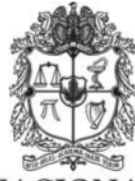

# Cosmological Features of Primordial Magnetic Fields

Héctor Javier Hortúa Orjuela

Universidad Nacional de Colombia
Facultad de Ciencias, Departamento de Física
Bogotá, Colombia
2018

# Cosmological Features of Primordial Magnetic Fields

Héctor Javier Hortúa Orjuela

Thesis submitted in fulfillment of the requirements for the degree of
**Doctor en Ciencias Física**

Line of Research:
Theoretical Physics
Group of Research:
Gravitation and Cosmology Group

Universidad Nacional de Colombia
Facultad de Ciencias, Departamento de Física
Bogotá, Colombia
2018

# Declaration

I hereby declare that the work reported in this thesis entitled **Cosmological Features of Primordial Magnetic Fields** has been carried out at the Department of Physics of Universidad Nacional of Colombia, under the supervision of Professor Leonardo Castañeda. The work reported in this thesis is original and It has not been submitted earlier for any degree to any university.

\_\_\_\_\_\_\_\_\_\_\_\_\_\_\_\_\_\_\_\_\_\_\_\_\_\_\_\_\_\_\_\_\_\_  \_\_\_\_\_\_\_\_\_\_\_\_\_\_\_\_\_\_\_\_\_\_\_\_\_\_\_\_\_\_

Prof. Leonardo Castañeda  Héctor Javier Hortúa
(Thesis Supervisor)  (The Candidate)

Gravitation & Cosmology Group
Universidad Nacional de Colombia
Bogotá D.C, Colombia, 2018

**To my Family**

> Life is short, break the rules, forgive quickly, kiss slowly, love truly, laugh uncontrollably, and never regret anything that made you smile. Twenty years from now you will be more disappointed by the things you didn't do than by the ones you did.
>
> Mark Twain

# Acknowledgments (in spanish)

Me gustaría que estas líneas sirvieran para expresar mi más profundo y sincero agradecimiento a todas aquellas personas que con su ayuda han colaborado en la realización de esta tesis, en especial al Profesor Leonardo Castañeda, director de esta investigación. Gracias totales profesor Leonardo por compartir su gran conocimiento con nosotros y ayudarnos a surgir en este maravilloso mundo académico, y sobretodo por ser mi apoyo en momentos difíciles y mi amigo.

Agradezco también a mis compañeros del Grupo de Gravitación y Cosmología (Joel, Luz Angela, Jonathan, Leandro, Oscar, Fercho(el bueno y el "lumpe"), William, Daniel, Alejo, Sebastian, los muchachos de pregrado y todos aquellos que están y que pasaron) por su ayuda y su compañia en esta desafiante etapa de mi vida. A los profesores de la Universidad por sus enseñanzas, sobretodo al profesor Angelo Fazio por su colaboración en cuestiones académicas y por sus excelentes consejos. Además, agradezco a mi querida Alma Mater, la Universidad Nacional de Colombia por darme la oportunidad de crecer, aprender y llegar a ser cada día mejor. Gracias a la convocatoria 647 de Becas de doctorado de Colciencias por permitir la financiación de mis estudios.

Por otro lado, un apoyo diferente al académico es también vital para alcanzar los objetivos propuestos. Quiero agradecer a mis compañeros Carlos Ávila y Tatiana, Manuel, Angela, Karl R., Roger, Edilson, Jorge, Alexander, Alejo Duitama, Ricardo y el profe Luis, por sus consejos y su apoyo en todo momento. A mis amigos de toda la vida Christian Martínez, Jhon Sánchez, Oscar Reyes, nano y Carito, Lucho, por estar siempre conmigo y brindar con ellos unas buenas cervezas. Quiero expresar un agradecimiento muy especial para Marina García, Karen Acosta, y a mi princesa Sasha por ser parte de mi vida; ustedes son y serán un motivo más para salir siempre adelante. Gracias por tanto hermosas.

Por supuesto, mi familia ha sido una parte vital para el desarrollo de esta tesis. Quiero agradecer a mi familia de Chipaque, mi papá, tios, tias, primos y primas por todo su apoyo y cariño, me siento muy afortunado de ser parte de esta excelente familia y a mi abuelita por estar conmigo presente siempre en mi corazón. A mi familia por parte de mamá, gracias por compartir y estar conmigo en todo momento, también agradecer a mi padrastro.

Finalmente, esta tesis es dedicada a tres personas que son el motor en mi vida y que sin ellos habría sido imposible la culminación de este documento, Mi madre y mis dos hermanos: Alejandro y Adrián.

# List of publications

The work presented in this thesis has been already published, the details of which are as follows:

1. Power spectrum of post-inflationary primordial magnetic fields, **Héctor J. Hortúa**, L. Castañeda, *Phys. Rev.**D90**, 123520 (2014)* [1]

2. Contrasting formulations of cosmological perturbations in a magnetic FLRW cosmology, **Héctor J. Hortúa**, L. Castañeda, *Class. Quantum Grav., 32 235026 (2015)* [2]

3. Reduced bispectrum seeded by helical primordial magnetic fields, **Héctor J. Hortúa**, L. Castañeda, *Journal of Cosmology and Astroparticle Physics **06** 020, (2017)* [3]

4. Parity odd CMB power spectrum via helical magnetic fields, **Héctor J. Hortúa**, L. Castañeda, *Proceedings 28th Texas Symposium on Relativistic Astrophysics* Geneva, Switzerland – December 13-18, (2015) [4]

5. Primordial Magnetic Fields and the CMB, **Héctor J. Hortúa**, L. Castañeda, *Cosmology, IntechOpen*, DOI: 10.5772/intechopen.81853 (2018) [5]

# Abstract

Recently, it has been found that our Universe holds magnetic fields in almost all scales probed so far. The fields in galaxies and galaxy clusters have strength of a few $\mu$Gauss and they are correlated up to Kpc scales. Furthermore, new observational evidence suggests the existence of magnetic fields of $10^{-16}$Gauss in the intergalactic medium with a correlation length of Mpc. However, the origin of these large scale magnetic fields is one of the most puzzling topics in cosmology and astrophysics. It is assumed that the observed magnetic fields result from the amplification of an initial field produced in the early Universe. Indeed, if those primordial fields were generated in early stages of the Universe, they could have left a distinctive signature on the Cosmic Microwave Background anisotropies (CMB). Thus, one of the most appealing ways of detecting those primordial magnetic fields is through temperature and polarization CMB observations. Therefore, the aim of this thesis is to study the effects on the CMB anisotropy due to primordial magnetic fields and to analyze some favorable scenarios of magnetogenesis constrained by those signatures, including limits on the amplitude of the fields from bounds on CMB non-Gaussianity and background models. In fact, we found out that helicity in the fields plays an important role in the analysis PMFs origin, by generating significant features in the cross-correlation polarization pattern and the increasing of the signal in the reduced CMB bispectrum. In the latter case, we reported that non-causal fields (mainly generated during the inflation epoch) are the most favourable models constrained by CMB observations. Moreover, we have studied the presence of an IR cutoff in the spectra and bispectra finding appealing unique features from primordial magnetic fields. Another important result shown in this thesis, is the equivalence between different approaches of cosmological perturbation theory in the magnetized context. In fact, assuming a magnetized Universe and building gauge invariant quantities in both approaches: the $1+3-$covariant and the gauge invariant; we found out that those invariants represent the same physical meaning. Besides, we define gauge invariant related to the electromagnetic potentials which in future works, could help us to study magnetogenesis models on perturbed scenarios.

# Resumen

Recientemente se ha encontrado que el Universo contiene campo magnéticos en todas las escalas observadas hasta ahora. Campos magnéticos en galaxias y los cúmulos de galaxias tienen amplitudes del orden de $\mu-$Gauss que son coherentes con escalas de hasta Kpc. Incluso, recientes observaciones sugieren la existencia de campos magnéticos del orden de $10^{-16}$Gauss en el medio intergaláctico coherente en las escalas de Mpc. Sin embargo, el origen de estos campos magnéticos con largas escalas de coherencia constituye en uno de los más grandes problemas abiertos en Cosmología y Astrofísica. Se cree que estos campos fueron el resultado de la amplificación de un campo inicial producido en el Universo temprano. En particular, si estos campos primordiales fueron generados en épocas tempranas del Universo, ellos pudieron haber dejado un rasgo único sobre las anisotropías de la radiación de fondo cósmico (CMB). Así, una excelente forma de detectar estos campos magnéticos primordiales es a través de las observaciones en la temperatura y polarización del (CMB). Por lo tanto, el objetivo de esta tesis es estudiar los efectos sobre el CMB debido a la presencia de campos magnéticos primordiales y analizar algunos escenarios favorables de magnetogenesis restringidos por aquellas señales, incluyendo limites en la amplitud de estos campos por medio de cotas en la no gaussianidad del CMB y effectos sobre modelos de fondo. De hecho, se encontró que la helicidad juega un papel importante en el análisis del origen de los campos magnéticos cósmicos al generar rasgos significativos en el patrón de polarización del CMB y el incremento en la señal del bispectrum reducido. En este último caso, se ha reportado que los campos no causales (generados principalmente durante la era de inflación) son los modelos más favorables a través de las observaciones del CMB. Más aún, se ha estudiado la presencia de cortes infrarojos en el espectro y el bispectrum, encontrando rasgos únicos generados por estos campos magnéticos. Otro resultado importante mostrado en esta tesis, es la equivalencia entre diferentes formalismos de la teoría de perturbaciones cosmológicas en el contexto magnetizado. En efecto, asumiendo un Universo magnetizado y construyendo allí cantidades invariantes gauge en dos formalismos: $1+3$-covariante e invariante gauge; se encontró que estas cantidades tienen el mismo significado físico. Además, se construyeron invariantes gauge relacionados con los potenciales electromagnéticos los cuales en el futuro, nos ayudarán a estudiar modelos de generación de campo magnético sobre espacios perturbados.

**Palabras clave:** Campos Magnéticos Primordiales, Radiación de Fondo Cósmico, No-Gaussianidad, Teoría de Perturbaciones Cosmológicas

# Contents

# List of Figures

# List of Tables

# Notations and conventions

- We shall use natural units throughout this thesis $c = \hbar = 1$ and adopt the Einstein notation.

- Greeks letters are space-time indices, $\mu = 0, 1, 2, 3$ and latin letters are spatial indices, $i = 1, 2, 3$. For the metric we use the $(- + + +)$ signature.

- We denote the derivative with respect to the conformal time $\tau$ by a prime, and the derivative with respect to the cosmic time $t$ by a dot,

$$\frac{dX}{d\tau} \equiv X', \quad \frac{dX}{dt} \equiv \dot{X}.$$

- Cosmological quantities evaluated today are indexed by a "0" e.g. $a_0 = a(t_0)$.

# Abbreviations

| Abbreviations | Phrase |
|---|---|
| $PMFs$ | Primordial Magnetic Fields |
| $AGN$ | Active Galactic Nuclei |
| $QCD$ | Quantum Chromodynamics |
| $EW$ | Electroweak |
| $GUT$ | Grand Unified Theories |
| $CMB$ | Cosmic Microwave Background |
| $BBN$ | Big Bang Nucleosynthesis |
| $GWs$ | Gravitational Waves |
| $NG$ | Non-Gaussianities |
| $GR$ | General Relativity |
| $FLRW$ | Friedmann-Lemaître-Robertson-Walker |
| $CDM$ | Cold Dark Matter |
| $MHD$ | Magnetohydrodynamic |
| $EMT$ | Energy Momentum Tensor |
| $IR-UV$ | Infrared-Ultraviolet |

# 1. Introduction

Cosmic magnetic fields have become one of the most pervasive features of the Universe, they have been observed in almost all scales probed so far [13, 14]. Starting from small scales, we can find magnetic fields in planets like the Earth with strength of around Gauss(G), and most of the planets (and larger moons) of solar system have a field, or had one in the past [15]. In stars, magnetic fields play an important role in all stages of the evolution and the amplitude of the field in these objects can vary from 1kG (for instance, in small scale sunspots generated in the Sun) to 30kG detected in magnetic variable stars [16, 17]. Going beyond the small scales, magnetic fields are also detected in galaxies of all types, in fact, Fermi proposed by first time the existence of a large-scale magnetic field in the interstellar space in order to confine cosmic rays [18].

Now, galaxies are known for having magnetic fields that are partly coherent on the scale of the galaxy, with field strengths reaching a few-to-tens of $\mu$G and surprisingly, these values are independent of the redshift [19–21]. These fields have been also detected in galaxy clusters and superclusters, although magnetic fields in these structures are notoriously difficult to measure and so far, observational constraints have been reported only in a few galaxy clusters, either by observing their synchrotron radiation or by Faraday rotation measurements [22]. The strength of the magnetic field in those systems is typically of the order of $0,1 - 10\mu$G on scales as large as 1Mpc [23, 24]. Additionally, over recent years developments in $\gamma$-ray observations have provided lower bounds in the strength of magnetic fields coherent on scales larger than Mpc [25, 26]. The detection of this large scale magnetic field comes from the deficit of GeV $\gamma$-ray in the direction of TeV $\gamma$-ray photons produced by TeV Blazars. Basically, these TeV photons interact with the diffuse extragalactic background light producing a beam of electron-positron pairs. Then, these particles scatter off Cosmic Microwave Background radiation (CMB) via inverse Compton, emitting a secondary beam of GeV photons. However, the presence of a weak intergalactic magnetic field would deflect the intermediate electron-positron pairs, and as a result the secondary GeV photons do not contribute entirely with the original TeV photon source, producing thus a missing GeV photons. Observations of this effect give a quantitative estimation of the lower bound on any cosmologically magnetic field [24, 27]. Other effects can also emerge due to the presence of this magnetic field, such as time-delayed observations of the GeV signal with respect to the primary emision or the detection of an extended emission of the secondary GeV source (for a deeper discussion see Refs. [22, 28–31]).

Let us now comment briefly on the four main observational probes used to study astrophysi-cal magnetic fields, which are: polarization of optical starlight, synchotron radiation, Zeeman effect, and the Faraday rotation. Polarized light from stars can show the presence of magnetic fields. This is the result of extinction by elongated dust grains in the line of sight which are aligned in the interstellar magnetic field(Davis-Greenstein effect) [21]. This probe is often used to study the magnetic field in our galaxy and nearby galaxies. Further, starlight polarization yields the orientation of magnetic fields in the Milky way; but, since this method depends on extinction and the polarization of light can be also produced via scattering processes, the measure of fields through this technique turns out to be in some cases a strenuous task [13]. Equally important is the synchotron radiation produced by electrons spiralling around magnetic field lines; this method allows to estimate the total magnetic field perpendicular to the line of sight. Synchotron emission is also very useful for detections of magnetic fields in spiral galaxies. Indeed, unpolarized radio synchrotron emission traces isotropic magnetic fields which are strongest in the spiral arms and bars $20-30\mu$G, and in central starburt regions $50-100\mu$G; while polarized radio emission traces ordered fields around of $10-15\mu$G and are generally found in interarm regions(see Chapter 18 in Ref. [21]). Astronomers have also found ordered magnetic fields in elliptical and irregular galaxies via synchrotron emission with a strength comparable to the one found in spiral galaxies [14]. Next, we have the Zeeman effect, that is the effect of splitting a spectral line into several components in the presence of a magnetic field. It may be a good probe for determining the intensity of strong magnetic fields in different astrophysical objects. However, the Zeeman effect is extremely difficult to observe because thermal effects can readily induce a greater splitting (Doppler broadening) than Zeeman effect [13]. Finally, Faraday rotation is caused by the interaction between light and the magnetic field. It causes a rotation of the plane of polarization which is proportional to the component of the magnetic field in the direction of propagation. This technique is helpful for detecting large scale fields. In fact, some observations have revealed large-scale spiral patterns that can be described by superposition of azimuthal modes implying a regular field generated by dynamo mechanisms of type $\alpha-\omega$ [21, 32]. When the rotation angle is sensitive to the sign of the field direction, only regular fields give rise to Faraday rotation, while the Faraday rotation contributions of turbulent fields cancel along the line of sight [21]. Evidence for magnetic fields through Faraday rotation in radio-halo clusters has been equally strong. It has been found the rotation measure for 18 sources behind the Coma cluster whose strength of the field gives an estimate of $2,5(\frac{L}{10\text{kpc}})^{-1/2}\mu$G where $L$ is the typical scale over which the field reverses direction [13, 33]. On supercluster scales also, astronomers detected faint radio emission in the region between the Coma cluster and the cluster Abell 1367, finding large magnetic fields with a strength of about $0,2-0,6\mu$G [13, 34]; the reader will find further clarification and detailed discussion of the aforementioned techniques in Refs. [13, 14, 21], and Chapter 18 in Ref. [21].

Large scale magnetic fields have been a subject of intense study, and the wealthly observations allow to obtain valuable information about their main features and the significant role that these fields might have in various aspects of Cosmology, Astrophysics and Particle physics. Despite the efforts, the question about the origin of the large-scale magnetic fields observed in the Universe remains to date as an unsolved puzzle. According to the standard paradigm, magnetic fields in virialized structures of different sizes are produced during structure formation by amplification of an initial weaker magnetic field, either via adiabatic compression or different types of dynamo but, both mechanisms can act only if a pre-existing magnetic field *a seed* is present [32,35–38]. This seed might typically vary between $10^{-22}$ and $10^{-30}$G depending on the efficiency of the dynamo amplification or the cosmological model, but as we said above, the strength and origin of this seed is still unknown [24]. So, in order to overcome this problem, two main scenarios are set for the generation of these seed fields: the astrophysical and the cosmological one. In the astrophysical scenarios, the seeds are created during structure formation via some processes like Biermann battery effect, Active Galactic Nuclei (AGN) or during the first generation of stars [13, 39, 40].

On the other hand, the cosmological scenario states that the so-called primordial magnetic fields(PMFs) are created in the early Universe either prior to or during recombination epoch. These PMFs can be generated during cosmic inflation [41–56], or in epochs that involve out-of-thermal equilibrium condition and parity violation, well known examples are the cosmological phase transitions (for example, QCD, Electroweak (EW), and Grand Unified Theories (GUTs)) [57–65]; and lastly via charge separation effects (like, Harrison's mechanism) [66–68]. In spite of the amount of data collected related to magnetic fields in galaxies and galaxy clusters, we cannot provide a new insight on the origin of these seed fields. This is because the complexity and the uncertainties of the details of dynamo mechanisms operating in those structures, lead to magnetic field amplitudes that are largely independent of their initial conditions. As a result, regions where primordial magnetic fields have been not affected by dynamo mechanisms (these regions may be associated with the intergalactic medium, more precisely, the voids of large scale structure) would be excellent laboratories for determining the nature of the initial seed fields [24]. Consequently, the recent observations mentioned earlier as upper limits for the intergalactic magnetic field, and the widespread presence of magnetic fields at high redshifts with similar strength to the Milky Way, might tip the scale in favour of a primodial origin [69–71].

Now, if magnetic fields were originated in the early Universe, these fields could strongly influence several processes which happened during and after its generation, and therefore signature of PMFs can be found in many astrophysical and cosmological probes such as formation of super massive black holes and stars, reionization [72]; Lyman$-\alpha$ forest, weak lensing; structure formation [73, 74], Big Bang Nucleosynthesis (BBN) [75], Gravitational waves (GWs) [76, 77], and Cosmic Microwave Background (CMB) among others [78, 79]. Especially the latter, i.e. *effects of PMFs on the CMB*, has been extensively studied during the last years [80–88]. Indeed, CMB has played a key role in the development of modern cosmology and our understanding of the Universe and provides tools for testing several other models and cosmological scenarios [89]. Since PMFs affect the evolution of cosmological perturbations, these fields might leave significant signals on the CMB temperature and pola-rization patterns, and produce non-Gaussianities (NG). As a matter of fact, PMFs introduce scalar, vector and tensor pertubations that affect the CMB in many ways. For instance, the scalar mode generates magnetosonic waves which influence the acoustic peaks and change the baryon fraction; vector mode contributes notably in scales below the Silk damping, and tensor mode induces gravitational waves that affect large angular scales [90]. Further, helical PMFs produce parity odd cross correlation which would not arise in the standard cosmological scenario [62]. Recently enough, CMB experiments like Planck and Polarbear have presented new limits on the amplitude of PMFs using temperature and polarization measurements that offer the possibility of investigating the nature of PMFs, and it is expected with future CMB polarization experiments like CMB-S4, Simons Observatory among others, to improve significantly the constraints to the helicity of PMFs, NG and to be able to provide new insight into the early Universe [91].

Therefore, the goal of this thesis is to study the impact of primordial magnetic fields on the CMB observables and to discuss the main features and possibilities for obtaining information about the helical(non-helical) primordial magnetic field generation mechanisms from (NG) and the angular power spectra of CMB. Futhermore, we will review some promising inflationary mechanisms of magnetogenesis which could be compatible with CMB observations. The structure of this thesis is as follows. In Chapter 2 we summarize the standard model of Cosmology and describe the evolution of the Universe under the FLRW model. In Chapter 3 we present the cosmological pertubation theory in order to model small desviations from the FLRW metric. Then, we will apply this theory to a magnetized Universe and build the physical observables. Later, we will introduce two formalisms of the perturbation theory and find the equivalence between them. In order to describe the stochastic properties of the perturbations, we need to introduce the concept of random fields and the $n$-point correlation functions as a way to characterize the probability density function of these observables, these concepts will be discussed in chapter 4. Moreover, we will also describe in detail the presence of sharp cut-offs at the convolution integrals and as an original result of the thesis, it will be shown the integration scheme used for calculating the spectra and bispectra for any stochastic field. In the Chapter 5 it will be reviewed the inflationary mechanisms capable of generating the seed for explaining the observed cosmic magnetic fields. In Chapter 6 we will show the PMF power spectra and describe the effects of these field on the CMB. We will start decribing the statistics for PMFs and we used the results given in Chapter 4 in order to calculate the magnetic spectra affected by an infrared and an ultraviolet cut-off. Afterward, we shall implement the CMB theory in order to analyze the effect of PMFs on the temperature and polarization CMB spectra. NG signals driven by PMFs will be also discussed in Chapter 7. Here we are going to start calculating the magnetic bispectrum and through the Komatsu-sperger estimator and, we will report bound on smoothed amplitude of PMFs on scales of 1Mpc for different spectral indices which are related to the generation mechanisms of these fields. Finally, in Chapter 8 we conclude by summarize the main results.

# 2. Standard cosmology model

Cosmology is considered as a scientific discipline that studies our Universe in terms of models based on well understood and tested physics. Since Einstein proposed his theory of general relativity in 1915, we have had a predictive science with the capacity of describing our Universe [92]. This field has progressed remarkably in the last decades, with predictions followed by confirmations. For instance, Friedmann and Lemaître predicted the expansion of the Universe, which was later confirmed by Hubble in 1929. Gamow, Alpher, and Hermann predicted the existence of the Cosmic Microwave Background (CMB) in 1948 which was discovered accidentally by Penzias and Wilson in 1965 [93].
Now, given the rapid improving observational technology, Cosmology has become a precision science which allow to cosmologists to reach a wide consensus about the best description of our Universe, i.e., what we currently know as the standard cosmological Model. Nowadays, CMB observations are one of the most powerful experiments of this precision cosmology era. The Planck satellite, after COBE and WMAP missions, has boosted the precision of cosmological measurements getting the most detailed maps of temperature and polarization of CMB anisotropies confirming the standard model of cosmology, although forthcoming cosmological surveys and datasets will leave also room for new physics. In this chapter we present some elements of standard model of cosmology that will be used for the remainder of this thesis. For further details concerning the background Universe see Refs. [94–96]

## 2.1. Cosmological principle

Modern cosmology is based on three main assumptions: the Copernican principle; the energy content of the Universe modelled in terms of fluids with constant equation of state (photons, baryons, neutrinos, cold dark matter and dark energy); and the Einstein's General Relativity (GR) as a theory used to describe gravity on all scales. The Copernican principle states that there is not a special position in the Universe. If we assume that this principle is true and include the (almost-)isotropy probed by observations of galaxy distribution and CMB performed so far [97–99], we can argue that our Universe is also (almost-)homogeneous. Based on this, we can adopt the idea that on sufficiently large scales, the statistical properties of our Universe are both rotationally (isotropy) and translationally (homogeneity) invariant, known as the *cosmological principle*. Here, isotropy means that there are no preferred directions in the Universe, so it looks statistically the same no matter which direction you look at, while homogeneity means that all locations are equal, therefore our Universe has the same average physical properties no matter where you are. When we talk about sufficiently large scales we are refering to scales larger than $60 - 70$Mpc which have been tested by galaxy distribution in the Sloan Digital Sky Survey Data or by galaxy sample of BOSS spectroscopic survey [100–104]. However, It is important to keep in mind that homogeneity cannot be established directly by observations given the inherent limitations of lightcone-based data, for this reason we need to link the observational test of isotropy with the Copernican principle to require homogeneity (assuming isotropy at all given points in space does imply homogeneity) [105, 106].

## 2.2. The Friedmann equations

We assume that GR is the correct description of gravity in the Universe, therefore the spacetime is regarded as a four dimensional pseudo-Riemannian manifold $\mathcal{M}$ with a metric $g$ that encodes the geometry of our Universe. According to the cosmological principle, the spacetime admits a slicing into space-like hypersurfaces of constant time which are homogeneous and isotropic, while a preferred time-like geodesics labeled with the coordinate $t$ called cosmic time defines the threading. These three spatial hypersurfaces of constant time are maximally symmetric with constant curvature $k$ and the metric can be written in the form [94]

$$ds^2 = g_{\mu\nu}dx^\mu dx^\nu = -dt^2 + a^2(t)\gamma_{ij}dx^i dx^j, \tag{2-1}$$

where $a(t)$ is called the scale factor, and $\gamma_{ij}$ is the three metric of the 3-D space with constant curvature $k$. In several applications it is convenient to use the conformal time $\tau$ defined as

$$\tau \equiv \int \frac{dt}{a(t)}. \tag{2-2}$$

With this definition the metric then becomes

$$ds^2 = g_{\mu\nu}dx^\mu dx^\nu = a^2(\tau)(-d\tau^2 + \gamma_{ij}dx^i dx^j). \tag{2-3}$$

The metric $\gamma$ has different forms [94]

$$\gamma_{ij}dx^i dx^j = \frac{\delta_{ij}dx^i dx^j}{\left(1+\frac{kx^2}{4}\right)^2}, \tag{2-4}$$

$$\gamma_{ij}dx^i dx^j = dr^2 + \chi^2(r)(d\theta^2 + \sin^2(\theta)d\varphi^2), \tag{2-5}$$

where

$$\chi(r) = \begin{cases} r, & \text{for Euclidean case } k = 0, \\ \frac{1}{\sqrt{k}}\sin(\sqrt{k}r), & \text{for spherical case } k > 0, \\ \frac{1}{\sqrt{|k|}}\sinh(\sqrt{|k|}r), & \text{for hyperbolic case } k < 0, \end{cases} \quad \text{and} \quad \delta_{ij} = \begin{cases} 1, & \text{if } i = j, \\ 0, & \text{otherwise.} \end{cases} \tag{2-6}$$

Recent results from Planck collaboration [107] constrain the spatial curvature to be negligible, suggesting an Universe with flat geometry. We shall assume $k = 0$ for the rest of the thesis, and in this way we write the line element as

$$ds^2 = g_{\mu\nu}dx^\mu dx^\nu = a^2(\tau)(-d\tau^2 + \delta_{ij}dx^i dx^j), \tag{2-7}$$

this is the flat Friedmann-Lemaître-Robertson-Walker (FLRW) metric. The matter content of the Universe describing by the energy-momentum tensor is assumed to be like a perfect fluid

$$T_{\mu\nu} = (\rho + P)u_\mu u_\nu + P g_{\mu\nu}, \tag{2-8}$$

where $\rho$ is the energy density of each species, that is, $\rho = \sum_i \rho^i$, where $i$ corresponds to photons(r), baryons(b), neutrinos($\nu$), cold dark matter(CDM), dark energy($\Lambda$), total matter(m=CDM+b) or magnetic fields ($B^2$); $P$ is its pressure, and $u^\mu$ the fluid four velocity satisfying the constraint $u^\mu u_\mu = -1$. Since the fluid elements will be comoving in the cosmological rest frame, the normalized four velocity becomes

$$u^\mu = \frac{1}{a}(1,0), \quad u_\mu = -a(1,0). \tag{2-9}$$

We shall assume barotropic fluids, i.e., their equation of state can be parametrized as

$$P = w\rho, \quad \text{where:} \quad w = \begin{cases} 0, & \text{for dust,} \\ \frac{1}{3}, & \text{for radiation,} \\ -1, & \text{for cosmological constant.} \end{cases} \tag{2-10}$$

The Einstein equations relate the Einstein tensor $G_{\mu\nu}$ to the energy momentum tensor $T_{\mu\nu}$ through the following equation (see Appendix A for details)

$$G_{\mu\nu} \equiv R_{\mu\nu} - \frac{1}{2}R g_{\mu\nu} = 8\pi G T_{\mu\nu}, \tag{2-11}$$

being $G = 6{,}673 \times 10^{-8} \text{cm}^3 \text{g}^{-1} \text{s}^{-2}$ the gravitational constant. Substituting the FLRW metric (2-1) and the energy momentum tensor (2-8) into the Einstein equations (2-11), we get

$$\left(\frac{a'}{a}\right)^2 = \frac{8\pi G}{3} a^2 \rho, \tag{2-12}$$

$$\left(\frac{a'}{a}\right)' = -\frac{4\pi G}{3} a^2 (\rho + 3P) = -\frac{4\pi G}{3} a^2 \rho(1 + 3w), \tag{2-13}$$

where $w$ is the equation of state parameter and $a' \equiv da/d\tau$. The above expressions are commonly called the Friedmann equation and acceleration equation respectively. The evolution of the species is determined by the conservation of the energy momentum tensor

$$\nabla_\mu T^\mu_{\ \nu} = \partial_\mu T^\mu_{\ \nu} + \Gamma^\mu_{\alpha\mu} T^\alpha_{\ \nu} - \Gamma^\alpha_{\mu\nu} T^\mu_{\ \alpha} = 0, \tag{2-14}$$

due to matter variables are time dependent, the only equation different from zero $\nu = 0$ is

$$\rho' = -3\frac{a'}{a}(\rho + P) = -3\frac{a'}{a}\rho(1+w), \tag{2-15}$$

finding that

$$\left(\frac{\rho}{\rho_0}\right) = \left(\frac{a_0}{a}\right)^{3(1+w)}, \tag{2-16}$$

where quantities indexed by a '0' like $a_0$ are evaluated today $T_0 \equiv T(\tau_0)$. By inserting the above equation into the Friedmann equation (2-12) and solving for $a(\tau)$ we have

$$\left(\frac{a}{a_0}\right) = \left(\frac{\tau}{\tau_0}\right)^{2/(1+3w)}. \tag{2-17}$$

We usually normalize the scale factor such that $a_0 = 1$ and $\tau_0$ is the value of the conformal time today. The quantity

$$H(t) \equiv \frac{\dot{a}}{a} = \frac{a'}{a^2} \equiv \frac{\mathcal{H}}{a}, \tag{2-18}$$

is called the Hubble parameter, and $\mathcal{H}$ is the conformal Hubble parameter which represents the homogeneous expansion rate, the value of this parameter today is

$$H_0 = 67{,}8 \pm 0{,}9 \text{km s}^{-1} \text{Mpc}^{-1},$$

from Planck TT+TE+EE+lowP at 95 % confidence level [107]. Then, we can use Eqs.(2-16),(2-17),(2-18) in order to analitically infer the evolution of variables given the equation of state for each species

$$\begin{aligned}
a &\propto \tau^2, & \mathcal{H} &\propto \frac{2}{\tau}, & \rho &\propto a^{-3}, & w &= 0, & \text{(dust)}, \\
a &\propto \tau, & \mathcal{H} &\propto \frac{1}{\tau}, & \rho &\propto a^{-4}, & w &= 1/3, & \text{(radiation)}, \\
a &\propto \frac{1}{|\tau|}, & |\mathcal{H}| &\propto \frac{1}{|\tau|}, & \rho &\propto \text{constant}, & w &= -1, & \text{(cosmological constant)}.
\end{aligned} \tag{2-19}$$

We can also define a critical density $\rho_{cr}$ which is associated with a flat universe (setting $k = 0$) as

$$\rho_{cr} \equiv \frac{3H^2}{8\pi G}, \tag{2-20}$$

and its present-day value can be easily computed in terms of the actual Hubble constant [94]

$$\rho_{cr,0} = 8{,}637 \times 10^{-27} \text{kg m}^{-3}.$$

With these expressions, It is standard to define the present day density parameter $\Omega_i$ for various species $i$ as the dimensionless ratio

$$\Omega_r \equiv \Omega_{r,0} = \frac{\rho_{r,0}}{\rho_{cr,0}}, \quad \Omega_m \equiv \Omega_{m,0} = \frac{\rho_{m,0}}{\rho_{cr,0}}, \quad \Omega_\Lambda \equiv \Omega_{\Lambda,0} = \frac{\rho_{\Lambda,0}}{\rho_{cr,0}}, \quad \Omega_b \equiv \Omega_{b,0} = \frac{\rho_{b,0}}{\rho_{cr,0}}, \quad (2\text{-}21)$$

some values reported by Planck are $\Omega_\Lambda = 0{,}685 \pm 0{,}013$, $\Omega_m = 0{,}315 \pm 0{,}013$ from Planck TT+lowP at 68 % confidence level [107] and $\Omega_r = 9{,}236 \times 10^{-5}$ assuming relativistic primordial neutrinos and a current CMB temperature of $T_0 = 2{,}7255\text{K}$ [94]. Using the information on the equations of state of the various species in Eq.(2-19), the Friedmann equation can be recast in terms of the density parameters

$$H^2 = H_0^2 \left( \frac{\Omega_r}{a^4} + \frac{\Omega_m}{a^3} + \Omega_\Lambda \right). \tag{2-22}$$

Due to the fact that the energy density of each component scales differently with the scale factor (see Eq.(2-19)), we could think that for most of Its history the Universe was presumably dominated by a specific component. Going back in time, the dynamics of the early Universe was set by radiation since the main contribution comes from relativistic particles. Then, with cooling from the expansion, the Universe entered into a matter era dominated by nonrelativistic fluids. This transition between both eras took place at $a_{eq} = \frac{\Omega_r}{\Omega_m} = 2{,}98 \times 10^{-4}$. Later, because the energy density of dark energy remains constant while matter and radiation density drop as the Universe expands, the dark energy eventually starts to dominate over others at late times around $a_\Lambda = \frac{\Omega_m}{\Omega_\Lambda} = 0{,}47$, and keeps the Universe expanding forever. The three stages of evolution are shown in Figure **2-1**, where we can see the values for $a_{eq}$ and $a_\Lambda$ through the intersection of the vertical lines with the x-axis. In Figure **2-1**a the evolution of the energy density $\rho_i$ for different species as a function of the conformal time are reported. Here, we show how radiation dominates for values $\tau < \tau_{eq}$ being $\tau_{eq} \sim 100\text{Mpc}$. Afterwards, our Universe is dominated by nonrelativistic matter between $\tau_{eq} < \tau < \tau_\Lambda$ with $\tau_\Lambda \sim 10^4\text{Mpc}$, and finally it enters a stage of acelerating expansion when $\tau > \tau_\Lambda$. On the other hand, the evolution of the density parameter $\Omega_i$ for each $i$ species is illustrated in Figure **2-1**b. In this plot we also include the presence of neutrinos. In late times, the massive neutrinos (following the Normal Hierarchy scheme $M_\nu = \{0{,}0\text{eV} - 0{,}01\text{eV} - 0{,}05\text{eV}\}$ [7]) track the matter, because they become nonrelativistic when the energy of the Universe falls below the neutrino rest mass and, Its density is enhanced in that time. We can also study analytically the transition between radiation and matter solving the Friedmann equation in terms of $a_{eq}$. In order to do that, let us suppose that content of the Universe is only dust and radiation

$$\rho = \rho_m + \rho_r = \frac{\rho_{0,m}}{a^3} + \frac{\rho_{0,r}}{a^4},$$

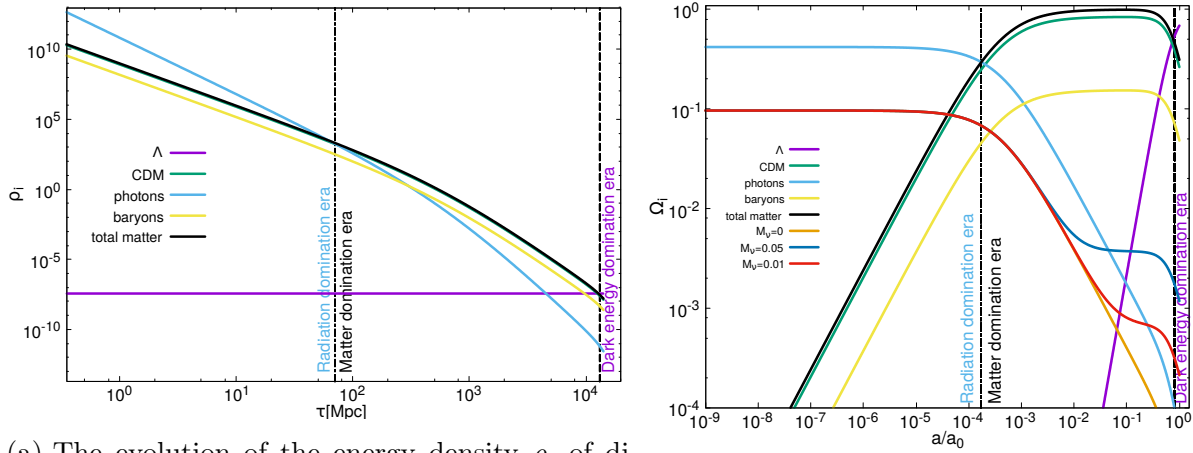

(a) The evolution of the energy density $\rho_i$ of different species as a function of the conformal time. The present value of the conformal time is around $\tau_0 = 14200$ Mpc.

(b) The evolution of the density parameter $\Omega_i$ of different species as a function of the logarithm of $(a/a_0)$.

**Figure 2-1**.: The evolution of the density parameter $\Omega_i$ and energy density $\rho_i$ of radiation(photons), baryons, Cold dark matter, total matter (CDM+b) cosmological constant ($\Lambda$) and (non-)relativistic neutrinos ($\nu$). The intersections between the matter, radiation and $\Lambda$, naturally split the cosmic history in three epochs: the radiation domination era, the matter domination era, and the dark energy, these periods are separated by the vertical lines. Here the neutrino mass is measured in eV. These plots were derived using the CLASS code [6] and adapted from [7].

and inserting this equation into Eq.(2-12) we will have

$$\left(\frac{da}{d\tau}\right)^2 = \frac{8\pi G a^4}{3}\left(\frac{\rho_{0,m}}{a^3} + \frac{\rho_{0,r}}{a^4}\right) = H_0^2\left(a\Omega_m + \Omega_r\right)$$
$$= H_0^2 \Omega_m \left(a + \frac{\Omega_r}{\Omega_m}\right) = H_0^2 \Omega_m \left(a + a_{eq}\right), \tag{2-23}$$

whose solution is

$$a(\tau) = a_{eq}\left(\left(\frac{\tau}{\tau^\star}\right)^2 + 2\left(\frac{\tau}{\tau^\star}\right)\right), \quad \text{where:} \quad \tau^\star = \frac{2\sqrt{\Omega_r}}{\Omega_m H_0} \sim 285 \text{Mpc}. \tag{2-24}$$

If we evaluate the expression above in $\tau = \tau_{eq}$, we found that $\tau_{eq} = (\sqrt{2}-1)\tau^\star$. Notice that for $\tau \ll \tau^\star \sim \tau_{eq}$, we get $a \sim \tau$ corresponding to the radiation era, while for $\tau \gg \tau^\star \sim \tau_{eq}$, we recover the matter domination era $a \sim \tau^2$.

## 2.3.  Cosmic inflation

The ΛCDM model described above, has been successful in explaining the expansion of the Universe observed by Edwin Hubble in 1929, the existence of the CMB and the abundance of the light elements formed during nucleosynthesis, among other things. Despite Its successes, this model had several open important observational and theoretical issues which we list below [94, 108]

- *Horizon problem*, the CMB radiation coming from areas of the universe that were never in causal contact are observed to have the same temperature with a precision of one part in ten thousand regardless of the direction of observation.

- *Flatness Problem*, the present energy density of the universe is close to Its critical value. This means that, for the Universe being so close to flat today, It requires an extreme fine-tuning of curvature parameter close to zero (within one part in a hundred trillion) in the past.

- *Density perturbations*, what mechanism produced the initial seed fluctuations for the cosmic structures and the CMB anisotropies we see today?

- *Magnetic Monopoles* they are likely produced in the early Universe, but any evidence for its existence has not been found.

In the 80's, Guth, Starobinsky, Albrecht, Steinhardt, and Linde postulated the existence of an epoch of accelerated expansion in the early Universe called inflation, which solved the aforementioned cosmological problems [109, 110]. This section summarizes some basic concepts concerning inflation using a simple toy model driven by a scalar field. For an extensive review on this topic see Refs. [10, 108, 111] and reference therein. Let us consider the action of a homogeneous scalar field $\phi(\tau)$ called inflaton, and minimally coupled with gravity as [9]

$$S = \int d^4x \sqrt{-g}\left(\frac{1}{2}(8\pi G)^{-1/2}R - \frac{1}{2}g^{\mu\nu}\partial_\mu\phi\partial_\nu\phi - V(\phi)\right), \tag{2-25}$$

where $R$ denotes the Ricci scalar, $V(\phi)$ is the potential. The energy momentum tensor and the equation of motion for the field are given by

$$T^{(\phi)}_{\mu\nu} = -\frac{2}{\sqrt{-g}}\frac{\delta S}{\delta g^{\mu\nu}} = \partial_\mu\phi\partial_\nu\phi - g_{\mu\nu}(\frac{1}{2}\partial^\alpha\phi\partial_\alpha\phi + V(\phi)), \tag{2-26}$$

$$\frac{\delta S}{\delta \phi} = \frac{1}{\sqrt{-g}}\partial_\mu(\sqrt{-g}g^{\mu\nu}\partial_\mu\phi) - \frac{dV(\phi)}{d\phi} = 0. \tag{2-27}$$

From last expressions we infer that

$$\rho_\phi = -T^{0(\phi)}_0 = \frac{1}{2a^2}\phi'^2 + V(\phi), \quad P_\phi = \frac{1}{3}T^{i(\phi)}_i = \frac{1}{2a^2}\phi'^2 - V(\phi), \tag{2-28}$$

and the well-known Klein-Gordon equation

$$\phi'' + 2\mathcal{H}\phi' + a^2 \frac{dV(\phi)}{d\phi} = 0. \tag{2-29}$$

The equation of state for the scalar field

$$w_\phi = \frac{\frac{1}{2a^2}\phi'^2 - V(\phi)}{\frac{1}{2a^2}\phi'^2 + V(\phi)}, \tag{2-30}$$

shows that if $V(\phi)$ dominates over the kinetic energy, $w_\phi < -1/3$ and using Eq.(2-13), it is found that the Universe would experience an accelerated expansion. In the limit of zero kinetic energy, the scalar field mimics the standard cosmological model, and the expansion would be exponential (in the cosmic time), this is called De Sitter expansion [112]. So, we require that

$$\frac{1}{2a^2}\phi'^2 \ll V(\phi), \tag{2-31}$$

in order to hold the accelerated expansion over an extended period. Moreover, this acceleration is kept stable thus, we must impose that the time-derivative of this condition

$$\frac{-\mathcal{H}}{a^2}\phi'^2 + \frac{1}{a^2}\phi'\phi'' \ll \frac{dV(\phi)}{d\phi}\phi', \quad \rightarrow \quad |-\mathcal{H}\phi' + \phi''| \ll a^2 \left|\frac{dV(\phi)}{d\phi}\right|. \tag{2-32}$$

Using this condition in the Klein Gordon equation (2-29), we get

$$a^2 \frac{dV(\phi)}{d\phi} \approx -3\mathcal{H}\phi', \quad \rightarrow \quad |\phi''| \ll 2\mathcal{H}|\phi'|. \tag{2-33}$$

Eqs.(2-31),(2-33) are the slow-roll conditions. When these two inequalities hold, the Friedmann and Klein-Gordon equations take the form

$$3\mathcal{H}^2 \approx 8\pi G a^2 V(\phi), \quad \phi' \approx \frac{-a^2}{3\mathcal{H}} \frac{dV(\phi)}{d\phi}. \tag{2-34}$$

Replacing the above equations into Eq.(2-31), we found that both slow-roll conditions can be rewritten as conditions on the flatness of the potential

$$\left(\frac{\frac{dV(\phi)}{d\phi}}{V(\phi)}\right)^2 \ll 48\pi G, \quad \left|\frac{\frac{d^2V(\phi)}{d\phi^2}}{V(\phi)}\right| \ll 48\pi G, \tag{2-35}$$

where the second expression is obtained by taking the time derivative of the first one. If we introduce the following dimensionless slow-roll parameters [94]

$$\epsilon_1 \equiv \frac{1}{16\pi G}\left(\frac{\frac{dV(\phi)}{d\phi}}{V(\phi)}\right)^2, \quad \epsilon_2 \equiv \frac{-1}{24\pi G}\left(\frac{\frac{d^2V(\phi)}{d\phi^2}}{V(\phi)}\right). \tag{2-36}$$

We see from Eqs. (2-35) that the slow-roll conditions yield

$$\epsilon_1 \ll 1, \quad |\epsilon_2| \ll 1, \tag{2-37}$$

and inflation ends when $\epsilon_1$ approaches unity. Planck collaboration [113] has reported that temperature and polarization data are consistent with the spatially flat $\Lambda$CDM model described above. Furthermore, primordial perturbations are Gaussian and adiabatic with a spectrum described by a power law, as predicted by the simplest inflationary models. Some models like $V(\phi) \sim \phi^2$, natural or chaotic inflation are disfavoured compared to models such as $V(\phi) \sim (1 - \exp^{(-\sqrt{2/3}\phi/M_{pl})})^2$ inflation[1]. For further discussion, see Ref. [113]. So far, we have treated our Universe as perfectly homogeneous and isotropic; however, to understand the formation and evolution of the cosmic structures, we have to introduce inhomogeneities. In what follows, we will use the concepts introduced in this Section in order to build a more realistic model that allows us to understand better our Universe.

---

[1] Here $M_{pl} = 2{,}435 \times 10^{18}$ GeV is the reduced Planck mass.

# 3. Cosmological perturbation theory

Cosmological perturbation theory has become a standard tool in modern cosmology to understand the formation of large scale structure in the universe, and also to calculate the fluctuations in the Cosmic Microwave Background (CMB) [114]. The first treatment of perturbation theory within General Relativity was developed by Lifshitz [115], where the evolution of structures in a perturbed Friedmann-Lemaître-Robertson-Walker universe (FLRW) under synchronous gauge was addressed. Later, the covariant approach of perturbation theory was formulated by Hawking [116] and followed by Olson [117] where perturbation in the curvature was worked rather than on metric variables. Then, based on early works by Gerlach and Sengupta [118], Bardeen [119] introduced a full gauge invariant approach to first order in cosmological perturbation theory. In his work, he built a set of gauge invariant quantities related to density perturbations commonly known as Bardeen potentials (see also Kodama & Sasaki [120] for an extensive review).

However, alternative representations of previous formalisms were appearing due to the gauge-problem [121]. This issue arises in cosmological perturbation theory due to the fact that splitting all metric and matter variables into a homogeneous and isotropic spacetime plus small desviations of the background, is not unique. Basically, peturbations in any quantity are defined by choosing a correspondence between a fiducial background spacetime and the physical universe. But, given the general covariance in perturbation theory, which states that there is not a preferred correspondence between these spacetimes[1], a freedom in the way how to identify points between two manifolds appears [122]. This arbitrariness generates a residual degree of freedom, which would imply that variables might not have a physical interpretation.

Following the research mentioned above, two main formalisms have been developed to study the evolution of matter variables and to deal with the gauge-problem, and will be reviewed in this chapter. The first is known as *1+3 covariant gauge invariant* presented by Ellis & Bruni [123]. This approach is based on earlier works of Hawking and Stewart & Walker [124] and consists in building covariantly variables such that they vanish in the background. Therefore, they can be considered as gauge invariant under gauge transformation in accordance with the Stewart-Walker lemma [125]. These gauge-invariant variables manage the gauge ambiguities and acquire a physical interpretation. Since the covariant variables do not as-sume linearization, exact equations are found for their evolution. On the other hand, the second approach considers arbitrary order perturbations in a geometrical perspective, it has been deeply discussed by Bardeen [119], Kodama & Sasaki [120], Mukhanov, Feldman & Brandenberger [126], and Bruni [127] and it is known as *gauge invariant* approach. Here, perturbations are descomposed into the so-called scalar, vector and tensor parts and the gauge invariants are found with both gauge transformations and the Stewart-Walker lemma. The gauge transformations are generated by arbitrary vector fields, defined on the background spacetime and associated with a one-parameter family of diffeomorphisms. This approach allows to find the conditions for the gauge invariance of any tensor field, although at high order sometimes appears unclear. As alternative description of the latter approach, it is important to comment the work done by Nakamura [128] where he splits the metric perturbations into a gauge invariant and gauge variant part, and thus, the equations of motion are written in terms of gauge invariant quantities. Given the importance and advantage of these two approaches, it is nessesary to find equivalences between them. Some authors have compared different formalisms, for example [129] discussed the invariant quantities found by Bardeen with the ones built on the 1+3 covariant gauge invariant in a specific coordinate system, also the authors in [130] found a way to reformulate the Bardeen approach in a covariant scenario and the authors in [131] constrasted the non-linear approach described by Malik et al. [132] with the Nakamura's approach.

---
[1] The only restriction is that perturbation must be small respect to its value in the background, even so, it does not help to specify the map in a unique way.

The aim of this chapter is twofold. Firstly, we want to present the perturbation theory formalism, its application to cosmology and specifically, to the study of magnetic fields in curves spacetimes and the early Universe. Secondly, we develop a way for contrasting the approaches of perturbation theory mentioned above: 1+3 covariant gauge invariant and gauge invariant. Here we will follow the methodology used in Refs. [129, 133] where a comparation of gauge invariant quantities built in each approach were done. Let us start with some background material introducing cosmological perturbation theory in the gauge invariant approach, our discussion in this section closely follows [127, 134]. Then, we are going to apply this formalism to Cosmology and describe in detail the perturbation under a FLRW Universe along with some gauge invariant quantities. We also show another example of this theory in the magnetized context where we write the equation of motion for the electromagnetic field in curve spacetimes and define the electromagnetic potentials in this scenario. Finally, we will review the 1+3 covariant approach and through magnetic gauge invariants, we will demostrate the equivalence of these formalisms. This chapter is based on the work published in [2].

## 3.1. Taylor expansion of tensor fields on manifolds and the perturbation concept

Cosmological perturbation theory help us to find approximated solutions of the Einstein field equations through small desviations from an exact solution. Let us start establishing some general results about the Taylor expansion of tensors on a manifold $\mathcal{M}$ [122, 127, 129]. For functions $f$ on $\mathbb{R}^n$, the Taylor expansion allows to write approximately the value of the function at some point in terms of its value, and the value of all its derivatives at the neighborhood points. For tensor fields on a manifold $\mathcal{M}$, this treatment cannot be done directly because tensors evaluated in different points belong to different spaces. Therefore, we must introduce a map between these tensors at different points on $\mathcal{M}$ through a one-parameter family of diffeomorphisms $\Psi_{\lambda \in \mathbb{R}} : \mathbb{R} \times \mathcal{M} \to \mathcal{M}$. In the case where $\Psi_\lambda := \phi_\lambda$, we can write the expansion of the tensor field $T$ around $\lambda = 0$ as [127]

$$\phi_\lambda^* T = \sum_{k=0}^{\infty} \frac{\lambda^k}{k!} \mathcal{L}_\xi^k T, \tag{3-1}$$

where $\xi$ is the vector field generating the flow $\phi_\lambda$, and $\mathcal{L}_\xi^k$ is just the Lie derivative of $T$ with respect to $\xi$. This flow has the property that $\phi_{\lambda=0}(p) = p$, $\forall p \in \mathcal{M}$ for any $\lambda \in \mathbb{R}$. The notation $\phi_\lambda^*$ refers to the fact that we are making correspondences between cotangent spaces (pull-back). In the case where $\Psi_\lambda := \phi_{\lambda^2/2}^{(2)} \circ \phi_\lambda^{(1)}$, where $\phi^{(1)}$ and $\phi^{(2)}$ are flows generated by vector fields $\xi_{(1)}$ and $\xi_{(2)}$ respectively, this one-parameter family of diffeomorphisms displaces a point $p$ an interval $\lambda$ along the integral curve of $\xi_{(1)}$, and then and interval $\lambda^2/2$ along the integral curve $\xi_{(2)}$. Finally, we can generalize this concept to the case in which $n$ independent vector fields $\xi_{(1)}, .., \xi_{(n)}$ are present generating their corresponding flows $\phi^{(1)}, .., \phi^{(n)}$ via $\Psi_\lambda := \phi_{\lambda^n/n!}^{(n)} \circ .. \circ \phi_{\lambda^2/2}^{(2)} \circ \phi_\lambda^{(1)}$ more commonly known as knight diffeomorphisms of rank $n$. Now, let us calculate the expansion around $\lambda = 0$ of the pull-back $\Psi_\lambda^* T$

$$\begin{aligned} \Psi_\lambda^* T &= \phi_\lambda^{(1)*} \phi_{\lambda^2/2}^{(2)*} \cdots \phi_{\lambda^n/n!}^{(n)*} \cdots T = \sum_{l_1=0}^{+\infty} \frac{\lambda^{l_1}}{l_1!} \mathcal{L}_{\xi_{(1)}}^{l_1} \left( \phi_{\lambda^2/2}^{(2)*} \cdots \phi_{\lambda^n/n!}^{(n)*} T \right) \\ &= \sum_{l_1=0}^{+\infty} \sum_{l_2=0}^{+\infty} \frac{\lambda^{l_1+2l_2}}{2^{l_2} l_1! l_2!} \mathcal{L}_{\xi_{(1)}}^{l_1} \mathcal{L}_{\xi_{(2)}}^{l_2} \left( \phi_{\lambda^3/3!}^{(3)*} \cdots \phi_{\lambda^n/n!}^{(n)*} T \right) \\ &\vdots \quad \text{repeating successively Eq.(3-1)} \Rightarrow \\ &= \sum_{l_1=0}^{+\infty} \sum_{l_2=0}^{+\infty} \cdots \sum_{l_n=0}^{+\infty} \frac{\lambda^{l_1+2l_2+\cdots+nl_n}}{2^{l_2} \cdots (n!)^{l_n} l_1! l_2! \cdots l_n!} \mathcal{L}_{\xi_{(1)}}^{l_1} \mathcal{L}_{\xi_{(2)}}^{l_2} \cdots \mathcal{L}_{\xi_{(n)}}^{l_n} T , \end{aligned} \tag{3-2}$$

where Eq.(3-1) was repeatedly used. Here, the vector fields $\xi_{(1)}, .., \xi_{(n)}$ of the one parameter family $\Psi_\lambda$ represent the direction along which the Taylor expansion is carried out. In fact, $\xi_{(1)}$ is the first-order approximation of the expansion, and $\xi_{(n)}$ represents the $n$th-order correction to this flow. Eq.(3-2) up to second order in $\lambda$ is given by

$$\Psi_\lambda^* T = T + \lambda \mathcal{L}_{\xi_{(1)}} T + \frac{\lambda^2}{2} \left( \mathcal{L}_{\xi_{(1)}}^2 + \mathcal{L}_{\xi_{(2)}} \right) T + \mathcal{O}(\lambda^3). \tag{3-3}$$

Now, a simple way to interpret the representation of the Taylor expansion is introducing a chart $(\mathcal{U}, x^\mu)$ of $\mathcal{M}$, with $\mathcal{U} \in \mathcal{M}$ an open set and $x^\mu : p \in \mathcal{M} \to x^m(p) \in \mathbb{R}^m$, $\forall p \in \mathcal{U}$. Then, let us define a new chart $(\Psi_\lambda(\mathcal{U}), y^\mu)$ with $y^\mu := x^\mu \circ \Psi_\lambda^{-1}$. With this choice, the coordinates of the point $q := \Psi_\lambda(p)$ in the new coordinates $y$ are the same as the ones of $p$ in the old coordinates $x$, i.e. $y^\mu(q) = x^\mu(p)$ thus, we can use Eq.(3-3) to derive

$$x^\mu(q) = (\Psi_\lambda^* x^\mu)(p) = x^\mu(p) + \lambda \xi_{(1)}^\mu(x(p)) + \frac{\lambda^2}{2}\left(\xi_{(1)}^\nu \partial_\nu \xi_{(1)}^\mu + \xi_{(2)}^\mu\right)|_{x(p)} + \mathcal{O}(\lambda^3). \tag{3-4}$$

The previous expresion states a transformation between different points $p$ and $q$ taking into account the same chart $(\mathcal{U}, x^\mu)$. In other words, the transformation of any quantity is evaluated in the same coordinate point, this is called the active view. Now we are going to describe another view, so by definition we have

$$y^\mu(q) := x^\mu(p) = x^\mu(q) - \lambda \xi_{(1)}^\mu(x(p)) - \frac{\lambda^2}{2}\left(\xi_{(1)}^\nu \partial_\nu \xi_{(1)}^\mu + \xi_{(2)}^\mu\right)|_{x(p)} - \mathcal{O}(\lambda^3), \tag{3-5}$$

using the first terms in Eq.(3-4) we have

$$x^\mu(q) = x^\mu(p) + \lambda \xi_{(1)}^\mu(x(p)) \tag{3-6}$$

making a Taylor expansion for $\xi_{(1)}$

$$\xi_{(1)}^\mu(x(p)) = \xi_{(1)}^\mu(x(q) - \lambda \xi_{(1)}^\mu(x(p))) \approx \xi_{(1)}^\mu(x(q)) - \lambda \partial_\nu \xi_{(1)}^\mu(x(q))\xi_{(1)}^\nu(x(q)), \tag{3-7}$$

and finally by substituting Eq.(3-7) into Eq.(3-5) gives as result

$$y^\mu(q) = x^\mu(q) - \lambda \xi_{(1)}^\mu(x(q)) + \frac{\lambda^2}{2}\left(\xi_{(1)}^\nu \partial_\nu \xi_{(1)}^\mu + \xi_{(2)}^\mu\right)|_{x(q)} + \mathcal{O}(\lambda^3). \tag{3-8}$$

This relation provides a transformation between the coordinates at the same point $q$ in two different charts $(\mathcal{U}, x^\mu)$ and $(\Psi_\lambda(\mathcal{U}), y^\mu)$. This is called passive view and the quantities in this approach transform at the same physical point. It is important to notice that the choice of a chart $(\mathcal{U}_\alpha \in \mathcal{M}, x_\alpha^\mu)$ which introduces a coordinate system on a single manifold $\mathcal{M}$ is called a first kind gauge choice. By considering another open set $(\mathcal{U}_\beta \in \mathcal{M}, x_\beta^\mu)$, we have another choice such that $x_\beta^\mu$ assigns coordinates in $x^\mu(\mathcal{U}_\beta) \in \mathbb{R}^m$ to points in the neighbourhood $\mathcal{U}_\beta \in \mathcal{M}$. If these two open sets satisfy $\mathcal{U}_\alpha \cap \mathcal{U}_\beta \neq \emptyset$, a coordinate transformation is induced $x_\alpha^\mu \circ (x_\alpha^\mu)^{-1} : x_\alpha^\mu(\mathcal{U}_\alpha \cap \mathcal{U}_\beta) \in \mathbb{R}^m \to x_\beta^\mu(\mathcal{U}_\alpha \cap \mathcal{U}_\beta) \in \mathbb{R}^m$, this diffeomorphism is usually called gauge transformation of the first kind. In order to formalize the previous ideas and describe the notion of perturbations in relativistic theories, let us introduce two different spacetimes, one is the perturbed spacetime $(\mathcal{M}, g_{\alpha\beta})$ which describes the real universe and the other is the background spacetime $(\mathcal{M}_0, g_{\alpha\beta}^{(0)})$ which is an idealization and it is taken as reference to generate the real spacetime. Then, the perturbation of any quantity $\Gamma$ (e.g., energy density $\mu(x,t)$, 4-velocity $u^\alpha(x,t)$, magnetic field $B^i(x,t)$ or metric tensor $g_{\alpha\beta}$) is the difference between the value that the quantity $\Gamma$ takes in the real spacetime and the value in the background at a given point. In order to determinate the perturbation in $\Gamma$, we must have a way to compare $\Gamma$ (tensor on the real spacetime) with $\Gamma^{(0)}$ (being $\Gamma^{(0)}$ the value on $\mathcal{M}_0$). This requires the assumption to identify points of $\mathcal{M}$ with those of $\mathcal{M}_0$. This is accomplished by assigning a mapping between these spacetimes called second kind gauge choice[2] given by a function $\mathcal{X}: \mathcal{M}_0(p) \longrightarrow \mathcal{M}(\bar{p})$ for any point $p \in \mathcal{M}_0$ and $\bar{p} \in \mathcal{M}$, which generates a pull-back

$$\mathcal{X}^*: \begin{array}{c} \mathcal{M} \\ T^*(\bar{p}) \end{array} \longrightarrow \begin{array}{c} \mathcal{M}_0 \\ T^*(p) \end{array}, \tag{3-9}$$

thus, points on the real and background spacetime can be compared through of $\mathcal{X}$. Now, the perturbation for $\Gamma$ can be defined as

$$\delta \Gamma(p) = \Gamma(\bar{p}) - \Gamma^{(0)}(p), \tag{3-10}$$

here we see that the perturbation $\delta \Gamma$ is completely dependent on the gauge choice because the mapping determines the representation on $\mathcal{M}_0$ of $\Gamma(\bar{p})$. In fact, $\Gamma(\bar{p})$ can be seen as a spatial average over each time slide but, since this slicing is dependent on the coordinates, the value of the pertubation will be also coordinate dependent. Futhermore, one can also choose another correspondence $\mathcal{Y}$ between these spacetimes so that $\mathcal{Y}: \mathcal{M}_0(q) \to \mathcal{M}(\bar{p}), (p \neq q)$.[3] In the literature a change of this identification map is called second kind gauge transformation. The freedom to choose between different correspondences is due to the general covariance in General Relativity, which states that there is no preferred coordinate system in nature [121, 122]. Hence, this freedom will generate an arbitrariness in the value of $\delta \Gamma$ at any spacetime point $p$, which is called *the gauge problem* in the general relativistic perturbation theory and it has been treated in Refs. [119, 123, 127, 128]. This problem generates unphysical degree of freedom to the solutions in the theory and, therefore one should fix the gauge or build up non-dependent quantities of the gauge. We should point out that second kind gauge transformation is not a transformation coordinate itself, because the perturbation in $\Gamma(\bar{p})$ is modified by changing the mapping between both spacetimes while the chart introduced on $\mathcal{M}_0$ remains fixed. Since second kind gauge choice only appears in perturbation theories, from now on will be focus on this second kind of gauges.

### 3.1.1. Gauge transformations and gauge invariant variables

To define the perturbation to a given order, let us consider a family of four-dimensional submanifolds $\mathcal{M}_\lambda$ with $\lambda \in \mathbb{R}$, embedded in a 5-dimensional manifold $\mathcal{N} = \mathcal{M} \times \mathbb{R}$ [127, 128, 131]. Each submanifold in the family represents a perturbed spacetime and the background spacetime is represented by the manifold $\mathcal{M}_0$ ($\lambda = 0$). On these manifolds we consider that the Einstein field or Maxwell's equations are satisfied

$$\mathsf{E}[g_\lambda, T_\lambda] = 0 \quad \text{and} \quad M[F_\lambda, J_\lambda] = 0; \tag{3-11}$$

---

[2] hereafter a gauge choice

[3] This is the active approach where transformations of the perturbed quantities are evaluated at the same coordinate point.

and each tensor field $\Gamma_\lambda$ on a given manifold $\mathcal{M}_\lambda$ is extended to all manifold $\mathcal{N}$ through $\Gamma(p, \lambda) \equiv \Gamma_\lambda(p)$ to any $p \in \mathcal{M}_\lambda$, likewise the above equations are extended to $\mathcal{N}$.[4] In order to make the operation in the right side of Eq. (3-10) meaningful at all, one has to introduce an one-parameter group of diffeomorphisms $\mathcal{X}_\lambda$ which identifies points in the background with points in the real spacetime labeled with the value $\lambda$. Each $\mathcal{X}_\lambda$ is a member of a flow $\mathcal{X}$ on $\mathcal{N}$ and it specifies a vector field $X$ with the property $X^4 = 1$ everywhere (transverse to the $\mathcal{M}_\lambda$)[5], then points which lie on the same integral curve of $X$ have to be regarded as the same point [128]. According to the above, one gets a definition for the tensor perturbation

$$\Delta\Gamma_\lambda \equiv \mathcal{X}_\lambda^*\Gamma|_{\mathcal{M}_0} - \Gamma_0. \tag{3-12}$$

At higher orders the Taylor expansion is given by [127],

$$\Delta^{\mathcal{X}}\Gamma_\lambda = \sum_{k=0}^{\infty} \frac{\lambda^k}{k!} \delta_{\mathcal{X}}^{(k)}\Gamma - \Gamma_0 = \sum_{k=1}^{\infty} \frac{\lambda^k}{k!} \delta_{\mathcal{X}}^{(k)}\Gamma, \tag{3-13}$$

where

$$\delta_{\mathcal{X}}^{(k)}\Gamma = \left[\frac{d^k \mathcal{X}_\lambda^*\Gamma}{d\lambda^k}\right]_{\lambda=0,\mathcal{M}_0}. \tag{3-14}$$

Now, rewriting eq. (3-12) we get

$$\mathcal{X}_\lambda^*\Gamma|_{\mathcal{M}_0} = \Gamma_0 + \lambda \delta_{\mathcal{X}}^{(1)}\Gamma + \frac{\lambda^2}{2} \delta_{\mathcal{X}}^{(2)}\Gamma + \mathcal{O}(\lambda^3), \tag{3-15}$$

Notice in the Eqs.(3-14),(3-15) the representation of $\Gamma$ on $\mathcal{M}_0$ is splitting in the background value $\Gamma_0$ plus $\mathcal{O}(k)$ perturbations in the gauge $\mathcal{X}_\lambda$ so, the $k$-th order $\mathcal{O}(k)$ in $\Gamma$ depends on gauge $\mathcal{X}$. The first term in Eq.(3-12) admits an expansion around $\lambda = 0$ described in Eq.(3-1) as

$$\mathcal{X}_\lambda^*\Gamma|_{\mathcal{M}_0} = \sum_{k=0}^{\infty} \frac{\lambda^k}{k!} \mathcal{L}_X^k \Gamma|_{\mathcal{M}_0} = \exp(\lambda \mathcal{L}_X)\Gamma|_{\mathcal{M}_0}, \tag{3-16}$$

where $\mathcal{L}_X\Gamma$ is the Lie derivative of $\Gamma$ with respect to a vector field $X$ that generates the flow $\mathcal{X}$. If we define $\mathcal{X}_\lambda^*\Gamma|_{\mathcal{M}_0} \equiv \Gamma_\lambda^{\mathcal{X}}$ and proceeding in the same way for another gauge choice $\mathcal{Y}$, using Eqs.(3-12)-(3-16), the tensor fields $\Gamma_\lambda^{\mathcal{X},\mathcal{Y}}$ can be written as

$$\Gamma_\lambda^{\mathcal{X}} = \sum_{k=0}^{\infty} \frac{\lambda^k}{k!} \delta_{\mathcal{X}}^{(k)}\Gamma = \sum_{k=0}^{\infty} \frac{\lambda^k}{k!} \mathcal{L}_X^k \Gamma|_{\mathcal{M}_0}, \tag{3-17}$$

---

[4] In Eq. (3-11), $g_\lambda$ and $T_\lambda$ are the metric and the matter fields on $\mathcal{M}_\lambda$, similarly $F_\lambda$ and $J_\lambda$ are the electromagnetic field and the four-current on $\mathcal{M}_\lambda$.

[5] Here we introduce a coordinate system $x^\alpha$ through a chart on $\mathcal{M}_\lambda$ with $\alpha = 0, 1, 2, 3$, thus, giving a vector field on $\mathcal{N}$, which has the property that $X^4 = \lambda$ in this chart, while the other components remain arbitrary.

$$\Gamma^{\mathcal{Y}}_{\lambda} = \sum_{k=0}^{\infty} \frac{\lambda^k}{k!} \delta^{(k)}_{\mathcal{Y}} \Gamma = \sum_{k=0}^{\infty} \frac{\lambda^k}{k!} \mathcal{L}^k_Y \Gamma |_{\mathcal{M}_0}, \tag{3-18}$$

if $\Gamma^{\mathcal{X}}_{\lambda} = \Gamma^{\mathcal{Y}}_{\lambda}$ for any arbitrary gauge $\mathcal{X}$ and $\mathcal{Y}$, from here it is clear that $\Gamma$ is totally gauge invariant. It is also clear that $\Gamma$ is gauge invariant to order $n \geqslant 1$ if only if satisfy $\delta^{(k)}_{\mathcal{Y}} \Gamma = \delta^{(k)}_{\mathcal{X}} \Gamma$, or in other way

$$\mathcal{L}_X \delta^{(k)} \Gamma = 0, \tag{3-19}$$

for any vector field $X$ and $\forall k < n$. To first order ($k = 1$) any scalar that is constant in the background or any tensor that vanished in the background are gauge invariant. This result is known as Stewart-Walker Lemma [124], i.e., Eq.(3-19) generalizes this Lemma. However, when $\Gamma$ is not gauge invariant and there are two gauge choices $\mathcal{X}_{\lambda}, \mathcal{Y}_{\lambda}$, the representation of $\Gamma|_{\mathcal{M}_0}$ is different depending on the gauge used. In order to transform the representation from a gauge choice $\mathcal{X}^*_{\lambda}\Gamma|_{\mathcal{M}_0}$ to another $\mathcal{Y}^*_{\lambda}\Gamma|_{\mathcal{M}_0}$, we use the map $\Phi_{\lambda} : \mathcal{M}_0 \to \mathcal{M}_0$ given by

$$\Phi_{\lambda} \equiv \mathcal{X}_{-\lambda} \circ \mathcal{Y}_{\lambda} \Rightarrow \Gamma^{\mathcal{Y}}_{\lambda} = \Phi^*_{\lambda} \Gamma^{\mathcal{X}}_{\lambda}, \tag{3-20}$$

as a consecuence the diffeomorphism $\Phi_{\lambda}$ induces a pull-back $\Phi^*_{\lambda}$ which changes the representation $\Gamma^{\mathcal{X}}_{\lambda}$ of $\Gamma$ in a gauge $\mathcal{X}_{\lambda}$ to the representation $\Gamma^{\mathcal{Y}}_{\lambda}$ of $\Gamma$ in other gauge $\mathcal{Y}_{\lambda}$. Now, following [135] and using the Baker-Campbell-Haussdorf formula [136], it is possible to generalize Eq.(3-16) to write $\Phi^*_{\lambda} \Gamma^{\mathcal{X}}_{\lambda}$ in the following way

$$\Phi^*_{\lambda} \Gamma^{\mathcal{X}}_{\lambda} = \exp\left(\sum_{k=1}^{\infty} \frac{\lambda^k}{k!} \mathcal{L}_{\xi_k}\right) \Gamma^{\mathcal{X}}_{\lambda}, \tag{3-21}$$

where $\xi_k$ is any vector field on $\mathcal{M}_{\lambda}$. Substituting Eq.(3-21) in Eq.(3-20), we have explicitly that

$$\Gamma^{\mathcal{Y}}_{\lambda} = \Gamma^{\mathcal{X}}_{\lambda} + \lambda \mathcal{L}_{\xi_1} \Gamma^{\mathcal{X}}_{\lambda} + \frac{\lambda^2}{2} \left(\mathcal{L}^2_{\xi_1} + \mathcal{L}_{\xi_2}\right) \Gamma^{\mathcal{X}}_{\lambda} + \mathcal{O}(\lambda^3). \tag{3-22}$$

Replacing Eqs.(3-17),(3-18) into Eq.(3-22), the relations to first and second order perturbations of $\Gamma$ in two different gauge choices are given by

$$\delta^{(1)}_{\mathcal{Y}} \Gamma - \delta^{(1)}_{\mathcal{X}} \Gamma = \mathcal{L}_{\xi_1} \Gamma_0, \tag{3-23}$$

$$\delta^{(2)}_{\mathcal{Y}} \Gamma - \delta^{(2)}_{\mathcal{X}} \Gamma = 2\mathcal{L}_{\xi_1} \delta^{(1)}_{\mathcal{X}} \Gamma_0 + \left(\mathcal{L}^2_{\xi_1} + \mathcal{L}_{\xi_2}\right) \Gamma_0, \tag{3-24}$$

where the generators of the gauge transformation $\Phi$ are

$$\xi_1 = Y - X \quad \text{and} \quad \xi_2 = [X, Y]. \tag{3-25}$$

This vector field can be split in their time and space part

$$\xi_\mu^{(r)} \to \left(\alpha^{(r)}, \partial_i\beta^{(r)} + d_i^{(r)}\right), \tag{3-26}$$

here $\alpha^{(r)}$ and $\beta^{(r)}$ are arbitrary scalar functions, and $\partial^i d_i^{(r)} = 0$. The function $\alpha_{(r)}$ determines the choice of constant time hypersurfaces, while $\partial_i\beta^{(r)}$ and $d_i^{(r)}$ fix the spatial coordinates within these hypersurfaces. The choice of coordinates is arbitrary and the definitions of perturbations are thus gauge dependent. The gauge transformations given by Eqs.(3-23),(3-24) are quite general. To first order $\Gamma$ is gauge invariant if $\mathcal{L}_{\xi_1}\Gamma_0 = 0$, while to second order one must have other conditions $\mathcal{L}_{\xi_1}\delta_\mathcal{X}^{(1)}\Gamma_0 = \mathcal{L}_{\xi_1}^2\Gamma_0 = 0$ and $\mathcal{L}_{\xi_2}\Gamma_0 = 0$, and so on at high orders.

## 3.2. The perturbed FLRW spacetime

We will apply the formalism described above to the Robertson-Walker metric. We consider the perturbations about a FLRW background, so the metric tensor is given by

$$g_{00} = -a^2(\tau)\left(1 + 2\sum_{k=1}^\infty \frac{1}{k!}\psi^{(k)}\right), \tag{3-27}$$

$$g_{0i} = a^2(\tau)\sum_{k=1}^\infty \frac{1}{k!}\omega_i^{(k)}, \tag{3-28}$$

$$g_{ij} = a^2(\tau)\left[\left(1 - 2\sum_{k=1}^\infty \frac{1}{k!}\phi^{(k)}\right)\gamma_{ij} + \sum_{k=1}^\infty \frac{\chi_{ij}^{(k)}}{k!}\right]. \tag{3-29}$$

where the variables $\psi$, $\omega_i$, $\phi$ and $\chi_{ij}$ are functions of spacetime coordinates and they vanish on the FLRW background[6]. $\gamma_{ij}$ is the metric tensor for a 3-space of uniform spatial curvature and it can be used to lower (or raise) indices of spatial 3-vectors and tensors. Since we will assume a vanishing spatial curvature of the Universe, we can choose Cartesian coordinates such that $\gamma_{ij} = \delta_{ij}$. The vector perturbation can be split into longitudinal and transverse parts

$$\omega_i^{(k)} = \partial_i\omega^{(k)\|} + \omega_i^{(k)\perp}, \tag{3-30}$$

with $\partial^i\omega_i^{(k)\perp} = 0$ and similarly for $\chi_{ij}^{(k)}$

$$\chi_{ij}^{(k)} = D_{ij}\chi^{(k)\|} + \partial_i\chi_j^{(k)\perp} + \partial_j\chi_i^{(k)\perp} + \chi_{ij}^{(k)\top}, \tag{3-31}$$

with $\partial^i\chi_{ij}^{(k)\top} = 0$, $\partial^i\chi_i^{(k)\perp} = 0$, $\chi_i^{(k)i} = 0$ and $D_{ij} \equiv \partial_i\partial_j - \frac{1}{3}\delta_{ij}\partial_k\partial^k$. As a consequence, we have four scalar modes $\phi$, $\psi$, $\omega^\|$ and $\chi^\|$ which are invariant under rotations and each

---

[6]Here $^{(k)}$ represents the $k-$th order perturbation of the variable.

one has 1 degree of freedom (spin-0), two vector modes $\omega_i^\perp$ and $\chi_i^\perp$ which behave like a spin-1 field under spatial rotations and each having 2 degrees of freedom, and one tensor mode $\chi_{ij}^\top$ physically representing gravitational radiation with two degrees of freedom. So, the perturbed variables contain $4 + 4 + 2 = 10$ independent components (physical degrees of freedom of the metric) as expected from a symmetric tensor. At first order, the metric tensor transforms as

$$\delta \widetilde{g}_{\mu\nu}^{(1)} = \delta g_{\mu\nu}^{(1)} + \partial_\lambda g_{\mu\nu}^{(0)} \xi_{(1)}^\lambda + g_{\mu\lambda}^{(0)} \partial_\nu \xi_{(1)}^\lambda + g_{\nu\lambda}^{(0)} \partial_\mu \xi_{(1)}^\lambda, \tag{3-32}$$

where we use Eq.(3-23). With this formula we can get the transformation of the metric variables which are given by

$$\widetilde{\psi}_{(1)} = \psi_{(1)} + \alpha'_{(1)} + \frac{a'}{a} \alpha_{(1)}, \tag{3-33}$$

$$\widetilde{\omega}_i^{(1)} = \omega_i^{(1)} - \partial_i \alpha^{(1)} + \partial_i \beta^{(1)\prime} + d_i^{(1)\prime}, \tag{3-34}$$

$$\widetilde{\phi}_{(1)} = \phi_{(1)} - \frac{1}{3} \nabla^2 \beta_{(1)} - \frac{a'}{a} \alpha_{(1)}, \tag{3-35}$$

$$\widetilde{\chi}_{ij}^{(1)} = \chi_{ij}^{(1)} + 2 D_{ij} \beta^{(1)} + \partial_j d_i^{(1)} + \partial_i d_j^{(1)}, \tag{3-36}$$

here we simply use a tilde to denote quantities in the new gauge. We can also see that $\chi_{ij}^{(1)\top}$ is gauge invariant. The energy density, pressure and velocity are expanded as

$$\rho = \rho^{(0)} + \sum_{k=1}^{\infty} \frac{1}{k!} \delta \rho^{(k)}, \quad P = P^{(0)} + \sum_{k=1}^{\infty} \frac{1}{k!} \delta P^{(k)}, \quad u^\alpha = \frac{1}{a} \left( \delta_0^\alpha + \sum_{k=1}^{\infty} \frac{1}{k!} v_{(k)}^\alpha \right), \tag{3-37}$$

where the fluid four velocity is subject to the constraint $u_\mu u^\mu = -1$ and it has the contravariant and covariant components expanded up to second order as

$$\begin{aligned}
u^i &= \frac{1}{a} \left( v_{(1)}^i + \frac{1}{2} v_{(2)}^i \right), u^0 = \frac{1}{a} \left( 1 - \psi^{(1)} - \frac{1}{2} \psi^{(2)} + \frac{3}{2} \psi_{(1)}^2 + \omega_i^{(1)} v_{(1)}^i + \frac{1}{2} v_i^{(1)} v_{(1)}^i \right), \\
u_0 &= a \left( -1 - \psi^{(1)} - \frac{1}{2} \psi^{(2)} + \frac{1}{2} \psi_{(1)}^2 - \frac{1}{2} v_i^{(1)} v_{(1)}^i \right), \\
u_i &= a \left( \omega_i^{(1)} + v_i^{(1)} - \psi^{(1)} \omega_i^{(1)} + \frac{1}{2} \left( \omega_i^{(2)} + v_i^{(2)} \right) - 2 v_i^{(1)} \phi_{(1)} + v_{(1)}^j \chi_{ij}^{(1)} \right)
\end{aligned} \tag{3-38}$$

Futhermore it can also be split as

$$v_{(k)}^i = \partial^i v_{(k)}^{\|} + v_\perp^{i\,(k)}. \tag{3-39}$$

Thus, by using the velocity along with the expansions of the energy density and pressure, we obtain the components of the energy momentum tensor up to second order

$$T^0_0 = -(\rho^{(0)} + \delta\rho^{(1)} + \frac{1}{2}\delta\rho^{(2)}) - (\rho^{(0)} + P^{(0)})(v^i_{(1)} + \omega^i_{(1)})v^{(1)}_i \tag{3-40}$$

$$\begin{aligned}T^0_i &= (\rho^{(0)} + P^{(0)})(v^{(1)}_i + \omega^{(1)}_i + \frac{1}{2}(v^{(2)}_i + \omega^{(2)}_i)) \\ &\quad - \psi^{(1)}(v^{(1)}_i + \omega^{(1)}_i) + 2\phi^{(1)}v^{(1)}_i + \chi^{(1)}_{ij}v^j_{(1)}) + (\delta\rho^{(1)} + \delta P^{(1)})(v^{(1)}_i + \omega^{(1)}_i)\end{aligned} \tag{3-41}$$

$$T^i_j = (P^{(0)} + \delta P^{(1)} + \frac{1}{2}\delta P^{(2)})\delta^i_j + (\rho^{(0)} + P^{(0)})(v^{(1)}_j + \omega^{(1)}_j)v^i_{(1)} + \Pi^{i\,(1)}_j + \frac{1}{2}\Pi^{i\,(2)}_j, \tag{3-42}$$

where $\Pi^i_j$ is the transverse and tracefree part of the anisotropic stress having five degrees of freedom, and may be decomposed as in Eq.(3-31) into scalar, vector, and tensor parts. Again, we have four scalar modes $\delta\rho$, $\delta P$, $v^{\|}$ and $\Pi^{\|}$, two vector modes $v^{\perp}_i$ and $\Pi^{\perp}_i$, and one tensor mode $\Pi^{\top}_{ij}$ for a total of $4+4+2=10$ degrees of freedom of the matter perturbations. Now We will look at the transformation for the matter variables

$$\tilde{\delta\rho}_{(1)} = \delta\rho_{(1)} + \rho'_{(0)}\alpha_{(1)}, \tag{3-43}$$

$$\tilde{v}^0_{(1)} = v^0_{(1)} - \frac{a'}{a}\alpha_{(1)} - \alpha'_{(1)}, \tag{3-44}$$

$$\tilde{v}^i_{(1)} = v^i_{(1)} - \partial^i\beta'_{(1)} - d^{i\prime}_{(1)}, \tag{3-45}$$

and the tensor component $\Pi^i_j$ is gauge invariant. As we mentioned earlier, the perturbed quantities would be plagued by non-physical modes due to the gauge problem. To deal with this issue, we can follow two routes: Firstly, It is viable to get gauge invariant quantities and express the equations of motions in terms of these quantities. In fact, since we have four scalars in the metric and two scalars $\alpha^{(1)}$ and $\beta^1$ which determine the gauge transformation Eq.(3-26), one can find the scalar *gauge invariant variables* at first order given by [119]

$$\Psi^{(1)} \equiv \psi^{(1)} + \frac{1}{a}\left(\mathcal{S}^{\|}_{(1)}a\right)', \tag{3-46}$$

$$\Phi^{(1)} \equiv \phi^{(1)} + \frac{1}{6}\nabla^2\chi^{(1)} - H\mathcal{S}^{\|}_{(1)}, \tag{3-47}$$

$$\tag{3-48}$$

with $\mathcal{S}^{\|}_{(1)} \equiv \left(\omega^{\|(1)} - \frac{(\chi^{\|(1)})'}{2}\right)$ the scalar contribution of the shear. The functions $\Psi$ and $\Phi$ are known as the Bardeen's potentials [137]. Other gauge invariant related to the density are

$$\Delta^{(1)} \equiv \delta\rho_{(1)} + (\rho_{(0)})'\mathcal{S}^{\|}_{(1)}, \tag{3-49}$$

$$\Delta^{(1)}_P \equiv \delta P_{(1)} + (P_{(0)})'\mathcal{S}^{\|}_{(1)}. \tag{3-50}$$

For the vector modes we can define

$$v^i_{(1)} \equiv v^i_{(1)} + \left(\chi^i_{\perp(1)}\right)', \tag{3-51}$$

$$\vartheta^{(1)}_i \equiv \omega^{(1)}_i - \left(\chi^{\perp(1)}_i\right)', \tag{3-52}$$

$$\mathcal{V}^i_{(1)} \equiv \omega^i_{(1)} + v^i_{(1)}. \tag{3-53}$$

with $\mathcal{S}^{\|}_{(1)}$ the scalar contribution of the shear (associated with $\alpha^{(1)}$). Therefore, we may write the equation of motion of perturbed variables like energy density or velocity in terms of these quantities

$$\left(\Delta^{(1)}\right)' + 3H\left(\Delta^{(1)}_P + \Delta^{(1)}\right) - 3\left(\Phi^{(1)}\right)'\left(P_{(0)} + \rho_{(0)}\right) + \left(P_{(0)} + \rho_{(0)}\right)\nabla^2 v^{(1)}$$
$$- 3H\left(P_{(0)} + \rho_{(0)}\right)'\mathcal{S}^{\|}_{(1)} - \left(\left(\rho_{(0)}\right)'\mathcal{S}^{\|}_{(1)}\right)'$$
$$+ \left(P_{(0)} + \rho_{(0)}\right)\left(-\frac{1}{2}\nabla^2\chi^{(1)} + 3H\mathcal{S}^{\|}_{(1)}\right)' - \left(P_{(0)} + \rho_{(0)}\right)\nabla^2\left(\frac{1}{2}\chi^{\|(1)}\right)' = 0, \tag{3-54}$$

$$\left(\mathcal{V}^{(1)}_i\right)' + \frac{\left(\mu_{(0)} + P_{(0)}\right)'}{\left(\mu_{(0)} + P_{(0)}\right)}\mathcal{V}^{(1)}_i - 4H\mathcal{V}^{(1)}_i + \partial_i\Psi^{(1)} - \frac{\partial_i\left(\Delta^{(1)}_P - \left(P_{(0)}\right)'\mathcal{S}^{\|}_{(1)}\right) + \partial_l\Pi^{(1)l}_i}{\left(\mu_{(0)} + P_{(0)}\right)} = \frac{\partial_i\left(\mathcal{S}^{\|}_{(1)}a\right)'}{a} \tag{3-55}$$

Secondly, one can fix the gauge i.e. eliminate these degrees of freedom by imposing gauge conditions. In any case, It is possible to show that both views should not make any difference. Now, let us concentrate on the second method and show the definitions of various gauges commonly used throughout the literature.

### 3.2.1. Longitudinal and Poisson gauge

The longitudinal gauge is defined by diagonal metric scalar perturbations

$$\omega^{(r)}_i = \chi^{(r)}_{ij} = 0, \tag{3-56}$$

this allows to express the line element at first order as

$$ds^2 = a^2(\tau)\left(-(1 + \psi^{(1)})d\tau^2 + (1 - 2\phi^{(1)})\delta_{ij}dx^i dx^j\right). \tag{3-57}$$

Using Eqs.(3-34),(3-36), the vector $\xi^{(1)}$ that determines the gauge transformation is fully specified

$$\alpha^{(1)} = \left(\omega^{\|(1)} - \frac{\left(\chi^{\|(1)}\right)'}{2}\right), \quad \beta^{(1)} = -\frac{\chi^{\|}_{(1)}}{2}, \tag{3-58}$$

and by substituting the above expression into Eqs.(3-33),(3-35), we obtain the two Bardeen potentials defined in Eq.(3-46). Thus, not unphysical modes are present in this gauge and the only two variables here are related to the Bardeen ones $\psi^{(1)} = \Psi^{(1)}$ and $\phi^{(1)} = \Phi^{(1)}$. This gauge is widely used in the literature to analyze the evolution of scalar perturbations and it gives equations of motion closer to the Newtonian ones. Some extensions could include vector and tensors such as

$$\partial^i \omega_i^{(r)} = \partial^i \chi_{ij}^{(r)} = 0, \rightarrow \omega^{\|(r)} = \chi_i^{(r)\perp} = \chi^{(r)\|} = 0, \tag{3-59}$$

this is called the Poisson gauge. Note that in the Poisson gauge there are two scalar potentials $\phi$, $\psi$, one transverse vector mode $\omega_i^\perp$, and one transverse-traceless tensor mode $\chi_{ij}^\top$. So, this gauge condition fixes the temporal and spatial gauge function as

$$\alpha^{(1)} = \omega_{(1)}^\| + \beta'_{(1)}, \quad \beta^{(1)} = -\frac{\chi^{\|(1)}}{2}, \quad d_i^{(1)} = -\chi_i^{\perp(1)}. \tag{3-60}$$

### 3.2.2. Synchronous gauge

The synchronous gauge constrains the metric perturbations only to the spatial part

$$\omega_i^{(r)} = \psi^{(r)} = 0, \tag{3-61}$$

so that the proper time for observers at fixed spatial coordinates coincides with cosmic time in the FLRW background [132]. The line element at first order is given by

$$ds^2 = a^2(\tau)\left(-d\tau^2 + (\delta_{ij} - 2\phi^{(1)}\delta_{ij} + \chi_{ij}^{(1)})dx^i dx^j\right). \tag{3-62}$$

If we define

$$h^{(1)} \equiv -\frac{1}{6}\phi^{(1)}, \quad D_{ij}\chi^{(1)\|} \equiv h_{ij}^\|, \quad (\partial_j \chi_i^{(1)\perp} + \partial_i \chi_j^{(1)\perp}) \equiv h_{ij}^\perp,$$

$$\chi_{ij}^{(1)\top} \equiv h_{ij}^\top, \quad \eta^{(1)} \equiv \phi^{(1)} + \frac{1}{6}\nabla^2 \chi_{(1)}^\|, \tag{3-63}$$

we can rewrite the line element as

$$ds^2 = a^2(\tau)\left(-d\tau^2 + (\delta_{ij} + h_{ij}^{(1)})dx^i dx^j\right), \tag{3-64}$$

with $h_{ij}^{(1)} = h^{(1)}\delta_{ij}/3 + h_{ij}^{(1)\|} + h_{ij}^{(1)\perp} + h_{ij}^{(1)\top}$. However, it is well known that synchronous gauge conditions do not fix the spacetime coordinates completely, because of the freedom to redefine the perturbed time-slicing and relabelling the spatial coordinates within these hypersurfaces [138]. These conditions correspond to a gauge transformation where

$$\alpha^{(1)} = \frac{1}{a}\left(\int a\psi^{(1)}d\tau - \hat{C}_a(x^i)\right), \quad \beta^{(1)} = \int(\alpha^{(1)} - \omega^\|)d\tau + \hat{C}_b(x^i),$$

$$d_i^{(1)} = -\int \omega_i^\perp d\tau + \hat{C}_i^c(x^i), \tag{3-65}$$

where $\hat{C}_a(x^i)$, $\hat{C}_b(x^i)$ and $\hat{C}_i^c(x^i)$ are constants of integration which leaves the choice of the initial spatial hypersurface free, and so the synchronous gauge has residual freedom in the form of one scalar $\hat{C}_a(x^i)$ and one transverse vector $\partial_i \hat{C}_b(x^i) + \hat{C}_i^c(x^i)$. A commonly used method for removing this gauge mode is setting the perturbation in the 3-velocity of a pressureless fluid to zero, indeed this fluid could be associated with cold dark matter (CDM) fluid $v_{CDM} = 0$. Then

$$\tilde{v}^{(1)} = 0 = v^{(1)} - \beta'_{(1)} \rightarrow \hat{C}_a(x^i) = a(v^{(1)} + \omega^{\|}_{(1)}) + \int a\psi^{(1)} d\tau, \tag{3-66}$$

where we use Eq.(3-45). The momentum conservation equation for a pressureless fluid ensures that $a(v^{(1)} + \omega^{\|}_{(1)}) =$ is constant, thus

$$\hat{C}_a(x^i) = a(v^{(1)} + \omega^{\|}_{(1)})|_{CDM}. \tag{3-67}$$

This gauge is very useful for numerical studies, and used in Boltzmann codes.

### 3.2.3. Comoving gauge

Finally, the comoving gauge is defined by setting spatial coordinates such that the 3-velocity of the fluid vanishes and the time slices are orthogonal to the fluid 4-velocity (see Eq.(3-38)), then

$$v^i = 0, \quad v_i + \omega_i = 0. \tag{3-68}$$

From transformation equations, this implies that

$$\alpha^{(1)} = \omega^{\|}_{(1)} + v^{\|}_{(1)}, \quad \beta^{(1)} = \int v^{\|}_{(1)} d\tau + \hat{C}_d(x^i), \quad d_i^{(1)} = \int v_i^{(1)\perp} d\tau + \hat{C}_i^e(x^i), \tag{3-69}$$

being $\hat{C}_d(x^i)$ and $\hat{C}_i^e(x^i)$ constants of integration corresponding to a shift of the spatial coordinates. By defining the variable

$$\phi_{\mathcal{R}}^{(1)} \equiv \phi^{(1)} - \frac{1}{6}\nabla^2 \chi^{\|}_{(1)} \tag{3-70}$$

we can build two gauge invariant quantities

$$\begin{aligned}
\mathcal{R} &\equiv \tilde{\phi}_{\mathcal{R}}^{(1)} = \tilde{\phi}^{(1)} - \frac{1}{6}\nabla^2 \tilde{\chi}^{\|}_{(1)} = \phi_{\mathcal{R}}^{(1)} - \frac{a'}{a}\alpha = \phi_{\mathcal{R}}^{(1)} - \frac{a'}{a}(\omega^{\|}_{(1)} + v^{\|}_{(1)}), \\
\Delta_{\mathcal{R}} &\equiv \Delta^{(1)} = \delta\rho_{(1)} + (\rho_{(0)})'\alpha = \delta\rho_{(1)} + (\rho_{(0)})'(\omega^{\|}_{(1)} + v^{\|}_{(1)}).
\end{aligned} \tag{3-71}$$

The first gauge invariant represents the spatial curvature perturbation on the initial comoving hypersurface and, since it is conserved on super Hubble scales for adiabatic initial conditions, it can be useful to parametrize the primordial spectrum. The second one was introduced by first time in [119], and it is related to the density perturbation on the comoving orthogonal hypersurfaces.

### 3.2.4. Transformation laws between longitudinal and synchronous gauges

Transforming from the synchronous to the longitudinal or Poisson gauge can be easy from Eqs.(3-33)-(3-36), where the non-tilde refers to synchronous gauge ($S$) and tilde refers to whether longitudinal ($l$) or Poisson gauge ($P$) [127, 138]

$$\alpha_{(1)} = \beta'_{(1)}, \tag{3-72}$$

$$\omega^{(1)}_{(P)i} = d_i^{(1)\prime}, \quad \text{and for longitudinal gauge:} \to d_i^{(1)}(\tau, x^i) = d_i^{(1)}(x^i) \tag{3-73}$$

$$D_{ij}(2\beta^{(1)} + \chi_S^{(1)\|}) = 0 \quad \to h_{ij}^{(1)\|} = -2D_{ij}\beta^{(1)}, \tag{3-74}$$

$$\partial_j \chi_{(S)i}^{(1)\perp} + \partial_i \chi_{(S)j}^{(1)\perp} + \partial_j d_i^{(1)} + \partial_i d_j^{(1)} = 0 \quad \to h_{ij}^{(1)\perp} = -\partial_j d_i^{(1)} - \partial_i d_j^{(1)}, \tag{3-75}$$

$$\psi^{(1)}_{P,l} = \alpha'_{(1)} + \frac{a'}{a}\alpha_{(1)} = \beta''_{(1)} + \frac{a'}{a}\beta'_{(1)}, \tag{3-76}$$

$$\phi^{(1)}_{P,l} = \phi^{(1)}_S - \frac{1}{3}\nabla^2\beta_{(1)} - \frac{a'}{a}\alpha_{(1)} = -\frac{1}{6}h_{(1)} - \frac{1}{3}\nabla^2\beta_{(1)} - \frac{a'}{a}\beta'_{(1)}, \tag{3-77}$$

$$\chi^{(1)\top}_{(P)ij} = \chi^{(1)\top}_{(S)ij} = h^{(1)\top}_{ij}, \tag{3-78}$$

$$\delta\rho^{(1)}_{l,P} = \delta\rho^{(1)}_S + \rho'_{(0)}\alpha^{(1)} = \delta\rho^{(1)}_S + \rho'_{(0)}\beta'_{(1)}, \tag{3-79}$$

$$\delta P^{(1)}_{l,P} = \delta P^{(1)}_S + P'_{(0)}\alpha^{(1)} = \delta P^{(1)}_S + P'_{(0)}\beta'_{(1)}, \tag{3-80}$$

$$v^{(1)}_{l,P} = v^{(1)}_S - \beta'_{(1)}, \tag{3-81}$$

$$\Pi^{(1)\top}_{(P)ij} = \Pi^{(1)\top}_{(S)ij}. \tag{3-82}$$

We usually work in the Fourier space $k$ to solve the linearized Einstein equations, and it is convenient in this framework to find the transformations between geometrical potentials $\phi$, $\psi$ in the longitudinal gauge and $h$, $\eta$ in the synchronous one. In order to do that, let us assume only scalar pertubation at first order and start writing the Fourier transform for the gravitational potentials

$$\phi(\tau, x^i) = \frac{1}{(2\pi)^3} \int \exp^{(-i\mathbf{k}\cdot\mathbf{x})} \phi(\tau, \mathbf{k}) d^3k, \quad \chi^{\|}(\tau, x^i) = \frac{1}{(2\pi)^3} \int \frac{1}{k^2} \exp^{(-i\mathbf{k}\cdot\mathbf{x})} \chi^{\|}(\tau, \mathbf{k}) d^3k,$$

$$\beta(\tau, x^i) = \frac{1}{(2\pi)^3} \int \exp^{(-i\mathbf{k}\cdot\mathbf{x})} \beta(\tau, \mathbf{k}) d^3k, \quad \alpha(\tau, x^i) = \frac{1}{(2\pi)^3} \int \exp^{(-i\mathbf{k}\cdot\mathbf{x})} \alpha(\tau, \mathbf{k}) d^3k \quad (3\text{-}83)$$

where $\mathbf{k} = \hat{k}k$. We can now use the previous relations to express the perturbations which appear in line element of Eq.(3-62) in the Fourier space

$$-2\phi(\tau, x^i)\delta_{ij} + D_{ij}\chi^{\|}(\tau, x^i) = \frac{1}{(2\pi)^3} \int \exp^{(-i\mathbf{k}\cdot\mathbf{x})}(-2\phi(\tau, \mathbf{k})\delta_{ij} - \hat{k}_i\hat{k}_j\chi^{\|}(\tau, \mathbf{k}) + \chi^{\|}(\tau, \mathbf{k})\frac{\delta_{ij}}{3})d^3k.$$
$$(3\text{-}84)$$

Using the definitions presented in Eq.(3-63), we have

$$-2\phi(\tau, x^i)\delta_{ij} + D_{ij}\chi^{\|}(\tau, x^i) = h_{ij}^{(S)}(\tau, x^i) = \frac{1}{3}h(\tau, x^i)\delta_{ij} + h_{ij}^{\|}(\tau, x^i), \tag{3-85}$$

and therefore we obtain

$$h_{ij}^{(S)}(\tau, x^i) = \frac{1}{(2\pi)^3} \int \exp^{(-i\mathbf{k}\cdot\mathbf{x})}(\frac{1}{3}h(\tau, \mathbf{k})\delta_{ij} - \hat{k}_i\hat{k}_j\chi^{\|}(\tau, \mathbf{k}) + \frac{1}{3}\chi^{\|}(\tau, \mathbf{k})\delta_{ij})d^3k,$$

$$= \frac{1}{(2\pi)^3} \int \exp^{(-i\mathbf{k}\cdot\mathbf{x})}(\hat{k}_i\hat{k}_j(h(\tau, \mathbf{k}) + 6\eta(\tau, \mathbf{k})) - 2\eta(\tau, \mathbf{k})\delta_{ij})d^3k,$$

$$= \frac{1}{(2\pi)^3} \int \exp^{(-i\mathbf{k}\cdot\mathbf{x})}(\hat{k}_i\hat{k}_j h(\tau, \mathbf{k}) + (\hat{k}_i\hat{k}_j - \frac{1}{3}\delta_{ij})6\eta(\tau, \mathbf{k}))d^3k, \tag{3-86}$$

this is the Fourier transform of the scalar mode $h_{ij}$ corresponding to the synchronous gauge. With the above result, let us try to determine $D_{ij}\chi^{\|}(\tau, x^i)$

$$D_{ij}\chi^{\|}(\tau, x^i) = h_{ij}^{(S)}(\tau, x^i) + 2\phi(\tau, x^i)\delta_{ij} = h_{ij}^{(S)}(\tau, x^i) - \frac{1}{3}h(\tau, x^i)\delta_{ij}$$

$$= \frac{1}{(2\pi)^3} \int \exp^{(-i\mathbf{k}\cdot\mathbf{x})} \left(\hat{k}_i\hat{k}_j - \frac{1}{3}\delta_{ij}\right)(h(\tau, \mathbf{k}) + 6\eta(\tau, \mathbf{k}))\, d^3k. \tag{3-87}$$

According to the Fourier transform for $\beta(\tau, x^i)$, we have

$$-2D_{ij}\beta(\tau, x^i) = \frac{2}{(2\pi)^3} \int \exp^{(-i\mathbf{k}\cdot\mathbf{x})} k^2 \left(\hat{k}_i\hat{k}_j - \frac{1}{3}\delta_{ij}\right)\beta(\tau, \mathbf{k})\, d^3k, \tag{3-88}$$

thus, inserting Eqs.(3-87),(3-88) into Eq.(3-75), the expression for $\beta(\tau, x^i)$ can be phrased in terms of the geometrical potentials $h$ and $\eta$

$$\beta(\tau, x^i) = \frac{1}{(2\pi)^3} \int \frac{1}{2k^2} \exp^{(-i\mathbf{k}\cdot\mathbf{x})} (h(\tau, \mathbf{k}) + 6\eta(\tau, \mathbf{k}))\, d^3k, \tag{3-89}$$

finding that

$$\beta(\tau, \mathbf{k}) = \frac{1}{2k^2} \left( h(\tau, \mathbf{k}) + 6\eta(\tau, \mathbf{k}) \right). \tag{3-90}$$

To finish off, Eq.(3-90) can be inserted into Eqs.(3-76),(3-77) obtaining a formula to relate the geometrical portentials in both gauges

$$\psi(\tau, \mathbf{k}) = \frac{1}{2k^2} \left( h''(\tau, \mathbf{k}) + 6\eta''(\tau, \mathbf{k}) + \frac{a'}{a} \left( h'(\tau, \mathbf{k}) + 6\eta'(\tau, \mathbf{k}) \right) \right). \tag{3-91}$$

$$\phi(\tau, \mathbf{k}) = \eta(\tau, \mathbf{k}) - \frac{1}{2k^2}\frac{a'}{a} \left( h'(\tau, \mathbf{k}) + 6\eta'(\tau, \mathbf{k}) \right). \tag{3-92}$$

## 3.3. Weakly magnetized FLRW-background

We can also allow the presence of a weak magnetic field into our FLRW space-time with the property $B_{(0)}^2 \ll \mu_{(0)}$ and it must be sufficiently random to satisface $\langle B_i \rangle = 0$ and $\langle B_{(0)}^2 \rangle \neq 0$ to ensure that symmetries and the evolution of the background remain unaffected (we assume that at zero order the magnetic field has been generated by some random process which is statistically homogeneous so that $B_{(0)}^2$ just time depending and $\langle .. \rangle$ denotes the expectation value) [139]. We work under magnetohydrodynamic(MHD) approximation, thus, in large scales the plasma is globally neutral, charge density is neglected and the electric field with the current should be zero, thus the only zero order magnetic variable is $B_{(0)}^2$. Some gauge invariant variables are the 3-current $J$, the charge density $\varrho$ and the electric and magnetic fields, because they vanish in the background. At first order, the electric field and the current become nonzero and assuming the ohmic current is not neglected, we find the Ohm's law

$$J_i^{(1)} = \sigma \left[ E_i^{(1)} + \left( \mathcal{V}^{(1)} \times B^{(0)} \right)_i \right]. \tag{3-93}$$

In order to study the evolution of magnetic field in large-scales we must write Maxwell's equations in the perturbed FLRW. The deduction of the following equations is shown in Ref. [67]. At first order the Maxwell's equations are expressed as

$$\partial_i E_{(1)}^i = a\varrho_{(1)}, \tag{3-94}$$
$$\partial_i B_{(1)}^i = 0, \tag{3-95}$$
$$\epsilon^{ilk}\partial_l B_k^{(1)} = (E_{(1)}^i)' + 2HE_{(1)}^i + aJ_{(1)}^i, \tag{3-96}$$
$$(B_{(1)}^i)' + 2HB_{(1)}^i = -\epsilon^{ilk}\partial_l E_k^{(1)}, \tag{3-97}$$

these equations represent the evolution of fields in a total invariant way. Furthermore, the energy density of the magnetic field is the unique variable which is gauge dependent and evolves under MHD approximation as $\sim a^{-4}$ and transforms at first order as

$$\tilde{B}_{(1)}^2 = B_{(1)}^2 + \left( B_{(0)}^2 \right)' \alpha^{(1)}. \tag{3-98}$$

30                                                                     3 Cosmological perturbation theoryAt second order, the Maxwell's equations are given by [67]

$$\partial_i E^i_{(2)} = -4E^i_{(1)}\partial_i\left(\psi^{(1)} - 3\phi^{(1)}\right) + a\varrho^{(2)}, \tag{3-99}$$

$$\left(\nabla \times B^{(2)}\right)^i = 2E^i_{(1)}\left(2\left(\psi^{(1)}\right)' - 6\left(\phi^{(1)}\right)'\right) + \left(E^i_{(2)}\right)' + 2HE^i_{(2)}$$
$$+2\left(\nabla\left(2\psi^{(1)} - 6\phi^{(1)}\right) \times B_{(1)}\right)^i + aJ^i_{(2)}, \tag{3-100}$$

$$\frac{1}{a^2}\left(a^2 B^{(2)}_k\right)' + \left(\nabla \times E_{j(2)}\right)_k = 0, \tag{3-101}$$

$$\partial_i B^{i(2)} = 0. \tag{3-102}$$

dependent on gauge choice. The magnetic gauge dependent variables transform as

$$\tilde{E}^{(2)}_i = E^{(2)}_i + 2\left[\frac{\left(a^2 E^{(1)}_i \alpha^{(1)}\right)'}{a^2} + \left(\xi'_{(1)} \times B^{(1)}\right)_i + \xi^l_{(1)}\partial_l E^{(1)}_i + E^{(1)}_l \partial_i \xi^l_{(1)}\right], \tag{3-103}$$

$$\tilde{B}^{(2)}_i = B^{(2)}_i + 2\left[\frac{\alpha^{(1)}}{a^2}\left(a^2 B^{(1)}_i\right)' + \left(\nabla \times \left(B^{(1)} \times \xi^{(1)}\right) + E^{(1)} \times \nabla \alpha^{(1)}\right)_i\right], \tag{3-104}$$

here $\varrho^{(2)}$ and $J^i_{(2)}$ transform according to equations (80),(81) in [67]. The energy density at second order evolves as equation (117) in [67] and it transforms

$$\tilde{B}^2_{(2)} = B^2_{(2)} + B^{2\prime}_{(0)}\alpha_{(2)} + \alpha_{(1)}\left(B^{2\prime\prime}_{(0)}\alpha_{(1)} + B^{2\prime}_{(0)}\alpha'_{(1)} + 2B^{2\prime}_{(1)}\right) + \xi^i_{(1)}\left(B^{2\prime}_{(0)}\partial_i \alpha^{(1)} + 2\partial_i B^2_{(1)}\right) \tag{3-105}$$

Fixing the gauge, we find out the gauge invariant variables related to the electromagnetic fields. Finally, applying the divergence to (3-100) and using Eq.(3-99), we obtain the conservation's equations up to second order for the charge given by

$$\varrho' + 3H\varrho + \nabla \cdot J = 0. \tag{3-106}$$

Here, at first order approximation, the equation is completly invariant, but at second order the involved variables are gauge dependent and transform according to (80),(81) in Ref. [67].

### 3.3.1. Electromagnetic potentials

In order to study the behavior of electromagnetic fields in scenarios such as inflation, vector-tensor theories [140, 141] or quantization of gauge theories in nontrivial spacetimes [142], it is more convenient to write the Maxwell's equations in terms of a four-potential. Therefore, in this section we will apply the gauge invariant approach to scenarios where the prensence of electromagnetic four-potential becomes relevant. The covariant form of the Maxwell's equations (see homogeneous equation (3-126)) reflects the existence of a four-potential [143]. This means that we can define the four potential as $A_\mu = (-\varphi, A_i)$ with the antisymmetric condition given by $F_{\mu\nu} = \partial_\nu A_\mu - \partial_\mu A_\nu$. At first order, the four-potential is gauge invariant (because they are null at the background)[7]. Using the homogeneous Maxwell's equations, It is plausible to define the fields in terms of four-vector potentials

$$B_i^{(1)} = (\nabla \times A^{(1)})_i \quad \text{and} \quad E_i^{(1)} = -(A_i^{(1)\prime} + 2HA_i^{(1)} + \partial_i \varphi^{(1)}). \tag{3-107}$$

Therefore the inhomogeneous Maxwell's equations could be reduced to two invariant equations

$$\nabla^2 \varphi^{(1)} + \frac{1}{a^2}\frac{\partial}{\partial t}\left(\nabla \cdot (a^2 A^{(1)})\right) = -a\varrho_{(1)} \tag{3-108}$$

$$\nabla^2 A_i^{(1)} - \frac{1}{a^2}\frac{\partial^2}{\partial t^2}(a^2 A_i^{(1)}) - \partial_i \left(\nabla \cdot A^{(1)} + \frac{1}{a^2}\frac{\partial}{\partial t}(a^2 \varphi^{(1)})\right) = -aJ_i^{(1)}. \tag{3-109}$$

The latter equations although are written in terms of gauge invariant quantities, they have an arbitrariness in the potentials known in electrodynamics given by the transformations $\widetilde{A}_i^{(1)} = A_i^{(1)} + \partial_i \Lambda$ and $\widetilde{\varphi}^{(1)} = \varphi^{(1)} - \frac{1}{a^2}\frac{\partial}{\partial t}(a^2 \Lambda)$, being $\Lambda$ some scalar function of same order that potentials and, where the fields are left unchanged under this transformation. It is commonly known in the literature that the freedom given by this transformation implies we can choose the set of potentials satisfying the Lorenz conditions which in this case is

$$\nabla \cdot A^{(1)} + \frac{1}{a^2}\frac{\partial}{\partial t}(a^2 \varphi^{(1)}) = 0. \tag{3-110}$$

Therefore, we can arrive at an uncoupled set of equations for the potentials, which are equivalents to the standard Maxwell's equations

$$\nabla^2 \varphi^{(1)} - \frac{1}{a^2}\frac{\partial^2}{\partial t^2}(a^2 \varphi^{(1)}) = -a\varrho_{(1)} \tag{3-111}$$

$$\nabla^2 A_i^{(1)} - \frac{1}{a^2}\frac{\partial^2}{\partial t^2}(a^2 A_i^{(1)}) = -aJ_i^{(1)}. \tag{3-112}$$

At second order the procedure is more complex given the gauge dependence of the potentials. Using the antisymmetrization and the gauge transformation equation (3-24), we have found that the four-portential transforms as

$$\widetilde{\varphi}^{(2)} = \varphi^{(2)} + 2\left[\frac{\alpha_{(1)}}{a^2}(a^2 \varphi_{(1)})' + \xi_{(1)}^i \partial_i \varphi^{(1)} + \alpha'_{(1)}\varphi_{(1)} - \xi_i^{(1)\prime} A_{(1)}^i\right], \tag{3-113}$$

$$\widetilde{A}_i^{(2)} = A_i^{(2)} + 2\left[\frac{\alpha_{(1)}}{a^2}(a^2 A_i^{(1)})' + \partial_l A_i^{(1)} \xi_{(1)}^l - \varphi_{(1)} \partial_i \alpha^{(1)} + A_{(1)}^l \partial_i \xi_l^{(1)}\right]. \tag{3-114}$$

Applying the curl operator at vector potential $\mathcal{A}_i^{(2)}$ and after a long but otherwise straightforward algebra, we obtain the transformation of magnetic field given by (3-104) and the vector potential can expressed as

$$\widetilde{B}_i^{(2)} = (\nabla \times \widetilde{A}^{(2)})_i, \tag{3-115}$$

---

[7] The magnetic potential is null at the background while that electric potential at most is a constant, but due to Stewart-Walker lemma it is gauge invariant.

being a original result of this paper. Similarly, we can use the induction equation (3-100) and with some algebra we find that the scalar potential is described in terms of electric field equation (3-103) via

$$\partial_i \tilde{\varphi}^{(2)} = -\tilde{E}_i^{(2)} - \frac{1}{a^2}\left(a^2 \tilde{A}_i^{(2)}\right)', \tag{3-116}$$

again the four-potential at this order has a freedom mediated by some scalar function $\Lambda$ with same order and under similar transformations showed at first order, the fields $E_i^{(2)}$ and $B_i^{(2)}$ are left unchanged. Let us continue with Maxwell's equations at second order written in terms of the four-potential. For this purpose, we substitute the Eqs.(3-115),(3-116) in the inhomogeneous Maxwell Eqs.(3-99), (3-100) obtaining a coupling set of equations given by

$$\nabla^2 \varphi^{(2)} + \frac{1}{a^2}\frac{\partial}{\partial t}\left(\nabla \cdot (a^2 A^{(2)})\right) - 4\left(\frac{1}{a^2}(a^2 A_i^{(1)})' + \partial_i \varphi^{(1)}\right) \times$$
$$\partial^i(\psi^{(1)} - 3\phi^{(1)}) = -a\varrho^{(2)}, \tag{3-117}$$

$$\nabla^2 A_i^{(2)} - \frac{1}{a^2}\frac{\partial^2}{\partial t^2}(a^2 A_i^{(2)}) - \partial_i\left(\nabla \cdot A^{(2)} + \frac{1}{a^2}\frac{\partial}{\partial t}(a^2 \varphi^{(2)})\right)$$
$$-4\left(\frac{1}{a^2}(a^2 A_i^{(1)})' + \partial_i \varphi^{(1)}\right)(\psi'_{(1)} - 3\phi'_{(1)}) + 4\left(\nabla^2 A_i^{(1)} - \partial_i(\nabla \cdot A^{(1)})\right) \times$$
$$\left(\psi_{(1)} - 3\phi_{(1)}\right) = -aJ_i^{(2)}. \tag{3-118}$$

in a dependent gauge way. The gravitational potentials $\psi$ and $\phi$ transform via Eqs.(3-33), (3-35). With these equations we can see a strong dependence between the electromagnetic fields and the gravitational effects with first order couplings between these variables. The Maxwell's equations found above, are still gauge dependent due to the fact that electromagnetic and gravitational potentials have a freedom in the choice of $\xi^\nu$, the gauge vector. Thus fixing the value of $\xi^\nu$, the variables might take their given meaning. For example, assuming that

$$\tilde{\psi}^{(1)} - 3\tilde{\phi}^{(1)} = 0, \tag{3-119}$$

in order to have the same expression gotten in the first order case, and using Eqs.(3-33), (3-35), an important constraint for the vector part of the gauge dependence is found

$$-\nabla^2 \beta^{(1)} = \psi^{(1)} - 3\phi^{(1)} + 4H\alpha^{(1)} + \alpha'_{(1)}. \tag{3-120}$$

With this choice, the conservation's equation given by expression (3-106) reads as

$$\Delta_\varrho^{(2)\prime} + 3H\Delta_\varrho^{(2)} + \partial_i \mathcal{J}_{(2)}^i + 2\varrho_{(1)}(\Psi'_{(1)} - 3\Phi'_{(1)}) + 2J_{(1)}^i \partial_i(\Psi_{(1)} - 3\Phi_{(1)}) = 0, \tag{3-121}$$

which is gauge invariant and equivalent to the equation (B2) in [67]. We can also use the Lorenz condition by fixing the freedom of the fields

$$\nabla \cdot A^{(2)} + \frac{1}{a^2}\frac{\partial}{\partial t}(a^2 \varphi^{(2)}) = 0, \tag{3-122}$$

obtaining the Maxwell's equation in terms of the potential and written in a invariant way.

## 3.4. The 1+3 Covariant approach: Preliminaries

We first review the Ellis & Bruni [123] covariant formalism and the extension of it with magnetic field described by Tsagas & Barrow [143, 144] briefly. The average motion of matter in the universe defines a future-directed timelike four-velocity $u^\alpha$, corresponding to a fundamental observer ($u_\alpha u^\alpha = -1$), and generates a unique splitting of spacetime into the tangent 3-spaces orthogonal to $u_\alpha$. The second order rank symmetric tensor $h_{\alpha\beta}$ written as

$$h_{\alpha\beta} = g_{\alpha\beta} + u_\alpha u_\beta, \tag{3-123}$$

is the projector tensor which defines the spatial part of the local rest frame of the fundamental observes ($h^\beta{}_\alpha u_\beta = 0$). The proper time derivative along the fluid-flow lines and spatial derivative in the local rest frame for any tensorial quantity $T^{\alpha\beta..}{}_{\gamma\delta..}$ are given by

$$\dot{T}^{\alpha\beta..}{}_{\gamma\delta..} = u^\lambda \nabla_\lambda T^{\alpha\beta..}{}_{\gamma\delta..} \quad \text{and} \quad D_\lambda T^{\alpha\beta..}{}_{\gamma\delta..} = h^\epsilon{}_\lambda h^\omega{}_\gamma h^\tau{}_\delta h^\alpha{}_\mu h^\beta{}_\nu \nabla_\epsilon T^{\mu\nu..}{}_{\omega\tau..} \tag{3-124}$$

respectively[8]. The operator $D_\lambda$ is the covariant derivative operator orthogonal to $u_\alpha$. The kinematic variables are introduced by splitting the covariant derivative of $u_\alpha$ into it's spatial and temporal parts, thus

$$\nabla_\alpha u_\beta = \sigma_{\beta\alpha} + \omega_{\beta\alpha} + \frac{\Theta}{3} h_{\beta\alpha} - a_\beta u_\alpha, \tag{3-125}$$

where, the variable $a_\alpha = u^\beta \nabla_\beta u_\alpha$ is the acceleration ($a_\alpha u^\alpha = 0$), $\Theta = \nabla_\alpha u^\alpha$ is the volume expansion, $\sigma_{\beta\alpha} = D_{(\alpha} u_{\beta)} - \frac{\Theta}{3} h_{\beta\alpha}$ is the shear ($\sigma_{\alpha\beta} u^\alpha = 0, \sigma^\alpha{}_\alpha = 0$) and $\omega_{\beta\alpha} = D_{[\alpha} u_{\beta]}$ is the vorticity ($\omega_{\alpha\beta} u^\alpha = 0, \omega^\alpha{}_\alpha = 0$). Also, on using the totally antisymmetric Levi-Civita tensor $\epsilon_{\alpha\beta\gamma\delta}$, one defines the vorticity vector $\omega^\alpha = \frac{1}{2} \omega_{\mu\nu} \epsilon^{\alpha\mu\nu\beta} u_\beta$. A length scale factor $a$ is introduced along the fluid flow of $u_\alpha$ by means of $H = \frac{\dot{a}}{a} = \frac{\Theta}{3}$, with $H$ the local Hubble parameter. Now, we summarize some results of the covariant studies of electromagnetic fields. The Maxwell's equations in their standard tensor form are written as

$$\nabla_{[\alpha} F_{\beta\gamma]} = 0 \quad \text{and} \quad \nabla^\beta F_{\alpha\beta} = j_\alpha. \tag{3-126}$$

These equations are covariantly characterized by the antisymmetric electromagnetic tensor $F_{\alpha\beta}$ and where $j_\alpha$ is the four-current that sources the electromagnetic field [145]. Using the four-velocity, the electromagnetic fields can be expressed as a four-vector electric field $E_\alpha$ and magnetic field $B_\alpha$ as

$$E_\alpha = F_{\alpha\beta} u^\beta \quad \text{and} \quad B_\alpha = \frac{1}{2} \epsilon_{\alpha\beta\gamma\delta} F^{\gamma\delta} u^\beta. \tag{3-127}$$

By definition, the electromagnetic four-vectors must be purely spatial and orthogonal to four-velocity ($E_\alpha u^\alpha = B_\alpha u^\alpha = 0$). We can write the electromagnetic tensor in terms of the electric and magnetic fields

$$F_{\alpha\beta} = u_\alpha E_\beta - E_\alpha u_\beta + B^\gamma \epsilon_{\alpha\beta\gamma\delta} u^\delta. \tag{3-128}$$

---

[8]The notation showed here, will be only used in this Section.

The electromagnetic tensor determines the energy-momentum tensor of the field which is given by

$$T^{(EM)}_{\alpha\beta} = -F_{\alpha\gamma}F^{\gamma}_{\beta} - \frac{1}{4}g_{\alpha\beta}F_{\gamma\delta}F^{\gamma\delta}. \tag{3-129}$$

Using the four-vector $u_\alpha$ and the projection tensor $h_{\alpha\beta}$, one can decompose the Maxwell's equations (3-126) into a timelike and a spacelike component, getting the following set of equations [144]

$$h^{\alpha}_{\beta}\dot{E}^{\beta} = \left(\sigma^{\alpha}_{\beta} + \omega^{\alpha}_{\beta} - \frac{2}{3}\Theta\delta^{\alpha}_{\beta}\right)E^{\beta} + \epsilon^{\alpha\beta\delta\gamma}B_{\delta}\dot{u}_{\beta}u_{\gamma} + \operatorname{curl} B^{\alpha} - J^{\alpha}, \tag{3-130}$$

$$h^{\alpha}_{\beta}\dot{B}^{\beta} = \left(\sigma^{\alpha}_{\beta} + \omega^{\alpha}_{\beta} - \frac{2}{3}\Theta\delta^{\alpha}_{\beta}\right)B^{\beta} - \epsilon^{\alpha\beta\delta\gamma}E_{\delta}\dot{u}_{\beta}u_{\gamma} - \operatorname{curl} E^{\alpha}, \tag{3-131}$$

$$\mathrm{D}^{\alpha}E_{\alpha} = \varrho - 2\omega^{\alpha}B_{\alpha}, \tag{3-132}$$

$$\mathrm{D}^{\alpha}B_{\alpha} = 2\omega^{\alpha}E_{\alpha}. \tag{3-133}$$

Where the curl operator is defined as $\operatorname{curl} E^{\alpha} = \epsilon^{\beta\alpha\delta\gamma}u_{\delta}\nabla_{\beta}E_{\gamma}$ and the four-current $j_\alpha$ splits along and orthogonal to $u^\alpha$ [143], then

$$\varrho = -j_{\alpha}u^{\alpha} \quad \text{and} \quad J_{\beta} = h^{\alpha}_{\beta}j_{\alpha} \quad \text{with} \quad J_{\alpha}u^{\alpha} = 0. \tag{3-134}$$

Finally, using the antisymmetric electromagnetic tensor together with Maxwell's equations (3-126), one arrives at a covariant form of the charge density conservation law

$$\dot{\varrho} = -\Theta\,\varrho - \mathrm{D}^{\alpha}J_{\alpha} - \dot{u}^{\alpha}J_{\alpha}. \tag{3-135}$$

In this approach, Ellis & Bruni [123] built gauge invariant quantities associated with the orthogonal spatial gradients of the energy density $\mu$, pressure $P$ and fluid expansion $\Theta$. Assuming that the unperturbed background universe is represented by a FLRW metric, the following basic variables are considered

$$X_{\alpha} = \kappa\, h^{\beta}_{\alpha}\nabla_{\beta}\mu, \quad Y_{\alpha} = \kappa\, h^{\beta}_{\alpha}\nabla_{\beta}P \quad \text{and} \quad Z_{\alpha} = \kappa\, h^{\beta}_{\alpha}\nabla_{\beta}\Theta, \tag{3-136}$$

where $\kappa = 8\pi G$. In fact, the variables such as pressure or energy density are usually nonzero in the FLRW background and so are not gauge invariant. However the spatial projection of these variables defined in (3-136) vanishes in the background so, they are gauge invariant and covariantly defined in the physical universe. Also, it is important to define quantities which are more easy to measure, thus is defined the fractional density gradient

$$\mathcal{X}_{\alpha} = \frac{X_{\alpha}}{\kappa\mu} \quad \text{and} \quad \mathcal{Y}_{\alpha} = \frac{Y_{\alpha}}{\kappa P}. \tag{3-137}$$

In the same way we can define the gauge invariant for magnetic fields $\mathcal{B}_\alpha$ in a magnetized universe [146]. For instance, the comoving fractional magnetic energy density distributions and the magnetic field vector can be defined as follows

$$\mathcal{B}_\alpha = D_\alpha B^2, \tag{3-138}$$

$$\mathcal{B} = \frac{a^2}{B^2} D^\alpha \mathcal{B}_\alpha, \tag{3-139}$$

$$\mathcal{M}_{\alpha\beta} = a D_\beta B_\alpha, \tag{3-140}$$

with $B^2$ the local density of the magnetic field. As it has been argued by Tsagas et.al. [143], they describe the spatial variation in the magnetic energy density and the spatial inhomogeneites in the distribution of the vector field $B_i$, as measured by a pair of neighbouring fundamental observers (which represent the motion of typical observers in the Universe being the four-velocity its vector tangent) in a gauge-invariant way. A further discussion of fundamental observers and the meaning of these gauge invariant respect to these observers is given in section 6.3.1 of [147].

## 3.5. Equivalence between two approaches

In this section we present the method to find the equivalence between both approaches mentioned above. For developing this, we compare the gauge invariant quantities built in each approach, similar to the one used by [129, 133]. Let us start defining a gauge invariant quantity related to the energy density of the magnetic field in the gauge invariant approach by substituting the value of $\xi^\mu$ from (3-69) in (3-98) obtaining

$$\Delta^{(1)}_{mag} := \tilde{B}^2_{(1)} = B^2_{(1)} + \left(B^2_{(0)}\right)' (v^\parallel_{(1)} + \omega^\parallel_{(1)}), \quad \to \textbf{Comoving Gauge.} \tag{3-141}$$

Now, we start expanding the equation (3-138), where we use the projector defined in (3-124) and the four-velocity given by (3-38); at first order we obtain

$$\mathcal{B}_0 = D_0 B^2_{(1)} = 0, \tag{3-142}$$

for the temporal part. For spatial part we get

$$\mathcal{B}_i = D_i B^2_{(1)} = \partial_i \left( B^2_{(1)} + \left(B^2_{(0)}\right)' \left(v^\parallel_{(1)} + \omega^\parallel_{(1)}\right) \right), \tag{3-143}$$

where both equations correspond to the gauge invariant in the $1+3$-covariant approach. If we compare the latter equation with the gauge invariant quantity corresponding to energy density of magnetic field (see Eq.(3-141)), we have finally

$$\mathcal{B}_i = D_i B^2_{(1)} \equiv \partial_i \Delta^{(1)}_{mag}. \tag{3-144}$$

The authors in [133] found similar results for the matter density case. For describing the equivalence at second order, we will make use $\tilde{u}_i = 0$ again (comoving condition), thus checking Eq.(3-38) we found that

$$\frac{1}{2}\left(\tilde{\omega}^{(2)}_i + \tilde{v}^{(2)}_i\right) - \tilde{\omega}^{(1)}_i \tilde{\psi}^{(1)} - 2\tilde{v}^{(1)}_i \tilde{\phi}^{(1)} + \tilde{v}^j_{(1)} \tilde{\chi}^{(1)}_{ij} = 0. \tag{3-145}$$

On the other hand, we can use Eq.(3-24) to obtain the gauge transformation to second order for $\omega_i$ and $v_i$ [127]

$$\begin{aligned}
\tilde{\omega}_i^{(2)} &= \omega_i^{(2)} - \partial_i \alpha^{(2)} + \xi_i^{(2)\prime} + \xi_{(1)}^j \left( 2\partial_j \omega_i^{(1)} - \partial_i \partial_j \alpha^{(1)} + \partial_j \xi_i^{(1)\prime} \right) \\
&+ \alpha^{(1)} \left[ 2 \left( \omega_i^{(1)\prime} + 2H\omega_i^{(1)} \right) - \partial_i \alpha^{(1)\prime} + \xi_i^{(1)\prime\prime} - 4H \left( \partial_i \alpha^{(1)} - \xi_i^{(1)\prime} \right) \right] \\
&+ \alpha'_{(1)} \left( 2\omega_i^{(1)} - 3\partial_i \alpha^{(1)} + \xi_i^{(1)\prime} \right) + \xi_{(1)}^{j\prime} \left( -4\phi^{(1)} \delta_{ij} + 2\chi_{ij}^{(1)} + 2\xi_{j,i}^{(1)} + \xi_{i,j}^{(1)} \right) \\
&+ \xi_{(1),i}^j \left( 2\omega_j^{(1)} - \partial_j \alpha^{(1)} \right) - 4\psi^{(1)} \partial_i \alpha^{(1)},
\end{aligned} \quad (3\text{-}146)$$

$$\begin{aligned}
\tilde{v}_i^{(2)} &= v_i^{(2)} - \xi_i^{(2)\prime} + \alpha_{(1)} \left[ 2 \left( v_i^{(1)\prime} - H v_i^{(1)} \right) - \left( \xi_i^{(1)\prime\prime} - 2H \xi_i^{(1)\prime} \right) \right] \\
&+ \xi_{(1)}^j \partial_j \left( 2v_i^{(1)} - \xi_i^{(1)\prime} \right) - \partial_j \xi_i^{(1)} \left( 2v_{(1)}^j - \xi_{(1)}^{j\prime} \right) + \xi_i^{(1)\prime} \left( 2\psi_{(1)} + \alpha'_{(1)} \right).
\end{aligned} \quad (3\text{-}147)$$

Substituting the above equations along with gauge transformations shown in Eqs.(3-33)-(3-36) and the Eq.(3-45) into Eq.(3-145), we obtain the temporal gauge dependence $\alpha^{(2)}$ written in the comoving gauge given by

$$\begin{aligned}
\partial_i \alpha^{(2)} &= \omega_i^{(2)} + v_i^{(2)} - 4\psi^{(1)} \left( \omega_i^{(1)} + v_i^{(1)} \right) + 2v_i^{(1)} \left( \psi^{(1)} - 2\phi^{(1)} \right) \\
&+ \left( \omega_\parallel^{(1)} + v_\parallel^{(1)} \right) \left( \omega_i^{(1)} + v_i^{(1)} \right)' - \left( \omega_\parallel^{(1)} + v_\parallel^{(1)} \right)' \left( \omega_i^{(1)} + v_i^{(1)} \right) \\
&+ \partial_i \xi_j^{(1)} \left( \omega_{(1)}^j + v_{(1)}^j \right) + 2\chi_{ij} v^j + \xi_{(1)}^j \partial_j \left( \omega_i^{(1)} + v_i^{(1)} \right),
\end{aligned} \quad (3\text{-}148)$$

As an alternative way, we can use the equation (A12) in Ref. [67] and transforms it from Poisson to comoving gauge. We can also define a gauge invariant quantity related to the energy density of the magnetic field in the gauge invariant approach at second order fixing the value of $\alpha^{(2)}$ from Eq.(3-148) and $\xi_i^{(1)}$ from Eq.(3-69) into Eq.(3-105) yields

$$\Delta_{mag}^{(2)} := \tilde{B}_{(2)}^2, \quad \rightarrow \textbf{Comoving Gauge.} \quad (3\text{-}149)$$

On the other hand, expanding (3-138) at second order (which comes from 1+3 covariant approach), the temporal part corresponds to

$$\mathcal{B}_0 = D_0 B_{(2)}^2 = -v_{(1)}^i B_{(0)}^{2\prime} \left( v_i^{(1)} + \omega_i^{(1)} \right) - v_{(1)}^i \partial_i B_{(1)}^2, \quad (3\text{-}150)$$

where is the same result found in (3-143) times $v_{(1)}^i$, therefore the temporal part is zero and give us an important constraint for our work. For the spatial part we found out the following

$$\begin{aligned}
\mathcal{B}_i = D_i B^2 &= \frac{1}{2} \partial_i B_{(2)}^2 + \left( \omega_i^{(1)} + v_i^{(1)} \right) B_{(1)}^{2\prime} + B_{(0)}^{2\prime} \left( \frac{1}{2} \left( \omega_i^{(2)} + v_i^{(2)} \right) \right. \\
&\left. - 2\omega_i^{(1)} \psi_{(1)} - 2v_i^{(1)} \phi_{(1)} - \psi^{(1)} v_i^{(1)} + \chi_{ij}^{(1)} v_{(1)}^j \right)
\end{aligned} \quad (3\text{-}151)$$

Now, applying the gradient operator $\partial_i$ to $\Delta^{(2)}_{(mag)}$ showed in (3-149), which is an invariant quantity associated with energy density at second order, we get

$$\partial_i \Delta^{(2)}_{(mag)} = \partial_i B^2_{(2)} + \partial_i \alpha^{(2)} B^{2\prime}_{(0)} + 2\alpha_{(1)} \partial_i \alpha^{(1)} B^{2\prime\prime}_{(0)} + B^{2\prime}_{(0)} \left( \alpha^{(1)\prime} \partial_i \alpha^{(1)} + \alpha^{(1)} \partial_i \alpha'_{(1)} \right)$$
$$+ 2 B^{2\prime}_{(1)} \partial_i \alpha^{(1)} + 2\alpha^{(1)} \partial_i B^{2\prime}_{(1)} + \partial_i \xi^j_{(1)} \partial_j \alpha^{(1)} B^{2\prime}_{(0)} + \xi^j_{(1)} \partial_i \partial_j \alpha^{(1)} B^{2\prime}_{(0)}$$
$$+ 2 \partial_i \xi^j_{(1)} \partial_j B^{2\prime}_{(1)} + 2 \xi^j_{(1)} \partial_j \partial_i B^2_{(1)} \quad (3\text{-}152)$$

Thus, substituting Eq.(3-69) and using Eq.(3-150) in the latter equation, we obtain

$$\partial_i \Delta^{(2)}_{(mag)} = \frac{1}{2} \partial_i B^2_{(2)} + \left( \omega^{(1)}_i + v^{(1)}_i \right) B^{2\prime}_{(1)} + B^{2\prime}_{(0)} \left( \frac{1}{2} \left( \omega^{(2)}_i + v^{(2)}_i \right) \right.$$
$$\left. - 2\omega^{(1)}_i \psi_{(1)} - 2 v^{(1)}_i \phi_{(1)} - \psi^{(1)} v^{(1)}_i + \chi^{(1)}_{ij} v^j_{(1)} \right), \quad (3\text{-}153)$$

which is the expression found in (3-151). Therefore we have obtained the desired result, an equivalence between the invariants of the two approaches up to second order

$$\mathcal{B}_i = D_i B^2 \equiv \partial_i \Delta^{(2)}_{mag}. \quad (3\text{-}154)$$

For the gauge invariant vector field defined in (3-140) we have

$$\mathcal{M}_{00} = 0. \quad (3\text{-}155)$$
$$\mathcal{M}_{[0\,i]} = \left( a B_{i(2)} \right)' + a v^j_{(1)} \partial_{[j} B^{(1)}_{i]}. \quad (3\text{-}156)$$
$$\mathcal{M}_{[i\,j]} = a \left( \partial_{[j} B^{(2)}_{i]} + B^{(1)\prime}_{[i} \mathcal{V}^{(1)}_{j]} \right). \quad (3\text{-}157)$$
$$\mathcal{M}_i{}^i = a \left( \partial_i B^i_{(2)} - \frac{1}{a} B^i_{(1)} (a \mathcal{V}_{i(1)})' - 3 B^i_{(1)} \partial_i \phi \right). \quad (3\text{-}158)$$

If we consider neither the magnetic field nor vorticity in linear perturbation theory in (3-158), we get the usual equation of divergence of the magnetic field (which confirms a claim in Ref. [148]). Making the antisymmetric product between the 4-acceleration equation $a_\mu = u^\nu \nabla_\nu u_\mu$ with the magnetic field, gives an equation of the type

$$a^{(1)}_{[i} B^{(1)}_{j]} = B^{(1)}_{[i} \mathcal{V}^{(1)\prime}_{j]} + B^{(1)}_{[i} \partial_{j]} \psi^{(1)} + H B^{(1)}_{[i} \mathcal{V}^{(1)}_{j]}, \quad (3\text{-}159)$$

where we use the 4-velocity expressed in (3-38) and where the temporal part is zero. If we contract the indices in (3-159) and we use Eq.(3-119), we get a consistency condition with Eq.(3-158) under a null electric field condition. Therefore a magnetic field with no accompanying electric field and currents, provides the relation

$$a^{(1)}_{[\alpha} B^{(1)}_{\beta]} = \mathcal{M}_{[\alpha\,\beta]}, \quad (3\text{-}160)$$

establishing an important relation between the gradient of the magnetic field with a kinematic quantity, as it has been argued by [143]. Taking the curl of equation (3-104) and using the Maxwell's equation (3-100), we find out that

$$(\nabla \times \tilde{H}^{(2)})_i \equiv (\nabla \times \tilde{B}^{(2)})_i = a(\nabla \times B^{(2)})_i. \quad (3\text{-}161)$$

where the electric field and vorticity (this assumption will be reflected as $\epsilon^{kij}\partial_i\xi_j^{(1)} = 0$) have been ignored. Here $\tilde{H}_i^{(2)}$ is the gauge invariant quantity related to the magnetic field vector in the gauge invariant approach. Therefore, by means of equations (3-157) and (3-161) allows us to find the vector equivalence up to second order given as

$$\epsilon^{kij}\mathcal{M}_{[ij]} = (\nabla \times \tilde{H})^k, \tag{3-162}$$

where we use the package VEST (Vector Einstein Summation Tools) to obtain this expression [149]. This result can be described as the variations of the magnetic field vector. In short, assuming a magnetized universe we have verified the equivalence of both approaches by finding connections among their gauge invariant quantities via Eqs.(3-144), (3-154) for scalar and Eq.(3-162) for the tensor case. To conclude, relativistic perturbation theory has been an important tool in theoretical cosmology to link scenarios of the early universe with cosmological data such as CMB-fluctuations. However, there is an issue in the treatment of this theory, which is called gauge problem. Due to the general covariance, a gauge degree of freedom, arises in cosmological perturbations theory. If the correspondence between a real and background space-time is not completely specified, the evolution of the variables will have unphysical modes. Different approaches have been developed to overcome this problem, among them, 1+3 covariant gauge invariant and the gauge invariant approaches, which were studied in the present chapter. Following some results shown in [129] and [133], we have contrasted these formalisms comparing their gauge invariant variables defined in each case. Using a magnetic scenario, we have shown a strong relation between both formalisms, indeed, we found that gauge invariant defined by 1+3 covariant approach is related to spatial variations of the magnetic field energy density (variable defined in the invariant gauge formalism) between two closed fundamental observers as it is noticed in Eqs.(3-144), (3-154) and (3-162). Moreover, we have also derived the gauge transformations for electromagnetic potentials, Eqs.(3-113) and (3-114), which are relevant in the study of evolution of primordial magnetic fields in scenarios such as inflation or later phase transitions. With the description of the electromagnetic potentials, we have expressed the Maxwell's equations in terms of these ones, finding again an important coupling with the gravitational potentials.

# 4. Statistical description of perturbations and random fields

Most of physical observables in cosmology are viewed as continuous functions of spatial coordinates and time. The temporal dependence of these variables are well determined by the Einstein field equations but, since we only can get information from our past light cone, we cannot have access to the complete space at fixed time so, the perturbation of the physical variables should be treated as random fields. Their statistical properties are obtained averaging over ensembles of these fields [150]. However, we live in only one Universe, i.e., one realization of the ensemble, therefore we must average over many disjoint regions by considering that ensamble averages are equal to spatial averages, that is, supposing that our fields are ergodic [94]. We presently believe that the structure we observe in the galaxy distribution and the CMB, was possibly generated by those field fluctuations during the inflationary epoch. Thus, the purpose of this chapter is to describe the stochastic properties of cosmological perturbations by introducing the mathematical concept of random fields and, we shall discuss the Fourier expansion as an important tool for analyzing the statistics of these fields. We follow [151–155] in detail for the discussion of the stochastic nature of perturbations and finally, we conclude showing the integration domain for the spectra when we consider the presence of a cutoff, which is one of the original results of this thesis and It was used as a basis for the work published in [1, 3, 4].

## 4.1. Random variables

There are some observables in cosmology for which we cannot predict its properties completely like the precise location of any galaxy or the exact CMB temperature of the sky at a specific point. From a mathematical point of view, these observables can be described through random variables. A random variable takes different values with distinct probabilities. There are two types of random variables: discrete variables which only can take integral values (like the number of heads when tossing two coins) and continuous variable which can take any value within a certain range (like height of a group of people). Associated with any random variable there will be a corresponding probability function which tell us the probability that the variable takes a particular value [151]. It is also convenient to use the cumulative distribution function which is the probability that one random variable takes a value less than or equal to some particular value $\beta$. For example, if $X$ is the sum of the numbers on two dice, the probability $P$ of obtaining 2 or 3 and its cumulative probability function $P_c(\beta) \equiv \sum_{u \leq \beta} P(u)$ is

$$P(X=2) = \frac{1}{36}, \quad P(X=3) = \frac{2}{36}; \quad P_c(X=2) = \frac{1}{36}, \quad P_c(X=3) = \frac{3}{36},$$

for a discrete variable. For the case of a continuous variable, the probability that the random variable lies between two values $x_1$ and $x_2$ is

$$P(x_1 < x \leq x_2) = \int_{x_1}^{x_2} f(u)du,$$

where $f(x)$ is the probability density function, and its cumulative density function is

$$P(x \leq \beta) = \int_{-\infty}^{\beta} f(u)du.$$

These concepts can be extended to the joint distribution of several random variables, i.e., the probability $P(x,y)$ that each of the random variables falls in any particular range or discrete set of values specified for that variable. We can also calculate the probabilities of various values of one variable regardless of the values taken by the others [151]

$$P_{m_1}(x) = P(x) = \sum_y P(x,y), \quad P_{m_2}(y) = P(y) = \sum_x P(x,y) \rightarrow \text{discrete},$$

$$f_{m_1}(x) = \int f(x,y)dy, \quad f_{m_2}(y) = \int f(x,y)dx \rightarrow \text{continuous},$$

these distributions are sometimes called the marginal distribution of $x$ and $y$ respectively. The condicional probability that one random variable takes the value $x$ given that the other random variable has the fixed value $y$ is

$$P(x|y) = \frac{P(x,y)}{P_{m2}(y)} \rightarrow \text{discrete}, \quad f(x|y) = \frac{f(x,y)}{f_{m2}(y)} \rightarrow \text{continuous}.$$

If these random fields are statistical independent, the conditional probability will be equal to the overall probability distribution

$$P(x,y) = P_{m_1}(x)P_{m_2}(y), \quad f(x,y) = f_{m_1}(x)f_{m_2}(y).$$

### 4.1.1. Absolute moments

It is often useful to talk about the statistical moments which are parameters that describe the main properties of a distribution such as its center, dispersion, symmetry and peakedness. There is a commonly measure for the average of a set of observations which is called the mean $\bar{x}$ given by

$$\bar{x} \equiv \sum_{i=1}^{n} x_i/n = \sum_x xp(x) \xrightarrow{\text{more observations}} \sum_x xP(x), \qquad (4\text{-}1)$$

where $n$ is the number of observations $x_1, x_2, .., x_n$; $p(x) = n(x)/n$ is the proportion of times on which the value $x$ has occured, and in the last term we use the fact that $p(x) \to P(x)$ when the number of observations increases. You may also notice that as the number of data increases, $\bar{x}$ tends towards the parameter $\mu$ which represents the mean of the corresponding probability distribution [151]. In the case of continuous distribution the mean of the distribution is given by

$$\mu = \int x f(x) dx, \tag{4-2}$$

evaluate over the range of values of the random variable, this is also called the first moment of the distribution. Another feature of a random variable is its dispersion. The variance can describe this dispersion which is given by

$$m_2 \equiv \frac{1}{n} \sum_{i=1}^{n} (x_i - \bar{x})^2 = \sum_x (x - \bar{x})^2 p(x) \xrightarrow{\text{more observations}} \sum_x (x - \mu)^2 P(x) \equiv \sigma^2, \tag{4-3}$$

where $m_2$ is also called the second moment around the mean. Furthermore, the variance of a continuous probability distribution is

$$\sigma^2 = \int (x - \mu)^2 f(x) dx. \tag{4-4}$$

Now, for distribution which have only one peak (unimodal) we can measure its degree of lack of symmetry. We can define the skew of a set of observation as

$$m_3 \equiv \frac{1}{n} \sum_{i=1}^{n} (x_i - \bar{x})^3 = \sum_x (x - \bar{x})^3 p(x) \xrightarrow{\text{more observations}} \sum_x (x - \mu)^3 P(x) \equiv B, \tag{4-5}$$

which is also called the third statistic moment around the mean. A negative skew indicates that the tail on the left side of the probability function is longer than the right side. Conversely, positive skew indicates that the tail on the right side is longer than the left side and a zero value represents the case of a symmetric distribution [151]. However, this skew is dimensional so, we can define an adimensional quantity called the skewness

$$s \equiv m_3 / m_2^{3/2} \xrightarrow{\text{more observations}} B/\sigma^3, \tag{4-6}$$

where a zero skewness stands for a symmetrical distribution and an asymmetrical distribution will have a positive or negative skewness depending on this skew to is on the right or left respectively. We can extend again this definition for a continuous distribution as

$$B = \int (x - \mu)^3 f(x) dx. \tag{4-7}$$

Finally, we can write the fourth moment around the mean as

$$m_4 \equiv \frac{1}{n} \sum_{i=1}^{n} (x_i - \bar{x})^4 = \sum_x (x - \bar{x})^4 p(x) \xrightarrow{\text{more observations}} \sum_x (x - \mu)^4 P(x) \equiv T, \tag{4-8}$$

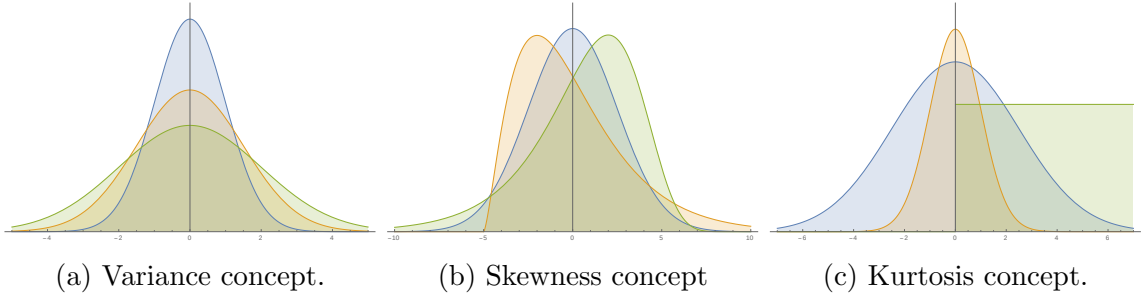

(a) Variance concept.     (b) Skewness concept     (c) Kurtosis concept.

**Figure 4-1**.: Graphic description of the statistical moments used to summarize the shape of a probability distribution. The blue curve in each plot corresponds to the Gaussian distribution with mean zero and variance 1. In the left plot we observe the variance as a measure of dispersion for a probability distribution, here the orange and green curves refer to $\sigma^2 = 2$ and $\sigma^2 = 3$ respectively. In the middle plot we show the skewness as a measure of asymmetry for a probability distribution, here the orange and green curves refer to $s = +1{,}31$ and $s = -1{,}26$ respectively, while for Gaussian distribution we have $s = 0$. Finally, in the right plot we show the kurtosis as a measure of peakedness for a probability distribution, here the orange and green (the uniform distribution) curves refer to $t = 3{,}9$ and $t = 1{,}8$ respectively, while for the Gaussian distribution we have $t = 3$.

or rather the dimensionless quantity

$$t \equiv m_4/m_2^2 \xrightarrow{\text{more observations}} T/\sigma^4. \tag{4-9}$$

This quantity is sometimes used to measure the degree of peakedness and is usually known as kurtosis of the distribution [153]. For continuous distribution we have that

$$T = \int (x - \mu)^4 f(x) dx. \tag{4-10}$$

In Figure **4-1** we show a heuristic description of these statistical moments.

### 4.1.2. Expected value and the characteristic function

The mean of a probability distribution function is often called the expected value denoted by $\langle .. \rangle$, since It is the value of the average over the ensemble (large collection of identical systems). The expected value of a random variable $X$ is given by [151]

$$\langle X \rangle = \sum_x x P(x) \ \to \text{discrete}; \quad \langle X \rangle = \mu = \int x f(x) dx \ \to \text{continuous}, \tag{4-11}$$

and for any function $\phi(X)$ its expectation value can be obtained from

$$\langle \phi(X) \rangle = \sum_x \phi(X) P(x) \ \to \text{discrete}; \quad \langle \phi(X) \rangle = \int \phi(X) f(x) dx \ \to \text{continuous}. \tag{4-12}$$

Let us review some properties of expected values [151]

$$\langle a + bX \rangle = a + b\langle X \rangle, \quad \text{for } a, b \text{ constants}, \tag{4-13}$$

$$\langle X + Y \rangle = \langle X \rangle + \langle Y \rangle, \quad \text{for } X, Y \text{ any pair of random variables}, \tag{4-14}$$

$$\langle XY \rangle = \langle X \rangle \langle Y \rangle, \quad \text{if } X, Y \text{ are independently distributed}. \tag{4-15}$$

Now, we shall consider a function which help us to find all moments of the probability distribution. Let us define the moment generating function $\psi(t)$ of a random variable $X$ by

$$\psi(t) = \langle \exp^{tX} \rangle, \tag{4-16}$$

being $t$ a parameter. Expanding the exponential function we have that

$$\exp^{tX} = 1 + tX + \frac{t^2 X^2}{2!} + \frac{t^3 X^3}{3!} + \ldots \rightarrow \psi(t) = \langle \exp^{tX} \rangle = \sum_{l=0}^{\infty} \frac{\langle X^l \rangle}{l!} t^l, \tag{4-17}$$

hence, if we differentiate $\psi(t)$ $k$ times with respect to $t$ and then set $t = 0$, we get

$$\left. \frac{d^k \psi(t)}{dt^k} \right|_{t=0} = \langle X^k \rangle, \tag{4-18}$$

which is the $k-$th moment about a mean zero $\mu = 0$. Using the properties of expected values mentioned above, we can find some results

$$\langle \exp^{t(X+Y)} \rangle = \langle \exp^{tX} \rangle \langle \exp^{tY} \rangle = \psi_1(t)\psi_2(t), \quad \text{if } X, Y \text{ are independently distributed}.$$

$$\langle \exp^{t(a+bX)} \rangle = \langle \exp^{at} \exp^{btX} \rangle = \exp^{at} \langle \exp^{btX} \rangle = \exp^{at} \psi(bt), \quad \text{for } a, b \text{ constants}. \tag{4-19}$$

If we assume $a = -\mu \neq 0$ and $b = 1$ in the last expression, we find that $\exp^{-\mu t} \psi(t)$ is the moment generating function of $X - \mu$ that generates the moments about the mean $\mu$. Replacing $t$ by $it$ in Eq.(4-16) we obtain the characteristic function of the stochastic variable $X$ in the average

$$\psi(it) \equiv C_X(t) = \langle \exp^{itX} \rangle = \int f(x) \exp^{itx} dx, \tag{4-20}$$

which coincides with the Fourier transform of the probability density function. The McLaurin expansion of the characteristic function reads as

$$C_X(t) = 1 + \sum_{l=1}^{\infty} \frac{(it)^l}{l!} \langle X^l \rangle, \tag{4-21}$$

and the $k-$th moment of $X$ about a zero mean $\mu = 0$, when It exists, can be obtained from the characteristic function by differentiation

$$\left. \frac{d^k C_X(t)}{dt^k} \right|_{t=0} = i^k \langle X^k \rangle. \tag{4-22}$$

If we expand $\ln C_X(t)$ in a power series of $it$

$$\begin{aligned}\ln C_X(t) &= \ln\left(1+\sum_{l=1}^{\infty}\frac{(it)^l}{l!}\langle X^l\rangle\right) = \sum_{l=1}^{\infty}\frac{(it)^l}{l!}\langle X^l\rangle - \frac{1}{2}\left(\sum_{l=1}^{\infty}\frac{(it)^l}{l!}\langle X^l\rangle\right)^2,\\ &- \frac{1}{3}\left(\sum_{l=1}^{\infty}\frac{(it)^l}{l!}\langle X^l\rangle\right)^3 - \ldots \\ &= \langle X\rangle(it) + (\langle X^2\rangle - \langle X\rangle^2)\frac{(it)^2}{2!} + (\langle X^3\rangle + 2\langle X\rangle^3 - 3\langle X\rangle\langle X^2\rangle)\frac{(it)^3}{3!}\\ &+ (\langle X^4\rangle - 3\langle X^2\rangle^2 - 4\langle X^3\rangle\langle X\rangle + 12\langle X^2\rangle\langle X\rangle^2 - 6\langle X\rangle^4)\frac{(it)^4}{4!}\ldots\end{aligned} \quad (4\text{-}23)$$

and defining the cumulants $k_n$ of $X$ through the relation [154]

$$\ln C_X(t) = \sum_{n=1}^{\infty}\frac{(it)^n}{n!}k_n, \rightarrow k_n = \frac{1}{i^n}\left.\frac{d^n \ln C_X(t)}{dt^n}\right|_{n=0}, \quad (4\text{-}24)$$

we can express the moments in terms of the cumulants by comparing Eq.(4-23) and Eq.(4-24) getting [153]

$$\begin{aligned}k_1 &= \langle X\rangle = \mu, \quad k_2 = \langle X^2\rangle - \langle X\rangle^2 = \sigma^2, \quad k_3 = \langle X^3\rangle + 2\langle X\rangle^3 - 3\langle X\rangle\langle X^2\rangle = B,\\ k_4 &= \langle X^4\rangle - 3\langle X^2\rangle^2 - 4\langle X^3\rangle\langle X\rangle + 12\langle X^2\rangle\langle X\rangle^2 - 6\langle X\rangle^4 = T - 3\sigma^4.\end{aligned} \quad (4\text{-}25)$$

To find the cumulants of a linear function $a + bX$, we have from Eq.(4-19)

$$\ln C_{a+bX}(t) = \ln(\exp^{ait}\psi(ibt)) = ait + \ln\psi(ibt) = ait + \ln C_{bX}(t) = ait + \sum_{n=1}^{\infty}\frac{b^n(it)^n}{n!}k_n, \quad (4\text{-}26)$$

where we see that the first cumulant of $a + bX$ is $a + bk_1$, and that of order $n > 1$ is $b^n k_n$. Therefore, we see from Eqs.(4-21),(4-24) that the whole set of cumulants (or moments) determines the probability density function of a random variable completely. As an example, let us calculate the cumulants for the Gaussian distribution

$$C_X(t) = \frac{1}{\sqrt{2\pi}\sigma}\int_{-\infty}^{\infty}\exp^{itx}\exp^{\frac{-(x-\mu)^2}{2\sigma^2}}dx = \exp^{i\mu t - \frac{1}{2}\sigma^2 t^2} \rightarrow \ln C_X(t) = \mu(it) + \sigma^2\frac{(it)^2}{2!}.$$

So we show that $k_1 = \mu$ the mean, $k_2 = \sigma^2$ the variance, and the $k_3 = k_4 = \ldots = 0$ higher-order cumulants vanish (related to skewness, kurtosis and so on).

## 4.2. Random fields

A $n-$dimensional random field, $\mathcal{R}(\mathbf{x})$, is a set of random variables one for each point $\mathbf{x}$ in a $n-$dimensional real space, characterised by a functional probability $\mathcal{P}(\hat{\mathcal{R}}(\mathbf{x}))$ which specifies the probability for ocurrence of one of the possible outcomes of the random field (realisation) $\hat{\mathcal{R}}$ over the ensemble [152]. Such a probability distribution is normalized by requiring that

$$\int \prod_x d\hat{\mathcal{R}}(\mathbf{x})\mathcal{P}(\hat{\mathcal{R}}(\mathbf{x})) = 1. \tag{4-27}$$

Consequently, we can generalize the expectation value of any functional $\mathcal{F}(\mathcal{R}(\mathbf{x}))$ as

$$\langle \mathcal{F}(\mathcal{R}(\mathbf{x})) \rangle = \int \prod_x d\hat{\mathcal{R}}(\mathbf{x})\mathcal{P}(\hat{\mathcal{R}}(\mathbf{x}))\mathcal{F}(\hat{\mathcal{R}}(\mathbf{x})). \tag{4-28}$$

Instead of dealing with the probability functions, we often work with the correlators, that is the expectation values of the product of $n$ functions in different positions at the same time

$$\langle \mathcal{R}(\mathbf{x_1})\ldots\mathcal{R}(\mathbf{x_n}) \rangle = \int \prod_x d\hat{\mathcal{R}}(\mathbf{x})\mathcal{P}(\hat{\mathcal{R}}(\mathbf{x}))\mathcal{R}(\mathbf{x_1})\ldots\mathcal{R}(\mathbf{x_n}), \tag{4-29}$$

this is called the $n-$point correlation functions and It specifies fully the random field. In analogy with (4-21), we can introduce the functional partition [154]

$$\mathcal{Z}(f(\mathbf{x})) = \langle \exp^{\left[i\int d\mathbf{x}'\mathcal{R}(\mathbf{x}')f(\mathbf{x}')\right]} \rangle = \int \prod_x d\mathcal{R}(\mathbf{x})\mathcal{P}(\mathcal{R}(\mathbf{x}))\exp^{\left[i\int d\mathbf{x}'\mathcal{R}(\mathbf{x}')f(\mathbf{x}')\right]}, \tag{4-30}$$

where $f(\mathbf{x})$ is a realisation. With the partition functional we can define the disconnected $n-$point correlation functions as the coefficients of its McLaurin expansion as we have seen in Eq.(4-22)

$$\langle \mathcal{R}(\mathbf{x_1})\ldots\mathcal{R}(\mathbf{x_n}) \rangle = \frac{\delta}{\delta f(\mathbf{x_1})}\ldots\frac{\delta}{\delta f(\mathbf{x_n})}\mathcal{Z}(f(\mathbf{x}))\bigg|_{f=0}, \tag{4-31}$$

which are the generalisation of the moments mentioned in the previous section. Similarly, we can use the coefficients of its McLaurin of the logarithm of partition functional to define the connected correlation functions

$$\langle \mathcal{R}(\mathbf{x_1}),\ldots,\mathcal{R}(\mathbf{x_n}) \rangle_c = \frac{\delta}{\delta f(\mathbf{x_1})}\ldots\frac{\delta}{\delta f(\mathbf{x_n})}\ln\mathcal{Z}(f(\mathbf{x}))\bigg|_{f=0}, \tag{4-32}$$

which are the generalisation of the cumulants. In Cosmology, It is convenient to use the connected functions because they vanish if at least two points belong to causally disconnected regions, i.e., each disconnected region behaves as a realisation within the realisation. Additionally, random fields are usually supposed to be statistically homogeneous and isotropic. Homogeneity means that the probability attached to a realization $\hat{\mathcal{R}}(\mathbf{x})$ is the same that the one attached to $\hat{\mathcal{R}}(\mathbf{x}+\mathbf{s})$ for each fixed $\mathbf{s}$, therefore by allowing $\mathbf{s}$ takes on arbitrary values, we can generate the complete random field [155]. This property is transferred to the correlation functions, for instance, $\langle \mathcal{R}(\mathbf{x_1})\mathcal{R}(\mathbf{x_2}) \rangle$ would be function only of $\mathbf{x_2}-\mathbf{x_1}$.

Isotropy instead means that the probability attached to a realization $\hat{\mathcal{R}}(\mathbf{x})$ is the same as the one attached to $\hat{\mathcal{R}}(\tilde{\mathbf{x}})$, where $\tilde{\mathbf{x}}$ are rotated coordinates [152]. In this case $\langle \mathcal{R}(\mathbf{x_1})\mathcal{R}(\mathbf{x_2})\rangle$ depends only on $|\mathbf{x_2} - \mathbf{x_1}|$. A random field which is homogeneous and isotropic is usually called stationary. Finally, in Cosmology one can never measure expectation values because we have only one Universe so, observations refer to a single realization of the random field. Therefore, we must assume that spatial averages over large scales in a single realization are equal to expectations over the ensemble, i.e., ergodicity. This assumption holds when we can average over many independent volumes, which consist of well separated regions of size smaller than the Hubble horizon, the size of the observable Universe; this is a fair sample of the Universe. However, when the scales become similar to the Hubble horizon, we cannot longer average over many independent volumes and hence, the volume average is quite far from the ensemble average, this difference is called cosmic variance [94].

## 4.3. Fourier transform

Fourier transform is a powerful tool for analyzing stochastic properties. It is often convenient to consider a random field as the linear superposition of many modes. For a flat geometry, the natural tool for achieving this, is via Fourier analysis. For non-flat models, plane waves are not a complete set and one should use instead, the eigenfunctions of the wave equation in a curved space. We consider a random field evaluated at some particular time $f(\mathbf{x})$. Working in a box of comoving size $L$ much bigger than the region of interest, we can write the Fouier expansion of the random field and its inverse as

$$f(\mathbf{x}) = \frac{1}{L^3} \sum_n f_n \exp^{-i\mathbf{k}_n \cdot \mathbf{x}}, \quad f_n = \int d^3x f(\mathbf{x}) \exp^{i\mathbf{k}_n \cdot \mathbf{x}}, \tag{4-33}$$

so that the wave-vector $\mathbf{k}_n$ form a cubic lattice with spacing $2\pi/L$ and

$$\int d^3x \exp^{i(\mathbf{k}_n - \mathbf{k}_m) \cdot \mathbf{x}} = L^3 \delta_{mn}. \tag{4-34}$$

In the limit of $L \to \infty$ we have that Fourier transformation for $f(\mathbf{x})$ is written as

$$f(\mathbf{x}) = \frac{1}{(2\pi)^3} \int d^3k f(\mathbf{k}) \exp^{-i\mathbf{k}\cdot\mathbf{x}}, \quad f(\mathbf{k}) = \int d^3x f(\mathbf{x}) \exp^{i\mathbf{k}\cdot\mathbf{x}}, \tag{4-35}$$

with these conventions we have

$$\delta^3(\mathbf{k} - \mathbf{k}') = \frac{1}{(2\pi)^3} \int d^3x \exp^{i(\mathbf{k}-\mathbf{k}')\cdot\mathbf{x}}, \quad \int d^3k \delta(\mathbf{k}) \exp^{-i\mathbf{k}\cdot\mathbf{x}} = 1, \quad f(-\mathbf{k}) = f^*(\mathbf{k}), \tag{4-36}$$

being the last expression the reality condition, and the Fourier transform of a product is given as (covolution theorem)

$$(fg)(\mathbf{k}) = \frac{1}{(2\pi)^3} \int d^3p f(\mathbf{p}) g(\mathbf{k} - \mathbf{p}) = \frac{1}{(2\pi)^3} \int d^3p d^3q f(\mathbf{p}) g(\mathbf{q}) \delta^3(\mathbf{k} - \mathbf{p} - \mathbf{q}). \tag{4-37}$$

## 4.4. Gaussian random fields

Gaussian random fields are ubiquitous in cosmology. For instance, the simplest inflationary scenario predicts very nearly the Gaussian primordial perturbations with properties consistent with the observed structure. Moreover, cosmological magnetic fields are also modeled as a stochastic background since It agrees with an homogeneous and isotropic background. A Gaussian random field, $\mathcal{G}((x))$, with vanishing mean value is given by the probability functional distribution [152]

$$\mathcal{P}[\mathcal{G}] = \sqrt{det K} \exp^{(-\frac{1}{2}\int d^3x d^3y \mathcal{G}(\mathbf{x})K(\mathbf{x},\mathbf{y})\mathcal{G}(\mathbf{y}))}, \tag{4-38}$$

where $K(\mathbf{x},\mathbf{y})$ is an invertible and symmetric operator which satifies the following condition

$$\int d^3y K(\mathbf{x},\mathbf{y})K^{-1}(\mathbf{y},\mathbf{z}) = \delta^3(\mathbf{x}-\mathbf{y}), \quad \text{with} \quad \langle \mathcal{G}(\mathbf{x})\mathcal{G}(\mathbf{y})\rangle = K^{-1}(\mathbf{x},\mathbf{y}). \tag{4-39}$$

As we mentioned above, the Gaussian random fields are completely characterized by their two-point connected correlation function, while their all higher connected moments are zero. Let us now consider this Gaussian field like a set of finite points rather than a full realisation. Thus, the probability of measuring the values $\{\alpha_1, \cdots \alpha_n\}$ in the space points $\{x_1, \cdots x_n\}$ is given by the multivariate Gaussian distribution [152]

$$\mathcal{P}_n(\alpha_1, \cdots \alpha_n)d\alpha_1, \cdots d\alpha_n = \frac{1}{\sqrt{2\pi det M}} \exp^{-\frac{1}{2}\alpha_i M_{ij}^{-1}\alpha_j} d\alpha_1, \cdots d\alpha_n, \tag{4-40}$$

with $M_{ij} = \langle \alpha_i \alpha_j \rangle$ the covariance matrix. In the Fourier space we can express the Gaussian random field as [154]

$$\mathcal{G}(\mathbf{k}) = a(\mathbf{k}) + ib(\mathbf{k}) = |\mathcal{G}(\mathbf{k})|\exp^{i\varphi(\mathbf{k})}, \tag{4-41}$$

being $\varphi(\mathbf{k})$ a phase, and where $a(\mathbf{k}) = a(-\mathbf{k})$ and $b(\mathbf{k}) = b(-\mathbf{k})$ in order to get $\mathcal{G}(\mathbf{x})$ real. It leads to

$$\langle a(\mathbf{k})a(\mathbf{k}')\rangle = \langle b(\mathbf{k})b(\mathbf{k}')\rangle = P(k)\delta^3(\mathbf{k}-\mathbf{k}'), \quad \langle a(\mathbf{k})b(\mathbf{k}')\rangle = 0, \tag{4-42}$$

where $P(k)$ is the power spectrum of $\mathcal{G}$. As a result, the multivariate Gaussian distribution of $a(\mathbf{k})$ and $b(\mathbf{k})$ are written for each mode $\mathbf{k}$ as

$$\mathcal{P}(\alpha(\mathbf{k}))d\alpha(\mathbf{k}) = \frac{1}{\sqrt{2\pi P(k)}} \exp^{-\frac{\alpha^2(\mathbf{k})}{2P(k)}} d\alpha(\mathbf{k}). \tag{4-43}$$

Therefore, for a given wavevector $\mathbf{k}$, the distribution function in terms of $|\mathcal{G}|$ and its phase $\varphi(\mathbf{k})$ is given by

$$\mathcal{P}(|\mathcal{G}(\mathbf{k})|, \varphi(\mathbf{k}))d|\mathcal{G}(\mathbf{k})|d\varphi(\mathbf{k}) = \exp^{-\frac{|\mathcal{G}(\mathbf{k})|^2}{4P(k)}} \frac{|\mathcal{G}(\mathbf{k})|}{2}\frac{d|\mathcal{G}(\mathbf{k})|}{P(k)}\frac{d\varphi(\mathbf{k})}{2\pi}. \tag{4-44}$$

Thus, for a Gaussian field its real and imaginary parts of individual modes are mutually independent, and this, in turn, implies that the phases $\varphi(\mathbf{k})$ of different modes are also mutually independent and have random distribution [154]. We can then go back to Eq.(4-30) and write the characteristic function $\mathcal{P}_n(\alpha_1, \cdots \alpha_n)$ of Gaussian fields as

$$C_{\mathcal{G}}(\mathbf{b}) = \exp^{(\frac{1}{2} b_i M_{ij} b_j)}, \tag{4-45}$$

and the joint cumulants $k_n$ of the Gaussian variables can be obtained from Eq.(4-24)

$$k_n(x_1, \cdots, x_n) = \frac{\partial}{\partial b_1} \cdots \frac{\partial}{\partial b_n} \ln C_{\mathcal{G}}(\mathbf{b})\big|_{b=0}. \tag{4-46}$$

Since the characteristic function is quadratic in $\mathbf{b}$, the only non-zero cumulant is $k_n(x_i, x_j) = M_{ij}$ [152]. It follows that any higher order moments of the Gaussian distribution can be written in terms of the two-point correlation function alone. Precisely, this is done by using the Wick's theorem [152, 156]

$$\langle \prod_i^n \mathcal{R}_i \rangle = \sum_{p \in \mathbb{P}} \prod_{b \in p} \langle b \rangle_c, \tag{4-47}$$

where $\mathbb{P}$ is the set of all possible ways to partition $\{\mathcal{R}_i, ..., \mathcal{R}_n\}$ and the product goes over each block $b$ of the considered partition. Writing a script in Mathematica [8], we can calculate the n-point correlation function for a random field via Eq.(4-47). For instance, up to n = 4 we have

$$\langle \mathcal{R}_1 \rangle = \langle \mathcal{R}_1 \rangle_c, \tag{4-48}$$

$$\langle \mathcal{R}_1 \mathcal{R}_2 \rangle = \langle \mathcal{R}_1, \mathcal{R}_2 \rangle_c + \langle \mathcal{R}_1 \rangle_c \langle \mathcal{R}_2 \rangle_c, \tag{4-49}$$

$$\langle \mathcal{R}_1 \mathcal{R}_2 \mathcal{R}_3 \rangle = \langle \mathcal{R}_1, \mathcal{R}_2, \mathcal{R}_3 \rangle_c + \langle \mathcal{R}_1, \mathcal{R}_3 \rangle_c \langle \mathcal{R}_2 \rangle_c + \langle \mathcal{R}_2, \mathcal{R}_3 \rangle_c \langle \mathcal{R}_1 \rangle_c$$
$$+ \langle \mathcal{R}_1, \mathcal{R}_2 \rangle_c \langle \mathcal{R}_3 \rangle_c, \tag{4-50}$$

$$\langle \mathcal{R}_1 \mathcal{R}_2 \mathcal{R}_3 \mathcal{R}_4 \rangle = \langle \mathcal{R}_1, \mathcal{R}_2, \mathcal{R}_3, \mathcal{R}_4 \rangle_c + \langle \mathcal{R}_1, \mathcal{R}_2 \rangle_c \langle \mathcal{R}_3 \rangle_c \langle \mathcal{R}_4 \rangle_c + \langle \mathcal{R}_1, \mathcal{R}_3 \rangle_c \langle \mathcal{R}_2 \rangle_c \langle \mathcal{R}_4 \rangle_c$$
$$+ \langle \mathcal{R}_1, \mathcal{R}_4 \rangle_c \langle \mathcal{R}_3 \rangle_c \langle \mathcal{R}_2 \rangle_c + \langle \mathcal{R}_2, \mathcal{R}_3 \rangle_c \langle \mathcal{R}_1 \rangle_c \langle \mathcal{R}_4 \rangle_c + \langle \mathcal{R}_2, \mathcal{R}_4 \rangle_c \langle \mathcal{R}_1 \rangle_c \langle \mathcal{R}_3 \rangle_c$$
$$+ \langle \mathcal{R}_3, \mathcal{R}_4 \rangle_c \langle \mathcal{R}_1 \rangle_c \langle \mathcal{R}_2 \rangle_c + \langle \mathcal{R}_1, \mathcal{R}_2 \rangle_c \langle \mathcal{R}_3, \mathcal{R}_4 \rangle_c + \langle \mathcal{R}_1, \mathcal{R}_3 \rangle_c \langle \mathcal{R}_2, \mathcal{R}_4 \rangle_c$$
$$+ \langle \mathcal{R}_1, \mathcal{R}_4 \rangle_c \langle \mathcal{R}_2, \mathcal{R}_3 \rangle_c. \tag{4-51}$$

For a Gaussian random field (zero mean) we have

$$\langle \mathcal{R}_1 \rangle = 0, \tag{4-52}$$

$$\langle \mathcal{R}_1 \mathcal{R}_2 \rangle = \langle \mathcal{R}_1, \mathcal{R}_2 \rangle_c, \tag{4-53}$$

$$\langle \mathcal{R}_1 \mathcal{R}_2 \mathcal{R}_3 \rangle = 0, \tag{4-54}$$

$$\langle \mathcal{R}_1 \mathcal{R}_2 \mathcal{R}_3 \mathcal{R}_4 \rangle = \langle \mathcal{R}_1, \mathcal{R}_2 \rangle_c \langle \mathcal{R}_3, \mathcal{R}_4 \rangle_c + \langle \mathcal{R}_1, \mathcal{R}_3 \rangle_c \langle \mathcal{R}_2, \mathcal{R}_4 \rangle_c$$
$$+ \langle \mathcal{R}_1, \mathcal{R}_4 \rangle_c \langle \mathcal{R}_2, \mathcal{R}_3 \rangle_c, \tag{4-55}$$

$$\langle \mathcal{R}_1 \mathcal{R}_2 \mathcal{R}_3 \mathcal{R}_4 \mathcal{R}_5 \mathcal{R}_6 \rangle = \langle \mathcal{R}_1, \mathcal{R}_2 \rangle_c \langle \mathcal{R}_3, \mathcal{R}_4 \rangle_c \langle \mathcal{R}_5, \mathcal{R}_6 \rangle_c + \langle \mathcal{R}_1, \mathcal{R}_3 \rangle_c \langle \mathcal{R}_2, \mathcal{R}_4 \rangle_c \langle \mathcal{R}_5, \mathcal{R}_6 \rangle_c$$
$$+ \langle \mathcal{R}_1, \mathcal{R}_2 \rangle_c \langle \mathcal{R}_3, \mathcal{R}_5 \rangle_c \langle \mathcal{R}_4, \mathcal{R}_6 \rangle_c + \cdots 12 \text{ terms more} \cdots, \tag{4-56}$$

where all connected correlations vanish except the covariance one.

## 4.5. Spectrum and bispectrum

As we pointed out before, the properties of the stochastic fields are completely defined by $n-$point correlation functions. Furthermore, we found that for gaussian perturbations their properties are determined only by their two point correlation function, this means that any non-zero primordial signal in the high correlators, will be a direct evidence of non-Gaussianity in the primordial perturbations, and will provide a powerful probe of new physics of the early Universe.

### 4.5.1. Power spectrum

Consider the expectation value for the product of two Fourier modes of a random field $\mathcal{R}$

$$\langle \mathcal{R}(\mathbf{k_1})\mathcal{R}^*(\mathbf{k_2})\rangle = \int d^3x d^3y \exp^{i(\mathbf{k_1}\cdot\mathbf{x}-\mathbf{k_2}\cdot\mathbf{y})}\langle \mathcal{R}(\mathbf{x})\mathcal{R}(\mathbf{y})\rangle. \tag{4-57}$$

Let us introduce a new variable which we will denote by $\mathbf{r} \equiv \mathbf{x} - \mathbf{y}$ which corresponds to the separation bewtween two points. Given the homogeneity, the two point function depends only on the separation $\mathbf{r}$, i.e., the statistical character of the fluctuations does not vary with location, so $\langle \mathcal{R}(\mathbf{y})\mathcal{R}(\mathbf{y}+\mathbf{r})\rangle \equiv \xi(\mathbf{r})$ getting

$$\begin{aligned}\langle \mathcal{R}(\mathbf{k_1})\mathcal{R}^*(\mathbf{k_2})\rangle &= \int d^3r \exp^{i\mathbf{k_1}\cdot\mathbf{r}} \int d^3y \langle \mathcal{R}(\mathbf{y})\mathcal{R}(\mathbf{r}+\mathbf{y})\rangle \exp^{i(\mathbf{k_1}-\mathbf{k_2})\cdot\mathbf{y}} \\ &= \int d^3r \xi(\mathbf{r}) \exp^{i\mathbf{k_1}\cdot\mathbf{r}} (2\pi)^3 \delta^3(\mathbf{k_1}-\mathbf{k_2}) = (2\pi)^3 \delta^3(\mathbf{k_1}-\mathbf{k_2})P(\mathbf{k_1}),\end{aligned} \tag{4-58}$$

where the $\langle \cdots \rangle$ operator only acts on the stochastic variables and we have defined the power spectrum $P(\mathbf{k})$ as the Fourier transform of the two point correlation function

$$P(\mathbf{k}) \equiv \int d^3r \xi(\mathbf{r}) \exp^{i\mathbf{k}\cdot\mathbf{r}}. \tag{4-59}$$

This is known as Wiener-Khintchine Theorem, which states that the two point correlation function of a homogeneous field has a spectral decomposition given by the power spectrum [152, 154]. We can also see that different Fourier modes are completely uncorrelated as a consecuence of the assumed statistical homogeneity of the field. By taking the inverse Fourier transform of power spectrum we obtain the two point correlator

$$\xi(\mathbf{r}) = \frac{1}{(2\pi)^3} \int d^3k P(\mathbf{k}) \exp^{-i\mathbf{k}\cdot\mathbf{r}}. \tag{4-60}$$

We can also invoke isotropy $\xi(\mathbf{r}) \to \xi(r)$, $P(\mathbf{r}) \to P(r)$ and reduce the integration to one dimension

$$\begin{aligned}\xi(r) &= \int_0^{2\pi} \frac{d\varphi}{(2\pi)^3} \int_0^\infty dk k^2 P(k) \int_{-1}^1 d\cos\theta \exp^{-ikr\cos\theta} = \frac{1}{2\pi^2} \int_0^\infty dk k^2 \frac{\sin(kr)}{kr} P(k), \\ P(k) &= \int_0^{2\pi} d\varphi \int_0^\infty dr r^2 \xi(r) \int_{-1}^1 d\cos\theta \exp^{-ikr\cos\theta} = 4\pi \int_0^\infty dr r^2 \frac{\sin(kr)}{kr} \xi(r).\end{aligned} \tag{4-61}$$

In the limit where $r \to 0$, the two point correlation reduces to the variance of the field

$$\lim_{r \to 0} \xi(r) = \langle \mathcal{R}(\mathbf{x})^2 \rangle = \sigma^2 = \frac{1}{2\pi^2} \int_0^\infty dk k^2 P(k), \tag{4-62}$$

where we use that $\lim_{r \to 0} \sin(kr)/(kr) = 1$. From this equation we see that power spectrum quantifies the power or the contribution to the variance of the field per unit $k-$space volume, due to the modes between $k$ and $k + dk$. The dimensionless counterpart is obtained by multiplying it by $k^3$

$$\mathcal{P}(k) = \frac{d\sigma^2}{d \ln k} = \frac{1}{2\pi^2} k^3 P(k) \to \sigma^2 = \int_0^\infty d \ln k \mathcal{P}(k), \tag{4-63}$$

which gives the contribution to the variance of the random field per bin of $\ln k$. A value of $\mathcal{P}(k) = 1$ means that the Fourier modes in a unit logarithmic bin around wavenumber $k$ generate fluctuations of order unity.

**Power law model**

Many physical systems generate fields whose power spectrum can be scale as a power law

$$P(k) = Ak^n, \tag{4-64}$$

where $A$ is the amplitude of the spectrum and $n$ is the spectral index. For values of $n = 0$ we have uncorrelated fields at different places and It is called white spectrum, while for $n > 0$ is called blue spectrum and $n < 0$ are often said to be red. We have a scale invariant spectrum when $\mathcal{P}(k) \sim constant$, that is, when $n = -3$. In fact, for $n = -D$ being $D$ the dimension of the space, we obtain a scale invariant spectrum. Sometimes in Cosmology we call to $n = 1$ the Zeldovich spectrum, i.e., this value renders the amplitude of the fluctuations for density perturbations constant on all scales [157]. Some examples for different spectral indices are shown in Figure **4-2**.

**Cut-off on the spectrum**

In order to guarantee the convergence of the integral Eq.(4-67), It is sometimes convenient to introduce a cut-off on the power spectrum. For instance, some people introduce the presence of a minimal cut-off in order to explain the lack of large-angle correlations at angles $> 60^o$ on the CMB [158]. Another example arises in the matter density perturbation where its power spectrum increases with $k$ up to a cut-off which corresponds to a coherence length [155]. In words, the range of correlation of the fields is limited so, It is appreciable only within some scale that we called coherence length. This cut-off is equivalent to smooth the perturbations on some scale bigger than this coherence length and It can be done by introducing the window function $W(y)$ in the smoothed perturbation $f(\mathbf{x})$ [155]

$$f(R, \mathbf{x}) = \frac{\int d^3 y W(|\mathbf{y} - \mathbf{x}|/R) f(\mathbf{y})}{\int d^3 x W(\mathbf{x}/R)}, \tag{4-65}$$

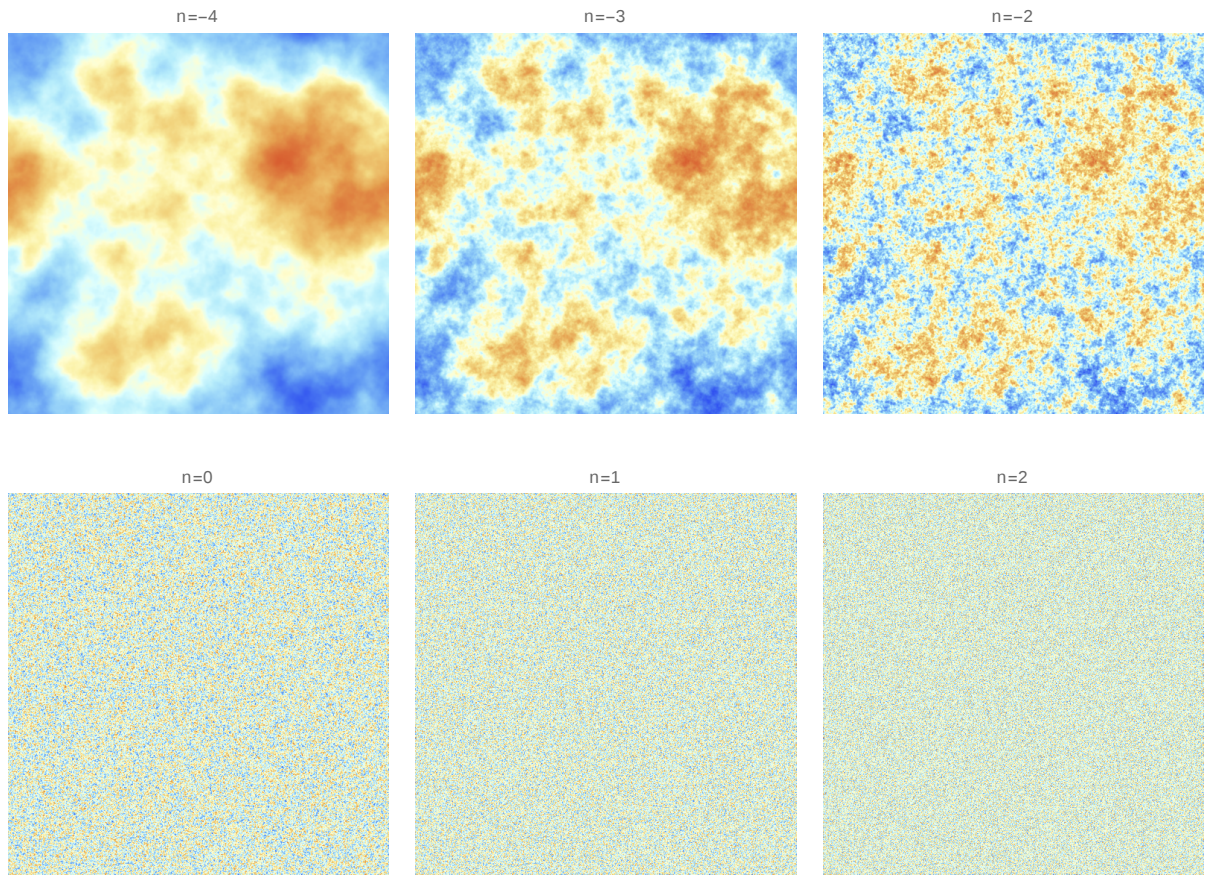

**Figure 4-2**.: Two-dimensional Gaussian random fields with a power spectrum modeled as a power law, whose spectral index is labeled as $n$. At first, a white noise Gaussian random field was generated, then It is smoothed with a small Gaussian filter on scale $R = 1$, and finally we applied the inverse Fourier transform. For $n = 0$ we have the white noise which roughly speaking, corresponds to throwing down in space, a large number of particles at random. $n = 1$ corresponds to the well-known Zeldovich spectrum used in Cosmology, and for $n = -2$ the spectrum is scale-invariant. These plots were generated in Mathematica software [8].

where this window function falls off for $y > 1$. One choice for this window function is the gaussian $W(kR) = \exp^{-k^2R^2/2}$, such is equal to 1 at $k = 0$ and falls off rapidly at $kR > 1$. Therefore the variance of this quantity is given by

$$\sigma^2(R) = \int_0^{R^{-1}} d\ln k \mathcal{P}(k) \sim \mathcal{P}(R^{-1}), \tag{4-66}$$

being the variance of the order of the spectrum evaluated at the cut-off [155]. For a power spectrum scale independent (like primodial spectrum), the integral is logarithmically divergent so, at small scales, It is necesary to smooth the perturbation and for large scales we need to impose a minimal cut-off

$$\sigma^2(R) = \int_{L^{-1}}^{R^{-1}} d\ln k \mathcal{P}(k) \sim \mathcal{P} \int_{L^{-1}}^{R^{-1}} d\ln k = \mathcal{P} \ln(L/R). \tag{4-67}$$

Finally, in the cosmological magnetic fields context the introduction of a cut-off is important to mimic the damping of these fields on small scales [159].

### 4.5.2. Bispectrum

The expectation value for the product of three Fourier modes reads as

$$\langle \mathcal{R}(\mathbf{k_1})\mathcal{R}(\mathbf{k_2})\mathcal{R}(\mathbf{k_3}) \rangle = \int d^3x d^3y d^3z \exp^{i(\mathbf{k_1}\cdot\mathbf{x}+\mathbf{k_2}\cdot\mathbf{y}+\mathbf{k_3}\cdot\mathbf{z})} \langle \mathcal{R}(\mathbf{x})\mathcal{R}(\mathbf{y})\mathcal{R}(\mathbf{z}) \rangle. \tag{4-68}$$

Introducing the variables $\mathbf{s} = \mathbf{y} - \mathbf{x}$ and $\mathbf{t} = \mathbf{z} - \mathbf{x}$, and defining the three point correlation as $\langle \mathcal{R}(\mathbf{x})\mathcal{R}(\mathbf{s}+\mathbf{x})\mathcal{R}(\mathbf{t}+\mathbf{x}) \rangle \equiv \xi(\mathbf{s},\mathbf{t})$ we obtain [152]

$$\begin{aligned}
\langle \mathcal{R}(\mathbf{k_1})\mathcal{R}(\mathbf{k_2})\mathcal{R}(\mathbf{k_3}) \rangle &= \int d^3x \exp^{i(\mathbf{k_1}+\mathbf{k_2}+\mathbf{k_3})\cdot\mathbf{x}} \int d^3t d^3s \exp^{i(\mathbf{k_2}\cdot\mathbf{s}+\mathbf{k_3}\cdot\mathbf{t})} \langle \mathcal{R}(\mathbf{x})\mathcal{R}(\mathbf{y})\mathcal{R}(\mathbf{z}) \rangle \\
&= \int d^3x \exp^{i(\mathbf{k_1}+\mathbf{k_2}+\mathbf{k_3})\cdot\mathbf{x}} \int d^3t d^3s \exp^{i(\mathbf{k_2}\cdot\mathbf{s}+\mathbf{k_3}\cdot\mathbf{t})} \xi(\mathbf{s},\mathbf{t}) \\
&= (2\pi)^3 \delta^3(\mathbf{k_1}+\mathbf{k_2}+\mathbf{k_3})B(\mathbf{k_2},\mathbf{k_3}),
\end{aligned} \tag{4-69}$$

where we have defined the bispectrum $B(\mathbf{k_2},\mathbf{k_3})$ as the Fourier transform of $\xi(\mathbf{s},\mathbf{t})$

$$B(\mathbf{k_2},\mathbf{k_3}) \equiv \int d^3t d^3s \exp^{i(\mathbf{k_2}\cdot\mathbf{s}+\mathbf{k_3}\cdot\mathbf{t})} \xi(\mathbf{s},\mathbf{t}). \tag{4-70}$$

Due to isotropy and the presence of the Dirac distribution (invariance under translations), we can parametrize the bispectrum in terms of the magnitude of all wavenumbers as

$$\langle \mathcal{R}(\mathbf{k_1})\mathcal{R}(\mathbf{k_2})\mathcal{R}(\mathbf{k_3}) \rangle = (2\pi)^3 \delta^3(\mathbf{k_1}+\mathbf{k_2}+\mathbf{k_3})B(k_1,k_2,k_3), \tag{4-71}$$

and so, the three point correlation is readily obtained by taking the inverse Fourier transform

$$\langle \mathcal{R}(\mathbf{x})\mathcal{R}(\mathbf{s}+\mathbf{x})\mathcal{R}(\mathbf{t}+\mathbf{x}) \rangle = \frac{1}{(2\pi)^3} \int d^3k_2 d^3k_3 \exp^{-i(\mathbf{k_2}\cdot\mathbf{s}+\mathbf{k_3}\cdot\mathbf{t})} B(k_1,k_2,k_3). \tag{4-72}$$

In the limit where $s, t \to 0$, we get

$$\langle \mathcal{R}(\mathbf{x})^3 \rangle = \frac{1}{(2\pi)^3} \int d^3k_1 d^3k_2 B(k_1, k_2, k_3), \tag{4-73}$$

with this relation we can obtain the skewness $s = \langle \mathcal{R}(\mathbf{x})^3 \rangle / (\langle \mathcal{R}(\mathbf{x})^2 \rangle)^{3/2}$ defined in Eq.(4-6). It is also often to define the reducen bispectrum $\mathcal{B}$ as [155]

$$B(k_1, k_2, k_3) \equiv \mathcal{B}(k_1, k_2, k_3)[P(k_1)P(k_2) + \text{ cyclic permutations}], \tag{4-74}$$

thus, if $\mathcal{B}(k_1, k_2, k_3)$ and $\mathcal{P}(k_1)$ are scale independent, the skewness reads as $s \sim \mathcal{B}\mathcal{P}^{1/2}$. Finally, we can generalise to $n$-point connected correlation functions of a homogeneous field $\mathcal{R}$ as [152]

$$\langle \mathcal{R}(\mathbf{k_1}) \cdots \mathcal{R}(\mathbf{k_n}) \rangle = (2\pi)^3 \delta^3(\mathbf{k_1} + \cdots + \mathbf{k_n}) S(\mathbf{k_2}, \cdots, \mathbf{k_n}), \tag{4-75}$$

where the polyspectrum $S(\mathbf{k_2}, \cdots, \mathbf{k_n})$ is defined as the Fourier transform of the $n$-point correlation function

$$S(\mathbf{k_2}, \cdots, \mathbf{k_n}) \equiv \int d^3r_2 \cdots d^3r_n \exp^{i(\mathbf{k_2} \cdot \mathbf{r_2} + \cdots + \mathbf{k_n} \cdot \mathbf{r_n})} \xi(\mathbf{r_2}, \cdots, \mathbf{r_n}). \tag{4-76}$$

## 4.6. Integration technique for the Fourier spectra with sharp cut-offs

When we consider the presence of a cut-off in the Fourier theory, we require a careful scheme for the straightforward calculations of the convolutions for the spectra. A sharp cut-off (infrared $k_m$ and ultraviolet $k_D$) enforces the following condition on the power spectrum

$$P(k) = \begin{cases} P(k), & \text{for } k_m \leq k \leq k_D \\ 0, & \text{otherwise,} \end{cases} \tag{4-77}$$

which introduces two conditions in the convolution when the power spectrum depends also on the modulus of separation between wavevectors

$$k_m \leq p \leq k_D, \quad k_m \leq |\mathbf{k} - \mathbf{p}| \leq k_D. \tag{4-78}$$

The previous conditions impose constraints on the angular integration and splits both the angular and radial components into multiple parts for convolutions with the form

$$\mathcal{T}(k) \sim \int d^3p P(p) P(|\mathbf{k} - \mathbf{p}|) \mathcal{O}(\gamma, \beta, \mu, ...), \tag{4-79}$$

where $d^3p = 2\pi p^2 dp d\gamma$, the variables $\{\gamma = \hat{p} \cdot \hat{k}, \beta, \mu, ..\}$ are the set of angle cosines, and $\mathcal{O}(\gamma, \beta, \mu, ...)$ is the angular component. Through a code developed in Mathematica [8], we found out that the spectra becomes non zero only for $0 < k < 2k_D$, and the conditions in Eq.(4-78) split the integrals in many parts for $k_m \neq 0$ (and in three parts for $k_m = 0$). Furthermore, as claimed in [83], the radial integrals need a further splitting for odd spectral indices (when we model the spectrum with a power law). A sketch of the integration is the following (we will show only the result for $k_D > 5k_m$ and $2k_m > k_D > k_m$)

**For** $\boxed{k_D > 5k_m}$, **we have:**

$k_m > k > 0$

$$\int_{k_m}^{k+k_m} d^3p_{(p>k)} \int_{-1}^{\frac{k^2+p^2-k_m^2}{2kp}} d\gamma + \int_{k_m+k}^{k_D-k} d^3p_{(p>k)} \int_{-1}^{1} d\gamma + \int_{k_D-k}^{k_D} d^3p_{(p>k)} \int_{\frac{k^2+p^2-k_D^2}{2kp}}^{1} d\gamma \quad (4\text{-}80)$$

$2k_m > k > k_m$

$$\int_{k_m}^{k} d^3p_{(k>p)} \int_{-1}^{\frac{k^2+p^2-k_m^2}{2kp}} d\gamma \;+\; \int_{k}^{k_m+k} d^3p_{(p>k)} \int_{-1}^{\frac{k^2+p^2-k_m^2}{2kp}} d\gamma$$

$$+ \int_{k+k_m}^{k_D-k} d^3p_{(p>k)} \int_{-1}^{1} d\gamma \;+\; \int_{k_D-k}^{k_D} d^3p_{(p>k)} \int_{\frac{k^2+p^2-k_D^2}{2kp}}^{1} d\gamma \quad (4\text{-}81)$$

$\dfrac{k_D - k_m}{2} > k > 2k_m$

$$\int_{k_m}^{k-k_m} d^3p_{(k>p)} \int_{-1}^{1} d\gamma \;+\; \int_{k-k_m}^{k} d^3p_{(k>p)} \int_{-1}^{\frac{k^2+p^2-k_m^2}{2kp}} d\gamma + \int_{k}^{k+k_m} d^3p_{(p>k)} \int_{-1}^{\frac{k^2+p^2-k_m^2}{2kp}} d\gamma$$

$$+ \int_{k_m+k}^{k_D-k} d^3p_{(p>k)} \int_{-1}^{1} d\gamma \;+\; \int_{k_D-k}^{k_D} d^3p_{(p>k)} \int_{\frac{k^2+p^2-k_D^2}{2kp}}^{1} d\gamma \quad (4\text{-}82)$$

$\dfrac{k_D}{2} > k > \dfrac{k_D - k_m}{2}$

$$\int_{k_m}^{k-k_m} d^3p_{(k>p)} \int_{-1}^{1} d\gamma + \int_{k+k_m}^{k_D} d^3p_{(p>k)} \int_{\frac{k^2+p^2-k_D^2}{2kp}}^{1} d\gamma + \int_{k}^{k_D-k} d^3p_{(p>k)} \int_{-1}^{\frac{k^2+p^2-k_m^2}{2kp}} d\gamma$$

$$+ \int_{k-k_m}^{k} d^3p_{(k>p)} \int_{-1}^{\frac{k^2+p^2-k_m^2}{2kp}} d\gamma + \int_{k_D-k}^{k+k_m} d^3p_{(p>k)} \int_{\frac{k^2+p^2-k_D^2}{2kp}}^{\frac{k^2+p^2-k_m^2}{2kp}} d\gamma \quad (4\text{-}83)$$

$$\frac{k_D}{2} < k < \frac{k_D + k_m}{2}$$

$$\int_{k_m}^{k-k_m} d^3 p_{(k>p)} \int_{-1}^{1} d\gamma + \int_{k+k_m}^{k_D} d^3 p_{(p>k)} \int_{\frac{k^2+p^2-k_D^2}{2kp}}^{1} d\gamma + \int_{k-k_m}^{k_D-k} d^3 p_{(k>p)} \int_{-1}^{\frac{k^2+p^2-k_m^2}{2kp}} d\gamma$$

$$+ \int_{k_D-k}^{k} d^3 p_{(k>p)} \int_{\frac{k^2+p^2-k_D^2}{2kp}}^{\frac{k^2+p^2-k_m^2}{2kp}} d\gamma + \int_{k}^{k+k_m} d^3 p_{(p>k)} \int_{\frac{k^2+p^2-k_D^2}{2kp}}^{\frac{k^2+p^2-k_m^2}{2kp}} d\gamma \qquad (4\text{-}84)$$

$$k_D - k_m > k > \frac{k_D + k_m}{2}$$

$$\int_{k_m}^{k_D-k} d^3 p_{(k>p)} \int_{-1}^{1} d\gamma + \int_{k_D-k}^{k-k_m} d^3 p_{(k>p)} \int_{\frac{k^2+p^2-k_D^2}{2kp}}^{1} d\gamma + \int_{k}^{k+k_m} d^3 p_{(p>k)} \int_{\frac{k^2+p^2-k_D^2}{2kp}}^{\frac{k^2+p^2-k_m^2}{2kp}} d\gamma$$

$$+ \int_{k-k_m}^{k} d^3 p_{(k>p)} \int_{\frac{k^2+p^2-k_D^2}{2kp}}^{\frac{k^2+p^2-k_m^2}{2kp}} d\gamma + \int_{k_m+k}^{k_D} d^3 p_{(p>k)} \int_{\frac{k^2+p^2-k_D^2}{2kp}}^{1} d\gamma \qquad (4\text{-}85)$$

$$k_D > k > k_D - k_m$$

$$\int_{k_m}^{k-k_m} d^3 p_{(k>p)} \int_{\frac{k^2+p^2-k_D^2}{2kp}}^{1} d\gamma + \int_{k}^{k_D} d^3 p_{(p>k)} \int_{\frac{k^2+p^2-k_D^2}{2kp}}^{\frac{k^2+p^2-k_m^2}{2kp}} d\gamma$$

$$+ \int_{k-k_m}^{k} d^3 p_{(k>p)} \int_{\frac{k^2+p^2-k_D^2}{2kp}}^{\frac{k^2+p^2-k_m^2}{2kp}} d\gamma \qquad (4\text{-}86)$$

$$k_D + k_m > k > k_D$$

$$\int_{k_m}^{k-k_m} d^3 p_{(k>p)} \int_{\frac{k^2+p^2-k_D^2}{2kp}}^{1} d\gamma \; + \; \int_{k-k_m}^{k_D} d^3 p_{(k>p)} \int_{\frac{k^2+p^2-k_D^2}{2kp}}^{\frac{k^2+p^2-k_m^2}{2kp}} d\gamma \qquad (4\text{-}87)$$

$$2k_D > k > k_D + k_m$$

$$\int_{k-k_D}^{k_D} d^3 p_{(k>p)} \int_{\frac{k^2+p^2-k_D^2}{2kp}}^{1} d\gamma. \qquad (4\text{-}88)$$

**For the case where** $\boxed{2k_m > k_D > k_m}$**, we have:**

$$\frac{k_D - k_m}{2} > k > 0$$

$$\int_{k_m}^{k+k_m} d^3 p_{(p>k)} \int_{-1}^{\frac{k^2+p^2-k_m^2}{2kp}} d\gamma + \int_{k_m+k}^{k_D-k} d^3 p_{(p>k)} \int_{-1}^{1} d\gamma + \int_{k_D-k}^{k_D} d^3 p_{(p>k)} \int_{\frac{k^2+p^2-k_D^2}{2kp}}^{1} d\gamma \quad (4\text{-}89)$$

$$k_D - k_m > k > \frac{k_D - k_m}{2}$$

$$\int_{k_m}^{k_D-k} d^3 p_{(p>k)} \int_{-1}^{\frac{k^2+p^2-k_m^2}{2kp}} d\gamma + \int_{k_D-k}^{k_m+k} d^3 p_{(p>k)} \int_{\frac{k^2+p^2-k_D^2}{2kp}}^{\frac{k^2+p^2-k_m^2}{2kp}} d\gamma + \int_{k+k_m}^{k_D} d^3 p_{(p>k)} \int_{\frac{k^2+p^2-k_D^2}{2kp}}^{1} d\gamma$$

$$(4\text{-}90)$$

$$k_m > k > k_D - k_m$$

$$\int_{k_m}^{k_D} d^3 p_{(p>k)} \int_{\frac{k^2+p^2-k_D^2}{2kp}}^{\frac{k^2+p^2-k_m^2}{2kp}} d\gamma \quad (4\text{-}91)$$

$$k_D > k > k_m$$

$$\int_{k_m}^{k} d^3 p_{(k>p)} \int_{\frac{k^2+p^2-k_D^2}{2kp}}^{\frac{k^2+p^2-k_m^2}{2kp}} d\gamma + \int_{k}^{k_D} d^3 p_{(p>k)} \int_{\frac{k^2+p^2-k_D^2}{2kp}}^{\frac{k^2+p^2-k_m^2}{2kp}} d\gamma \quad (4\text{-}92)$$

$$2k_m > k > k_D$$

$$\int_{k_m}^{k_D} d^3 p_{(k>p)} \int_{\frac{k^2+p^2-k_D^2}{2kp}}^{\frac{k^2+p^2-k_m^2}{2kp}} d\gamma \quad (4\text{-}93)$$

$k_m + k_D > k > 2k_m$

$$\int_{k-k_m}^{k_D} d^3p_{(k>p)} \int_{\frac{k^2+p^2-k_D^2}{2kp}}^{\frac{k^2+p^2-k_m^2}{2kp}} d\gamma + \int_{k_m}^{k-k_m} d^3p_{(k>p)} \int_{\frac{k^2+p^2-k_D^2}{2kp}}^{1} d\gamma \qquad (4\text{-}94)$$

$2k_D > k > k_m + k_D$

$$\int_{k-k_D}^{k_D} d^3p_{(k>p)} \int_{\frac{k^2+p^2-k_D^2}{2kp}}^{1} d\gamma. \qquad (4\text{-}95)$$

Finally, in the case where $\boxed{k_m = 0}$, we obtain the same results reported in [83]

$k_D/2 > k > 0$

$$\int_0^k d^3p_{(k>p)} \int_{-1}^{1} d\gamma + \int_k^{k_D-k} d^3p_{(k<p)} \int_{-1}^{1} d\gamma + \int_{k_D-k}^{k_D} d^3p_{(k<p)} \int_{\frac{k^2+p^2-k_D^2}{2kp}}^{1} d\gamma \qquad (4\text{-}96)$$

$k_D/2 < k < k_D$

$$\int_0^{k_D-k} d^3p_{(k>p)} \int_{-1}^{1} d\gamma + \int_{k_D-k}^{k} d^3p_{(k>p)} \int_{\frac{k^2+p^2-k_D^2}{2kp}}^{1} d\gamma + \int_k^{k_D} d^3p_{(k<p)} \int_{\frac{k^2+p^2-k_D^2}{2kp}}^{1} d\gamma \qquad (4\text{-}97)$$

$2k_D > k > k_D$

$$\int_{k-k_D}^{k_D} d^3p_{(k>p)} \int_{\frac{k^2+p^2-k_D^2}{2kp}}^{1} d\gamma. \qquad (4\text{-}98)$$

The integration domain above, generalizes the results obtained in [83, 84] and thus, they become an original result of this thesis. With this scheme, we can calculate the spectrum and bispectrum (under certain configurations) for any stochastic field. In following chapters we will use this integration scheme for the Fourier spectra in order to calculate the magnetic spectra and bispectra and thus, be able to find the effects of these fields on the CMB anisotropies. It is remarkable to note that those results are quite general and can be applied to numerous convolution integrals.

# 5. A theoretical framework for Magnetogenesis

Magnetic fields are ubiquitous in the Universe. Even if the origin of these fields is under debate, It is assumed that observed fields were originated from cosmological or astrophysical seed fields and then, amplified during structure formation via some astrophysical mechanism [70]. Currently, there is not a clear astrophysical process to generate magnetic fields on such large length scales. An appealing alternative we can think of, is to make use of inflation in the early Universe since it may be able to produce those coherent fields on large scales. It is well known that to create magnetic fields during inflation, the conformal invariance of the standard electrodynamics must be broken. One of the first models of inflamagnetogenesis was introduced by Ratra [41], where he proposed a conformal-breaking coupling between the scalar field (the inflaton) and the electromagnetic field. Many other mechanisms have been proposed following the same philosophy and several conditions were obtained in order to explain the observed large-scale magnetic fields. However, serious obstacles arise in those mechanisms such as the strong-coupling problem where the theory becomes uncontrollable [50]; the backreaction problem where an overproduction of the electric fields spoils inflation [50] and the curvature perturbation problem where the generation of both scalar and tensor curvature perturbations from PMFs would yield results in conflict with CMB observations [160].

Recently, different scenarios for inflationary magnetogenesis have been proposed to solve the above problems successfully. For instance, Membiela [161] showed that a bouncing cosmology with a contracting phase dominated by an equation of state with $P > -\rho/3$ can support magnetogenesis, evading the backreaction/strong-coupling problems; Ferreira et. al. [51] analyzed simple extensions of the minimal model which avoid both the strong coupling and back reaction problems; Caprini et. al [52] considered the pseudoscalar invariant $F_{\mu\nu}\tilde{F}^{\mu\nu}$ multiplied by a time dependent function; Domenech et. al., [162] pointed out that a successful inflationary magnetogenesis could be realised if the local U(1) gauge symmetry during inflation is broken; and a non-conformally invariant coupling between the inflaton and the photon in the minimal Lorentz violating standard model extension is compatible with observation as was claimed by Campanelli [163]. Futhermore, some authors have explored alternative possibilities, i.e., they avoided conformal triviality via a nonzero spatial curvature of the Universe [36], others have analyzed causality and the nature of the transition from inflation to reheating and then to the radiation era [164], and finally, other authors have considered deviations from the FLRW metric [44]. This latter situation is rather natural, since they do not require any modification of Maxwell electromagnetism, and they includes metric inhomogeneities which have been always presented at large scales. Therefore, finding inflationary magnetogenesis models in which the strong-coupling/backreaction and curvature perturbation problems are avoided; represents a serious challenge and It is still a work in progress. Only the cosmological scenario will be briefly discussed in this chapter. More complete treatments of this subject can be found in Refs. [13, 24, 36, 64, 165]

## 5.1. A primordial origin

Cosmological scenarios describe the generation of magnetic fields in the early Universe (so-called primordial magnetic field), approximately prior to or during Recombination, i.e. $T > 0,25\text{eV}$. At the same time, cosmological scenarios can be classified into two categories: inflationary and post-inflationary magnetogenesis. The first scenario generates PMFs correlated on very large scales during inflation, although the breaking of conformal invariance of the electromagnetic action is needed in order to obtain the suitable seed field. Besides, these kind of models also suffer from some problems such as backreaction and the strong coupling [163]. On the other hand, post-inflationary scenarios consider PMFs created after inflation via either cosmological phase transitions or during the Recombination era(Harrison's mechanism) [165]. However, these will lead to a correlation scale of the field smaller than the Hubble radius at that epoch, thus a suitable field cannot be generated unless we consider another dynamical effect, for instance helicity, which under certain conditions produces transference of energy from small to large scales required to explain the observational large-scale magnetic fields [24]. We will briefly summarize some models and properties of the cosmological scenario. Throughout this Section we will use units in which $c = \hbar = M_G = k_B = 1$, being $M_G = (8\pi G)^{-1/2}$ the reduced Planck mass.

### 5.1.1. Inflamagnetogenesis

A treatment of inflationary magnetic field which we follow in great detail in this Section, are given by K. Subramanian [90, 165], Martin et al., [166], Ferreira et al., [51], Caprini et al., [52] and Durrer et al., [167].
As mentioned earlier, inflation provides an interesting scenario for the generation of PMFs with large coherence scales. Let us start with the standard free electromagnetic(EM) action, given by

$$S_{EM} = \frac{-1}{4} \int \sqrt{-g} g^{\mu\alpha} g^{\nu\beta} F_{\mu\nu} F_{\alpha\beta} d^4x, \tag{5-1}$$

where $F_{\mu\nu} = \nabla_\mu A_\nu - \nabla_\nu A_\mu = \partial_\mu A_\nu - \partial_\nu A_\mu$ is conformally invariant and being $A_\mu$ the vector potential. By making a conformal transformation of the metric given by $g^*_{\mu\nu} = \Omega^2 g_{\mu\nu}$, the determinant $\sqrt{-g}$ and the contravariant metric change as

$$\sqrt{-g} \to \sqrt{-g}^* = \Omega^4\sqrt{-g}, \qquad g^{\mu\nu} \to g^{\mu\nu\,*} = \Omega^{-2}g^{\mu\nu}, \tag{5-2}$$

and the factors $\Omega^2$ cancel out the action, thus the action of the free EM is invariant under conformal transformations. Since the FLRW models are conformally flat i.e. $g_{\mu\nu}^{FLRW} = \Omega^2\eta_{\mu\nu}$, being $\eta_{\mu\nu}$ the Minkowski metric, one can transform the electromagnetic wave equation into its flat version. Therefore, It is not possible to amplify the EM field fluctuations and this leads to an adiabatic decay of the EM field as $\sim 1/a^2$ with the expansion of the universe. Hence, inflamagnetogenesis requires the breaking of conformal invariance of the EM action in order to amplify EM waves from vacuum fluctuations. A multitude of possibilities have been considered for this purpose and some of them are illustrated in the action

$$\begin{aligned}S &= \int \sqrt{-g}\left[-I^2(\phi,f(R))\left(\frac{1}{4}F^{\mu\nu}F_{\mu\nu} - \frac{\gamma_g}{8}\epsilon^{\mu\nu\alpha\beta}F_{\mu\nu}F_{\alpha\beta}\right) - RA^2\right.\\ &\quad \left. - \frac{\beta}{4m}\epsilon^{\mu\nu\alpha\beta}F_{\mu\nu}F_{\alpha\beta} - D_\mu\psi(D^\mu\psi)^*\right]d^4x - \int \sqrt{-g}\left[\frac{1}{2}g^{\mu\nu}\partial_\mu\phi\partial_\nu\phi + V(\phi)\right]d^4x.\end{aligned} \tag{5-3}$$

This action includes coupling of EM action to scalar fields ($\phi$) like the inflaton or dilaton and $V(\phi)$ its potential [166, 168]; coupling to curvature invariants ($R$) or a particular class of $f(R)$ theories [169]; coupling to a pseudo-scalar field like the axion ($\beta$) with a mass scale $m$ [170]; charged scalar fields ($\psi$) [171], and the presence of a constant $\gamma_g$ that leads to a magnetic field with a net helicity [52]; here $\epsilon^{\mu\nu\alpha\beta}$ is the totally anti-symmetric tensor in four dimensions with $\epsilon^{0123} = (-g)^{-1/2}$.

## 5.2. Scalar field plus a U(1) gauge field

For a pedagogical introduction of magnetogenesis during inflation, we consider in this Section only the first two terms in the previous action and we assume a dimensionless gauge coupling $I^2(\phi)$, being $\phi$ the field associated with the inflaton [166]

$$\begin{aligned}S[\phi, A_\mu] &= -\int \sqrt{-g}I^2(\phi)\left[\frac{1}{4}g^{\alpha\mu}g^{\beta\nu}F_{\alpha\beta}F_{\mu\nu} - \frac{\gamma_g}{8}\epsilon^{\mu\nu\alpha\beta}F_{\mu\nu}F_{\alpha\beta}\right]d^4x \\ &\quad - \int \sqrt{-g}\left[\frac{1}{2}g^{\mu\nu}\partial_\mu\phi\partial_\nu\phi + V(\phi)\right]d^4x.\end{aligned} \tag{5-4}$$

The equations of motion following from the above action are (see Appendix B for more details)

$$\frac{1}{\sqrt{-g}}\partial_\alpha\left[\sqrt{-g}I^2\left(F_{\mu\nu}g^{\beta\nu}g^{\alpha\mu} - \frac{\gamma_g}{2}\epsilon^{\alpha\beta\mu\nu}F_{\mu\nu}\right)\right] = 0, \tag{5-5}$$

$$\frac{1}{\sqrt{-g}}\partial_\alpha\left(\sqrt{-g}g^{\alpha\beta}\partial_\beta\phi\right) - \frac{\partial V}{\partial\phi} = \frac{1}{4}\frac{\partial I^2}{\partial\phi}\left(g^{\alpha\mu}g^{\beta\nu}F_{\alpha\beta}F_{\mu\nu} - \frac{\gamma_g}{2}\epsilon^{\mu\nu\alpha\beta}F_{\mu\nu}F_{\alpha\beta}\right). \tag{5-6}$$

In the following, we assume that EM field is a perturbative field which affects neither the scalar field evolution (2-29) nor the evolution of the background FLRW Universe (2-12). Hence, we can neglect the right hand side of the Klein-Gordon equation and work with the FLRW spacetime 2-7. In the Coulomb gauge $A_0(\eta, \mathbf{x}) = \partial_i A^i(\eta, \mathbf{x}) = 0$, the equation of motion becomes (see Appendix B)

$$A_i'' + 2I^{-1}\frac{\partial I}{\partial \phi}A_i' - a^2 \partial_j \partial^j A_i = 2a^2 \gamma_g I^{-1}\frac{\partial I}{\partial \phi}\epsilon_{ijk}\partial^j A^k, \quad \text{or:}$$

$$\bar{A}_i'' - I^{-1}\frac{\partial^2 I}{\partial \phi^2}\bar{A}_i' - a^2 \partial_j \partial^j \bar{A}_i = 2a^2 \gamma_g I^{-1}\frac{\partial I}{\partial \phi}\epsilon_{ijk}\partial^j \bar{A}^k, \tag{5-7}$$

being $\bar{A}_i \equiv IA_i$ and $\epsilon_{ijk}$ the three dimensional levi-civita symbol. Now, we are going to calculate the energy density of the EM fields by varying the action (5-4) with respect to the metric

$$T_{\mu\nu} \equiv -\frac{2}{\sqrt{-g}}\frac{\delta S}{\delta g^{\mu\nu}} = -I^2(\phi)g^{\alpha\beta}F_{\mu\alpha}F_{\beta\nu} - \frac{I^2(\phi)}{4}g_{\mu\nu}g^{\alpha\beta}g^{\gamma\delta}F_{\beta\delta}F_{\alpha\gamma}, \tag{5-8}$$

where the time-time component is given by

$$T_{00} = \underbrace{\frac{I^2(\phi)}{4}a^2 g^{ij}g^{k\ell}\left(\partial_j A_\ell - \partial_\ell A_j\right)\left(\partial_i A_k - \partial_k A_i\right)}_{\rho_B} + \underbrace{\frac{I^2(\phi)}{2}g^{ij}A_i'A_j'}_{\rho_E}, \tag{5-9}$$

here, the total energy density in the EM field can be written as a sum of the magnetic(first term) and electric(second term) contributions. In order to quantize the EM field in the FLRW background, we calculate the conjugate momentum $\pi_i$ associated to the gauge field $A_i$ by varying the action with respect to $A_i'$

$$\pi^i(\tau, x) = \frac{\delta S}{\delta A_i'} = I^2(\phi)a^2 g^{ij}\left(A_j'(\tau, x) + a^4 \gamma_g \epsilon_{lkj} g^{ln} g^{ks} \partial_n A_s(\tau, x)\right), \tag{5-10}$$

and postulate the standard commutation relation between $A^i(\tau, x)$ and $\pi_j(\tau, x)$

$$\begin{aligned}\left[A^i(\tau, x), \pi_j(\tau, y)\right] &= i\int \frac{d^3k}{(2\pi)^3}e^{ik\cdot(x-y)}\left(\delta^i{}_j - \delta_{j\ell}\frac{k^i k^\ell}{k^2}\right) \\ &= i\delta^{(3)i}_{\perp\;j}(x-y), \end{aligned} \tag{5-11}$$

with $\delta^{(3)}_{\perp ij}$ the transverse delta function [172]. Drawing on the experience from solutions of Klein-Gordon and Dirac equations, we may expand the vector potential $A_i(\tau, x)$ in Fourier space in an orthonormal basis (see Appendix B)

$$A_i(\tau, x) = \int \frac{d^3k}{(2\pi)^{3/2}}\sum_{\lambda=1}^{2}\epsilon_{i\lambda}(k)\left[b_\lambda(k)A(\tau, k)e^{ik\cdot x} + b_\lambda^\dagger(k)A^*(\tau, k)e^{-ik\cdot x}\right], \tag{5-12}$$

here the star denotes complex conjugation and this expression is written in terms of the annihilation and creation operators: $b_\lambda(k)$ and $b_\lambda^\dagger(k)$; the comoving wavenumber $k$ and, the transverse polarization vector $\epsilon_i^\lambda(k)$. We can also expand $A_i(\tau, x)$ in the helicity basis (see Appendix B)

$$A_i(\tau, x) = \int \frac{d^3k}{(2\pi)^{3/2}} \sum_{h=\pm} \left[ \epsilon_{ih}(k) b_h(k) A_h(\tau, k) e^{ik\cdot x} + \epsilon_{ih}^*(k) b_h^\dagger(k) A_h^*(\tau, k) e^{-ik\cdot x} \right], \tag{5-13}$$

which is helpful when we are studying helical fields.

### 5.2.1. Non-helical fields $\gamma_g = 0$

In the following, we will assume $\gamma_g = 0$ in the action 5-4. By substituting expression of $A_i(\tau, x)$ (5-12) into (5-7), the time-dependent Fourier amplitude obeys the following equation of motion

$$\mathcal{A}''(\tau, k) + \left( k^2 - \frac{I''}{I} \right) \mathcal{A}(\tau, k) = 0 \tag{5-14}$$

where $\mathcal{A}(\tau, k) \equiv a(\tau) \bar{A}(\tau, k)$. In order to satisfy (5-11), the creation and annihilation operators must obey

$$\left[ b_\lambda(k), b_{\lambda'}^\dagger(k') \right] = \delta^3(k - k') \delta_{\lambda\lambda'}, \quad [b_\lambda(k), b_{\lambda'}(k')] = \left[ b_\lambda^\dagger(k), b_{\lambda'}^\dagger(k') \right] = 0, \tag{5-15}$$

along with the normalization condition for the time-dependent amplitude $A(\tau, k)$

$$A(\tau, k) A'^*(\tau, k) - A'(\tau, k) A^*(\tau, k) = \frac{i}{I^2 a^2}. \tag{5-16}$$

Then, we substitute the Fourier expansion of the potential vector into the first term of Eq.(5-9) and take the expectation value in the vacuum state $|0>$ (defined by the condition $b_\lambda(k)|0>= 0$, for all $k$), obtaining the energy density for the magnetic field

$$\begin{aligned}\rho_B(\tau) &= -\left\langle 0 \left| T^{B\,0}{}_0 \right| 0 \right\rangle = \frac{I^2}{(2\pi)^3} \int d^3k \, |A(\tau, k)|^2 \frac{k^2}{a^2} \\ &= \frac{1}{2\pi^2} \int_0^{+\infty} \frac{dk}{k} k^5 I^2 \left| \frac{A(\tau, k)}{a(\tau)} \right|^2 \\ &= \frac{1}{2\pi^2} \int_0^{+\infty} \frac{dk}{k} k^5 \frac{1}{a^4(\tau)} |\mathcal{A}(\tau, k)|^2 \equiv \int_0^{+\infty} \frac{dk}{k} \frac{d}{d\ln k} \rho_B(\tau, k),\end{aligned} \tag{5-17}$$

where we have used the definition of energy density stored at a given scale in the last step, thus we get

$$\frac{d}{d\ln k} \rho_B(\tau, k) = \frac{1}{2\pi^2} \frac{k^5}{a^4(\tau)} |\mathcal{A}(\tau, k)|^2. \tag{5-18}$$

Finally, if we substitute the Fourier expansion of the potential vector into the second term of Eq.(5-9), we find the energy density for the electric field

$$\rho_E(\tau) = -\langle 0|T^{E0}{}_0|0\rangle = \frac{I^2}{(2\pi)^3}\int d^3k \frac{|(aA(\tau,k))'|^2}{a^4(\tau)}$$

$$= \frac{1}{2\pi^2}\int_0^{+\infty}\frac{dk}{k}\frac{k^3}{a^4(\tau)}\left|\left(\frac{\mathcal{A}(\tau,k)}{I}\right)'\right|^2 I^2 \equiv \int_0^{+\infty}\frac{dk}{k}\frac{d}{d\ln k}\rho_E(\tau,k), \quad (5\text{-}19)$$

and the energy density of the electric field stored at a given scale is

$$\frac{d}{d\ln k}\rho_E(\tau,k) = \frac{1}{2\pi^2}\frac{k^3}{a^4(\tau)}\left|\left(\frac{\mathcal{A}(\tau,k)}{I}\right)'\right|^2 I^2. \quad (5\text{-}20)$$

For illustrative purposes, we perform calculations of normal modes for a simple case when the scale factor and the coupling evolve with conformal time as

$$a(\tau) = a_0\left|\frac{\tau}{\tau_0}\right|^{1+\beta}, \quad I(\phi) \propto a^\alpha, \quad (5\text{-}21)$$

where $\alpha$ and $\beta$ are free indices and $\beta = -2$ corresponds to de Sitter spacetime. With the above expressions, the evolution of the mode function $\mathcal{A}(\tau,k)$ is given by

$$\mathcal{A}(\tau,k)'' + \left(k^2 - \frac{\gamma(\gamma-1)}{\tau^2}\right)\mathcal{A}(\tau,k) = 0, \quad \gamma \equiv \alpha(1+\beta). \quad (5\text{-}22)$$

with the general solution

$$\mathcal{A}(\tau,k) = (-k\tau)^{1/2}\left[C_1(k,\gamma)J_{\gamma-1/2}(-k\tau) + C_2(k,\gamma)J_{-\gamma+1/2}(-k\tau)\right], \quad (5\text{-}23)$$

where $C_1(k)$ and $C_2(k)$ are two scale-dependent coefficients which are constrained by the initial conditions. Using the asymptotic relation for the Bessel functions, we can determine the behavior of the mode function for the far remote past $-k\tau \gg 1$ and for late times $-k\tau \ll 1$, see Ref. [173]

$$J_n(s) = \begin{cases} \sqrt{\frac{2}{\pi s}}\cos(s - \frac{n\pi}{2} - \frac{\pi}{4}), & s \to \infty \\ \frac{1}{\Gamma(n+1)}\left(\frac{s}{2}\right)^n, & s \to 0 \end{cases}$$

Indeed, there is an ambiguity for choosing the vacuum state in a general spacetime, due to the fact that geometry of a curve spacetime could change notably across the spatial size of the wavepacket(localized particle), and therefore, plane waves could not a good definition of a particle with momentum $k$ [173]. However, if the Hubble scale exceeds the scale of the fluctuation $k/(aH) = -k\tau \to \infty$, the mode function is not affected significantly by gravity and it behaves as in Minkowski spacetime. Thus, we can neglect the $\tau^{-2}$ contribution compared with $k^2$ in Eq.(5-22) and the vacuum state corresponds to the positive-frecuency mode

$$\mathcal{A}(k,\tau) \to \frac{1}{\sqrt{2k}} e^{-ik\tau}, \tag{5-24}$$

these modes determine the *Bunch-Davis* vacuum which is annihilated by all operators $b_\lambda(k)$ [173]. As a consequence, the arbitrary coefficients $C_1$ and $C_2$ are given by

$$C_1(k,\gamma) = \sqrt{\frac{\pi}{k}} \frac{\exp^{i\pi\gamma/2}}{2\cos(\pi\gamma)}, \quad C_2(k,\gamma) = \sqrt{\frac{\pi}{k}} \frac{\exp^{-i\pi(\gamma-1)/2}}{2\cos(\pi(\gamma-1))}, \tag{5-25}$$

and they have the required asymptotic conditions to Eq.(5-24). On superhorizon scales $k/(aH) = -k\tau \to 0$, we use again the asymptotic behavior of the Bessel functions, and one gets

$$\begin{aligned}\mathcal{A}(k,\tau) &= \frac{\sqrt{\pi}}{2^{\gamma+1/2}} \frac{e^{i\pi\gamma/2}}{\Gamma(\gamma+1/2)\cos(\pi\gamma)} k^{-1/2}(k\tau)^\gamma \\ &+ \frac{\sqrt{\pi}}{2^{-\gamma+3/2}} \frac{e^{i\pi(1-\gamma)/2}}{\Gamma(-\gamma+3/2)\cos[\pi(1-\gamma)]} k^{-1/2}(k\tau)^{1-\gamma},\end{aligned} \tag{5-26}$$

here we see that the first term dominates for $\gamma < 1/2$, while the second one dominates for $\gamma > 1/2$. Then we can rewrite the last expression as

$$\mathcal{A}(k,\tau) = \sqrt{\mathcal{F}(\delta)} k^{-1/2}(k\eta)^\delta, \tag{5-27}$$

where the dimensionless function $\mathcal{F}(\delta)$ is defined as [166]

$$\delta = \begin{cases} \gamma, & \gamma < 1/2 \\ 1-\gamma, & \gamma > 1/2 \end{cases} \quad \text{and} \quad \mathcal{F}(\delta) \equiv \frac{\pi}{2^{2\delta+1}\Gamma^2(\delta+1/2)\cos^2(\pi\delta)}. \tag{5-28}$$

Therefore, by substituting (5-27) into (5-18) the magnetic spectrum is

$$\begin{aligned}\frac{d}{d\ln k}\rho_B(\tau,k) &= \frac{k^4}{2\pi^2}\mathcal{F}(\delta)\frac{1}{a^4}\left(\frac{k}{aH}\right)^{2\delta} \\ &\sim \frac{H^4}{2\pi^2}\mathcal{F}(\delta)\left(\frac{k}{aH}\right)^{2\delta+4},\end{aligned} \tag{5-29}$$

where in the last line we use the fact that the Hubble parameter $H$ varies very slowly during inflation. We can also calculate the electric spectrum by using the identities [90]

$$J'_\nu - \frac{\nu}{x}J_\nu = -J_{\nu+1}, \quad J'_\nu + \frac{\nu}{x}J_\nu = -J_{\nu-1}, \tag{5-30}$$

and be able to solve the term $\left(\frac{A}{I}\right)'$ in Eq.(5-20). After some calculations the magnetic spectrum takes the following form

$$\begin{aligned}\frac{d}{d\ln k}\rho_E(\tau,k) &= \frac{k^4}{2\pi^2}\mathcal{G}(\xi)\frac{1}{a^4}\left(\frac{k}{aH}\right)^{2\xi} \\ &\sim \frac{H^4}{2\pi^2}\mathcal{G}(\xi)\left(\frac{k}{aH}\right)^{2\xi+4},\end{aligned} \tag{5-31}$$

where the dimensionless function $\mathcal{G}(\xi)$ is defined as

$$\xi = \begin{cases} \gamma + 1, & \gamma < -1/2 \\ -\gamma, & \gamma > -1/2 \end{cases} \quad \text{and} \quad \mathcal{G}(\xi) \equiv \frac{\pi}{2^{2\xi+3}\Gamma^2(\xi + 3/2)\cos^2(\pi\xi)}. \tag{5-32}$$

If we consider that inflationary expansion is almost de Sitter, $\beta \sim -2$, then we have $\gamma = \alpha(1+\beta) \sim -\alpha$. For a de Sitter background, we have that $aH = -1/\eta$ and we can write the spectrum of the EM fields as

$$\frac{d}{d\ln k}\rho_B(\tau, k) \sim \frac{H^4}{2\pi^2}\mathcal{F}(\delta)(-k\tau)^{2\delta+4}, \quad \frac{d}{d\ln k}\rho_E(\tau, k) \sim \frac{H^4}{2\pi^2}\mathcal{G}(\delta)(-k\tau)^{2\xi+4}. \tag{5-33}$$

One can see that scale invariant for the magnetic field occurs for $\delta = -2$, i.e., $\gamma = -2(\alpha = 2)$ and $\gamma = 3(\alpha = -3)$; while the scale invariant for the electric field happens for $\xi = -2$, i.e., $\gamma = 2(\alpha = -2)$ and $\gamma = -3(\alpha = 3)$. Although most inflamagnetogenesis models can generate the magnetic fields currently observed in large-scales as we see above, they indeed suffer from some severe problems. Basically, constraints against an excessive production of EM energy (backreaction problem) and to avoid strong coupling between fermions and EM fields (strong coupling problem) turn these scenarios into a challenging theoretical problem. Let us now discuss both requirements that have to be imposed on any magnetogenesis model to become consistent. In order to avoid an excessive production of EM energy, we calculate the energy stored in the EM field at the end of inflation $\tau_f = (a_f H)^{-1}$, here the main contribution to the energy density comes from the scales larger than $H^{-1}$ because the contribution from the subhorizon scales is renormalized in the leading order [50]. Thus, using (5-33) one can obtain the following result for the electric energy [51]

$$\rho_E = \int_{Ha_i}^{Ha_f} \frac{dk}{k} \frac{d\rho_E}{d\ln k} = \frac{H^4}{2\pi^2} \int_{Ha_i}^{Ha_f} \frac{dk}{k} \times \begin{cases} \mathcal{G}(1-\alpha)(-k\tau_f)^{6-2\alpha}, & \alpha \geq 1/2 \\ \mathcal{G}(\alpha)(-k\tau_f)^{4+2\alpha}, & \alpha \leq 1/2 \end{cases} \tag{5-34}$$

$$\rho_E \approx \frac{H^4}{2\pi^2} \begin{cases} \mathcal{G}(1-\alpha) \begin{cases} \frac{1}{6-2\alpha} \begin{cases} 1, & \alpha < 3 \\ -\exp^{(2\alpha-6)N}, & \alpha > 3 \end{cases} \\ N, & \alpha = 3 \end{cases}, & \alpha \geq 1/2 \\ \mathcal{G}(\alpha) \begin{cases} \frac{1}{4+2\alpha} \begin{cases} 1, & \alpha > -2 \\ -\exp^{-(4+2\alpha)N}, & \alpha < -2 \end{cases} \\ N, & \alpha = -2 \end{cases}, & \alpha \leq 1/2 \end{cases} \tag{5-35}$$

where $a_i$, $a_f$ refer to scale factor at the begining and the end of inflation respectively, and number of e-folds is defined as $N \equiv \ln(a_f/a_i)$. For the magnetic energy one can get the following

$$\rho_B = \int_{Ha_i}^{Ha_f} \frac{dk}{k} \frac{d\rho_B}{d\ln k} = \frac{H^4}{2\pi^2} \int_{Ha_i}^{Ha_f} \frac{dk}{k} \times \begin{cases} \mathcal{F}(\alpha)(-k\tau_f)^{4-2\alpha}, & \alpha \geq -1/2 \\ \mathcal{F}(1+\alpha)(-k\tau_f)^{6+2\alpha}, & \alpha \leq -1/2 \end{cases} \tag{5-36}$$

$$\rho_B \approx \frac{H^4}{2\pi^2} \begin{cases} \mathcal{F}(\alpha) \begin{cases} \frac{1}{4-2\alpha} \begin{cases} 1, & \alpha < 2 \\ -\exp^{(2\alpha-4)N}, & \alpha > 2 \end{cases} \\ N, & \alpha = 2 \end{cases}, & \alpha \geq -1/2 \\ \mathcal{F}(1+\alpha) \begin{cases} \frac{1}{6+2\alpha} \begin{cases} 1, & \alpha > -3 \\ -\exp^{-(6+2\alpha)N}, & \alpha < -3 \end{cases} \\ N, & \alpha = -3 \end{cases}, & \alpha \leq -1/2 \end{cases} \quad (5\text{-}37)$$

The condition that EM backreaction on inflation is negligible is that [50, 51]

$$\rho_{EM} \leq H^2, \qquad (5\text{-}38)$$

taking $H^2 \approx 10^{-12} M_p^2$ required by primordial inhomogeneities [50]. So checking eq.(5-35), we observe that for the magnetic scale invariant spectrum values, $\alpha = 2$ does not lead to any backreaction while $\alpha = -3$ does, in particular, It is viable to observe that for $\alpha > 3$ and $\alpha < -2$ the electric energy rapidly grows. By using eq.(5-38) one can find the number of e-folds of inflation for these two divergent behaviors

$$\begin{aligned} H^2 &\geq H^4 \exp^{(2\alpha-6)N} \to N \leq \frac{6 \ln 10}{\alpha - 3}, \quad \text{for: } \alpha > 3, \\ H^2 &\geq H^4 \exp^{-(2\alpha+4)N} \to N \leq \frac{-6 \ln 10}{\alpha + 2}, \quad \text{for: } \alpha < -2. \end{aligned} \qquad (5\text{-}39)$$

For $\alpha = -3$, It is seen that $N \sim 13$ which is a low value respect to the standard one of 75 e-folds, thus we obtain that the contribution of the EM energy density spoils inflation! Although we have another value $\alpha = 2$ which corresponds to a magnetic scale invariant spectrum and it does not have any backreaction problem, due to the appearance of time dependence of the coefficient in the electromagnetic tensor in the action $(I^2(\phi))$, this value suffers of the strongly coupled problem. To illustrate this a little, we write the Lagrangian density of the EM fields coupled with a charged fermion

$$\mathcal{L} = -\frac{1}{4} F_{\mu\nu} F^{\mu\nu} + i\bar{\psi}\gamma^\mu(\partial_\mu + igA_\mu)\psi, \qquad (5\text{-}40)$$

where $g$ is the coupling constant. The vector potential can be rescaled to $A_\mu \to gA_\mu$, and we rewrite the lagrangian density as

$$\mathcal{L} = -\frac{1}{4g^2} F_{\mu\nu} F^{\mu\nu} + i\bar{\psi}\gamma^\mu(\partial_\mu + iA_\mu)\psi. \qquad (5\text{-}41)$$

If we compare the last equation with eq.(5-3) for the EM fields, we see that $(I^2(\phi))$ is associated with coupling constant $g$. Hence, in the beginning of inflation $I = \left(\frac{a_f}{a_i}\right)^\alpha = \exp^{(-\alpha N)} \sim \exp^{-150}$, and due to the fact that $I \sim g^{-1}$, we have a large coupling constant and therefore, the theory is not trustable at those scales, this is the strong coupling problem [50].

## 5.2.2. Helical fields $\gamma_g \neq 0$

By substituting expression of $A_i(\tau, x)$ (5-13) into (5-7), the time-dependent Fourier amplitude now obeys the following equation of motion

$$\mathcal{A}_h''(\tau, k) + \left(k^2 - \frac{I''}{I} + 2hk\gamma_g \frac{I'}{I}\right)\mathcal{A}_h(\tau, k) = 0 \tag{5-42}$$

where $\mathcal{A}_h(\tau, k) \equiv a(\tau)\bar{A}_h(\tau, k)$. In order to satisfy (5-11), the creation and annihilation operators must obey (see Appendix B)

$$\left[b_h(k), b_{h'}^\dagger(k')\right] = \delta^3(k - k')\delta_{hh'}, \quad [b_h(k), b_{h'}(k')] = \left[b_h^\dagger(k), b_{h'}^\dagger(k')\right] = 0, \tag{5-43}$$

along with the normalization condition for the time-dependent amplitude $A_h(\tau, k)$ (B-10). Again, we substitute (5-13) into the first term of Eq.(5-9) and take the expectation value in the vacuum state $|0>$ (this time defined by the condition $b_h(k)|0> = 0, \forall k$), obtaining the energy density for the magnetic field

$$\begin{aligned}\rho_B(\tau) &= -\langle 0|T^{B0}{}_0|0\rangle = \frac{1}{2\pi^2}\int_0^{+\infty}\frac{dk}{k}k^5 I^2\left[\left|\frac{A_+(\tau, k)}{a(\tau)}\right|^2 + \left|\frac{A_-(\tau, k)}{a(\tau)}\right|^2\right] \\ &= \frac{1}{2\pi^2}\int_0^{+\infty}\frac{dk}{k}k^5\left[\left|\frac{\mathcal{A}_+(\tau, k)}{a^2(\tau)}\right|^2 + \left|\frac{\mathcal{A}_-(\tau, k)}{a^2(\tau)}\right|^2\right] \equiv \int_0^{+\infty}\frac{dk}{k}\frac{d}{d\ln k}\rho_B(\tau, k),\end{aligned} \tag{5-44}$$

where we have used the definition of energy density stored at a given scale in the last step, thus we get

$$\frac{d}{d\ln k}\rho_B(\tau, k) = \frac{1}{2\pi^2}\frac{k^5}{a^4(\tau)}\left[|\mathcal{A}_+(\tau, k)|^2 + |\mathcal{A}_-(\tau, k)|^2\right]. \tag{5-45}$$

While the energy density for the electric field reads

$$\begin{aligned}\rho_E(\tau) &= -\langle 0|T^{E0}{}_0|0\rangle = \frac{I^2}{(2\pi)^3}\int d^3k \frac{[|(aA_+(\tau, k))'|^2 + |(aA_-(\tau, k))'|^2]}{a^4(\tau)} \\ &= \frac{I^2}{2\pi^2}\int_0^{+\infty}\frac{dk}{k}\frac{k^3}{a^4(\tau)}\left[\left|\left(\frac{\mathcal{A}_+(\tau, k)}{I}\right)'\right|^2 + \left|\left(\frac{\mathcal{A}_-(\tau, k)}{I}\right)'\right|^2\right] \\ &\equiv \int_0^{+\infty}\frac{dk}{k}\frac{d}{d\ln k}\rho_E(\tau, k),\end{aligned} \tag{5-46}$$

the energy density of the electric field stored at a given scale is

$$\frac{d}{d\ln k}\rho_E(\tau, k) = \frac{I^2}{2\pi^2}\frac{k^3}{a^4(\tau)}\left[\left|\left(\frac{\mathcal{A}_+(\tau, k)}{I}\right)'\right|^2 + \left|\left(\frac{\mathcal{A}_-(\tau, k)}{I}\right)'\right|^2\right]. \tag{5-47}$$

Finally, the magnetic helicity $\eta_h \equiv B_i A^i$ coupled to $I^2$ is given by

$$\begin{aligned}\langle 0 |\eta_h| 0 \rangle &= \frac{I^2}{(2\pi)^3} \int d^3k k \frac{[|A_+(\tau,k)|^2 - |A_-(\tau,k)|^2]}{a(\tau)} \\ &= \frac{I^2}{2\pi^2} \int_0^\infty \frac{dk}{k} k^4 \frac{[|A_+(\tau,k)|^2 - |A_-(\tau,k)|^2]}{a(\tau)} \equiv \int_0^{+\infty} \frac{dk}{k} \frac{d}{d\ln k} \eta_h(\tau,k),\end{aligned} \quad (5\text{-}48)$$

and magnetic helicity per logarithmic wave numben is

$$\frac{d}{d\ln k} \eta_h(\tau,k) = \frac{1}{2\pi^2} k^4 \frac{[|\mathcal{A}_+(\tau,k)|^2 - |\mathcal{A}_-(\tau,k)|^2]}{a^3(\tau)}. \quad (5\text{-}49)$$

Using (5-21), the evolution of the mode function $\mathcal{A}_h(\tau,k)$ under a purely de-sitter expansion now looks a bit more complicated

$$\mathcal{A}_h''(\tau,k) + \left( k^2 - \frac{\alpha(\alpha+1)}{\tau^2} - \frac{2\alpha\gamma_g}{\tau} hk \right) \mathcal{A}_h(\tau,k) = 0, \quad (5\text{-}50)$$

and redefining these variables $z \equiv 2ik\tau$, $\kappa \equiv i\alpha h\gamma_g$, $\dot{X} \equiv \partial_z X$ and $\mu \equiv \alpha + 1/2$, the previous expression can be recast as

$$\ddot{\mathcal{A}}_h + \left( \frac{\kappa}{z} + \frac{\frac{1}{4} - \mu^2}{z^2} - \frac{1}{4} \right) \mathcal{A}_h = 0, \quad (5\text{-}51)$$

whose solution is given by [11]

$$\mathcal{A}_h = C_1 W_{\kappa,\mu}(z) + C_2 W_{-\kappa,\mu}(-z), \quad (5\text{-}52)$$

being $W_{\kappa,\mu}(z)$ the Whittaker functions. The asymptotic representations of these functions are [11]

$$W_{\kappa,\mu}(z) \sim \begin{cases} \frac{\Gamma(2\mu)}{\Gamma(\frac{1}{2}+\mu-\kappa)} z^{\frac{1}{2}-\mu} + \frac{\Gamma(-2\mu)}{\Gamma(\frac{1}{2}-\mu-\kappa)} z^{\frac{1}{2}+\mu}, & z \to 0 \\ \exp^{-\frac{1}{2}z} z^\kappa. & z \to \infty \end{cases}$$

Again, in order to determine the coefficients $C_1$ and $C_2$, we have to match (5-52) with the Bunch-Davies vacuum (5-24) for $-k\tau \to \infty$, as a result we see that $C_2 = 0$ and $C_1$ becomes

$$\begin{aligned} C_1 W_{\kappa,\mu}(z) &\sim C_1 \exp^{-ik\tau}(2ik\tau)^{i\alpha h\gamma_g} = C_1 \exp^{-i(k\tau - \alpha h\gamma_g \ln(2k\tau))} \exp^{-\alpha h\gamma_g \pi/2} \\ &\sim C_1 \exp^{-ik\tau} \exp^{-\alpha h\gamma_g \pi/2} \to \frac{1}{\sqrt{2k}} \exp^{-ik\tau} \Rightarrow \\ C_1 &= \frac{1}{\sqrt{2k}} \exp^{\alpha h\gamma_g \pi/2}. \end{aligned} \quad (5\text{-}53)$$

For large $|-k\tau|$, the second line in the exponential, $\alpha h\gamma_g \ln(2k\tau)$ can be neglected with respect to $|-k\tau|$. At the end of the inflation, all the modes outside the horizon will be given by

$$\mathcal{A}_h = \frac{\exp^{\alpha h\gamma_g \pi/2}}{\sqrt{2k}} \left[ \underbrace{\frac{(-2i)^{-\alpha}\Gamma(2\alpha+1)}{\Gamma(\alpha+1-ih\gamma_g\alpha)}}_{C_3}(-k\tau)^{-\alpha} + \underbrace{\frac{(-2i)^{\alpha+1}\Gamma(-2\alpha-1)}{\Gamma(-\alpha-ih\gamma_g\alpha)}}_{C_4}(-k\tau)^{\alpha+1} \right]. \quad (5\text{-}54)$$

Without loss of generality we may assume maximal helicity $|A_+| = |A|$ and $|A_-| = 0$, so the spectra energy density of the field finally read

$$\frac{d}{d\ln k}\rho_B(\tau, k) = \frac{H^4}{2\pi^2} \exp^{\alpha\gamma_g \pi} \left[ |C_3|^2 \left(\frac{k}{aH}\right)^{-2\alpha+4} + |C_4|^2 \left(\frac{k}{aH}\right)^{2\alpha+6} \right], \quad (5\text{-}55)$$

$$\frac{d}{d\ln k}\rho_E(\tau, k) = \frac{H^4}{2\pi^2} \exp^{\alpha\gamma_g \pi} \left[ |C_4|^2 (1+2\alpha)^2 \left(\frac{k}{aH}\right)^{2\alpha+4} \right], \quad (5\text{-}56)$$

$$\frac{d}{d\ln k}\eta_h(\tau, k) = \frac{H^4}{2\pi^2} \exp^{\alpha\gamma_g \pi} \left[ |C_3|^2 \left(\frac{k}{aH}\right)^{-2\alpha+3} - |C_4|^2 \left(\frac{k}{aH}\right)^{2\alpha+5} \right]. \quad (5\text{-}57)$$

Notice that, we have two values for which the magnetic field is scale invariant $\alpha = 2, -3$. For the case $\alpha = -3$, a back-reaction arises because the electric field spectrum diverges as $(k/aH)^{-2}$ in the super horizon limit. For $\alpha = 2$, we can avoid the backreaction problem but, a kind of backreaction appears $(k/aH)^{-1}$ generated by the magnetic helicity. However, in Ref. [167] they reported that this effect works as a condition to constrain the value of coupling derivative to $I' < k_{max}$, being $k_{max}$ the UV cutoff of the modes under consideration. Finally, we can find that for $\alpha = 2$, the magnetic energy spectrum increases Its value by a factor of $\sim \exp^{4\pi\gamma_g}$ when we take into account helicity in the field. While typing this Chapter, the preprint [174] appeared where the computation of the magnetic spectrum with helicity during inflation was presented. Our calculations agree with those in Ref. [174]

## 5.3. Perturbations during Inflation

Another alternative to get amplification of EM quantum fluctuations during inflation, is not to modify Maxwell electromagnetism but the background metric. In [44], Maroto et. al. considered the presence of the inhomogeneous perturbations in the metric

$$g_{\mu\nu} = g^{(FLRW)}_{\mu\nu} + h_{\mu\nu}; \quad h_{\mu\nu}dx^\mu dx^\nu = 2\Phi(\eta, \mathbf{x})a^2(\eta)(d\eta^2 + \delta_{ij}dx^i dx^j), \quad (5\text{-}58)$$

where $\Phi$ is the Bardeen potential. However, although this possibility is more natural since the metric perturbations are taken into account in almost all stages of the Universe, the model does not produce high enough magnetic seeds to explain the fields observed in clusters of galaxies. Nevertheless, for scenarios where EM perturbations are generated, It is important to study the effect of PMFs on the primordial curvature perturbations and gravitational waves and be able to constrain bounds on the primordial spectrum [44, 160, 175].

## 5.4. Cosmological Phase Transitions

In the early Universe there have been at least two phase transitions: the cosmological QCD phase transition($\sim 250$MeV) and electroweak phase transition($\sim 125$GeV) [176]. If these are first order transitions, the Universe goes through an out of equilibrium process that generates bubble nucleation. As the Universe cools below critical temperature, bubbles nucleate and grow, the walls of these bubbles collide with the others generating turbulence, then dynamo mechanism create and amplify magnetic fields from this violent process that are concentrated later in the bubble walls [70]. Calculations of generation of magnetic fields during QCD phase transitions, have been carried out by several authors [59, 177, 178]. Quashnock et.al. [177] suggested different ingredients for the generation of PMFs at the QCD phase transition such as baryon asymmetry (which implies that there are more quarks than anti-quarks producing a net positive charge and It is compensated with an excess of negative charged leptonic component), the different equation of state and energy density between quarks and leptons and finally assuming a nonrelativistic limit of both fluids (quarks and leptons). As a consequence, an electric field (at $kT_c \sim 150$MeV) is generated with a strength of the order of

$$eE \sim 15 \left(\frac{\epsilon}{10\,\%}\right) \left(\frac{\delta}{10\,\%}\right) \left(\frac{kT_c}{150\text{MeV}}\right) \left(\frac{100\text{cm}}{L}\right) \frac{\text{keV}}{\text{cm}}, \tag{5-59}$$

where $\epsilon$ represents the ratio of the energy densities of the two fluids, $\delta = \frac{L\nabla P}{P}$ is the pressure gradient and $L$ is the average distance between nucleation sites. Using Maxwell's equation, the total current is $J \approx \frac{vE}{4\pi L}$, where $v$ is the fluid velocity, and neglecting the displacement current term, the magnetic field generated in that turbulent period is

$$B_L \approx \frac{v}{c} E \approx 5G, \quad \text{for: } L \sim 100 cm. \tag{5-60}$$

On scales larger than $L$, the physical mechanisms which generate the field are uncorrelated and the magnetic field amplitude on scales $l >> L$ is given by [179]

$$B_l = B_L \left(\frac{L}{l}\right)^{3/2}. \tag{5-61}$$

Finally, the magnetic field evolves as $B \sim a^{-2}$, where $a$ is the scale factor, and the magnetic field amplitude at Recombination is $\sim 2 \times 10^{-17}$G corresponding to $5 \times 10^{10}$cm. On the other hand, for temperatures below the $T_c \sim 125$GeV, the Higgs field with non-zero expectation value nucleates from unbroken $SU(2)_L \times U(1)_Y$ to the broken $U(1)_{em}$, forming bubbles which expand, collide among themselves and generate turbulence [176]. Some authors [70, 180] have shown that magnetic field strength at the present time generated by electroweak phase transitions is

$$B_L \approx 10^{-7} - 10^{-9} G, \quad \text{for: } L \sim 5 \times 10^{11} \text{cm}, \tag{5-62}$$

$$B_L \approx 10^{-17} - 10^{-20} G, \quad \text{for:} L \sim 5 \times 10^{19} \text{cm}. \tag{5-63}$$

As we can see, PMFs generated during phase transitions lead to a coherence length of the field smaller than the Hubble scale at that epoch and weaker fields on galactic scales are obtained. However, the presence of helical fields can undergo processes of inverse cascade that transfers power from small to large scales and thus, the result will be strong fields on very large scales [59].

### 5.4.1. Harrison's Mechanism

Other alternative for the production of PMFs arises during the radiation era in regions that have non-vanishing vorticity. The first attempt at such a model was done by Harrison [181], there, magnetic fields are created through vorticity generated by the velocity difference in the fluids present. We comment briefly on the phenomenology of this model but for a formal derivation of the mechanism, see ref [66, 67, 182]. At temperatures larger than the electron mass, the interactions between protons, electrons and photons are strong and they are locked together. This means that all the system has the same angular velocity and seed fields cannot be generated. For temperatures below $T < 230$eV, electrons and photons are tightly coupled through Thomson scattering while the coupling between protons and photons is weak in this stage. Protons and electrons are still tightly coupled through Coulomb scattering and so, the photon fluid drags the protons in its motion. Therefore, the difference of mass between electrons and protons will lead to non-zero electron and proton fluid angular velocities that give rise to currents and magnetic fields. Matarrese et. al [66], found that for comoving scales of $\sim 1$Mpc, the amplitude of PMFs generated via this mechanism today is given by

$$B \sim 10^{-29} G. \tag{5-64}$$

This value of the magnetic field generated by the differential rotational velocity of charged particles, is much smaller than those signals observed in clusters of galaxies. Now, if an initial vorticity is present during this epoch, magnetic fields may serve as seed for explaining the galactic fields, however, in the early Universe the vorticity decays rapidly due to expansion of the Universe and therefore, this mechanism cannot work efficiently [67].

# 6. Primordial magnetic fields and the CMB

Since PMFs affect the evolution of cosmological perturbations, these fields might leave significant signals on the CMB. Basically, PMFs add three contributions to the temperature and polarization of the CMB spectra: the scalar vector and tensor, which have been deeply studied [183–186]. For the scalar contribution, the shape of the temperature anisotropy (TT mode) presents an increase on large scales and It also shifts the acoustic peaks via fast magnetosonic waves, nevertheless the main effect of the scalar mode lies on large multipolar numbers, since the primary CMB is significanly suppressed by the Silk damping in these scales [184, 187]. Next, the vector contribution leaves an indistinguishable signal, because in stardard cosmology, vector contributions decay with time and do not affect the CMB anisotropies considerably [184]. Further, vector mode peaks where primary CMB is suppressed by Silk damping and so dominates over the scalar ones in small scales [188]. Vector modes are also very interesting in the polarization spectra, in particular, they induce B modes with amplitudes slightly larger than any other contribution, allowing us to constrain better PMFs in the next CMB polarization experiments [84].

Finally, tensor modes induce gravitational wave perturbations that lead to CMB temperature and polarization anisotropies on large angular scales, and the passive tensor modes (produced by the presence of PMFs before neutrino decoupling) generate the most significant magnetic contributions, so those modes become relevant to study the nature of PMFs [85, 86, 189]. Moreover, if helical PMFs are presented before Recombination, they affect drastically the parity-odd CMB cross correlations implying a strong feature of parity violation in the early Universe [190, 191]. This section will be dedicated to how to consider the effects of the PMFs on the CMB and we will introduce the PMF power spectrum concept. This chapter is based on the work published in [1, 4, 5].

## 6.1. Magnetic spectra and correlation functions

Two models have been proposed to model PMFs. The first one consists in describing PMFs as a homogeneous field such that $B^2$ is the local density of the field and where we must require an anisotropic background (like, Bianchi VII) to allow the presence of this field. Comparing those models with CMB quadrupole data, Barrow et. al, [192] reported an amplitude of PMFs of $B < 6,8 \times 10^{-9}(\Omega_m h^2)^{1/2}$G, there they used the most general flat and open anisotropic cosmologies containing expansion rate and 3-curvature anisotropies. However, they found that PMFs amplitude constraints are stronger than those imposed by nucleosynthesis and therefore, this description hardly agrees with other cosmological probes. On the other hand, PMFs can also be described by a stochastic test field where $B^2$ would be related to the average density of the field instead. This description does not break neither isotropy nor homogeneity of the background Universe, hence, this scheme allows to have a PMF model concordant to the current constraints. In consequence, we will consider a stochastic primordial magnetic field (PMF) generated in the very early Universe which could have been produced during inflation (non-causal field) or after inflation (causal field) throughout the thesis. The PMF power spectrum which is defined as the Fourier transform of the two point correlation can be written as

$$\langle B_l(\mathbf{k}) B_m^*(\mathbf{k}')\rangle = (2\pi)^3 \delta^3(\mathbf{k}-\mathbf{k}')\left(P_{lm}(k)P_B(k) + i\epsilon_{lmn}\hat{k}^n P_H(k)\right), \tag{6-1}$$

where $P_{lm}(k) = \delta_{lm} - \hat{k}_l\hat{k}_m$ is a projector onto the transverse plane[1], $\epsilon_{lmn}$ is the 3D Levi-Civita tensor and, $P_B(k)$, $P_H(k)$ are the symmetric/anti-symmetric parts of the power spectrum and represent the magnetic field energy density and absolute value of the kinetic helicity respectively [193]

$$\langle B_i(\mathbf{k}) B_i^*(\mathbf{k}')\rangle = 2(2\pi)^3 \delta^3(\mathbf{k}-\mathbf{k}') P_B(k), \tag{6-2}$$

$$-i\langle \epsilon_{ijl}\hat{k}^l B_i(\mathbf{k}) B_j^*(\mathbf{k}')\rangle = 2(2\pi)^3 \delta^3(\mathbf{k}-\mathbf{k}') P_H(k). \tag{6-3}$$

We assume that power spectrum scales as a simple power law

$$P_B(k) = A_B k^{n_B}, \quad P_H(k) = A_H k^{n_H}. \tag{6-4}$$

and we usually parametrize the fields through a convolution with a 3D-Gaussian window function smoothed over a sphere of comoving radius $\lambda$, $B_i(k) \to B_i(k) \times f(k)$, with $f(k) = e^{(-\lambda^2 k^2/2)}$ [184]. We also define $B_\lambda$ as the comoving PMF strength scaled to the present day on $\lambda$

$$\langle B^i(\mathbf{x}) B_i(\mathbf{x})\rangle |_\lambda \equiv B_\lambda^2 = \frac{1}{(2\pi)^6}\int\int d^3k d^3k' e^{(-i\mathbf{x}\cdot\mathbf{k}+i\mathbf{x}\cdot\mathbf{k}')}\langle B^i(\mathbf{k}) B_i^*(\mathbf{k}')\rangle |f(k)|^2,$$

$$= \frac{A_B}{(2\pi)^2}\frac{2}{\lambda^{n_B+3}}\Gamma\left(\frac{n_B+3}{2}\right), \tag{6-5}$$

and we define $\mathcal{B}_\lambda$ as the comoving kinetic helical PMF strength scaled to the present day on $\lambda$

$$\langle (\nabla\times\mathbf{B}(\mathbf{x}))_i B^i(\mathbf{x})\rangle |_\lambda \equiv \mathcal{B}_\lambda^2$$

$$= \frac{i\epsilon_{ilj}}{(2\pi)^6}\int\int d^3k d^3k' e^{(-i\mathbf{x}\cdot\mathbf{k}+i\mathbf{x}\cdot\mathbf{k}')}\langle k^j B^l(\mathbf{k}) B_i^*(\mathbf{k}')\rangle |f(k)|^2,$$

$$= \frac{|A_H|}{(2\pi)^2}\frac{2}{\lambda^{n_H+4}}\Gamma\left(\frac{n_H+4}{2}\right), \tag{6-6}$$

---

[1] This projector has the property $P_{lm}\hat{k}^m = 0$ with $\hat{k} = \frac{\mathbf{k}}{k}$.

with $\Gamma$ being the Gamma function. We then obtain the amplitudes as follows

$$A_B = \frac{B_\lambda^2 2\pi^2 \lambda^{n_B+3}}{\Gamma(\frac{n_B+3}{2})}, \quad A_H = \frac{H_\lambda^2 2\pi^2 \lambda^{n_H+3}}{\Gamma(\frac{n_H+4}{2})}, \quad \text{with} \quad n_B > -3, n_H > -4. \tag{6-7}$$

The more general case of the power spectrum for magnetic fields can be studied, if we assume that It is non-zero for $k_m \leq k \leq k_D$, being $k_m$ an infrared cut-off and $k_D$ an ultraviolet cut-off corresponding to damping scale of the field which can be written as [184]

$$k_D \approx \left(1{,}7 \times 10^2\right)^{\frac{2}{n+5}} \left(\frac{B_\lambda}{10^{-9}nG}\right)^{\frac{-2}{n+5}} \left(\frac{k_\lambda}{1 Mpc^{-1}}\right)^{\frac{n+3}{n+5}} h^{\frac{1}{n+5}} \frac{1}{Mpc}. \tag{6-8}$$

Hereafter we simply set this scale at $k_D \sim \mathcal{O}(10) \text{Mpc}^{-1}$ [184]. Given the Schwarz inequality [194],

$$\lim_{k' \to k} \langle \mathbf{B}(k) \cdot \mathbf{B}^*(k') \rangle \geq |\lim_{k' \to k} \langle (\hat{\mathbf{k}} \times \mathbf{B}(k)) \cdot \mathbf{B}^*(k') \rangle|, \tag{6-9}$$

an additional constraint is found for these fields

$$|A_H| \leq A_B k^{n_B - n_H}. \tag{6-10}$$

In the case where $A_H = A_B$ and $n_B = n_H$ we define the maximal helicity condition. We will also parametrize the infrared cut-off by a single constant parameter $\alpha$,

$$k_m = \alpha k_D, \quad 0 \leq \alpha < 1 \tag{6-11}$$

which in the case of inflationary scenarios would correspond to the wave mode that exits the horizon at inflation epoch and for causal modes would be important when this scale is larger than the wavenumber of interest (as claimed by Kim et.al. [195]). Thus, this infrared cut-off would be important in order to constrain PMF parameters and magnetogenesis models [195–198]. Equation (6-11) gives only an useful mathematical representation to constrain these cut-off values via cosmological datasets (for this case, the parameter space would be given by $(\alpha, k_D, B_\lambda, H_\lambda, n_H, n_B)$), and therefore we want to point out that latter expression does not state any physical relation between both wave numbers. In [196, 197], they showed constraints on the maximum wave number $k_D$ as a function of $n_B$ via big bang nucleosynthesis (BBN), and they considered the maximum and minimum wave numbers as independent parameters. In fact, we have found out that the integration scheme used for calculating the spectrum and bispectrum of PMFs is exactly the same if we parametrize $k_m$ as seen in (6-11), or if we consider $(k_m, k_D, B_\lambda, H_\lambda, n_H, n_B)$ as independent parameters.

Thus the inclusion of $k_m$ is done only for studying at a phenomenological level, and Its effects on the CMB bispectrum in the next section. At background level, we need only the energy density of the PMF which is given by $\rho_B = \langle B^2(\mathbf{x}) \rangle / (8\pi)$, therefore, by using Eqs.(6-4) and (6-5) we get (for the spatial dependence)

$$\rho_B = \frac{\langle B^2(\mathbf{x}) \rangle}{8\pi} = \frac{2}{8\pi} \int_{k_m}^{k_D} d^3k P_B(k) = \frac{\lambda^{n_B+3}}{8\pi} \frac{B_\lambda^2}{\Gamma\left(\frac{n_B+5}{2}\right)} \left[k_D^{n_B+3} - k_m^{n_B+3}\right], \tag{6-12}$$

where only the non-helical term contributes to the energy density of the PMF in the Universe. In Ref. [196] is also reported this equation, and we will study in more detail their effects on the CMB later. In order to study the impact of PMFs on cosmological perturbations, we start writing the energy momentum tensor (EMT) for the magnetic fields

$$T^0_0 = \frac{-1}{8\pi a^4}|B(\mathbf{x})|^2, \quad T^i_j = \frac{1}{4\pi a^4}\left(\delta^i_j\frac{|B(\mathbf{x})|^2}{2} - B_j(\mathbf{x})B^i(\mathbf{x})\right), \quad T^0_i = 0, \tag{6-13}$$

where, we can see that EMT of PMFs is quadratic in the fields [82]. Due to the high conductivity in the primordial Universe, the electric field is suppressed and the magnetic field is frozen into the plasma, therefore we have that $B_i(\mathbf{x}, \tau) = B_i(\mathbf{x})a^{-2}(\tau)$. Then, the spatial part of magnetic field EMT in Fourier space is given by

$$T^i_j(\mathbf{k}, \tau) = \frac{-1}{32\pi^4 a^4}\int d^3k' \left[B^i(\mathbf{k}')B_j(\mathbf{k}-\mathbf{k}') - \frac{1}{2}\delta^i_j B^l(\mathbf{k}')B_l(\mathbf{k}-\mathbf{k}')\right], \tag{6-14}$$

and the two-point correlation tensor related to the spatial dependence (6-14) gives

$$\begin{aligned}\langle T_{ij}(\mathbf{k})T^*_{lm}(\mathbf{p})\rangle &= \frac{1}{1024\pi^8}\int\int d^3p' d^3k' \langle B_i(\mathbf{k}')B_j(\mathbf{k}-\mathbf{k}')B^*_l(\mathbf{p}')B^*_m(\mathbf{p}-\mathbf{p}')\rangle \\ &+ \ldots\langle\ldots\rangle_{lm}\delta_{ij} + \ldots\langle\ldots\rangle_{ij}\delta_{lm} + \ldots\delta_{ij}\delta_{lm},\end{aligned} \tag{6-15}$$

where we can apply the Wick theorem Eq.(4-55), because the stochastic fields are Gaussianly distributed

$$\begin{aligned}\langle B_i(\mathbf{k}')B_j(\mathbf{k}-\mathbf{k}')B^*_l(\mathbf{p}')B^*_m(\mathbf{p}-\mathbf{p}')\rangle &= \langle B_i(\mathbf{k}')B_j(\mathbf{k}-\mathbf{k}')\rangle\langle B^*_l(\mathbf{p}')B^*_m(\mathbf{p}-\mathbf{p}')\rangle \\ &+ \langle B_i(\mathbf{k}')B^*_l(\mathbf{p}')\rangle\langle B_j(\mathbf{k}-\mathbf{k}')B^*_m(\mathbf{p}-\mathbf{p}')\rangle \\ &+ \langle B_i(\mathbf{k}')B^*_m(\mathbf{p}-\mathbf{p}')\rangle\langle B^*_l(\mathbf{p}')B_j(\mathbf{k}-\mathbf{k}')\rangle.\end{aligned} \tag{6-16}$$

On the other hand, the equations for the adimensional energy density of magnetic field and spatial part of the electromagnetic energy momentum tensor respectively written in Fourier space are given as

$$\rho_B(\mathbf{k}) \equiv \frac{1}{8\pi\rho_{\gamma,0}}\int \frac{d^3p}{(2\pi)^3}B_l(\mathbf{p})B^l(\mathbf{k}-\mathbf{p}),$$

$$\Pi_{ij}(\mathbf{k}) \equiv \frac{1}{4\pi\rho_{\gamma,0}}\int \frac{d^3p}{(2\pi)^3}\left[\frac{\delta_{ij}}{2}B_l(\mathbf{p})B^l(\mathbf{k}-\mathbf{p}) - B_i(\mathbf{p})B_j(\mathbf{k}-\mathbf{p})\right], \tag{6-17}$$

where we express each component of the energy momentum tensor in terms of photon energy density $\rho_\gamma = \rho_{\gamma,0}a^{-4}$, with $\rho_{\gamma,0}$ being its present value.[2] We can also see that using the previous definition, the EMT can be written as $T^i_j(\mathbf{k}, \tau) \equiv \rho_\gamma(\tau)\Pi^i_j(\mathbf{k})$. Since the spatial

---

[2] The adimensional energy density of magnetic field showed here is written with different notation in [199]: $\Omega_B \equiv \frac{B^2}{8\pi a^4 \rho_\gamma}$ and in [9, 200]: $\Delta_B \equiv \frac{B^2}{8\pi a^4 \rho_\gamma}$.

EMT is symmetric, we can decompose this tensor into two scalars ($\rho_B$, $\Pi^{(S)}$), one vector ($\Pi_i^{(V)}$) and one tensor ($\Pi_{ij}^{(T)}$) components

$$\Pi_{ij} = \frac{1}{3}\delta_{ij}\rho_B + (\hat{k}_i\hat{k}_j - \frac{1}{3}\delta_{ij})\Pi^{(S)} + (\hat{k}_i\Pi_j^{(V)} + \hat{k}_j\Pi_i^{(V)}) + \Pi_{ij}^{(T)} \tag{6-18}$$

which obey to $\hat{k}^i\Pi_i^{(V)} = \hat{k}^i\Pi_{ij}^{(T)} = \Pi_{ii}^{(T)} = 0$ [201, 202]. The components of this tensor are recovered by applying projector operators defined as

$$\begin{aligned}
\rho_B &= \delta^{ij}\Pi_{ij} \\
\Pi^{(S)} &= (\delta^{ij} - \frac{3}{2}P^{ij})\Pi_{ij} = \mathcal{P}^{ij}\Pi_{ij} \\
\Pi_i^{(V)} &= \hat{k}^{(j}P_i^{l)}\Pi_{lj} = \mathcal{Q}_i^{jl}\Pi_{lj} \\
\Pi_{ij}^{(T)} &= (P_i^{(a}P_j^{b)} - \frac{1}{2}P^{ab}P_{ij})\Pi_{ab} = \mathcal{P}_{ij}^{ab}\Pi_{ab},
\end{aligned} \tag{6-19}$$

where (..) in the indices denotes symmetrization [203]. The two-point correlation tensor related to Eq.(6-17) is

$$\begin{aligned}
\langle \Pi_{ij}(\mathbf{k})\Pi_{lm}^*(\mathbf{p}) \rangle &= \frac{1}{(4\pi\rho_{\gamma,0})^2}\delta^{(3)}(\mathbf{k}-\mathbf{p}) \int d^3k' \Big[ (P_B(k')P_B(|\mathbf{k}-\mathbf{k}'|)P_{il}(k')P_{jm}(|\mathbf{k}-\mathbf{k}'|) \\
&\quad - P_H(k')P_H(|\mathbf{k}-\mathbf{k}'|)\epsilon_{ilt}\epsilon_{jms}\hat{k}'_t\widehat{(\mathbf{k}-\mathbf{k}')}_s \\
&\quad + iP_B(k')P_H(|\mathbf{k}-\mathbf{k}'|)P_{il}(k')\epsilon_{jmt}\widehat{(\mathbf{k}-\mathbf{k}')}_t \\
&\quad + iP_B(k')P_H(|\mathbf{k}-\mathbf{k}'|)\epsilon_{ilt}P_{jm}(|\mathbf{k}-\mathbf{k}'|)\hat{k}'_t + (l \leftrightarrow m) \Big) \\
&\quad + \ldots\delta_{ij} + \ldots\delta_{lm} + \ldots\delta_{ij}\delta_{lm} \Big],
\end{aligned} \tag{6-20}$$

where we use Eqs.(6-1),(6-16). To determine the effect on cosmic perturbations It is necessary to compute the scalar, vector and tensor correlation functions of PMFs using the projector operators

$$\begin{aligned}
\langle \rho_B(\mathbf{k})\rho_B^*(\mathbf{k}') \rangle &= \delta^{ij}\delta^{lm}\langle \Pi_{ij}(\mathbf{k})\Pi_{lm}^*(\mathbf{k}') \rangle, \\
\langle \Pi^{(S)}(\mathbf{k})\Pi^{(S)*}(\mathbf{k}') \rangle &= \mathcal{P}^{ij}\mathcal{P}^{lm}\langle \Pi_{ij}(\mathbf{k})\Pi_{lm}^*(\mathbf{k}') \rangle, \\
\langle \Pi_i^{(V)}(\mathbf{k})\Pi_j^{(V)*}(\mathbf{k}') \rangle &= \mathcal{Q}_i^{lm}\mathcal{Q}_j^{ts}\langle \Pi_{ml}(\mathbf{k})\Pi_{st}^*(\mathbf{k}') \rangle, \\
\langle \Pi_{ij}^{(T)}(\mathbf{k})\Pi_{lm}^{(T)*}(\mathbf{k}') \rangle &= \mathcal{P}_{ij}^{ab}\mathcal{P}_{lm}^{cd}\langle \Pi_{ab}(\mathbf{k})\Pi_{cd}^*(\mathbf{k}') \rangle,
\end{aligned} \tag{6-21}$$

These convolutions can be written in terms of spectra as follows [193]

$$\langle \rho_B(\mathbf{k})\rho_B^*(\mathbf{k}') \rangle = (2\pi)^3 \left| \rho(k) \right|^2 \delta^{(3)}(\mathbf{k}-\mathbf{k}'), \tag{6-22}$$

$$\langle \Pi^{(S)}(\mathbf{k})\Pi^{(S)*}(\mathbf{k}') \rangle = (2\pi)^3 \left| \Pi^{(S)}(k) \right|^2 \delta^{(3)}(\mathbf{k}-\mathbf{k}'), \tag{6-23}$$

$$\langle \Pi_i^{(V)}(\mathbf{k})\Pi_j^{(V)*}(\mathbf{k}')\rangle = (2\pi)^3\delta^{(3)}(\mathbf{k}-\mathbf{k}')\left[P_{ij}\left|\Pi^{(V)}(k)\right|_S^2 + i\epsilon_{ijl}\hat{k}^l\left|\Pi^{(V)}(k)\right|_A^2\right], \tag{6-24}$$

$$\langle \Pi_{ij}^{(T)}(\mathbf{k})\Pi_{lm}^{(T)*}(\mathbf{k}')\rangle = (2\pi)^3\delta^{(3)}(\mathbf{k}-\mathbf{k}')\left[\mathcal{M}_{ijlm}\left|\Pi^{(T)}(k)\right|_S^2 + i\mathcal{A}_{ijlm}\left|\Pi^{(T)}(k)\right|_A^2\right], \tag{6-25}$$

being tensors $\mathcal{M}_{ijlm}$ and $\mathcal{A}_{ijlm}$ given by [189, 193]

$$\mathcal{M}_{ijlm} \equiv P_{il}P_{jm} + P_{im}P_{jl} - P_{ij}P_{lm}, \tag{6-26}$$

$$\mathcal{A}_{ijlm} \equiv \frac{\hat{k}_a}{2}\left(P_{il}\epsilon_{jma} + P_{im}\epsilon_{jla} + P_{jl}\epsilon_{ima} + P_{jm}\epsilon_{ila}\right). \tag{6-27}$$

We can summarize some properties for the previous tensors

$$\mathcal{M}_{ijlm}\mathcal{M}^{ijlm} = \mathcal{A}_{ijlm}\mathcal{A}^{ijlm} = 8; \quad \mathcal{M}_{ijij} = 4; \quad \mathcal{A}_{ijij} = 0; \quad P_{ij}P^{ij} = \epsilon_{ijl}\hat{k}^l\epsilon_{ijs}\hat{k}^s = 2;$$

$$\mathcal{M}_{ijlm}\mathcal{A}^{ijlm} = 0; \quad P_{ij}\epsilon^{ijl}\hat{k}_l = 0; \quad \mathcal{M}_{ijll} = \mathcal{A}_{ijll} = \mathcal{M}_{llij} = \mathcal{A}_{llij} = 0; \quad \delta_{ij}\delta^{ij} = 3. \tag{6-28}$$

Thus, using Eqs.(6-20)-(6-25), along with the Wick's theorem (6-16), the spectra take the form

$$|\rho(k)|^2 = \frac{1}{8(2\pi)^5\rho_{\gamma,0}^2}\int d^3p'\bigg((1+\mu^2)P_B(p')P_B(|\mathbf{k}-\mathbf{p}'|)$$
$$- 2\mu P_H(p')P_H(|\mathbf{k}-\mathbf{p}'|)\bigg), \tag{6-29}$$

$$\left|\Pi^{(S)}(k)\right|^2 = \frac{1}{8(2\pi)^5\rho_{\gamma,0}^2}\int d^3p'\bigg([4-3\gamma^2+\beta^2(-3+9\gamma^2)-6\beta\gamma\mu+\mu^2]P_B(p')P_B(|\mathbf{k}-\mathbf{p}'|)$$
$$- (6\beta\gamma-4\mu)P_H(p')P_H(|\mathbf{k}-\mathbf{p}'|)\bigg), \tag{6-30}$$

$$\left|\Pi^{(V)}(k)\right|_S^2 = \frac{1}{4(2\pi)^5\rho_{\gamma,0}^2}\int d^3p'\bigg([1-2\gamma^2\beta^2+\mu\gamma\beta]P_B(p')P_B(|\mathbf{k}-\mathbf{p}'|)$$
$$- (\gamma\beta-\mu)P_H(p')P_H(|\mathbf{k}-\mathbf{p}'|)\bigg), \tag{6-31}$$

$$\left|\Pi^{(V)}(k)\right|_A^2 = \frac{1}{8(2\pi)^5\rho_{\gamma,0}^2}\int d^3p'\bigg([\beta+\gamma\mu-2\gamma^2\beta]P_B(p')P_H(|\mathbf{k}-\mathbf{p}'|)$$
$$+ (\gamma+\beta\mu-2\beta^2\gamma)P_H(p')P_B(|\mathbf{k}-\mathbf{p}'|)\bigg), \tag{6-32}$$

$$\left|\Pi^{(T)}(k)\right|_S^2 = \frac{1}{16(2\pi)^5\rho_{\gamma,0}^2}\int d^3p'\bigg([(1+\gamma^2)(1+\beta^2)]P_B(p')P_B(|\mathbf{k}-\mathbf{p}'|)$$
$$+ 4\gamma\beta P_H(p')P_H(|\mathbf{k}-\mathbf{p}'|)\bigg), \tag{6-33}$$

$$\left|\Pi^{(T)}(k)\right|_A^2 = \frac{1}{8(2\pi)^5\rho_{\gamma,0}^2}\int d^3p'\bigg((\beta+\gamma^2\beta)P_B(p')P_H(|\mathbf{k}-\mathbf{p}'|)$$
$$+ (\gamma+\beta^2\gamma)P_H(p')P_B(|\mathbf{k}-\mathbf{p}'|)\bigg), \tag{6-34}$$

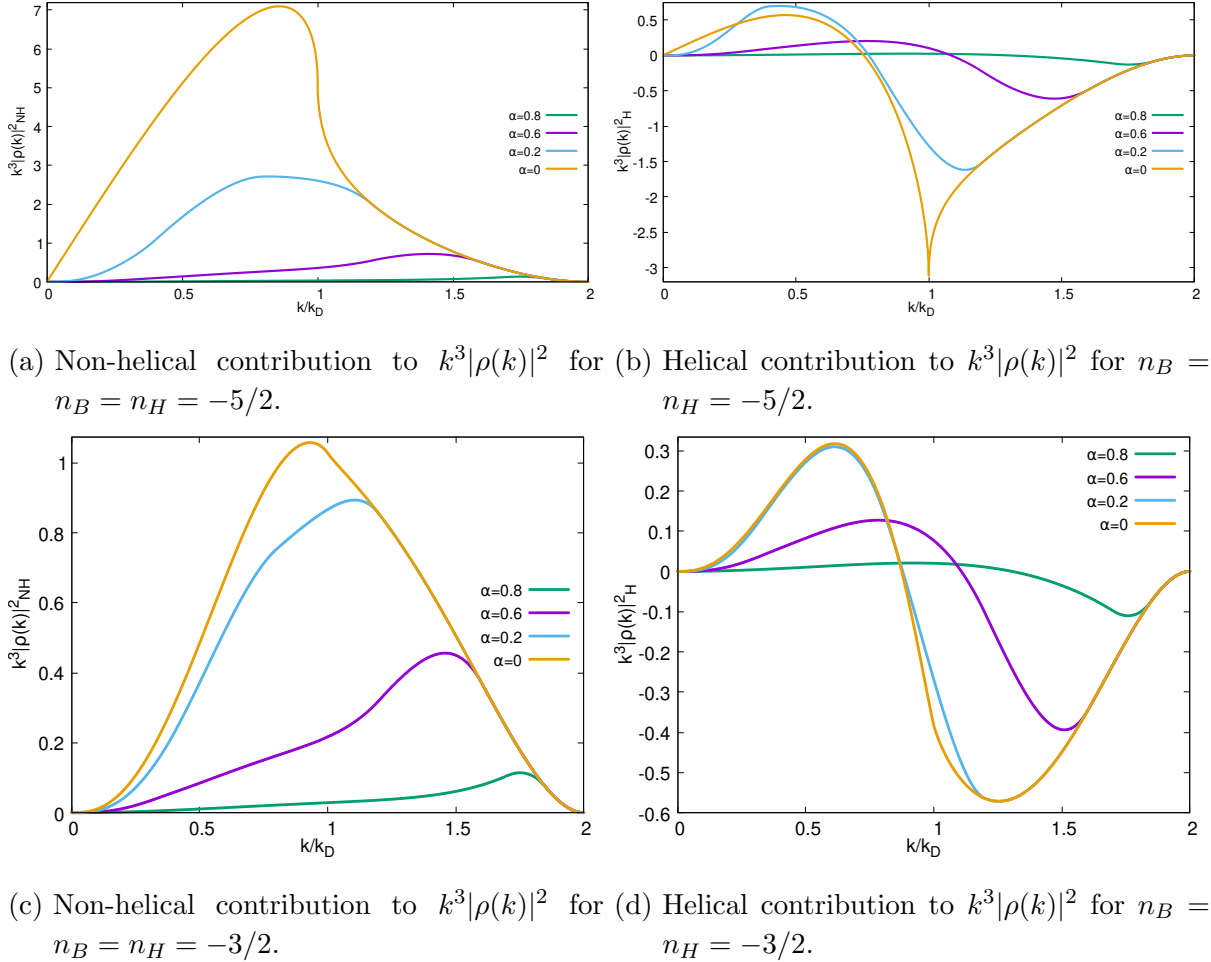

(a) Non-helical contribution to $k^3|\rho(k)|^2$ for $n_B = n_H = -5/2$.

(b) Helical contribution to $k^3|\rho(k)|^2$ for $n_B = n_H = -5/2$.

(c) Non-helical contribution to $k^3|\rho(k)|^2$ for $n_B = n_H = -3/2$.

(d) Helical contribution to $k^3|\rho(k)|^2$ for $n_B = n_H = -3/2$.

**Figure 6-1**.: (Non)-Helical contribution to $k^3|\rho(k)|^2$ for different spectral indices in units of $A^2_{(B),H}/(8(2\pi)^5\rho^2_{\gamma,0})$ versus $k/k_D$. Here we show the effect of an IR cut-off parametrized with $\alpha$ on the magnetic power spectrum.

where properties in Eqs.(6-28) It was necessary to find the symmetric (S)/anti-symmetric(A) terms and, the angular functions are defined as

$$\beta = \frac{\mathbf{k}\cdot(\mathbf{k}-\mathbf{k}')}{k\,|\mathbf{k}-\mathbf{k}'|}, \quad \mu = \frac{\mathbf{k}'\cdot(\mathbf{k}-\mathbf{k}')}{k'\,|\mathbf{k}-\mathbf{k}'|}, \quad \gamma = \frac{\mathbf{k}\cdot\mathbf{k}'}{kk'}. \tag{6-35}$$

The above relations and properties were obtained using the xAct software [204] and they agree with those reported in [193, 203]. Given these results, we are able to analyse the effects of PMFs on CMB by adding the previous contributions to the CMB angular power spectrum. Indeed, some authors [82–84,191,205] have added the above spectrum relations in Boltzmann codes like CAMB [206] or CMBeasy [207], while other authors [88,183,184,187] have analysed the effects of these fields through approximate solutions.

Using the integration scheme for the Fourier spectra reported in Sec.4 we obtain the solution for the magnetic spectra for different contributions. In Figs.**6-1**-**6-2** we show the total

```
```
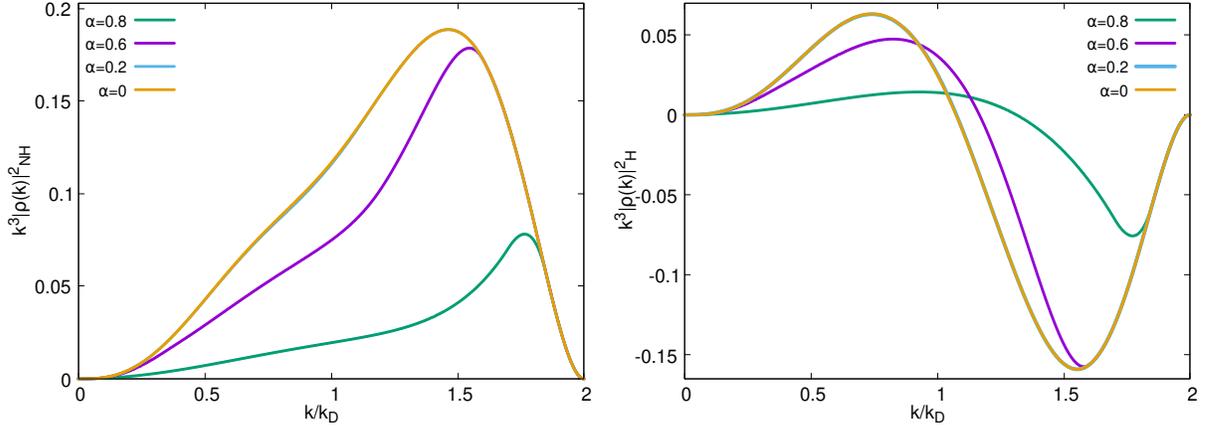

(a) Non-helical contribution to $k^3|\rho(k)|^2$ for $n_B = n_H = 1$.

(b) Helical contribution to $k^3|\rho(k)|^2$ for $n_B = n_H = 1$.

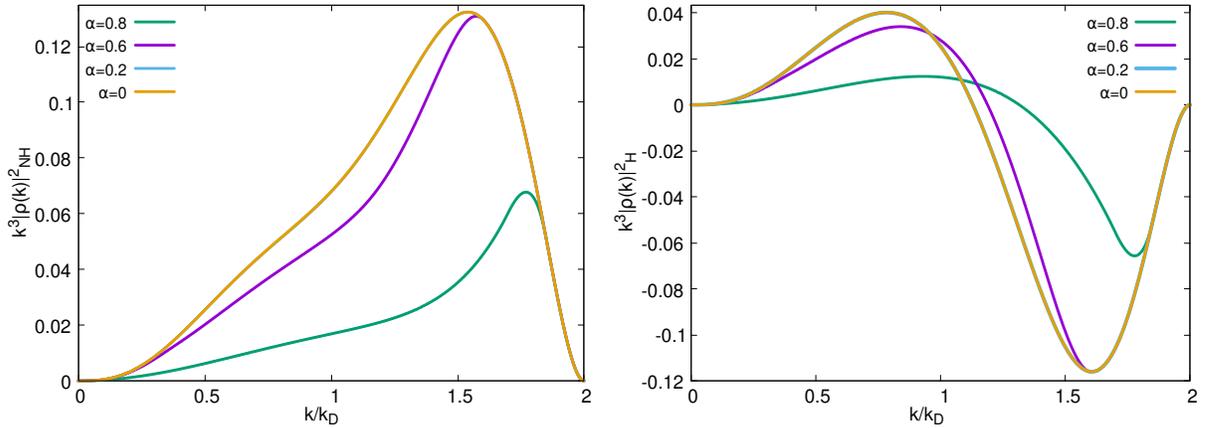

(c) Non-helical contribution to $k^3|\rho(k)|^2$ for $n_B = n_H = 2$.

(d) Helical contribution to $k^3|\rho(k)|^2$ for $n_B = n_H = 2$.

**Figure 6-2**.: (Non)-Helical contribution to $k^3|\rho(k)|^2$ for different spectral indices in units of $A^2_{(B),H}/(8(2\pi)^5\rho^2_{\gamma,0})$ versus $k/k_D$. Here we show the effect of an IR cut-off parametrized with $\alpha$ on the magnetic power spectrum.

6.1 Magnetic spectra and correlation functions     79

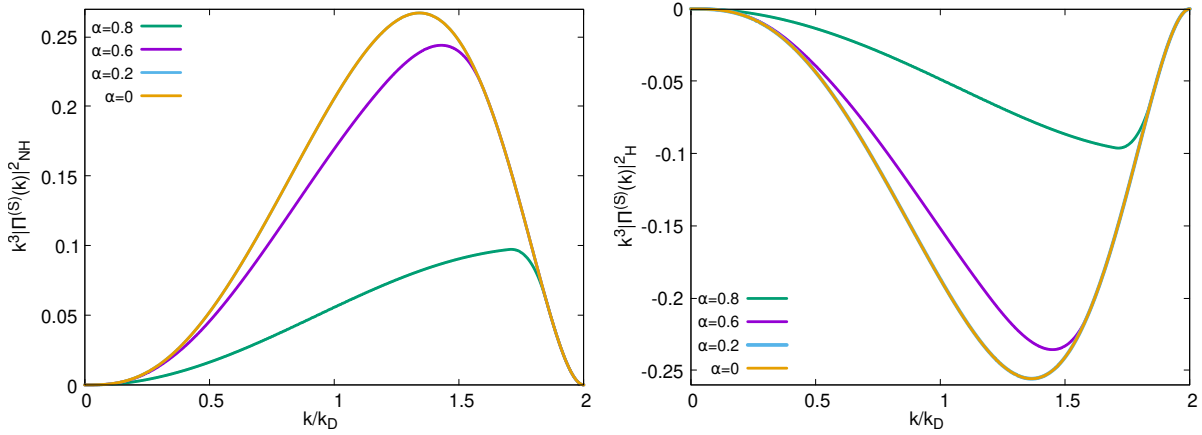

(a) Non-helical contribution to $k^3|\Pi^{(S)}(k)|^2$ for $n_B = n_H = 2$.

(b) Helical contribution to $k^3|\Pi^{(S)}(k)|^2$ for $n_B = n_H = 2$.

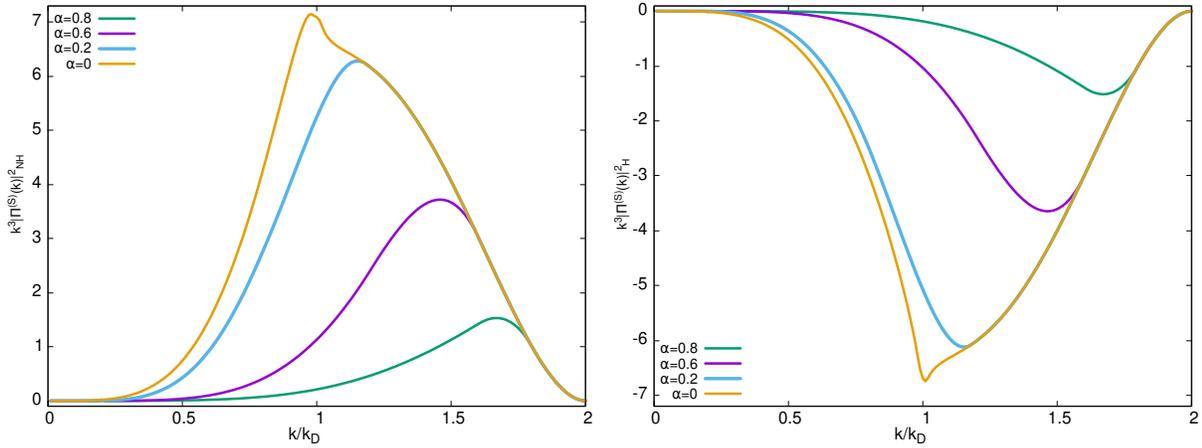

(c) Non-helical contribution to $k^3|\Pi^{(S)}(k)|^2$ for $n_B = n_H = -5/2$.

(d) Helical contribution to $k^3|\Pi^{(S)}(k)|^2$ for $n_B = n_H = -5/2$.

**Figure 6-3.**: (Non)-Helical contribution to $k^3|\Pi(k)^{(S)}|^2$ for different spectral indices in units of $A^2_{(B),H}/(8(2\pi)^5\rho^2_{\gamma,0})$ versus $k/k_D$. Here we show the effect of an IR cut-off parametrized with $\alpha$ on the magnetic power spectrum.

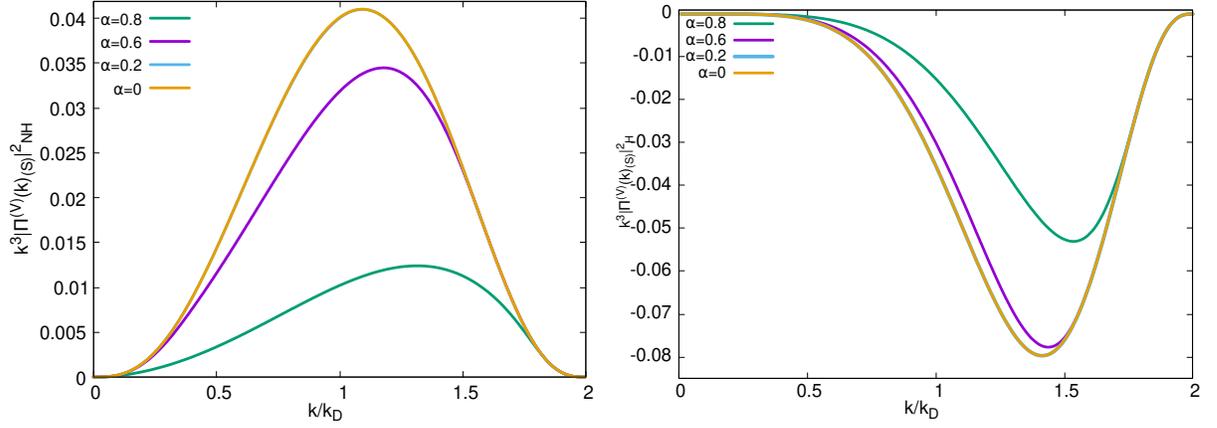

(a) Non-helical contribution to $k^3|\Pi^{(V)}(k)|^2_S$ for $n_B = n_H = 2$.

(b) Helical contribution to $k^3|\Pi^{(V)}(k)|^2_S$ for $n_B = n_H = 2$.

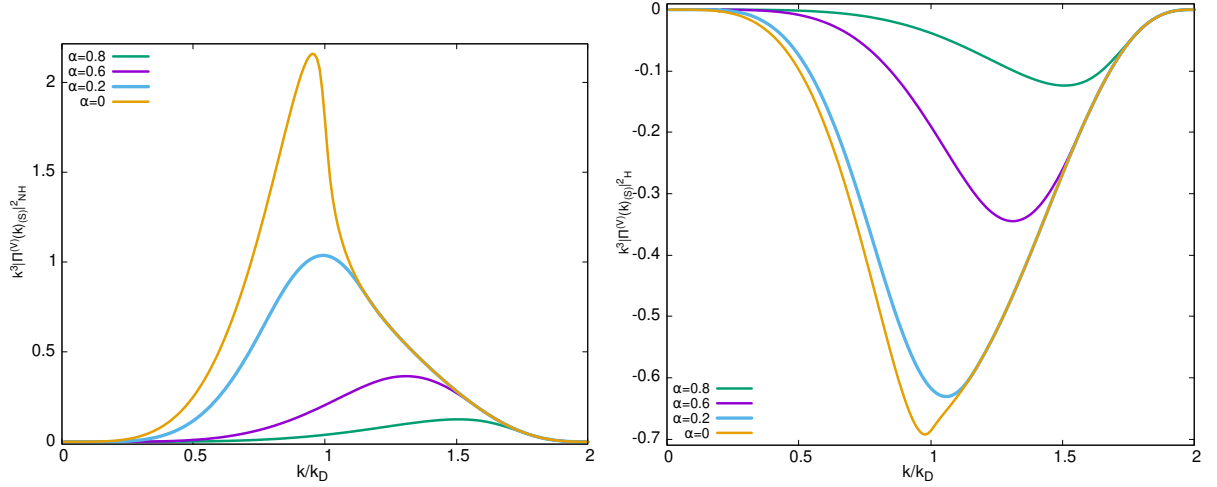

(c) Non-helical contribution to $k^3|\Pi^{(V)}(k)|^2_S$ for $n_B = n_H = -5/2$.

(d) Helical contribution to $k^3|\Pi^{(V)}(k)|^2_S$ for $n_B = n_H = -5/2$.

**Figure 6-4**.: (Non)-Helical contribution to $k^3|\Pi^{(V)}(k)|^2_S$ for different spectral indices in units of $A^2_{(B),H}/(4(2\pi)^5\rho^2_{\gamma,0})$ versus $k/k_D$. Here we show the effect of an IR cut-off parametrized with $\alpha$ on the magnetic power spectrum.

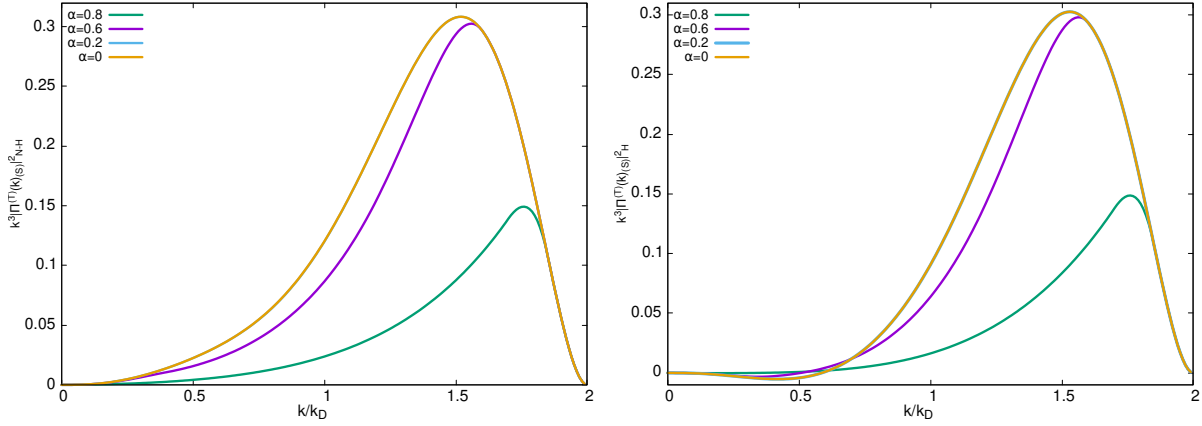

(a) Non-helical contribution to $k^3|\Pi^{(T)}(k)|^2_S$ for $n_B = n_H = 2$.

(b) Helical contribution to $k^3|\Pi^{(T)}(k)|^2_S$ for $n_B = n_H = 2$.

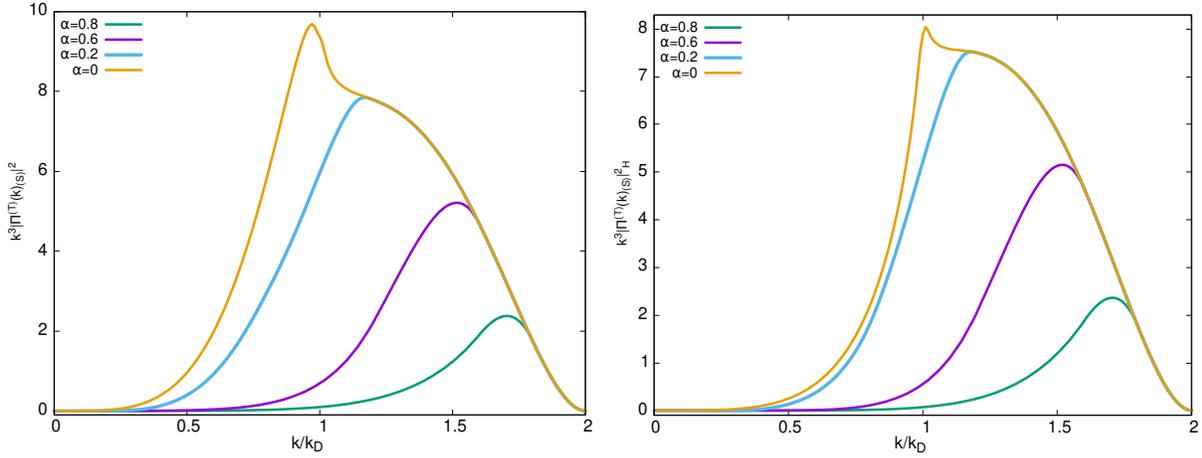

(c) Non-helical contribution to $k^3|\Pi^{(T)}(k)|^2_S$ for $n_B = n_H = -5/2$.

(d) Helical contribution to $k^3|\Pi^{(T)}(k)|^2_S$ for $n_B = n_H = -5/2$.

**Figure 6-5.**: (Non)-Helical contribution to $k^3|\Pi^{(T)}(k)|^2_S$ for different spectral indices in units of $A^2_{(B),H}/(16(2\pi)^5\rho^2_{\gamma,0})$ versus $k/k_D$. Here we show the effect of an IR cut-off parametrized with $\alpha$ on the magnetic power spectrum.

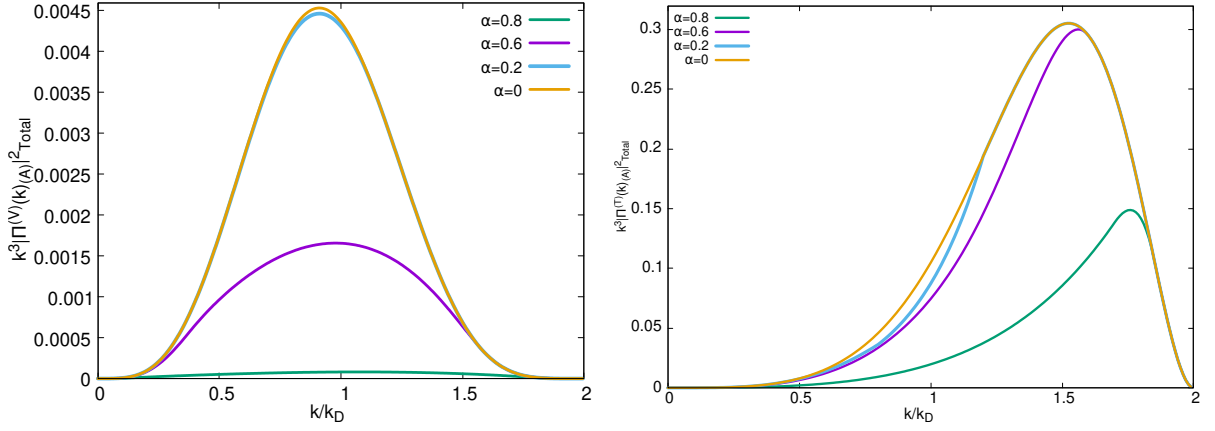

(a) Total contribution to $k^3|\Pi^{(V)}(k)|^2_A$ for $n_B = n_H = 2$.

(b) Total contribution to $k^3|\Pi^{(T)}(k)|^2_A$ for $n_B = n_H = 2$.

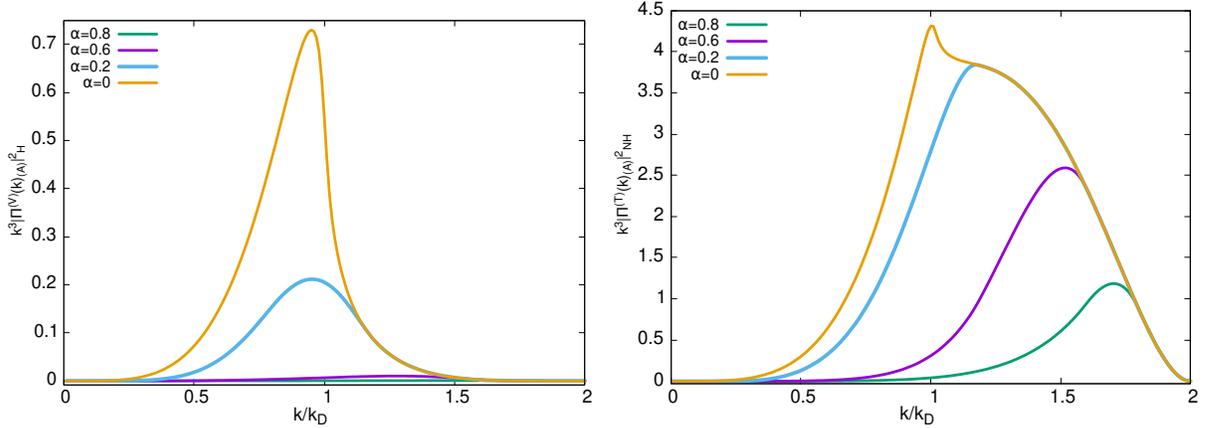

(c) Helical contribution to $k^3|\Pi^{(V)}(k)|^2_A$ for $n_B = n_H = -5/2$.

(d) Non-helical contribution to $k^3|\Pi^{(T)}(k)|^2_A$ for $n_B = n_H = -5/2$.

**Figure 6-6**.: (Non)-Helical contribution to $k^3|\Pi^{(V)}(k)|^2_A$ and $k^3|\Pi^{(T)}(k)|^2_A$ for different spectral indices in units of $A^2_{(B),H}/(8(2\pi)^5\rho^2_{\gamma,0})$ versus $k/k_D$. Here we show the effect of an IR cut-off parametrized with $\alpha$ on the magnetic power spectrum.

contribution for $k^3|\rho(k)|^2$ in the maximal helical case for several spectral indices and values of $\alpha$. Here we can see that for $n_B < 0$, the spectrum is red while for $n > 0$ the biggest contribution comes from large wavenumbers. In Fig.**6-3** is displayed the scalar part of the anisotropic stress and the effect of an IR cutoff on its spectrum. In Figs. **6-4** and **6-5** are shown the symmetric vector and tensor contribution respectively, while Fig.**6-6** displays the antisymmetric part for these contributions. The last figures are important because the magnetic source spectrum with an helical contribution will give non-vanishing odd CMB power spectra and may be used to break the degeneracy between the helical and non-helical contributions to CMB parity even correlators [193].

## 6.2. Magnetic contribution to CMB anisotropies

Using the total angular momentum formalism introduced by [208], the angular power spectrum of the CMB temperature anisotropy is given as

$$(2l+1)^2 C_l^{\mathcal{X}\mathcal{X}} = \frac{2}{\pi} \int \frac{dk}{k} \sum_{m=-2}^{2} k^3 \mathcal{X}_l^{(m)*}(\tau_0,k) \mathcal{X}_l^{(m)}(\tau_0,k), \tag{6-36}$$

where $m = 0, \pm 1, \pm 2$ are the scalar, vector and tensor perturbations modes and $\mathcal{X} = \{\Theta, E, B\}$. Here $\Theta_l^{(m)}(\tau_0, k)$ are the temperature fluctuation $\frac{\delta T}{T}$ multipolar moments, and $B, E$ represent the polarization electric and magnetic type respectively. In large scales, one can neglect the contribution on CMB temperature anisotropies by ISW effect in presence of a PMF [184]. Therefore, considering just the fluctuation via PMF perturbation, the temperature anisotropy multipole moment for $m = 0$ becomes [184]

$$\frac{\Theta_l^{(S)}(\tau_0,k)}{2l+1} \approx \frac{-8\pi G}{3k^2 a_{dec}^2} \rho_B(\tau_0,k) j_l(k\tau_0), \tag{6-37}$$

where $a_{dec}$ is the value of scalar factor at decoupling, $G$ is the Gravitational constant and $j_l$ is the spherical Bessel function. Substituting the last expression in (6-36) with $\mathcal{X} = \Theta$, the CMB temperature anisotropy angular power spectrum is given by

$$l^2 C_l^{\Theta\Theta\,(S)} = \frac{2}{\pi} \left(\frac{8\pi G}{3 a_{dec}^2}\right)^2 \int_0^\infty \frac{|\rho_B(\tau_0,k)|^2}{k^2} j_l^2(k\tau_0) l^2 dk, \tag{6-38}$$

here for our case, we should integrate only up to $2k_D$ since It is the range where energy density power spectrum is not zero. The result of the angular power spectrum induced by scalar magnetic perturbations given by (6-50) is shown in Fig.**6-7**. There, we plot the $\log l^2 C_l^{\Theta\Theta}$ in order to compare our results with those found by [184]. We calculate the angular power spectrum of CMB in units of $\frac{2}{\pi}\left(\frac{8\pi G}{3a_{dec}^2}\right)^2$. One of the important features of the CMB power spectrum (scalar mode) with PMFs, is that the distortion is proportional to strength of PMF and decreases with the spectral index and we must expect its greatest contribution at low multipoles.

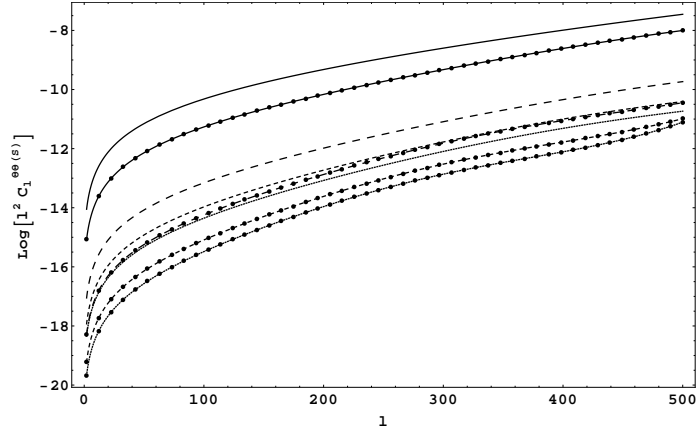

**Figure 6-7**.: Plot of the CMB temperature anisotropy angular power spectrum induced by scalar magnetic perturbations, where the lines with filled circles are for $n = 2$ and the other ones for $n = 5/2$. Here, the solid lines refer to $B_\lambda = 10$nG, large dashed lines for $B_\lambda = 8$nG, small dashed lines refer to $B_\lambda = 5$nG, and dotted lines for $B_\lambda = 1$nG.

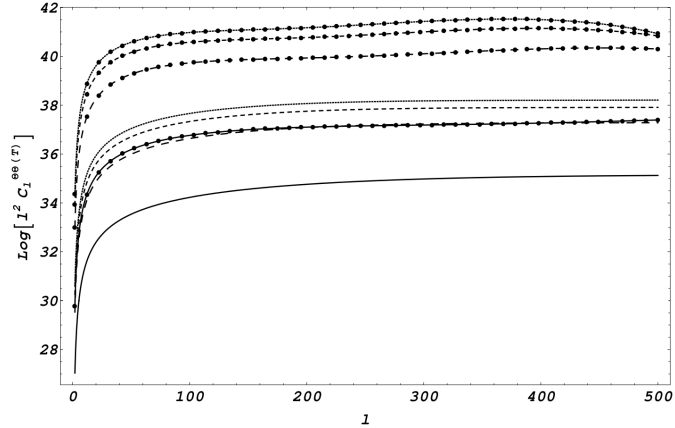

**Figure 6-8**.: Plot of the CMB temperature anisotropy angular power spectrum induced by tensor magnetic perturbations, where the lines with filled circles are for $n = 2$ and the other ones for $n = 4$. Here, the solid lines refer to $B_\lambda = 1$nG, large dashed lines for $B_\lambda = 5$nG, small dashed lines refer to $B_\lambda = 8$nG, and dotted lines for $B_\lambda = 10$nG.

In the case where $m \pm 2$ (tensor modes), the temperature anisotropy multipole moment is given by Eq.(5.22) of [184]

$$\frac{\Theta_l^{(T)}}{2l+1} \simeq -2\pi \sqrt{\frac{8(l+2)!}{3(l-2)!}} \left( G\tau_0^2 z_{eq} \ln\left(\frac{z_{in}}{z_{eq}}\right) \right) \times \Pi^{(T)}(k,\tau_0) \int_0^{x_0} \frac{j_2(x)}{x} \frac{j_l(x_0-x)}{(x_0-x)^2} dx, \quad (6\text{-}39)$$

where $z_{in}$ and $z_{eq}$ are the redshift when PMF was created and during equal matter-radiation era respectively and $x_0 = k\tau_0$. For the integral found in the last expression, we use the approximation made by [209]

$$\int_0^{x_0} \frac{j_2(x)}{x} \frac{j_l(x_0-x)}{(x_0-x)^2} dx \simeq \frac{7\pi}{25} \frac{\sqrt{l}}{x_0^3} J_{l+3}(x_0), \tag{6-40}$$

where $j_l(z) = \sqrt{\frac{\pi}{2z}} J_{l+\frac{1}{2}}(z)$, being $J_\nu(z)$ the Bessel functions of the first kind. With this approximation, the tensor CMB temperature anisotropy angular power spectrum induced by a PMF is given by

$$\begin{aligned} l^2 C_l^{\Theta\Theta(T)} &= \left( Gz_{eq} \ln\left(\frac{z_{in}}{z_{eq}}\right) \right)^2 \frac{l^4(l-1)(l+1)(l+2)}{(2l+1)^2 \tau_0^2} \\ &\times 1{,}25\pi^3 \int \frac{dk}{k^4} J_{l+3}^2(k\tau_0) \left|\Pi^{(T)}(k,\tau_0)\right|^2. \end{aligned} \tag{6-41}$$

The plot of CMB power spectra for tensor perturbations from a power law stochastic PMF with spectral index $n=2$ (lines with filled circles) and $n=4$ (without circles) for different amplitudes of the magnetic field is shown in Fig.**6-8**. Here we can see the same dependence of spectral index and amplitud of PMF as the scalar case. Here the spectra is in units of $\left( Gz_{eq} \ln\left(\frac{z_{in}}{z_{eq}}\right) \right)^2 \frac{1{,}25\pi^3}{\tau_0^2}$.

## 6.3. Infrared cutoff in the CMB spectra

Studying the effect of this lower cutoff of CMB spectra, we can constrain PMF generation models. In Fig.**6-2**c we plot the power spectrum of the energy density of PMF for different values of $k_m$. Here we can see the strong dependence of the power spectrum on this scale, basically the power spectrum does not change when $0{,}2k_D > k_m > 0$ with respect to the results of $k_m = 0$, but in the cases where $k_m > 0{,}2k_D$ (threshold described by dashed line) there is a significant variation with a null lower cutoff. Futhermore, for $k$ close to $2k_D$ the spectrum decays with the same slope, independent from lower cutoff, in this case for $n_B$, the slope of the energy density of PMF goes as $\sim k^{-3{,}2}$. Figs.**6-9**,**6-10** show the effects of PMF on the scalar mode of CMB spectra. Here we did a comparison between the Cls with a null cutoff respect to Cls generated by values of cutoff different from zero. The horizontal solid line shows the comparison with $k_m = 0$, $k_m = 0{,}001k_D$, $k_m = 0{,}1k_D$; no difference in effectiveness was found between these values. The dashed lines report a significant difference of the Cls for values of $k_m = 0{,}3k_D$, $k_m = 0{,}7k_D$, and $k_m = 0{,}9k_D$. In Fig.**6-5**a we show the dependence of the anisotropic trace-free tensor part power spectrum with the infrared cutoff. We observe again a strong dependence for values larger than $0{,}2k_D$ represented by the dashed line. In fact, from Figs.**6-11** and **6-12**, we find that tensor modes of the CMB spectra are distorted by values of $\alpha$ greater than 0.2. It is appropiate to remark that power spectrum of causal fields is a smooth function in the k-space without any sharp cutoff coming from the original mechanism, now, given the parametrization introduced here, we notice from Fig.(7) in [1] that for $\alpha$ very small, the calculations agree with previous work. It can be thinking as contribution of the super horizon modes is negligible and one would expect that scales as $\sim k^4$ for instance. But the results found here have demostrated that an infrared cutoff plays an important role in physical scenarios in other cases where $\alpha > 0{,}2$. Also, one of the characteristics of this dependence is the existence of a peak; indeed, for large values of $\alpha$ the peak moves to left as we see for instance with $\alpha = 0{,}4$ where the peak is in $l \sim 380$ while for $\alpha = 0{,}9$ the peak is shifted to $l \sim 200$.

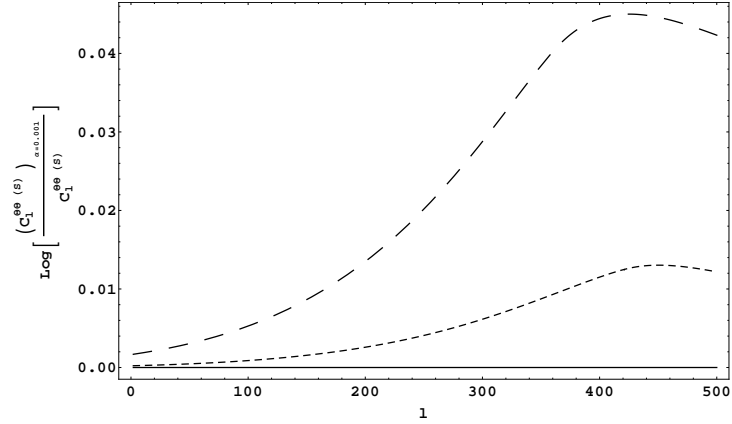

**Figure 6-9**.: Comparison between the CMB temperature anisotropy angular power spectrum induced by scalar PMF at $k_m = 0{,}001k_D$ lower cutoff, respect to the other ones with different values of infrared cutoff. Here, the solid horizontal line is for $k_m = 0{,}1k_D$; small and large dashed lines refer to $k_m = 0{,}3k_D$ and $k_m = 0{,}4k_D$ respectively.

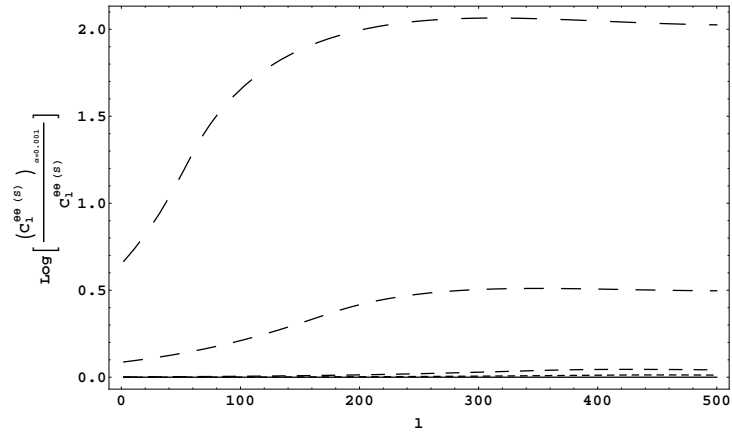

**Figure 6-10**.: This plot shows again a comparison between the CMB temperature anisotropy angular power spectrum induced by scalar PMF at $k_m = 0{,}001k_D$ lower cutoff, respect to the other ones with different values of infrared cutoff. Here, the solid horizontal line is for $k_m = 0{,}1k_D$. The dashed lines describe $k_m = 0{,}3k_D$, $k_m = 0{,}4k_D$, $k_m = 0{,}7k_D$, and $k_m = 0{,}9k_D$ from the small to the longest dashed lines respectively.

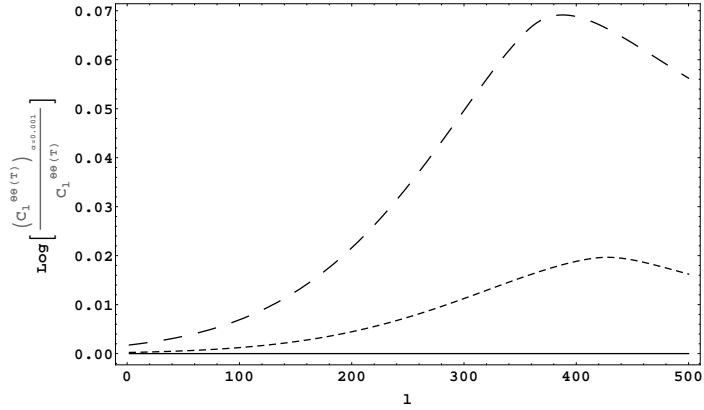

**Figure 6-11**.: Comparison between the CMB temperature anisotropy angular power spectrum induced by tensor magnetic perturbation at $k_m = 0{,}001 k_D$ lower cutoff, respect to the other ones with different values of infrared cutoff. Here, the solid horizon line is for $k_m = 0{,}1 k_D$; small and large dashed lines refer to $k_m = 0{,}3 k_D$ and $k_m = 0{,}4 k_D$ respectively.

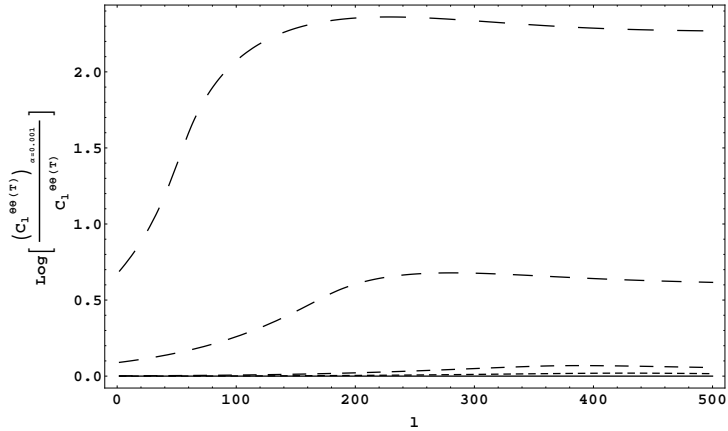

**Figure 6-12**.: This plot shows again, a comparison between the CMB temperature anisotropy angular power spectrum induced by tensor magnetic fluctuations at $k_m = 0{,}001 k_D$ lower cutoff, respect to the other ones with different values of infrared cutoff. Here, the solid horizon line is for $k_m = 0{,}1 k_D$. The dashed lines describe $k_m = 0{,}3 k_D$, $k_m = 0{,}4 k_D$, $k_m = 0{,}7 k_D$, and $k_m = 0{,}9 k_D$ from the small to the longest dashed lines respectively.

## 6.4. Infrared cutoff in the CMB cross-correlation

Using $m = \pm 1, \pm 2$ in Eq.(6-36), the cross-correlation power spectrum of the CMB is given by

$$(2l+1)^2 C_l^{\Theta B (X)} = \frac{4}{\pi} \int dk k^2 \Theta_l^{(X)*}(\tau_0, k) B_l^{(X)}(\tau_0, k), \qquad (6\text{-}42)$$

$$(2l+1)^2 C_l^{E B (X)} = \frac{4}{\pi} \int dk k^2 E_l^{(X)*}(\tau_0, k) B_l^{(X)}(\tau_0, k), \qquad (6\text{-}43)$$

Now, we use the approximate values for vector and tensor contribution of the E-type, B-type and temperature moments obtained in [184]

$$\frac{\Theta_l^{(V)}(\tau_0, k)}{2l+1} \approx \sqrt{\frac{l(l+1)}{2}} \frac{5\Pi_B^{(V)}(k)}{kL_\gamma(\rho_{\gamma_0} + P_{\gamma_0})} \frac{j_l(k\tau_0)}{k\tau_0}, \qquad (6\text{-}44)$$

$$\frac{B_l^{(V)}(\tau_0, k)}{2l+1} \approx -\sqrt{\frac{(l-1)(l+2)}{18}} \frac{5\Pi_B^{(V)}(k)}{(\rho_{\gamma_0} + P_{\gamma_0})} \frac{j_l(k\tau_0)}{k\tau_0}, \qquad (6\text{-}45)$$

$$\frac{E_l^{(V)}(\tau_0, k)}{2l+1} \approx -\sqrt{\frac{(l-1)(l+2)}{18}} \frac{5\Pi_B^{(V)}(k)}{(\rho_{\gamma_0} + P_{\gamma_0})} \left( (l+1) \frac{j_l(k\tau_0)}{(k\tau_0)^2} - \frac{j_{l+1}(k\tau_0)}{k\tau_0} \right), \qquad (6\text{-}46)$$

$$\frac{\Theta_l^{(T)}(\tau_0, k)}{2l+1} \approx -\frac{7(2\pi)^2}{50} \sqrt{\frac{l(l+2)!}{(l-2)!}} \left( G z_{eq} \tau_0^2 \ln\left(\frac{z_{in}}{z_{eq}}\right) \right) \Pi_B^{(T)}(k) \frac{J_{l+3}(k\tau_0)}{(k\tau_0)^3}, \qquad (6\text{-}47)$$

$$\frac{B_l^{(T)}(\tau_0, k)}{2l+1} \approx \frac{7(2\pi)^2}{100} \sqrt{l} \left( G z_{eq} \tau_0 \ln\left(\frac{z_{in}}{z_{eq}}\right) \right) \frac{\Pi_B^{(T)}(k)}{k} \left( l \frac{J_{l+3}(k\tau_0)}{k\tau_0} - J_{l+4}(k\tau_0) \right), \qquad (6\text{-}48)$$

$$\frac{E_l^{(T)}(\tau_0, k)}{2l+1} \approx -\frac{7(2\pi)^2}{100} \sqrt{l} \left( G z_{eq} \tau_0 \ln\left(\frac{z_{in}}{z_{eq}}\right) \right) \frac{\Pi_B^{(T)}(k)}{k} \left( J_{l+3}(k\tau_0) \left( 1 - \frac{l^2}{2(k\tau_0)^2} \right) + \frac{J_{l+4}(k\tau_0)}{k\tau_0} \right), \qquad (6\text{-}49)$$

being $j_l$ and $J_l$ the spherical Bessel and first kind Bessel functions respectively. Using the last expressions and Eqs.(6-42)-(6-43) one can derive the CMB cross-correlations power spectrum for vector and tensor modes

$$C_l^{\Theta B (V)} = -\frac{100}{\pi L_\gamma (\rho_{\gamma_0} + P_{\gamma_0})^2} \sqrt{\frac{l(l-1)(l+2)(l+1)}{32}} \int dk k \left|\Pi_B^{(V)}(k)\right|^2 \left(\frac{j_l(k\tau_0)}{(k\tau_0)}\right)^2, \quad (6\text{-}50)$$

$$C_l^{EB (V)} = -\frac{100(l-1)(l+2)}{18\pi(\rho_{\gamma_0} + P_{\gamma_0})^2} \int dk k^2 \left|\Pi_B^{(V)}(k)\right|^2 \frac{j_l(k\tau_0)}{k\tau_0} \left((l+1)\frac{j_l(k\tau_0)}{(k\tau_0)^2} - \frac{j_{l+1}(k\tau_0)}{k\tau_0}\right) \quad (6\text{-}51)$$

$$C_l^{\Theta B (T)} = -\frac{196(2\pi)^4}{5000\pi(l+1)} \sqrt{\frac{(l+2)!}{(l-2)!}} \left(Gz_{eq} \ln\left(\frac{z_{in}}{z_{eq}}\right)\right)^2 \times$$
$$\times \int dk k^{-2} \left|\Pi_B^{(T)}(k)\right|^2 J_{l+3}(k\tau_0) \left(l\frac{J_{l+3}(k\tau_0)}{k\tau_0} - J_{l+4}(k\tau_0)\right), \quad (6\text{-}52)$$

$$C_l^{EB (T)} = -\frac{196(2\pi)^4}{10000\pi(l+1)} \left(Gz_{eq}\tau_0 \ln\left(\frac{z_{in}}{z_{eq}}\right)\right)^2 \int dk \left|\Pi_B^{(T)}(k)\right|^2 \times$$
$$\times \left(J_{l+3}(k\tau_0)\left(1 - \frac{l^2}{2(k\tau_0)^2}\right) + \frac{J_{l+4}(k\tau_0)}{k\tau_0}\right) \left(l\frac{J_{l+3}(k\tau_0)}{k\tau_0} - J_{l+4}(k\tau_0)\right), \quad (6\text{-}53)$$

here for our case, the integration is only up to $2k_D$, since It is the range where power spectrum is different from zero as we can see in Fig.(**6-4**),(**6-5**),(**6-6**). The solutions are shown in Figs.(**6-13**a),(**6-13**b),(**6-13**c),(**6-13**d). Here, the colours denote different values of infrared cutoff of the PMF: Black($\alpha = 0,01$), Red( $\alpha = 0,3$), Blue($\alpha = 0,5$), Green($\alpha = 0,7$), Brown($\alpha = 0,8$) and Magenta($\alpha = 0,9$). We can observe that helical(antisymmetric) contribution generates nonzero values of the parity-odd correlators $C_l^{\theta B}$ and $C_l^{EB}$ as also pointed out in [184, 187, 190, 193]. We also observe the effects of IR cutoff on the parity odd spectra. In summary we are working on the assumption that after inflation a weak magnetic field, a seed, was created. This PMF is parametrized by its strength $B_\lambda$, smoothing length $\lambda$ and in accordance with the generation process, it also depends on $k_D$, $k_m$ and a blue spectral index $n_B$. In particular, $k_m$ is set by the size of the causal part of the Universe during its generation. Now, if this seed indeed is presented during late stages in the universe, this PMF prints a signal in the pattern on CMB spectra, signal that depends of the variables above mentioned, in particular $k_m$. If $\alpha$ tends to one, the effect of infrared cutoff must not be ignored, even in scenarios like inflation, this cutoff is also important (For a deeper discussion see [197]). Therefore, the feature of this signal which we found is strongly dependent on the infrared cutoff, will be useful for constraining PMF post inflation generation models. Besides this $k_m$ is important for studying the evolution of density perturbations and peculiar velocities due to primordial magnetic fields and effects on BBN [210], [210], [195], [211].

## 6.5. Effects of the background PMF on the CMB

The presence of energy density of the background PMF increases total radiation-like energy density $\rho_r$, and modifies the standard dynamics of the background Universe producing considerable effects on the primary temperature fluctuations of the CMB. In this Section we

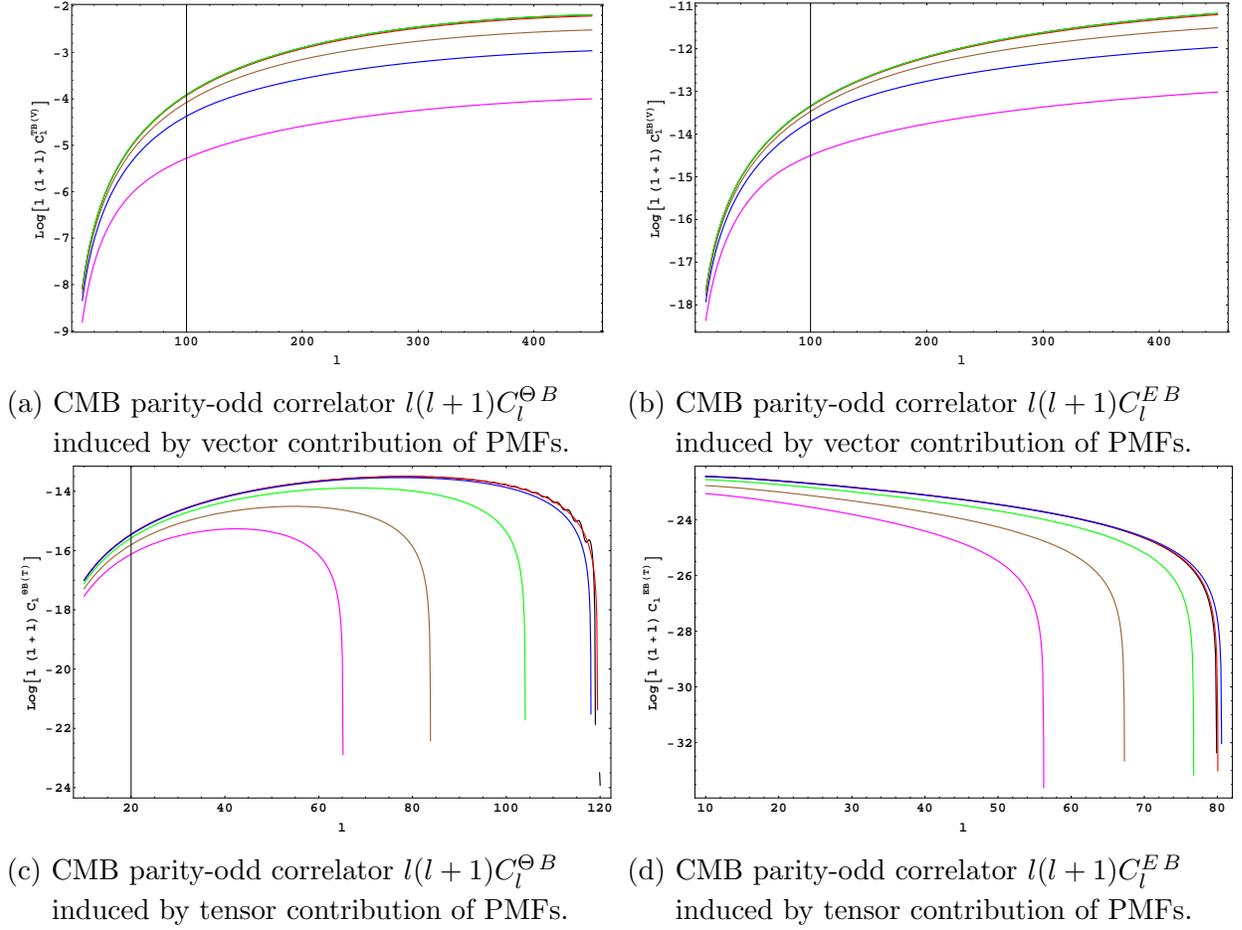

(a) CMB parity-odd correlator $l(l+1)C_l^{\Theta B}$ induced by vector contribution of PMFs.

(b) CMB parity-odd correlator $l(l+1)C_l^{EB}$ induced by vector contribution of PMFs.

(c) CMB parity-odd correlator $l(l+1)C_l^{\Theta B}$ induced by tensor contribution of PMFs.

(d) CMB parity-odd correlator $l(l+1)C_l^{EB}$ induced by tensor contribution of PMFs.

**Figure 6-13**.: Vector and Tensor temperature-polarization correlations $C_l^{\Theta B}$ and $C_l^{EB}$ in the CMB anisotropies.

will study the effects of background PMF on the CMB following the early work discussed in Ref. [196]. First, the total energy density and pressure are now written as $\rho = \rho^{(\Lambda CDM)} + \rho_B$, and $P = \sum_i w_i \rho_i^{(\Lambda CDM)} + P_B$ respectively (see Eq.(2-10)), modifying the solution of the Friedmann's equation (2-12)-(2-13). In order to study the effects of PMFs on the CMB, we include a background magnetic density given by Eq.(6-12) into the Boltzmann code CLASS [6]. As first shown by Ref. [196], the speed of sound in baryon fluid is described by $c_{s,eff}^2 = c_s^2 + \rho_B/\rho_t$, where $c_s^2$ is the speed of sound without magnetic field [94]

$$c_s^2 = \frac{1}{3(1+R)}, \quad \text{with:} \ R \equiv 3\rho_b/4\rho_r, \tag{6-54}$$

and $\rho_t$ is the total energy density. Including also this modification in the thermodynamic structure in the CLASS code, we obtain the spectrum of CMB temperature anisotropies shown in Fig.**6-14**. Here we can observe the effects of different parameters enclosed in Eq.(6-12) on the CMB spectrum. With a PMF in the primordial plasma, the time of matter-radiation density equality ($\rho_m = \rho_r$) increases as we can see in Eq.(2-24), enhancing the

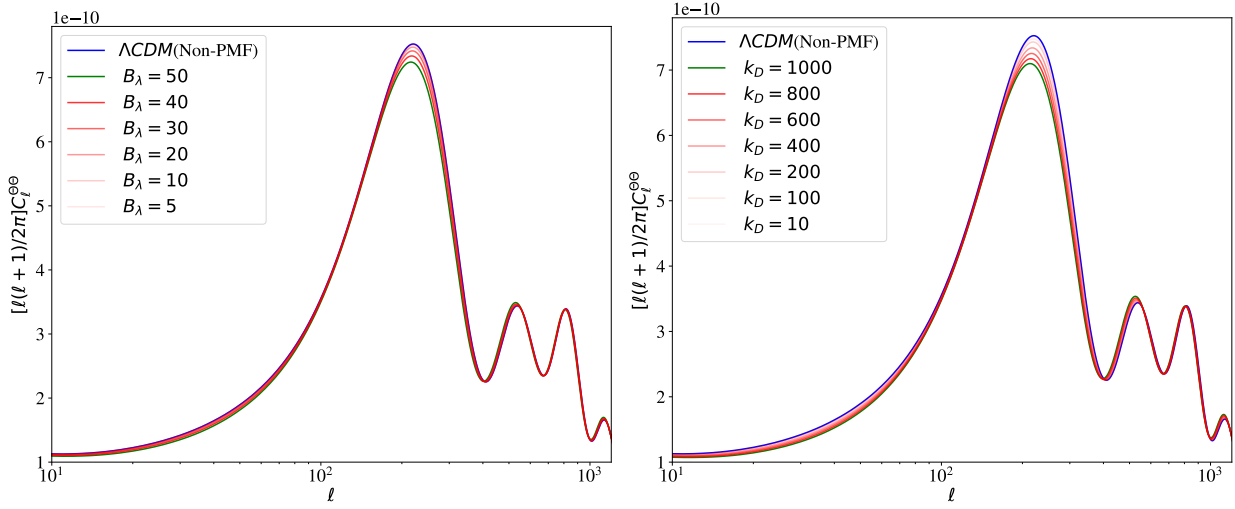

(a) $l(l+1)C_l$ with $k_D = 100$, $k_{min} = 0$, $n_B = -2$. The green line assumes $\rho_B/\rho_\gamma = 0{,}0041$.

(b) $l(l+1)C_l$ with $B_\lambda = 20$nG, $k_{min} = 0$, $n_B = -2$. The green line assumes $\rho_B/\rho_\gamma = 0{,}0065$.

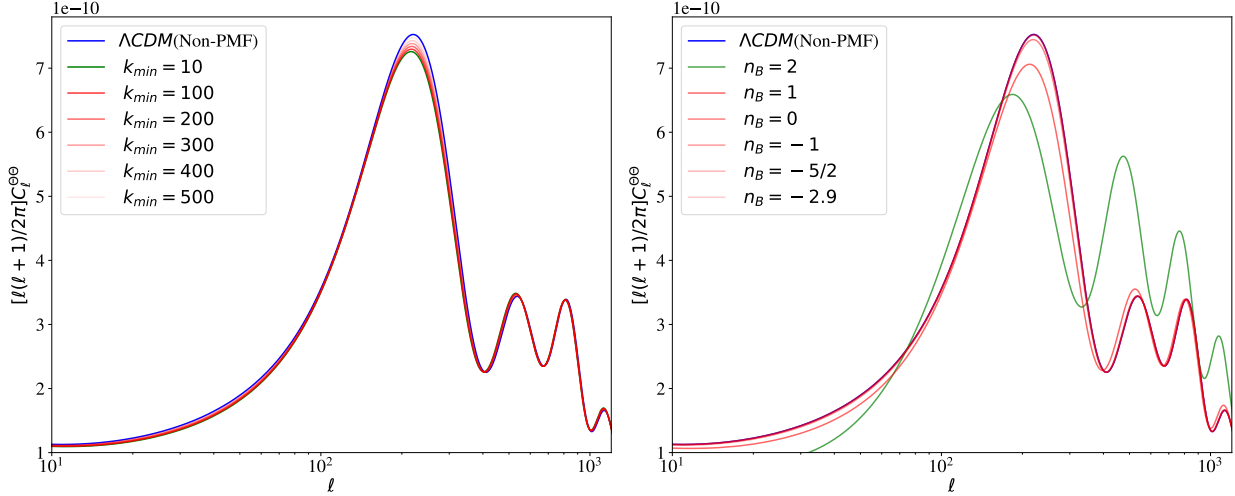

(c) $l(l+1)C_l$ with $B_\lambda = 20$nG, $k_D = 400$, $n_B = -2$. The green line assumes $\rho_B/\rho_\gamma = 0{,}0038$.

(d) $l(l+1)C_l$ with $B_\lambda = 20$nG, $k_{min} = 0$, $k_D = 400$. The green line assumes $\rho_B/\rho_\gamma = 0{,}044$.

**Figure 6-14**.: Spectrum of CMB temperature anisotropies with PMFs obtained numerically from CLASS code. Each plot display the effect of $B_\lambda$ (a), $k_D$ (b), $k_{min}$ (c), and $n_B$ (d) on the CMB spectrum. Here the blue line stands for the model without PMF and $B_\lambda$ is in units of nG, and $k$ in units of Mpc$^{-1}$.

amplitude of all peaks because there is not enough time to be supressed for the cosmic expansion. However, the contrast between odd and even peaks is reduced because It further depends on $(\rho_m/\rho_r)$ corresponding to the balance between gravity and the total radiation pressure [7]. Second, an important effect of PMFs come from increasing in sound speed $c_s$. In fact, the peaks location depends on the angle $\theta = d_s(\tau_{dec})/d_A(\tau_{dec})$, where $d_s(\tau_{dec})$ is the physical sound horizon at decoupling and $d_A(\tau_{dec})$ is the angular diameter distance at decoupling [7]

$$d_s = a \int_{\tau_{ini}}^{\tau} c_s d\tau, \quad d_A = \frac{1}{1+z} \int_0^z \frac{dz'}{H(z')}. \tag{6-55}$$

The angular diameter distance depends on the late history after decoupling $(\Omega_\Lambda, h)$, whereas physical sound horizon is further affected by the value of $c_s$. By adding PMFs to the primordial plasma, we increase the effective speed of sound which in turn increases the angle of the location of peaks, boosting the peaks to small $l's$[3], i.e., shifting the acoustic peaks to the left as we see in Fig.**6-15**a. Finally, since the value of the total radiation energy density is larger with PMFs, the gravitational potentials $\phi, \psi$ decay more quickly after their wavelengths become smaller than the sound horizon (see Fig.**6-15**a).

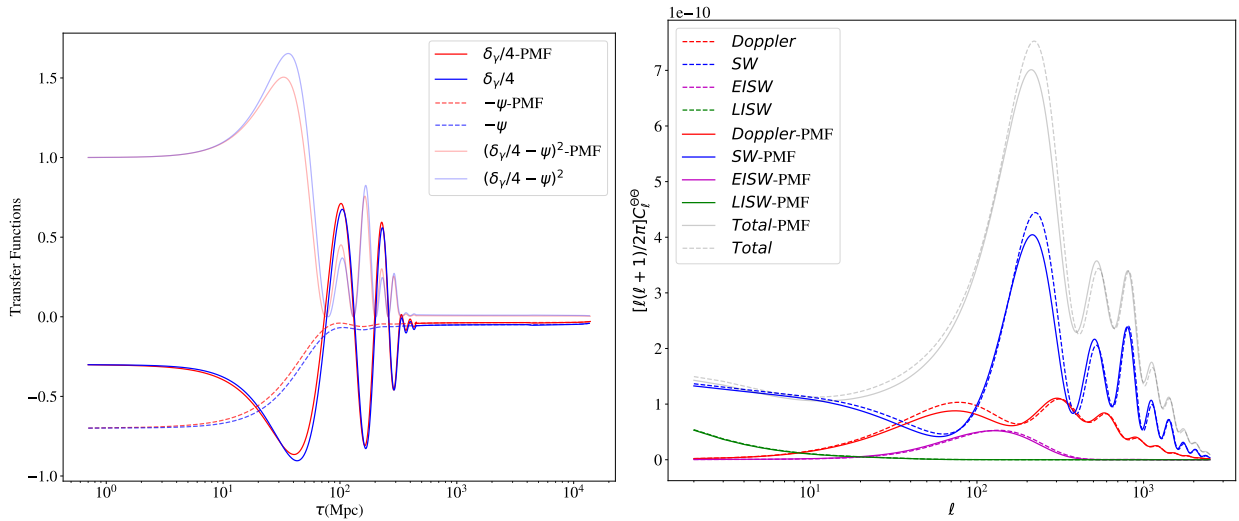

(a) Effect of PMF on the transfer functios for $k = 0,1 \text{Mpc}^{-1}$.

(b) Individual contribution of $l(l+1)C_l$ with PMFs.

**Figure 6-15**.: A snapshot of the transfer functios and $l(l+1)C_l$ temperature accounting for a PMF with $\rho_B/\rho_\gamma = 0,0041$. In left panel we plot the numerical solution obtained from CLASS code for $\delta_\gamma$ and $\psi$ with and without PMFs. In right panel we plot the CMB espectrum with and without PMFs showing the individual contributions explained in the text: Sachs-Wolf(SW), Doppler, early integrated Sachs-Wolfe(EISW) and late integrated Sachs-Wolfe (LISW). Note that the total spectrum labeled by the black line corresponds to all correlations.

In summary, accounting for a background PMF in our model modifies the shape of the temperature power spectrum significantly for large multipolar numbers, that is, the Sachs-Wolfe (SW), Doppler and early integrated Sachs-Wolfe (EISW) contributions are quite affected by the magnetic field. This fact can be noticed in Fig.**6-15**b, where we plot the features of PMFs ($\rho_B/\rho_\gamma = 0{,}0041$) for several contributions of the CMB spectrum. Since the late integrated Sachs-Wolfe (LISW) comes from interactions of the photons after last scattering, PMFs do not play a sizable role in this contribution. On the other hand, the EISW signal is shifted to small $l's$ because modes related to $\psi, \phi$ entered to sub-horizon scales earlier than if they had done without PMFs. This boost is also seen in the SW where the acoustic peak positions are shifted to larger scales. For $l > 100$, odd Doppler peaks are enhanced with respect to the ratio of baryon and radiation content [7] hence, PMFs produce supression in the amplitude for odd peaks while the even ones remain unaltered. These features are illustrated in Fig.**6-15**b.

---

[3]This can be understood geometrically using $\theta \sim \pi/l$, see Appendix E.

# 7. CMB Non-Gaussianity from PMFs

PMFs induce non-Gaussianity (NG) signals on the CMB anisotropies. This is because the Energy Momentum Tensor (EMT) of PMFs is quadratic in the fields and therefore the resulting fluctuations are non-Gaussian even if the PMF is a stochastic Gaussian field. Studies of such primordial NG and the CMB bispectrum and trispectrum have been deeply investigated in refs [199, 200, 203, 212]. Analyses of NG signals via bispectrum on CMB have found upper limits of PMFs around $2 - 22$nG derived from scalar magnetic modes, and $3,2 - 10$nG derived from vector-tensor magnetic modes smoothed on a scale of 1Mpc [199, 200, 212–217]. Planck Collaboration [218] also reported limits on the amplitude of $B_{1Mpc} < 3$nG for a spectral index of $n_B = -2,9$; $B_{1Mpc} < 0,07$nG for $n_B = -2$; and $B_{1Mpc} < 0,04$nG for $n_B = 2$ from compensated modes; and $B_{1Mpc} < 4,5$nG for $n_B = -2,9$ from the passive-scalar mode. PMFs have also been constrained by POLARBEAR experiment, where they reported that the PMF amplitude from the two-point correlation functions is less than 3,9nG at the 95 % confidence level [219]. Furthermore, distinct signatures on the parity-odd CMB cross correlations would carry valuable clues about a primordial magnetic helicity [193, 215]. In fact, helical contribution in the field (in the perfect conductivity limit) has been widely studied because it produces efficient transference of power from smaller to larger scales, and thus it is able to explain the actual observed magnetic fields [220]. Therefore, the study of helical fields via NG will give deeper comprehension of the magnetic field generation model and will help us to strengthen the constraints of PMF amplitude. Therefore, one of the aims of this chapter is to estimate limits on the amplitude of the field through bounds on non-Gaussianity via helical contributions of the field. This chapter is based on the work published in [3].

## 7.1. Shapes of Non-Gaussianity

In this section, we review some preliminary studies on NG based on Refs. [9, 10, 221]. The physics for generating the initial density fluctuations in the CMB and the large-scale structure is not complete understood. Most inflationary models predict that those seed perturbations are nearly Gaussian distributed, i,e., their statistical properties are completely characterized by the two-point correlation function. Consequently, any deviation from a Gaussian distribution or *non-Gaussianity* measured by the upcoming observational experiments, will enable us to move beyond the knowledge of the initial conditions and will open up new opportunities to study the physics of early Universe. The amoung of NG is described in terms of a dimensionless parameter $f_{NL}$ and from theoretical point of view, its value runs over a range of values depending on the inflation scenario. For instance, the simplest inflation models that are based on a slowly rolling scalar field predict an almost negligible level of NG [222]; but, for a very large class of general models such that multiple scalar fields, axion monodromy models, hybrid inflation or dubbed multi-brid inflation among others, generates substantially higher amounts of NG [223–225]. The measurement of the bispectrum is one of the most promising ways of constraining the value of $f_{NL}$. So far, Planck collaboration have obtained constraints on primordial NG and they have reported values of $f_{NL} = 5,8 \pm 5,0$ (68 % CL, statistical) [12]. Moreover, the presence of the delta function in the definition of the three-point function in Eq.(4-71), forms a triangle (see Figure **7-1**) in the three Fourier modes of the bispectrum whose shape becomes a powerful probe to distinguish between different scenarios of the physics of the primordial Universe. This shape can be parametrised as [10]

$$\mathcal{S}(k_1, k_2, k_3) \equiv N \times (k_1 k_2 k_3)^2 \times B(k_1, k_2, k_3), \tag{7-1}$$

where $N$ is a constant of normalisation and $B(k_1, k_2, k_3)$ is the bispectrum defined in Eq.(4-71). Following Ref. [221], the shape function $\mathcal{S}(k_1, k_2, k_3)$ can be expressed in terms of possible combinations of variables

$$\begin{aligned} K_p &= \sum_p (k_i)^p, \quad K_{pq} = \frac{1}{\Delta_{pq}} \sum_{i \neq j} (k_i)^p (k_j)^q, \quad K_{pqr} = \frac{1}{\Delta_{pqr}} \sum_{i \neq j \neq l} (k_i)^p (k_j)^q (k_l)^r, \\ \tilde{k}_{ip} &= K_p - 2k_p^i, \quad \Delta_{pq} = 1 + \delta_{pq}, \quad \Delta_{pqr} = \Delta_{pq}(\Delta_{qr} + \delta_{pr}) \text{ (no summation).} \end{aligned} \tag{7-2}$$

We will concentrate only in two representative shapes of the primordial bispectrum, local and equilateral.

### 7.1.1. Local type

One way to parametrise non-Gaussianity is expanding in Taylor series, the random field $\mathcal{R}$ around the Gaussian part $\mathcal{R}_\mathcal{G}$ [10]

$$\mathcal{R}(\mathbf{x}) = \mathcal{R}_\mathcal{G}(\mathbf{x}) + \frac{3}{5} f_{NL}^{local}(\mathcal{R}_\mathcal{G}(\mathbf{x})^2 - \langle \mathcal{R}_\mathcal{G}(\mathbf{x})^2 \rangle), \tag{7-3}$$

where $f_{NL}^{local}$ is the local-type parameter which parametrises the level of NG. Since this expression is localized at a given point in the real space, this parametrisation is commonly called local-type non-Gaussianity [9]. Let us now calculate the three point correlation of $\mathcal{R}$ in the real space keeping only terms linear in $f_{NL}^{local}$

$$\begin{aligned} \mathcal{R}(\mathbf{x})\mathcal{R}(\mathbf{y})\mathcal{R}(\mathbf{z}) &\approx \mathcal{R}_\mathcal{G}(\mathbf{x})\mathcal{R}_\mathcal{G}(\mathbf{y})\mathcal{R}_\mathcal{G}(\mathbf{z}) + \frac{3}{5} f_{NL}^{local} \bigg[ \Big(\mathcal{R}_\mathcal{G}(\mathbf{x})^2 - \langle \mathcal{R}_\mathcal{G}(\mathbf{x})^2 \rangle \Big) \mathcal{R}_\mathcal{G}(\mathbf{y})\mathcal{R}_\mathcal{G}(\mathbf{z}) \\ &+ \Big(\mathcal{R}_\mathcal{G}(\mathbf{y})^2 - \langle \mathcal{R}_\mathcal{G}(\mathbf{y})^2 \rangle \Big) \mathcal{R}_\mathcal{G}(\mathbf{x})\mathcal{R}_\mathcal{G}(\mathbf{z}) + \Big(\mathcal{R}_\mathcal{G}(\mathbf{z})^2 - \langle \mathcal{R}_\mathcal{G}(\mathbf{z})^2 \rangle \Big) \mathcal{R}_\mathcal{G}(\mathbf{x})\mathcal{R}_\mathcal{G}(\mathbf{y}) \bigg] \\ &= \mathcal{R}_\mathcal{G}(\mathbf{x})\mathcal{R}_\mathcal{G}(\mathbf{y})\mathcal{R}_\mathcal{G}(\mathbf{z}) + \frac{3}{5} f_{NL}^{local} \bigg[ \Big(\mathcal{R}_\mathcal{G}(\mathbf{x})^2 - \langle \mathcal{R}_\mathcal{G}(\mathbf{x})^2 \rangle \Big) \mathcal{R}_\mathcal{G}(\mathbf{y})\mathcal{R}_\mathcal{G}(\mathbf{z}) + 2\text{perms} \bigg]. \end{aligned} \tag{7-4}$$

Let us average this equation taking into account $\langle \mathcal{R}_\mathcal{G}(\mathbf{x})\mathcal{R}_\mathcal{G}(\mathbf{y})\mathcal{R}_\mathcal{G}(\mathbf{z}) \rangle = 0$ because all odd higher-point correlation functions vanish for Gaussian fluctuations, so

$$\begin{aligned}\langle \mathcal{R}(\mathbf{x})\mathcal{R}(\mathbf{y})\mathcal{R}(\mathbf{z}) \rangle &\approx \langle \mathcal{R}_\mathcal{G}(\mathbf{x})\mathcal{R}_\mathcal{G}(\mathbf{y})\mathcal{R}_\mathcal{G}(\mathbf{z}) \rangle + \frac{3}{5}f_{NL}^{local}\bigg[\langle \mathcal{R}_\mathcal{G}(\mathbf{x})\mathcal{R}_\mathcal{G}(\mathbf{x})\mathcal{R}_\mathcal{G}(\mathbf{y})\mathcal{R}_\mathcal{G}(\mathbf{z}) \rangle \\ &\quad - \langle \mathcal{R}_\mathcal{G}(\mathbf{x})\mathcal{R}_\mathcal{G}(\mathbf{x}) \rangle \langle \mathcal{R}_\mathcal{G}(\mathbf{y})\mathcal{R}_\mathcal{G}(\mathbf{z}) \rangle + 2\text{perms}\bigg] \\ &= \frac{6}{5}f_{NL}^{local}\bigg[\langle \mathcal{R}_\mathcal{G}(\mathbf{x})\mathcal{R}_\mathcal{G}(\mathbf{y}) \rangle \langle \mathcal{R}_\mathcal{G}(\mathbf{x})\mathcal{R}_\mathcal{G}(\mathbf{z}) \rangle + 2\text{perms}\bigg],\end{aligned} \quad (7\text{-}5)$$

in this instance, we used the properties in Eq.(4-13) along with the 4-point correlation function Eq.(4-55). Substituting this expression into (4-68) and using Eqs.(4-60),(4-58) we obtain

$$\langle \mathcal{R}(\mathbf{k_1})\mathcal{R}(\mathbf{k_2})\mathcal{R}(\mathbf{k_3}) \rangle = (2\pi)^3 \delta^3(\mathbf{k_1}+\mathbf{k_2}+\mathbf{k_3})\frac{6}{5}f_{NL}^{local}\bigg[P(k_1)P(k_2)+P(k_1)P(k_3)+P(k_2)P(k_3)\bigg], \quad (7\text{-}6)$$

if we compare this equation with (4-71), the bispectrum of local non-Gaussianity is written as

$$B(k_1, k_2, k_3)^{local} = \frac{6}{5}f_{NL}^{local}\bigg[P(k_1)P(k_2) + P(k_1)P(k_3) + P(k_2)P(k_3)\bigg]. \quad (7\text{-}7)$$

For a scale-invariant spectrum $P(k) = Ak^{-3}$, we have

$$B(k_1, k_2, k_3)^{local} = \frac{6}{5}f_{NL}^{local}A^2\bigg[\frac{1}{(k_1k_2)^3} + 2\,\text{perms}\bigg], \quad (7\text{-}8)$$

This is boosted in the squeezed limit $k_3 \ll k_1 \approx k_2$ obtaining

$$\lim_{k_3 \ll k_1 \approx k_2} B(k_1, k_2, k_3)^{local} = \frac{12}{5}f_{NL}^{local}A^2\bigg[\frac{1}{(k_1k_2)^3}\bigg], \quad (7\text{-}9)$$

and writing this configuration in terms of the shape function, we have

$$\mathcal{S}^{local} \sim \frac{K_3}{K_{111}}. \quad (7\text{-}10)$$

Shapes of the primordial bispectra are shown in Figure **7-1**, where we have defined the rescaled momenta as $x_2 \equiv k_2/k_1$ and $x_3 \equiv k_3/k_1$ and it is assumed the ordering given by $x_3 \leq x_2 \leq x_1$. Here we can observe that the signal for the local shape peaks at the squeezed limit, i.e., for $x_3 \sim 0$ and $x_2 \sim 1$. Single field inflationary models which satisfy the slow-roll inflation demand a local type bispectrum, however its amplitude is suppressed by slow-roll parameters $f_{NL} \sim \mathcal{O}(\epsilon_1, \epsilon_2)$ [10, 222].

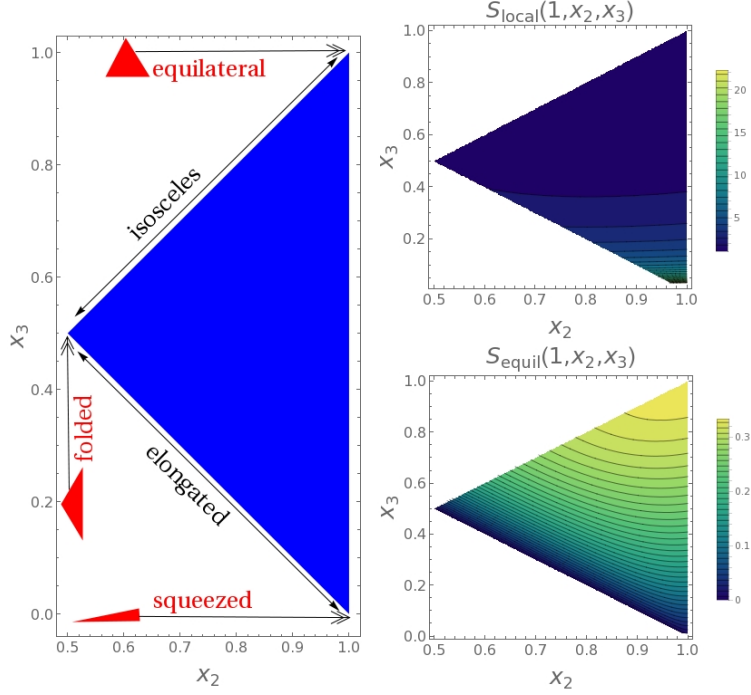

**Figure 7-1**.: Contour for the shape function of the bispectra. Left panel shows the shapes of non-Gaussianity in a region given by $1 - x_2 \leq x_3 \leq x_2$, with $x_2 = k_2/k_1, x_3 = k_3/k_1$. The top-right panel is the local form which diverges at squeezed configuration $x_3 \sim 0, x_2 \sim 1$, while the bottom-right panel is the equilateral one which peaks at $x_2 \sim x_3 \sim 1$. We use the normalization $\mathcal{S}^{local}(1,1,1) = 1$ in each panel. Figure adapted from [9, 10].

### 7.1.2. Equilateral type

The equilateral bispectrum is parametrized as [9]

$$B(k_1, k_2, k_3)^{equil} = \frac{36}{5} f_{NL}^{equil} \left[ \left( -\frac{1}{(k_1 k_2)^3} + 2\text{perms} \right) - \frac{2}{(k_1 k_2 k_3)^3} + \left( \frac{1}{k_1 k_2^2 k_3^2} + 5\text{perms} \right) \right], \quad (7\text{-}11)$$

whereas in terms of the shape function we have

$$\mathcal{S}^{equil} \sim \frac{\tilde{k}_1 \tilde{k}_2 \tilde{k}_3}{K_{111}}. \quad (7\text{-}12)$$

This shape of the primordial bispectra is shown in Figure **7-1**. This bispectrum has a peak at $x_3 \sim 1$ and $x_2 \sim 1$, namely, the equilateral limit $k_1 = k_2 = k_3$. Inflationary models with non-canonical kinetic terms for the scalar field or with higher-derivative interactions and non-trivial speeds of sound enforce this configuration [9,10,221]. Additionally, there are three intermediate cases: the elongated ($k_1 = k_2 + k_3$), isosceles ($k_1 > k_2 = k_3$) and folded ($k_1 = 2k_2 = 2k_3$), see Figures **7-1** and **7-3**. Finally, since primordial NGs are measurements of the interactions in inflation, these signals appear only through higher order effects. However, features of NG which comes from other contributions like magnetic fields can be generated even at the lowest order. Thus, the presence of these fields will lead to NG in the CMB anisotropies. So, in the next section we will investigate the imprints of these fields in the CMB bispectrum and infer some parameters which characterize those fields.

## 7.2. The magnetic bispectrum

Since the magnetic field is assumed as a Gaussianly-distributed stochastic helical field and the electromagnetic EMT is quadratic in the fields, the statistics must be NG and the bispectrum is non-zero as was claimed by [203]. Using (6-17) we have that three-point correlation function is expressed as

$$
\begin{aligned}
\langle \Pi_{ij}(\mathbf{k1})\Pi_{tl}(\mathbf{k2})\Pi_{mn}(\mathbf{k3})\rangle &= \frac{-1}{(4\pi\rho_{\gamma,0})^3} \int \frac{d^3p}{(2\pi)^3} \int \frac{d^3q}{(2\pi)^3} \int \frac{d^3s}{(2\pi)^3} \times \\
&\quad \langle B_i(\mathbf{p})B_j(\mathbf{k1}-\mathbf{p})B_l(\mathbf{k2}-\mathbf{q})B_n(\mathbf{k3}-\mathbf{s})B_t(\mathbf{q})B_m(\mathbf{s})\rangle \\
&\quad - \frac{\delta_{ij}}{2}\langle...\rangle - \frac{\delta_{tl}}{2}\langle...\rangle - \frac{\delta_{mn}}{2}\langle...\rangle + \frac{\delta_{ij}\delta_{mn}}{4}\langle...\rangle + \frac{\delta_{ij}\delta_{tl}}{4}\langle...\rangle \\
&\quad + \frac{\delta_{tl}\delta_{mn}}{4}\langle...\rangle - \frac{\delta_{ij}\delta_{mn}\delta_{tl}}{8}\langle...\rangle.
\end{aligned} \quad (7\text{-}13)
$$

Where $\langle...\rangle$ describes an ensemble average over six stochastic fields. We can use Wicks theorem to decompose the six point correlation function into products of the magnetic field power spectrum expressed in (6-2). Eight of fifteen terms contribute to the bispectrum and they are proportional to $\delta(\mathbf{k1}+\mathbf{k2}+\mathbf{k3})$ due to the homogeneity condition. In [202] they point out that expression (7-13) can be reduced to just one contribution, if the projection tensor used for extracting each one is symmetric in $(ij)$, $(tl)$ and $(mn)$. Therefore one can write the bispectrum as follows

$$
\begin{aligned}
\langle \Pi_{ij}(\mathbf{k1})\Pi_{tl}(\mathbf{k2})\Pi_{mn}(\mathbf{k3})\rangle &= \delta(\mathbf{k1}+\mathbf{k2}+\mathbf{k3}) \times \\
&\quad \frac{8}{(4\pi\rho_{\gamma,0})^3} \int \frac{d^3p}{(2\pi)^3} F_{it}(\mathbf{p})F_{jm}(\mathbf{k1}-\mathbf{p})F_{ln}(\mathbf{k2}+\mathbf{p}),
\end{aligned} \quad (7\text{-}14)
$$

being $F_{ij}(\mathbf{k}) = P_{ij}(\mathbf{k})P_B(\mathbf{k}) + i\epsilon_{ijm}\hat{\mathbf{k}}^m P_H(\mathbf{k})$. Wavevectors that appear in Eq.(7-14) generate a tetrahedron configuration (see Figure **7-2**) such that fifteen angles must be included for calculating the bispectrum. So, in order to a comparison with previos works, we use not only the geometry configuration for bispectrum but as well the notation of these angles defined

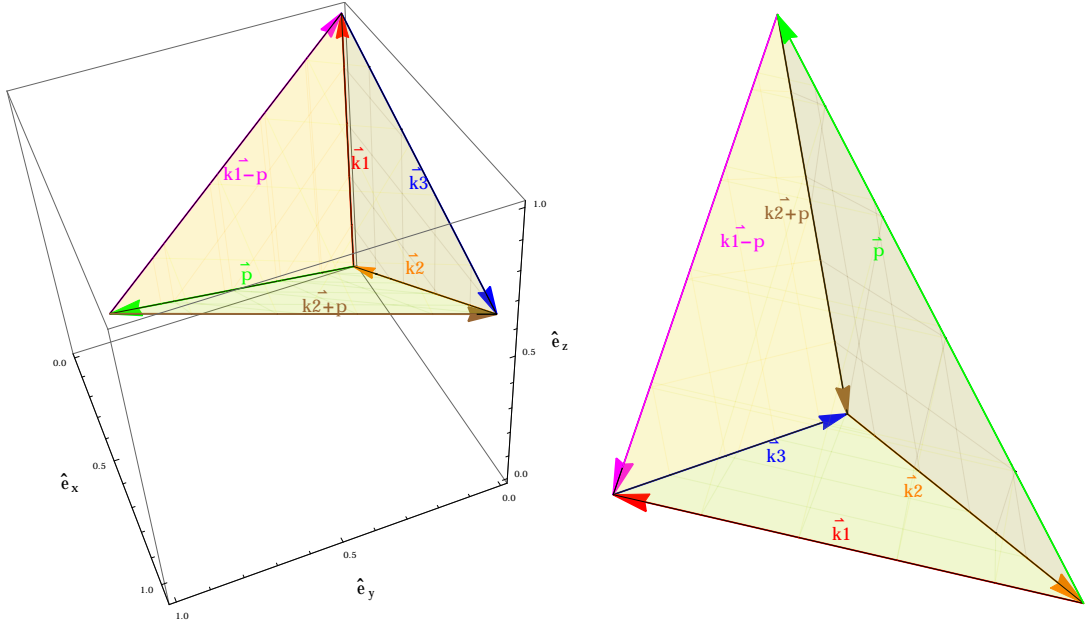

**Figure 7-2**.: Geometrical configuration for bispectrum. The wavevectors **k1**, **k2** and **k3** are free, while **p** is the integration mode.

in [202] given as

$$\begin{aligned}
\beta &= \hat{\mathbf{p}} \cdot \widehat{\mathbf{k1}-\mathbf{p}}, \quad \gamma = \hat{\mathbf{p}} \cdot \widehat{\mathbf{k2}+\mathbf{p}}, \quad \mu = \widehat{\mathbf{k1}-\mathbf{p}} \cdot \widehat{\mathbf{k2}+\mathbf{p}}, \quad \theta_{kp} = \hat{\mathbf{k1}} \cdot \hat{\mathbf{k2}} \\
\theta_{kq} &= \hat{\mathbf{k1}} \cdot \hat{\mathbf{k3}}, \quad \theta_{pq} = \hat{\mathbf{k2}} \cdot \hat{\mathbf{k3}}, \quad \alpha_k = \hat{\mathbf{k1}} \cdot \hat{\mathbf{p}}, \quad \alpha_p = \hat{\mathbf{k2}} \cdot \hat{\mathbf{p}}, \quad \alpha_q = \hat{\mathbf{k3}} \cdot \hat{\mathbf{p}} \\
\beta_k &= \hat{\mathbf{k1}} \cdot \widehat{\mathbf{k1}-\mathbf{p}}, \beta_p = \hat{\mathbf{k2}} \cdot \widehat{\mathbf{k1}-\mathbf{p}}, \quad , \beta_q = \hat{\mathbf{k3}} \cdot \widehat{\mathbf{k1}-\mathbf{p}}, \quad \gamma_k = \hat{\mathbf{k1}} \cdot \widehat{\mathbf{k2}+\mathbf{p}} \\
\gamma_p &= \hat{\mathbf{k2}} \cdot \widehat{\mathbf{k2}+\mathbf{p}}, \quad \gamma_q = \hat{\mathbf{k3}} \cdot \widehat{\mathbf{k2}+\mathbf{p}}.
\end{aligned} \quad (7\text{-}15)$$

As we will see, the bispectrum has two main contributions, first of them contains terms proportional to $A_B^3$ or $A_B A_H^2$ and this is called the even contribution denoted here with $B^{(S)}$. The second contribution is proportional to terms like $A_H^3$ or $A_B^2 A_H$ and It is called the odd contribution denoted as $\mathcal{B}^{(A)}$. Hence, we can define the three-point correlation for the scalar modes as

$$\langle Z_1(\mathbf{k1}) Z_2(\mathbf{k2}) Z_3(\mathbf{k3}) \rangle \equiv \delta\left(\sum_{i=1}^{3} \mathbf{ki}\right) \left(B^{(S)}_{Z_1 Z_2 Z_3} - i\epsilon_{ljk} \mathcal{B}^{(A)\, ljk}_{Z_1 Z_2 Z_3}\right), \quad (7\text{-}16)$$

where $Z_{\{1,2,3\}} = \{\rho_B, \Pi_B^{(S)}\}$. We begin calculating the bispectrum of the magnetic energy density. To do so, we shall apply the projector defined in (6-19) three times $\delta_{ij}\delta_{tl}\delta_{mn}$ onto (7-13) to obtain the following

$$\langle \Pi_{ij}(\mathbf{k1}) \Pi_{tl}(\mathbf{k2}) \Pi_{mn}(\mathbf{k3}) \rangle \delta_{ij}\delta_{tl}\delta_{mn} = \langle \rho_B(\mathbf{k1}) \rho_B(\mathbf{k2}) \rho_B(\mathbf{k3}) \rangle, \quad (7\text{-}17)$$

using Eq.(7-16), a straightforward calculation gives the following expression

$$
\begin{aligned}
B^{(S)}_{\rho_B \rho_B \rho_B} &= \frac{8}{(2\pi)^3 (4\pi \rho_{\gamma,0})^3} \int d^3p \, \big( P_B(p) P_B(|\mathbf{k1}-\mathbf{p}|) P_B(|\mathbf{p}+\mathbf{k2}|) F^1_{\rho\rho\rho} \\
&\quad + P_H(|\mathbf{p}+\mathbf{k2}|)(-P_B(p)P_H(|\mathbf{k1}-\mathbf{p}|)F^2_{\rho\rho\rho} + P_B(|\mathbf{k1}-\mathbf{p}|)P_H(p)F^3_{\rho\rho\rho}) \\
&\quad - P_B(|\mathbf{p}+\mathbf{k2}|)P_H(p)P_H(|\mathbf{k1}-\mathbf{p}|)F^4_{\rho\rho\rho} \big),
\end{aligned} \tag{7-18}
$$

for the even contribution. The values of $F_{\rho\rho\rho}$ are shown along with the odd contribution in appendix C. In orden to find the three-point cross-correlation between scalar anisotropic stress and magnetic energy density, we will apply the three projections $\delta_{ij}\delta_{tl}\mathcal{P}_{mn}$ defined in (6-19) onto (7-13) which gives us

$$
\langle \Pi_{ij}(\mathbf{k1})\Pi_{tl}(\mathbf{k2})\Pi_{mn}(\mathbf{k3})\rangle \delta_{ij}\delta_{tl}\mathcal{P}_{mn} = \langle \rho_B(\mathbf{k1})\rho_B(\mathbf{k2})\Pi^{(S)}_B(\mathbf{k3})\rangle, \tag{7-19}
$$

and using the three-point correlation Eq.(7-16), the even contribution yields

$$
\begin{aligned}
B^{(S)}_{\rho_B \rho_B \Pi^{(S)}_B} &= \frac{8}{2(2\pi)^3(4\pi\rho_{\gamma,0})^3} \int d^3p \, \Big( -P_B(p)P_B(|\mathbf{k1}-\mathbf{p}|)P_B(|\mathbf{p}+\mathbf{k2}|)F^1_{\rho\rho\Pi^{(S)}} \\
&\quad - P_B(p)P_H(|\mathbf{k1}-\mathbf{p}|)P_H(|\mathbf{p}+\mathbf{k2}|)F^2_{\rho\rho\Pi^{(S)}} \\
&\quad + P_H(p)\big(-P_H(|\mathbf{p}+\mathbf{k2}|)P_B(|\mathbf{k1}-\mathbf{p}|)F^3_{\rho\rho\Pi^{(S)}} \\
&\quad + P_B(|\mathbf{p}+\mathbf{k2}|)P_H(|\mathbf{k1}-\mathbf{p}|)F^4_{\rho\rho\Pi^{(S)}}\big) \Big).
\end{aligned} \tag{7-20}
$$

Other cross-bispectra is obtained by applying $\delta_{ij}\mathcal{P}_{tl}\mathcal{P}_{mn}$ on (7-13) and this gives

$$
\langle \Pi_{ij}(\mathbf{k1})\Pi_{tl}(\mathbf{k2})\Pi_{mn}(\mathbf{k3})\rangle \delta_{ij}\mathcal{P}_{tl}\mathcal{P}_{mn} = \langle \rho_B(\mathbf{k1})\Pi^{(S)}_B(\mathbf{k2})\Pi^{(S)}_B(\mathbf{k3})\rangle, \tag{7-21}
$$

as a result we found the following expression

$$
\begin{aligned}
B^{(S)}_{\rho_B \Pi^{(S)}_B \Pi^{(S)}_B} &= \frac{8}{4(2\pi)^3(4\pi\rho_{\gamma,0})^3} \int d^3p \, \Big( P_B(p)P_B(|\mathbf{k1}-\mathbf{p}|)P_B(|\mathbf{p}+\mathbf{k2}|)F^1_{\rho\Pi^{(S)}\Pi^{(S)}} \\
&\quad + P_B(p)P_H(|\mathbf{p}+\mathbf{k2}|)P_H(|\mathbf{k1}-\mathbf{p}|)F^2_{\rho\Pi^{(S)}\Pi^{(S)}} \\
&\quad - P_H(p)P_H(|\mathbf{p}+\mathbf{k2}|)P_B(|\mathbf{k1}-\mathbf{p}|)F^3_{\rho\Pi^{(S)}\Pi^{(S)}} \\
&\quad - P_H(p)P_H(|\mathbf{k1}-\mathbf{p}|)P_B(|\mathbf{p}+\mathbf{k2}|)F^4_{\rho\Pi^{(S)}\Pi^{(S)}} \Big).
\end{aligned} \tag{7-22}
$$

Finally the three-point correlation of scalar anisotropic stress is obtained by applying $\mathcal{P}_{ij}\mathcal{P}_{tl}\mathcal{P}_{mn}$ on (7-13) finding that

$$
\langle \Pi_{ij}(\mathbf{k1})\Pi_{tl}(\mathbf{k2})\Pi_{mn}(\mathbf{k3})\rangle \mathcal{P}_{ij}\mathcal{P}_{tl}\mathcal{P}_{mn} = \langle \Pi^{(S)}_B(\mathbf{k1})\Pi^{(S)}_B(\mathbf{k2})\Pi^{(S)}_B(\mathbf{k3})\rangle, \tag{7-23}
$$

thus, the expression for the bispectrum for that contribution is

$$
\begin{aligned}
B^{(S)}_{\Pi^{(S)}_B \Pi^{(S)}_B \Pi^{(S)}_B} &= \frac{1}{(2\pi)^3(4\pi\rho_{\gamma,0})^3} \int d^3p \, \Big( -P_B(p)P_B(|\mathbf{k1}-\mathbf{p}|)P_B(|\mathbf{p}+\mathbf{k2}|)F^1_{\Pi^{(S)}\Pi^{(S)}\Pi^{(S)}} \\
&\quad - P_B(p)P_H(|\mathbf{p}+\mathbf{k2}|)P_H(|\mathbf{k1}-\mathbf{p}|)F^2_{\Pi^{(S)}\Pi^{(S)}\Pi^{(S)}} \\
&\quad + P_H(p)P_B(|\mathbf{p}+\mathbf{k2}|)P_H(|\mathbf{k1}-\mathbf{p}|)F^3_{\Pi^{(S)}\Pi^{(S)}\Pi^{(S)}} \\
&\quad + P_H(p)P_B(|\mathbf{k1}-\mathbf{p}|)P_H(|\mathbf{p}+\mathbf{k2}|)F^4_{\Pi^{(S)}\Pi^{(S)}\Pi^{(S)}} \Big).
\end{aligned} \tag{7-24}
$$

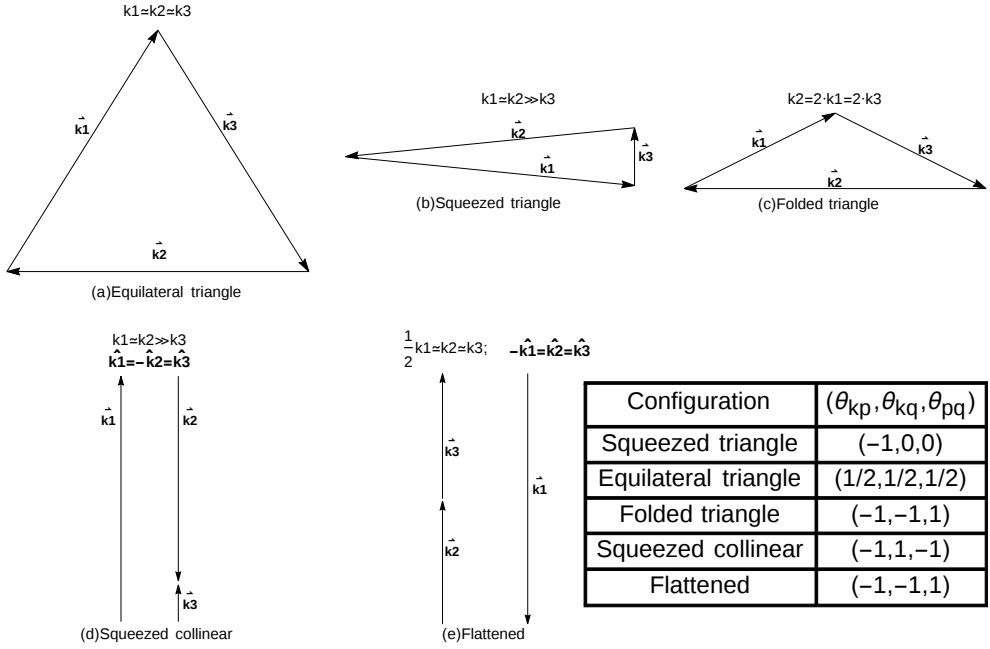

**Figure 7-3**.: Geometrical representations for the bispectrum. The figure shows a visual representation of the triangles and the collinear configuration of the bispectrum shape. The table in the bottom-right panel describes the values of the **p**-independent terms for each configuration.

Once more, the $F_{\Pi\Pi\Pi}$'s values can be checked along with the odd contribution in appendix C. In the case where the helicity of the field is not considered ($A_H = 0$), the only contribution that remains is sourced from $A_B^3$. The results of this contribution were reported in [202] and we have found the same expressions. Therefore, we have generalized those previous results to even and odd contributions of the PMFs bispectrum and thus these findings are the first results of the paper.

## 7.3. Full evaluation

With the derivation of the angular part of the three-point correlation for each component of the magnetic tensor, we proceed to make the evaluation of the above integrals. Due to the complexity of the calculation, we are going to follow the exact methodology proposed in Refs. [200, 213, 226]. The methodology considers two cases for finding solution of the correlator. In the first case, the terms dependent on the integration vector **p** are not considered in the evaluation. For the second case, the squeezed collinear configuration is used to make predictions. We will use five representative shapes of the bispectrum which are shown in the Figure **7-3**. The odd signal arising from $\mathcal{B}^{(A)}$ will not be considered here but it will be deferred for later work.

### 7.3.1. p-independent

For this case, the only terms which appear in the evaluation are those angles given in Eq.(7-15) independent of **p**; they are $(\theta_{kp}, \theta_{kq}, \theta_{pq})$. The values of these angles for each configuration are shown in Figure **7-3**. The $F$'s functions defined above take the following values under this approximation

$$F^1_{\rho\rho\rho} = \mu^2, \quad F^2_{\rho\rho\rho} = \mu, \quad F^3_{\rho\rho\rho} = F^4_{\rho\rho\rho} = 0, \tag{7-25}$$

$$F^1_{\rho\rho\Pi} = \mu^2 - 3, \quad F^2_{\rho\rho\Pi} = -\mu, \quad F^3_{\rho\rho\Pi} = F^4_{\rho\rho\Pi} = 0, \tag{7-26}$$

$$F^1_{\rho\Pi\Pi} = \mu^2 - 6 + 9\theta^2_{pq}, \quad F^2_{\rho\Pi\Pi} = (-7 + 9\theta^2_{pq})\mu, \quad F^3_{\rho\Pi\Pi} = F^4_{\rho\Pi\Pi} = 0, \tag{7-27}$$

$$\begin{aligned} F^1_{\Pi\Pi\Pi} &= -9 + 9\theta^2_{kp} - 27\theta_{kp}\theta_{kq}\theta_{pq} + 9\theta^2_{pq} + \mu^2 + 9\theta^2_{kq}, \quad F^3_{\Pi\Pi\Pi} = 0, \\ F^2_{\Pi\Pi\Pi} &= (-13 + 18\theta^2_{kp} + 9\theta^2_{kq} - 27\theta_{kp}\theta_{kq}\theta_{pq} + 9\theta^2_{pq})\mu, \quad F^4_{\Pi\Pi\Pi} = 0, \end{aligned} \tag{7-28}$$

where the result given for $F^1_{\Pi\Pi\Pi}$ is in agreement with the reported in [200][1]. The values of $F$ for each geometrical representation of the bispectrum are shown in Table **7-1** and $\mu = 0$ for all configuration except to squeezed configuration where it takes $\mu \sim -1$.

| Configuration | $F^1_{\rho\rho\rho}$ | $F^2_{\rho\rho\rho}$ | $F^1_{\rho\rho\Pi}$ | $F^2_{\rho\rho\Pi}$ | $F^1_{\rho\Pi\Pi}$ | $F^2_{\rho\Pi\Pi}$ | $F^1_{\Pi\Pi\Pi}$ | $F^2_{\Pi\Pi\Pi}$ |
|---|---|---|---|---|---|---|---|---|
| Squeezed triangle | 0 | 0 | -3 | 0 | -6 | 0 | 0 | 0 |
| Equilateral triangle | 0 | 0 | -3 | 0 | -15/4 | 0 | -45/8 | 0 |
| Folded triangle | 0 | 0 | -3 | 0 | 3 | 0 | -9 | 0 |
| Squeezed collinear | 1 | -1 | -2 | 1 | 4 | -2 | -8 | 4 |
| Flattened | 0 | 0 | -3 | 0 | 3 | 0 | -9 | 0 |

**Table 7-1**.: Values of $F$ for different geometrical configurations in the **p**-independent case.

### 7.3.2. Squeezed collinear configuration

In this approximation, the magnitude of one wavevector $(\hat{\mathbf{k3}})$ is small while the others have equal magnitudes but have opposing directions $(\hat{\mathbf{k1}} = -\hat{\mathbf{k2}})$ as shown in Figure **7-3**. With this assumption the angles can be reduced to

$$\begin{aligned} \beta &= \hat{\mathbf{p}} \cdot \widehat{\mathbf{k1} - \mathbf{p}} \sim -\hat{\mathbf{p}} \cdot \widehat{\mathbf{k1} - \mathbf{p}} \sim -\gamma, \quad \mu \sim -\widehat{\mathbf{k1} - \mathbf{p}} \cdot \widehat{\mathbf{k1} - \mathbf{p}} \sim -1, \\ \alpha_k &\sim -\alpha_p \sim \alpha_q, \quad \beta_k \sim -\beta_p \sim \beta_q \sim -\gamma_k \sim \gamma_p \sim -\gamma_q. \end{aligned} \tag{7-29}$$

---
[1] There is a difference with a minus sign because we are taking a different signature in the metric.

By using this approximation the values of the $F$'s are simplified to

$$F^1_{\rho\rho\rho} = 1+\beta^2, \quad F^2_{\rho\rho\rho} = -(1+\beta^2), \quad F^3_{\rho\rho\rho} = -2\beta, \quad F^4_{\rho\rho\rho} = 2\beta, \tag{7-30}$$

$$\begin{aligned}F^1_{\rho\rho\Pi} &= -2+3\alpha_k^2+\beta^2-6\alpha_k\beta\beta_k+3\beta_k^2+3\beta^2\beta_k^2, \quad F^2_{\rho\rho\Pi} = 1-3\alpha_k^2-2\beta^2+6\alpha_k\beta\beta_k,\\ F^3_{\rho\rho\Pi} &= -\beta(-1+3\beta_k^2), \quad F^4_{\rho\rho\Pi} = \beta(-1+3\beta_k^2),\end{aligned} \tag{7-31}$$

$$\begin{aligned}F^1_{\rho\Pi\Pi} &= -9\alpha_k\beta\beta_k^3+(2-3\beta_k^2)^2+\beta^2(1+3\beta_k^2)+\alpha_k^2(-3+9\beta_k^2), \quad F^3_{\rho\Pi\Pi} = 3\alpha_k\beta_k-\beta(4-3\beta_k^2),\\ F^2_{\rho\Pi\Pi} &= -2-2\beta^2+3\alpha_k\beta\beta_k+3\beta_k^2, \quad F^4_{\rho\Pi\Pi} = -6\alpha_k\beta_k+9\alpha_k\beta_k^3-\beta(-5+6\beta_k^2),\end{aligned} \tag{7-32}$$

$$\begin{aligned}F^1_{\Pi\Pi\Pi} &= -8+\beta^2+18\beta_k^2+3\beta^2\beta_k^2-9\beta_k^4+6\alpha_k\beta\beta_k(1-3\beta_k^2)+9\alpha_k^2(1-3\beta_k^2+3\beta_k^4),\\ F^2_{\Pi\Pi\Pi} &= 4-2\beta^2-3\beta_k^2+\alpha_k^2(-6+9\beta_k^2), \quad F^3_{\Pi\Pi\Pi} = F^4_{\Pi\Pi\Pi} = (2\beta-3\alpha_k\beta_k)(1-3\beta_k^2). \tag{7-33}\end{aligned}$$

Same results have been obtained in Ref. [200] for $F^1_{\Pi\Pi\Pi}$ (case where there is not helicity). The angular part of the integrals must be written in spherical coordinates $d^3p = p^2 dp d\alpha_k d\theta$, where $\theta$ is the azimuthal angle. Since we consider an upper cut-off $k_D$, we must introduce the $(k1, k2)$-dependence on the angular integration domain; this implies that we should split the integral domain in different regions such as $0 < k1, k2 < 2k_D$. The integration domains we use for calculating the integrals are shown in appendix D. By using the power spectrum expression for the magnetic fields (6-4) and the F's values for the $\mathbf{p}-$independent case given above, we get the causal magnetic bispectrum ($n_B = n_H = 2$) which is shown in the Figure **7-4**. We see that the biggest contribution for the bispectrum occurs when $k1 \sim k2$ and $\langle \Pi_B \Pi_B \Pi_B \rangle$ generates the largest value for scalar mode. Hence, we conclude that the shape of the NG associated with the PMF can be classified into the local-type configuration as was previously reported in Ref. [227] for a scale invariant shape. We also observe that effects from $A_B A_H^2$ contribution are smaller respect to $A_B^3$. Figures **7-4**g, **7-4**h show the cross-correlation of the bispectrum obtaining the same behavior and a large contribution (with respect to the energy density bispectrum).

On the other hand, in Figure **7-5** we present the results under the squeezed collinear configuration driven also by causal fields. Here, we see that the bispectrum for this configuration is less than the $\mathbf{p}-$independent case. We also notice that magnetic anisotropic stress change the sign under this configuration for wavenumbers larger than $k_D/2$, while the $A_B A_H^2$ contribution is practically negative. Note that all the palettes shown here are sequential colour palettes, so they are very well suited to show the amplitude of the bispectrum. Finally, for negative spectral indices, an approximate solution for the bispectrum can be found using the formula (5.13) in Ref. [199] (see also Eq.(D-6) in appendix D). Assuming $(n_B = n_H = n)$ along with $k2 < k1 < k_D$, the expression for bispectrum becomes

$$\begin{aligned}B^{(S)}_{\rho_B^{(S)}\rho_B^{(S)}\rho_B^{(S)}} &\sim \frac{16}{(2\pi)^3(4\pi\rho_{\gamma,0})^3}\left(\frac{nk1^n k2^{2n+3}}{(n+3)(2n+3)}+\frac{nk1^{3n+3}}{(2n+3)(3n+3)}+\frac{k_D^{3n+3}}{(3n+3)}\right)\\ &\quad \times \left(A_s^3 F^1_{\rho\rho\rho} + A_H^2 A_s \left(F^3_{\rho\rho\rho} - F^4_{\rho\rho\rho} - F^2_{\rho\rho\rho}\right)\right),\end{aligned} \tag{7-34}$$

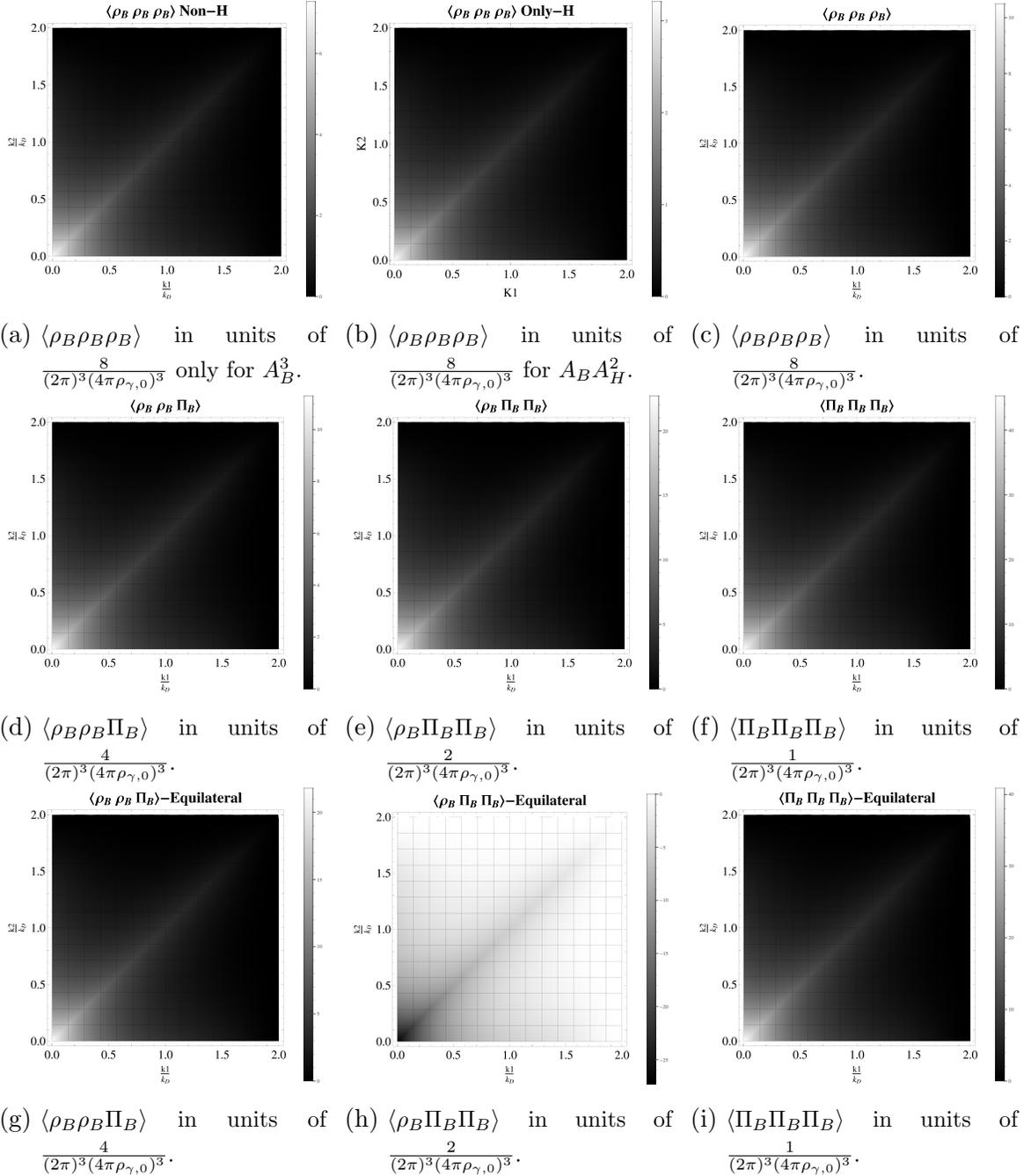

**Figure 7-4.**: Total contribution of three-point correlation for scalar modes using the **p**-independent approximation. Figures (a), (b) (c) show the correlation of the energy density of the magnetic field without, with $A_B A_H^2$ and full contribution respectively. Figures (d), (e) and (f) show the even cross correlation of the field. Finally, Figures (g), (h) (i) show the even cross three-point correlation field in the equilateral configuration, where the $\langle \rho_B \Pi_B \Pi_B \rangle$ has a negative contribution. We see that largest contribution to the bispectrum is obtained when $k1 \sim k2$. Also, the biggest contribution is given by $\langle \Pi_B \Pi_B \Pi_B \rangle$ into a squeezed configuration.

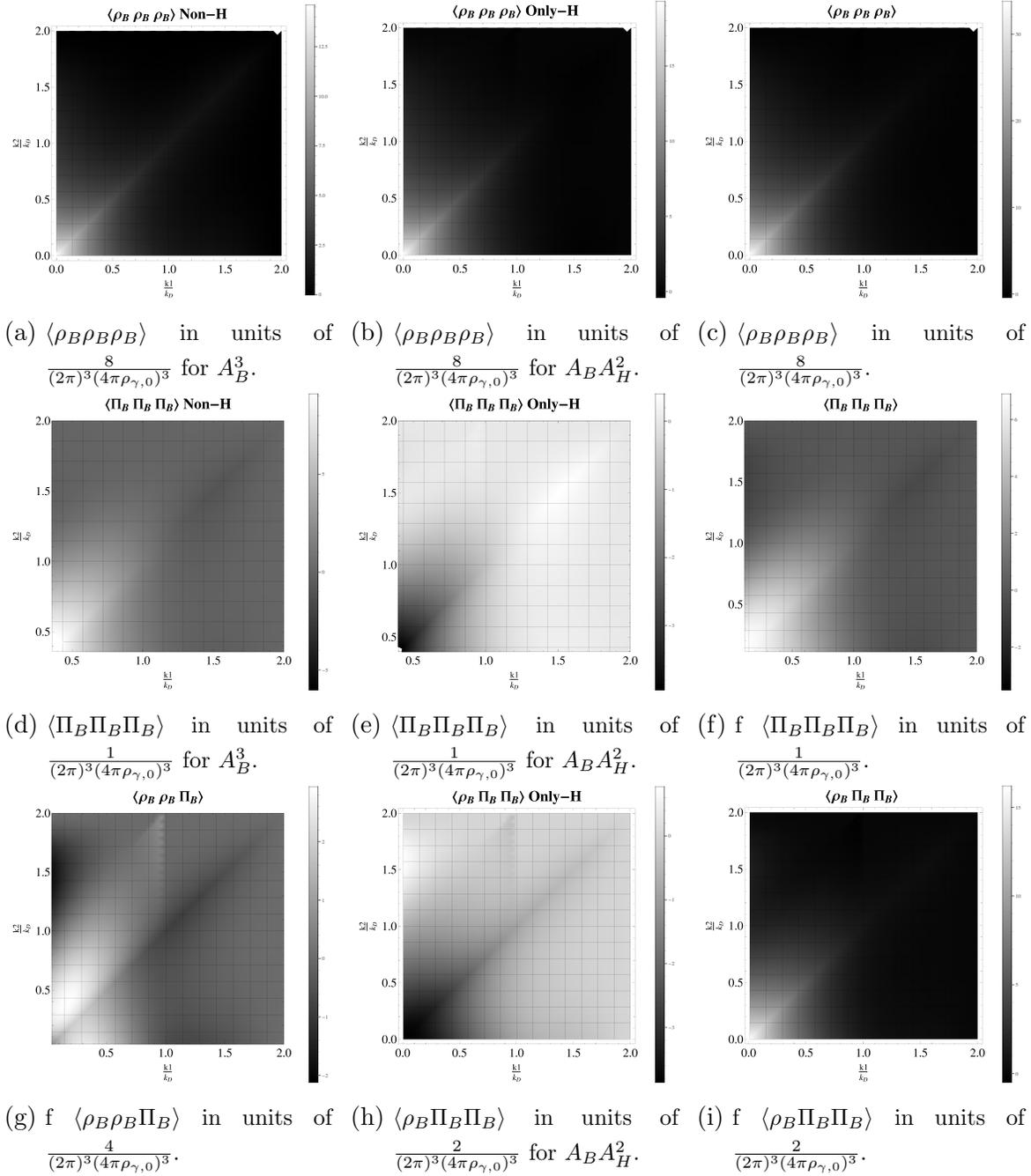

(a) $\langle\rho_B\rho_B\rho_B\rangle$ in units of $\frac{8}{(2\pi)^3(4\pi\rho_{\gamma,0})^3}$ for $A_B^3$.

(b) $\langle\rho_B\rho_B\rho_B\rangle$ in units of $\frac{8}{(2\pi)^3(4\pi\rho_{\gamma,0})^3}$ for $A_B A_H^2$.

(c) $\langle\rho_B\rho_B\rho_B\rangle$ in units of $\frac{8}{(2\pi)^3(4\pi\rho_{\gamma,0})^3}$.

(d) $\langle\Pi_B\Pi_B\Pi_B\rangle$ in units of $\frac{1}{(2\pi)^3(4\pi\rho_{\gamma,0})^3}$ for $A_B^3$.

(e) $\langle\Pi_B\Pi_B\Pi_B\rangle$ in units of $\frac{1}{(2\pi)^3(4\pi\rho_{\gamma,0})^3}$ for $A_B A_H^2$.

(f) f $\langle\Pi_B\Pi_B\Pi_B\rangle$ in units of $\frac{1}{(2\pi)^3(4\pi\rho_{\gamma,0})^3}$.

(g) f $\langle\rho_B\rho_B\Pi_B\rangle$ in units of $\frac{4}{(2\pi)^3(4\pi\rho_{\gamma,0})^3}$.

(h) $\langle\rho_B\Pi_B\Pi_B\rangle$ in units of $\frac{2}{(2\pi)^3(4\pi\rho_{\gamma,0})^3}$ for $A_B A_H^2$.

(i) f $\langle\rho_B\Pi_B\Pi_B\rangle$ in units of $\frac{2}{(2\pi)^3(4\pi\rho_{\gamma,0})^3}$.

**Figure 7-5**.: Contribution of three-point correlation of non-crossing scalar modes described in the text using the squeezed collinear configuration. Figures (a), (b) (c) show the correlation of the PMF energy density without and with $A_B A_H^2$; and full contribution respectively; while Figures (d), (e) and (f) show the correlation of the PMF anisotropic stress without, with $A_B A_H^2$ and full contribution respectively. Figures (g), (h) and (i) show the cross correlation of the field in this configuration. One important feature of the anisotropic stress mode is the negative contribution to the total bispectrum for $k > K_D/2$.

$$B^{(S)}_{\Pi^{(S)}_B \Pi^{(S)}_B \Pi^{(S)}_B} \sim \frac{2}{(2\pi)^3 (4\pi\rho_{\gamma,0})^3} \left( \frac{nk1^n k2^{2n+3}}{(n+3)(2n+3)} + \frac{nk1^{3n+3}}{(2n+3)(3n+3)} + \frac{k_D^{3n+3}}{(3n+3)} \right)$$
$$\times \left( A_s^3 F^1_{\Pi\Pi\Pi} + A_H^2 A_s \left( F^3_{\Pi\Pi\Pi} + F^4_{\Pi\Pi\Pi} - F^2_{\Pi\Pi\Pi} \right) \right). \tag{7-35}$$

### 7.3.3. Infrared cut-off in the bispectrum

Now, we analyze the effect of an infrared cut-off parametrized by $\alpha$ on the magnetic bispectrum (see Eq.(6-11)). We saw that the NG peaks at $k1 \sim k2$ under a squeezed configuration, so we compute the magnetic bispectrum using the strategy adopted in Sec. 4.6. In Figure **7-6**, the effect of this IR cut-off for causal fields is illustrated. The top Figures (**7-6**a) and (**7-6**b) show the effect of $\alpha$ on $k1^3 \langle \rho_B \rho_B \rho_B \rangle$ where the lines refer to different values of the cut-off, while the bottom Figures (**7-6**c) and (**7-6**d) show the effect of $\alpha$ on $k1^3 \langle \Pi_B \Pi_B \Pi_B \rangle$. What we read off these figures is how the peak of the bispectrum moves to high wavenumbers when we increase the value of $k_m$ in the same way that magnetic power spectrum (and its amplitude decreases too, due to reduction of the wavenumber space) and how the effects of the $A_B A_H^2$ contributions PMF are tiny compared with the non-helical case. On the other hand, Figure (**7-7**) shows the effect of the infrared cut-off when we are considering non-causal fields. The top panel shows the three-point correlation of the energy density of the magnetic field for $n_H = n_B = -5/2$ while the bottom panel shows the three-point correlation for $n_H = n_B = -1,9$. Here we can see that the contribution driven by $A_B A_H^2$ is bigger than $A_B^3$, which means that for negative spectral indices, the effect of helicity becomes relevant for our studies.

## 7.4. Reduced bispectrum from PMF

In this section, we estimate the reduced bispectrum and give a careful review the results of Refs. [200, 228, 229]. The CMB temperature perturbation at a direction of photon momentum $\hat{\mathbf{n}}$ can be expanded into spherical harmonics

$$\frac{\Delta T^{(Z)}}{T}(\hat{\mathbf{n}}) = \sum_{lm} a^{(Z)}_{lm} Y_{lm}(\hat{\mathbf{n}}), \tag{7-36}$$

where $Z = S, V, T$ refers to the contribution given by scalar, vector or tensor perturbations. The coefficient $a^{(Z)}_{lm}$ is written as [9]

$$a^Z_{lm} = 4\pi(-i)^l \int \frac{d^3\mathbf{k}}{(2\pi)^3} \Delta^Z_l(k) \sum_\lambda [sgn(\lambda)]^\lambda \xi^{(\lambda)}_{lm}(k),$$
$$\xi^{(\lambda)}_{lm}(k) = \int d^2\hat{\mathbf{k}} \xi^\lambda(k) {}_{-\lambda}Y^*_{lm}(\hat{\mathbf{k}}), \tag{7-37}$$

where $\lambda = 0, \pm 1, \pm 2$ describes the helicity of the scalar, vector, tensor mode; ${}_{-\lambda}Y^*_{lm}$ is the spin-weight spherical harmonics; $\xi^\lambda(k)$ is the primordial perturbation and $\Delta^Z_l(k)$ is the transfer

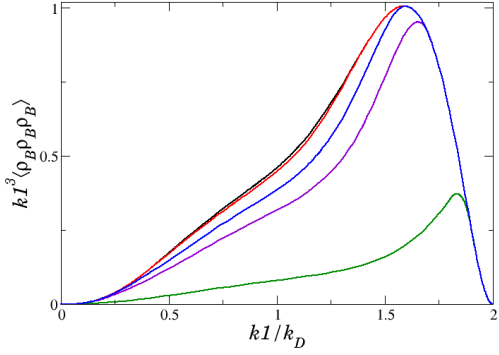
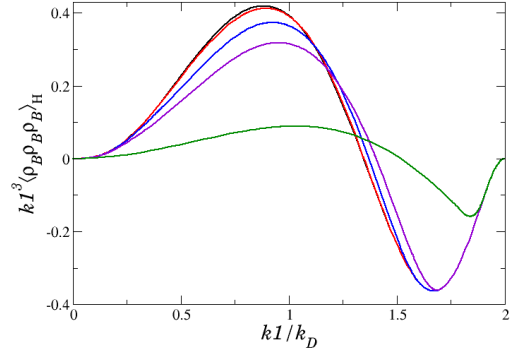

(a) Change of $k1^3\langle\rho_B\rho_B\rho_B\rangle$ respect to the infrared cut-off without $A_B A_H^2$.

(b) Change of $k1^3\langle\rho_B\rho_B\rho_B\rangle$ respect to the infrared cut-off only with $A_B A_H^2$.

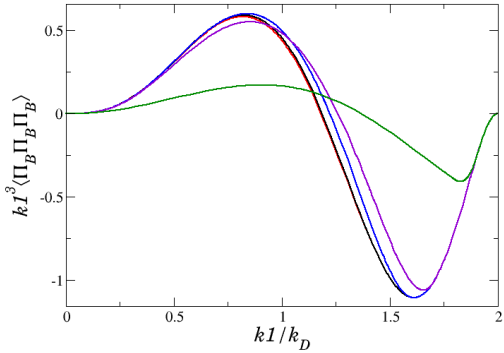
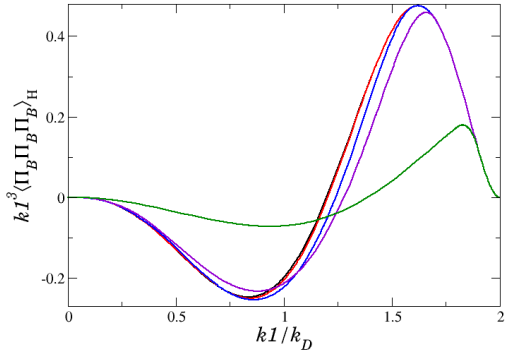

(c) Change of $k1^3\langle\Pi_B\Pi_B\Pi_B\rangle$ respect to the infrared cut-off without $A_B A_H^2$.

(d) Change of $k1^3\langle\Pi_B\Pi_B\Pi_B\rangle$ respect to the infrared cut-off only with $A_B A_H^2$.

**Figure 7-6**.: Effects of a lower cut-off at the three-point correlation of non-crossing scalar modes described in the text using the squeezed collinear configuration. Figures (a), (b) show the three-point correlation of the energy density of the magnetic field without and only with $A_B A_H^2$ respectively, while Figures (c) and (d) show the three-point correlation of the anisotropic stress of the magnetic field without and only with $A_B A_H^2$ respectively. The black, red, blue, violet and green lines refer to lower cut-off for $\alpha = 0,01$, $\alpha = 0,4$, $\alpha = 0,6$, $\alpha = 0,7$, $\alpha = 0,9$ respectively. Here the units are normalized respect to the values in Figure (a) and we use $n_B = n_H = 2$.

function. Let us define the CMB angular bispectrum as

$$B_{l_1\,l_2\,l_3}^{m_1\,m_2\,m_3} = \left\langle \prod_{n=1}^{3} a_{l_n\,m_n}^{(Z)} \right\rangle, \tag{7-38}$$

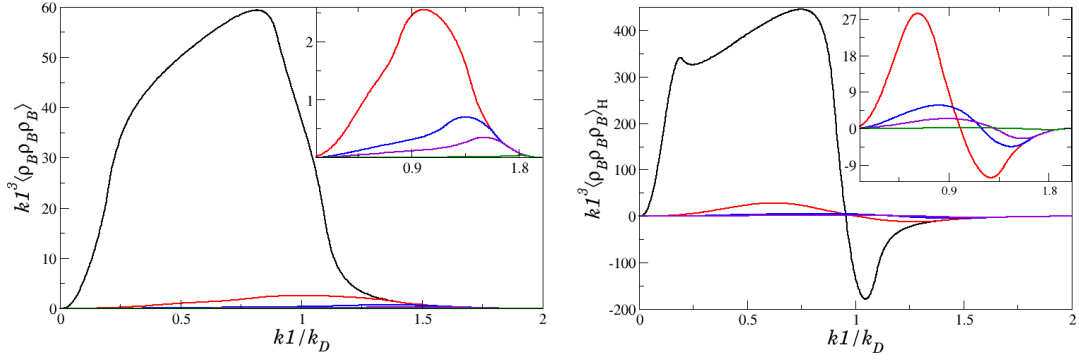

(a) Change of $k1^3 \langle \rho_B \rho_B \rho_B \rangle$ respect to the infrared cut-off without $A_B A_H^2$.

(b) Change of $k1^3 \langle \rho_B \rho_B \rho_B \rangle$ respect to the infrared cut-off only with $A_B A_H^2$.

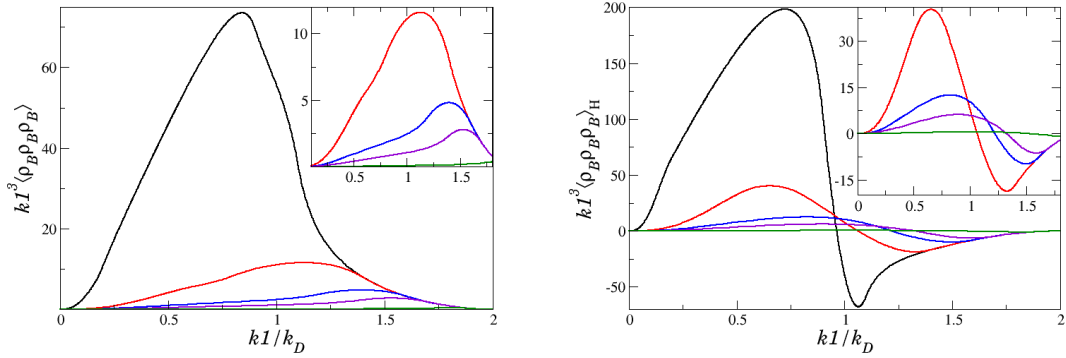

(c) Change of $k1^3 \langle \rho_B \rho_B \rho_B \rangle$ respect to the infrared cut-off without $A_B A_H^2$.

(d) Change of $k1^3 \langle \rho_B \rho_B \rho_B \rangle$ respect to the infrared cut-off only with $A_B A_H^2$.

**Figure 7-7**.: Effects of a lower cut-off at the three-point correlation of non-crossing scalar modes described in the text using the squeezed collinear configuration. Figures (a), (b) show the three-point correlation of the energy density of the magnetic field for $n_B = n_H = -5/2$ without and only with $A_B A_H^2$ respectively, while Figures (c) and (d) show the three-point correlation of the energy density of the magnetic field for $n_B = n_H = -1,9$ without and only with $A_B A_H^2$ respectively. The black, red, blue, violet and green lines refer to lower cut-off for $\alpha = 0,01$, $\alpha = 0,4$, $\alpha = 0,6$, $\alpha = 0,7$, $\alpha = 0,9$ respectively. Here the units are normalized respect to the values in Figure (**7-6**a).

where only scalar perturbations (Z=S) will be considered in the paper. Now, by substituting (7-37) into (7-38) we can find

$$\left\langle \prod_{n=1}^{3} a_{l_n m_n} \right\rangle = \left[ \prod_{n=1}^{3} 4\pi(-i)^{l_n} \int \frac{d^3 \mathbf{kn}}{(2\pi)^3} \Delta_{l_n}(kn) Y^*_{l_n m_n}(\hat{\mathbf{n}}) \right] \times \left\langle \prod_{n=1}^{3} \xi(kn) \right\rangle. \qquad (7\text{-}39)$$

We also consider a rough approximation for the transfer function that works quite well at large angular scales and for primordial adiabatic perturbations given by $\Delta_l(k) = \frac{1}{3}j_l(k(\tau_0 - \tau_*))$, where $j_l(x)$ the spherical Bessel function, $\tau_0 = 14{,}38$ Gpc the conformal time at present and $\tau_* = 284{,}85$ Mpc the conformal time at the recombination epoch [229]. This is the large scale Sachs Wolfe effect. Since we want to evaluate the contribution of the bispectrum by PMFs, the three-point correlator for the primordial perturbation must satisfy the relation

$$\left\langle \prod_{n=1}^{3} \xi(kn) \right\rangle = A_P \delta(\mathbf{k1}+\mathbf{k2}+\mathbf{k3}) B^{(S)}_{\Pi_B \Pi_B \Pi_B}. \tag{7-40}$$

Here, $A_P$ is a constant that depends on the type of perturbation (passive or compensated magnetic mode) and $B^{(S)}_{\Pi_B \Pi_B \Pi_B}$ is the magnetic bispectrum computed above. Using the following relations [228]

$$\delta(\mathbf{k1}+\mathbf{k2}+\mathbf{k3}) = \frac{1}{(2\pi)^3} \int_{-\infty}^{\infty} \exp^{i(\mathbf{k1}+\mathbf{k2}+\mathbf{k3})\cdot\mathbf{x}} d^3x, \tag{7-41}$$

$$\exp^{i\mathbf{k}\cdot\mathbf{x}} = 4\pi \sum_l i^l j_l(kx) \sum_m Y_{lm}(\hat{\mathbf{k}}) Y^*_{lm}(\hat{\mathbf{x}}), \tag{7-42}$$

with the Gaunt integral $\mathcal{G}^{l_1 l_2 l_3}_{m_1 m_2 m_3}$ defined by

$$\begin{aligned}\mathcal{G}^{l_1 l_2 l_3}_{m_1 m_2 m_3} &\equiv \int d^2\hat{\mathbf{n}} Y_{l_1 m_1}(\hat{\mathbf{n}}) Y_{l_2 m_2}(\hat{\mathbf{n}}) Y_{l_3 m_3}(\hat{\mathbf{n}}) \\ &= \sqrt{\frac{(2l_1+1)(2l_2+1)(2l_3+1)}{4\pi}} \begin{pmatrix} l_1 & l_2 & l_3 \\ 0 & 0 & 0 \end{pmatrix} \begin{pmatrix} l_1 & l_2 & l_3 \\ m_1 & m_2 & m_3 \end{pmatrix},\end{aligned} \tag{7-43}$$

and along with Eq.(7-40), the expression (7-39) takes the form

$$\begin{aligned}\left\langle \prod_{n=1}^{3} a_{l_n m_n} \right\rangle &= A_P \sqrt{\frac{(2l_1+1)(2l_2+1)(2l_3+1)}{4\pi}} \begin{pmatrix} l_1 & l_2 & l_3 \\ 0 & 0 & 0 \end{pmatrix} \begin{pmatrix} l_1 & l_2 & l_3 \\ m_1 & m_2 & m_3 \end{pmatrix} \times \\ &\quad \times \left[ \prod_{n=1}^{3} \frac{1}{3\pi^2} \int kn^2 \int j_{l_n}(kn\,x) j_{l_n}(kn(\tau_0-\tau_*)) dkn \right] B^{(S)}_{\Pi_B \Pi_B \Pi_B} x^2 dx.\end{aligned} \tag{7-44}$$

See Appendix E and Ref. [9] for more details about Wigner 3-j symbols. Given the rotational invariance, Komatsu-Spergel [228] defined a real symmetric function of $l_i$ called the reduced bispectrum $b_{l_1 l_2 l_3}$

$$\left\langle \prod_{n=1}^{3} a_{l_n m_n} \right\rangle \equiv \mathcal{G}^{l_1 l_2 l_3}_{m_1 m_2 m_3} b_{l_1 l_2 l_3}. \tag{7-45}$$

Checking the last two equations, the properties of the bispectrum generated by PMFs can be expressed via the reduced bispectrum as

$$b_{l_1 l_2 l_3} = A_P \left[ \prod_{n=1}^{3} \frac{1}{3\pi^2} \int kn^2 \int j_{l_n}(kn\,x) j_{l_n}(kn(\tau_0-\tau_*)) dkn \right] B^{(S)}_{\Pi_B \Pi_B \Pi_B} x^2 dx. \tag{7-46}$$

In order to calculate $A_P$, we must clarify the sources of primordial perturbations. Prior to neutrino decoupling ($\tau_\nu = 1\text{MeV}^{-1}$), the Universe is dominated by radiation and it is tightly coupled to baryons such that they cannot have any anisotropic stress contribution. Since we are also considering magnetic fields, they would be the only ones that develop anisotropic stress and therefore, at superhorizon scales the curvature perturbation depends on the primordial magnetic source [9]. But after neutrino decoupling, neutrinos generated anisotropic stress which compensates the one coming from PMF finishing the growth of the perturbations. Shaw-Lewis [85] showed that curvature perturbation is given by

$$\xi(k) \sim -\frac{1}{3}R_\gamma \ln\left(\frac{\tau_\nu}{\tau_B}\right)\Pi_B^{(S)}(k), \tag{7-47}$$

commonly known as passive mode, where $R_\gamma = \frac{\rho_\gamma}{\rho} \sim 0.6$ and $\tau_B$ is the epoch of magnetic field generation. Another contribution comes from the density-sourced mode with unperturbed anisotropic stresses, the magnetic compensated scalar mode, this is proportional to the amplitude of the perturbed magnetic density just as the magnetic Sachs Wolfe effect [85]. So, if the primordial perturbation is associated with the initial gravitational potential, in the limit on large-angular scales the compensated modes are expressed as [200, 230]

$$\xi(k) \sim \frac{1}{4}R_\gamma \rho_B(k). \tag{7-48}$$

Therefore if we use the passive mode contribution, $B^{(S)}_{\Pi_B\Pi_B\Pi_B}$ is given by $B^{(S)}_{\Pi_B^{(S)}\Pi_B^{(S)}\Pi_B^{(S)}}$ (see Eq.(7-24)) with $A_P = \left(-\frac{1}{3}R_\gamma \ln\left(\frac{\tau_\nu}{\tau_B}\right)\right)^3$, whilst compensated mode the primordial three-point correlation is described by $B^{(S)}_{\rho_B\rho_B\rho_B}$ (see Eq.(7-18)) with $A_P = \left(\frac{1}{4}R_\gamma\right)^3$. Since the magnetic bispectrum only depends on $(k1; k2)$, the $k3$ integral in the equation (7-46) gives $\frac{\pi}{2x^2}\delta(x - (\tau_0 - \tau_*))$ due to the closure relation [200], and integrating out the delta function one finally obtains

$$b_{l_1 l_2 l_3} = A_P \frac{\pi}{2}\left[\prod_{n=1}^{2}\frac{1}{3\pi^2}\int kn^2 j_{l_n}(kn(\tau_0 - \tau_*))^2 dkn\right]B^{(S)}_{\Pi_B\Pi_B\Pi_B}. \tag{7-49}$$

This is the master formula that we shall use in the following section in order to calculate the CMB reduced bispectrum.

## 7.5. Analysis

In this section we show the numerical results of the CMB reduced bispectrum produced by helical PMFs. In order to solve numerically Eq.(7-49), we use the adaptive strategy implemented in Mathematica called Levin-type rule which estimates the integral of an oscillatory function with a good accuracy [8].

### 7.5.1. Causal fields

Figure **7-8** presents the CMB reduced bispectrum generated by compensated PMFs modes under collinear configuration. In Figure (**7-8**a) we observe the signal produced by only the $A_B^3$ contribution, as well as the signal by the whole (**7-8**b). We found that $A_B A_H^2$ contribution (helical) is smaller than non-helical part $A_B^3$. Here we plot the change of the reduced bispectrum with regard to $l_1$, finding a large contribution for large values of $l_1$. We also see that helical contribution reaches a maximum around $l_2 \sim 400$ whilst non-helical contribution tends to increase at least until $l_2 \sim 500$. In Figures (**7-8**c) and (**7-8**d), the effect of an IR cut-off on the reduced bispectrum are shown. Each of these plots show the signal for different values of $l_1$. We see that signal is bigger for small $\alpha$ values (being the biggest contribution for spectrum without IR cut-off) similar to the one found with the power spectrum case [1].

In Figure (**7-9**) displays the CMB reduced bispectrum generated by PMFs passive modes under collinear configuration. Meanwhile, Figures (**7-9**a) and (**7-9**b) describe the signal for the non-helical and total contribution respectively. An interesting feature is that reduced bispectrum generated by helical contribution is totally negative, then, by using this unusual behavior we would have direct evidence of a helical component in the field.

Another important result of this paper is reported in Figs.**7-9**c and **7-9**d. Here we show again the effect of an IR cut-off on the reduced bispectrum and, we have found out that the biggest contribution of the bispectrum comes from an IR cut-off near $\alpha \sim 0,5$ instead of $\alpha = 0$. This peak might correspond to a type of dynamics in large scales and help us to determine the nature of PMFs (Since we are trying with causal fields, this infrared cut-off would correspond to the maximum scale in which magnetic fields may be generated at later times). Therefore the evidence of this cut-off in the bispectrum, would reveal an interesting signal from passive magnetic scalar mode. In addition, the change of the reduced bispectrum for helical PMFs in presence of an IR cut-off is shown in Fig.(**7-10**). Here we observe how the signal decreases when the IR cut-off increases for the compensated mode and how change the behavior for the passive case. We want to remark on some approximations used so far. Since we assume that the effects of PMFs are important for small multipolar numbers, we write the transfer functions in terms of spherical Bessel function. Previous papers have worked without this approximation. For instance, [216] computes the full radiation transfer function taking into account the effect of PMFs. The full numerical integration of the bispectrum can be done via second-order Einstein-Boltzmann codes like SONG [152] improving the estimation of the amplitude of PMFs. Moreover, we must note that [231] found a WMAP bound on non-helical passive mode for tensor temperature bispectrum of $B_{1Mpc} < 3,1$nG and the Planck paper [218] reported $B_{1Mpc} < 2,8$nG all of them for scale-invariant fields. Actually, the tensor mode is dominant in the passive mode, and can give quite tighter constraint of the PMF amplitude than the scalar mode. Thus, tensor mode contribution and a full transfer function determined by the presence of these fields will improve our results and for this reason, they will be interesting subjects of our future research.

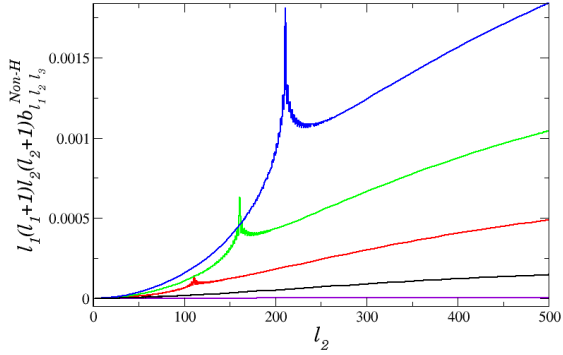
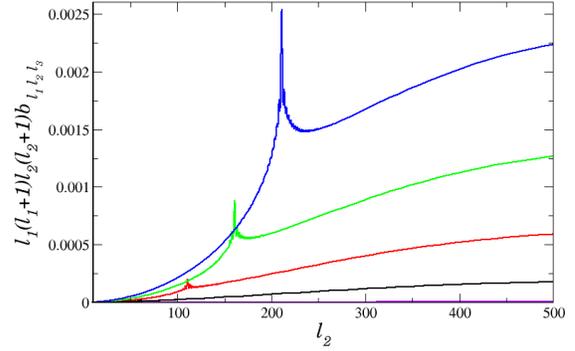

(a) Reduced bispectrum given by $A_B^3$ contribution of compensated PMFs.

(b) Reduced bispectrum given by total contribution of compensated PMFs.

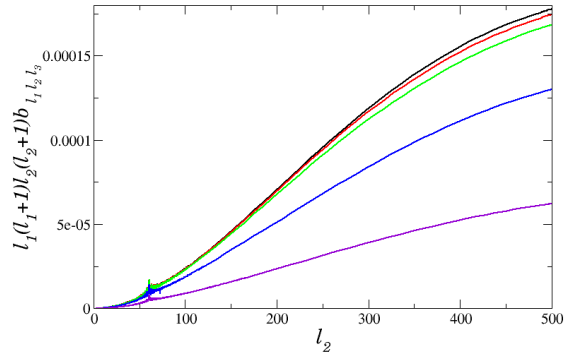
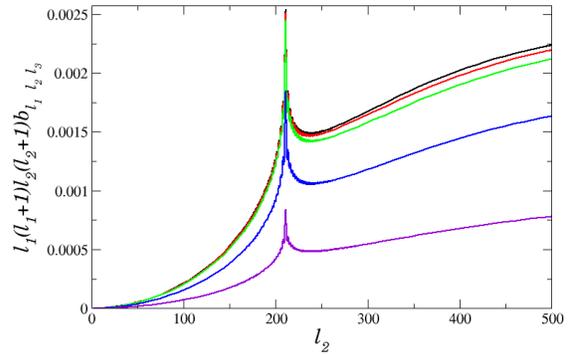

(c) Effects of infrared cut-off on the reduced bispectrum with $l_1 = 61$.

(d) Effects of infrared cut-off on the reduced bispectrum with $l_1 = 210$.

**Figure 7-8**.: Reduced bispectrum seeded by compensated PMFs with $n = 2$ using the squeezed collinear configuration. Figures (a) shows the reduced bispectrum of the magnetic field with only $A_B^3$, while Figure (b) shows the total contribution of the compensated mode; here the lines refers to different values of $l_1$, violet($l_1 = 11$), black($l_1 = 61$), red($l_1 = 110$), green($l_1 = 161$) and blue line($l_1 = 210$). Figures (c), (d) show the effects of an infrared cut-off on the reduced bispectrum for different values of multipolar numbers $l_1$. Black; red; green; blue; and violet lines refer to lower cut-off of $\alpha = 0{,}001$, $\alpha = 0{,}4$, $\alpha = 0{,}5$, $\alpha = 0{,}7$, $\alpha = 0{,}8$ respectively. The reduced bispectrum is in units of $4\pi 10^{-8} A_P/(8\pi^2 \rho_{\gamma,0})^3$.

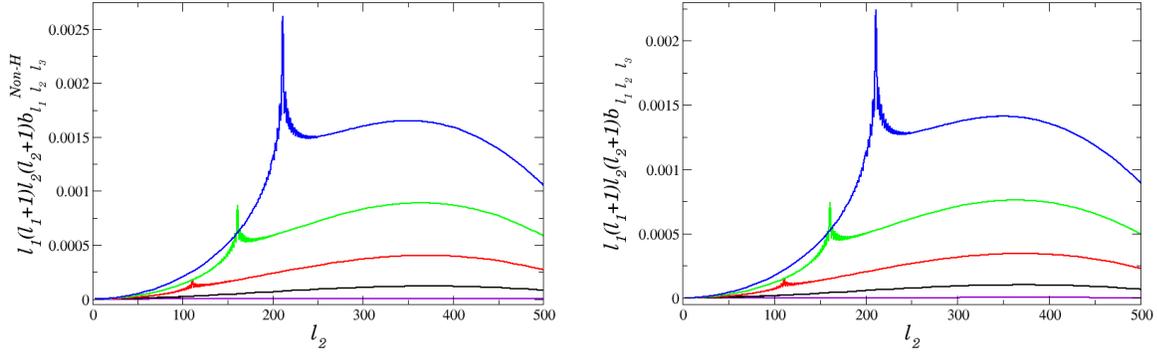

(a) Reduced bispectrum given by only $A_B^3$ contribution of passive PMFs.

(b) Reduced bispectrum given by total contribution of passive PMFs.

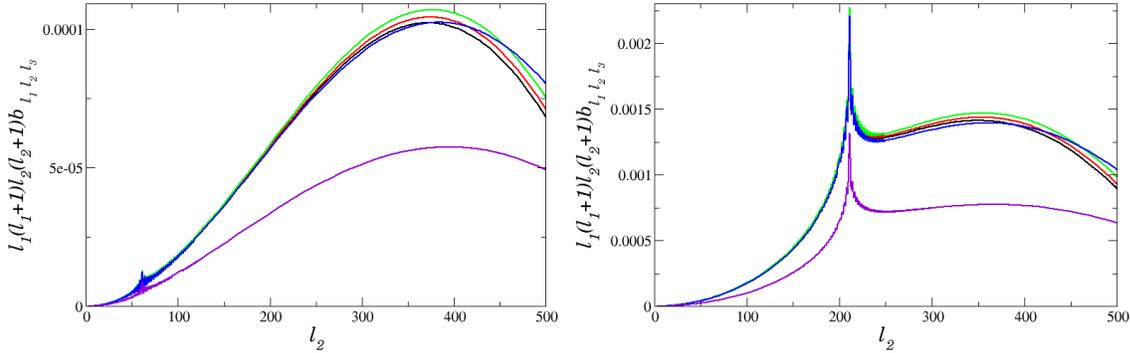

(c) Effects of infrared cut-off on the reduced bispectrum with $l_1 = 61$.

(d) Effects of infrared cut-off on the reduced bispectrum with $l_1 = 210$.

**Figure 7-9**.: Reduced bispectrum seeded by passive PMFs with $n_B = n_H = n = 2$ using the squeezed collinear configuration. Figure (a) shows reduced bispectrum of the magnetic field with only $A_B^3$, while Figure (b) shows the total contribution of the passive mode, here the lines refer to different values of $l_1$, violet($l_1 = 11$), black($l_1 = 61$), red($l_1 = 110$), green($l_1 = 161$) and blue line($l_1 = 210$). Figures (c) and (d) show the effects of an infrared cut-off on the reduced bispectrum for difference values of multipolar numbers $l_1$. Black; red; green; blue; and violet lines refer to lower cut-off of $\alpha = 0,001$, $\alpha = 0,4$, $\alpha = 0,5$, $\alpha = 0,7$, $\alpha = 0,8$ respectively. The reduced bispectrum is in units of $4\pi 10^{-8} A_P / 2(8\pi^2 \rho_{\gamma,0})^3$.

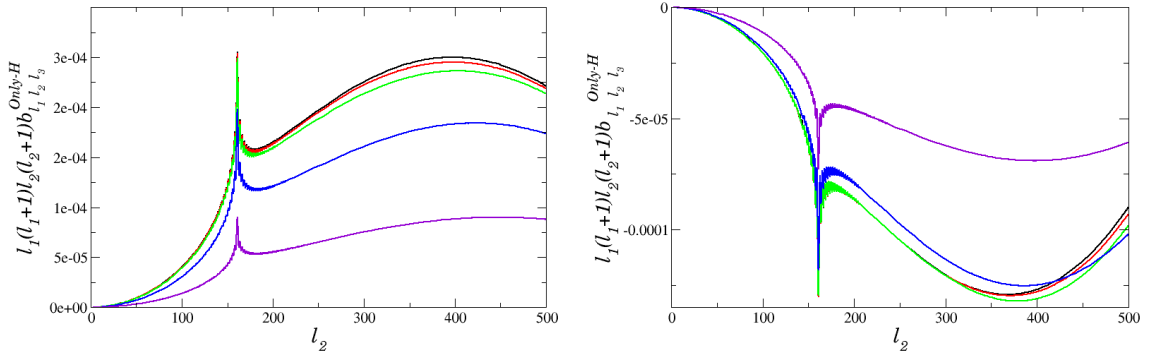

(a) Effects of infrared cut-off on the reduced bispectrum with $l_1 = 161$ seeded by total contribution of compensated helical PMFs.

(b) Effects of infrared cut-off on the reduced bispectrum with $l_1 = 161$ seeded by total contribution of passive helical PMFs.

**Figure 7-10**.: Reduced bispectrum seeded by compensated (a) and passive (b) helical PMFs with only $A_B A_H^2$ contribution using the squeezed collinear configuration. Black; red; green; blue; and violet lines refer to lower cut-off of $\alpha = 0,001$, $\alpha = 0,4$, $\alpha = 0,5$, $\alpha = 0,7$, $\alpha = 0,8$ respectively. The reduced bispectrum is in units of $4\pi 10^{-8} A_P/2(8\pi^2 \rho_{\gamma,0})^3$.

### 7.5.2. Non-causal fields

Let us consider a red magnetic spectrum. Fig.**7-11** shows the reduced bispectrum for compensated mode with $n_H = n_B = n = -5/2$. Again, we plot the non-helical **7-11**a and total **7-11**b contribution of the bispectrum while Figs. **7-11**c and **7-11**d correspond to the change of the signal due to an IR cut-off. Since the PMFs bispectrum is almost determined by the poles in each $k$, the value of it peaked for $l_1 = l_2 = l_3$ as we can observe in Figs.**7-11**, **7-12**, **7-13**. Additionally, Fig.**7-12** shows signals for $n_H = n_B = n = -3/2$ and $n_H = n_B = n = -1,9$. As a matter of fact, some mechanisms of inflationary magnetogenesis with parity violating terms which lead to helical magnetic field ($n = -1,9$), stand for the lower bound for which the field can satisfy the intensity of magnetic fields in the intergalactic medium; thereby the signal described in Fig.**7-12** constraints models for providing the seed for galactic magnetic fields [52]. On the other hand, Figure **7-13** shows the reduced bispectrum taking into account the p-independent approximation implemented in Sec.7.3.1. Due to complexity of the angular structure on the PMF bispectrum for passive mode, the numerical computation for reduced bispectrum requires a great deal of time. To avoid this problem we can use the p-independent approximation by reducing the PMF bispectrum to an independent angular form as we studied above. From Figs. **7-13**a and **7-13**b we observe an increase in the signal as we expected due to the lack of angular terms in the bispectrum. Finally, Figs.**7-13**c and **7-13**d show the total contribution of passive modes. Note that the amplitude is larger than the compensated mode. This behavior have also been reported in [200].

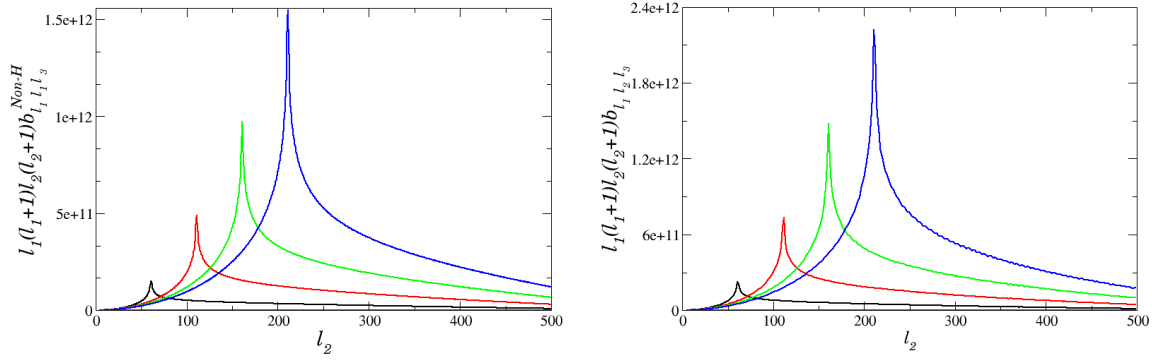

(a) Reduced bispectrum given by $A_B^3$ contribution of compensated PMFs.

(b) Reduced bispectrum given by total contribution of compensated PMFs.

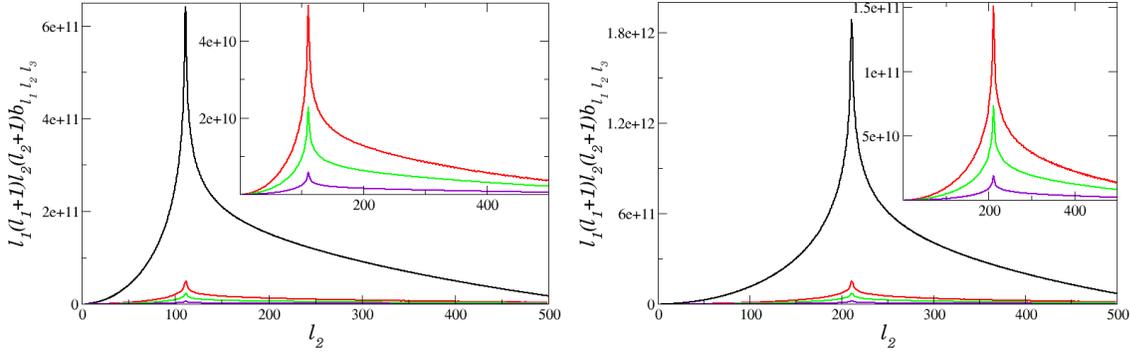

(c) Effects of infrared cut-off on the reduced bispectrum with $l_1 = 111$.

(d) Effects of infrared cut-off on the reduced bispectrum with $l_1 = 210$.

**Figure 7-11**.: Absolute value of reduced bispectrum seeded by compensated PMFs with $n = -5/2$ using the squeezed collinear configuration. Figure (a) shows reduced bispectrum of the magnetic field with only $A_B^3$ contribution, while Figure (b) shows the total contribution of the compensated mode; here the lines refers to different values of $l_1$, violet($l_1 = 11$), black($l_1 = 61$), red($l_1 = 110$), green($l_1 = 161$) and blue line($l_1 = 210$). Figures (c) and (d) show the effects of an infrared cut-off on the reduced bispectrum for difference values of multipolar numbers $l_1$. Black; red; green; and violet lines refer to lower cut-off of $\alpha = 0,001$, $\alpha = 0,4$, $\alpha = 0,6$, $\alpha = 0,8$ respectively. The reduced bispectrum is in units of $4\pi 10^{16} A_P/(8\pi^2 \rho_{\gamma,0})^3$.

This result is interesting because an estimation of $B_\lambda$ through a local-type primordial NG in curvature perturbation generates constraints stronger than the compensated ones. Finally, if we compare the results reported in this section with the ones shown in above section for causal fields, we can observe that effect of $k_m$ is more significant in negative spectral indices, specially for nearly scale invariant scale fields. We conclude that $k_m$ plays an important role

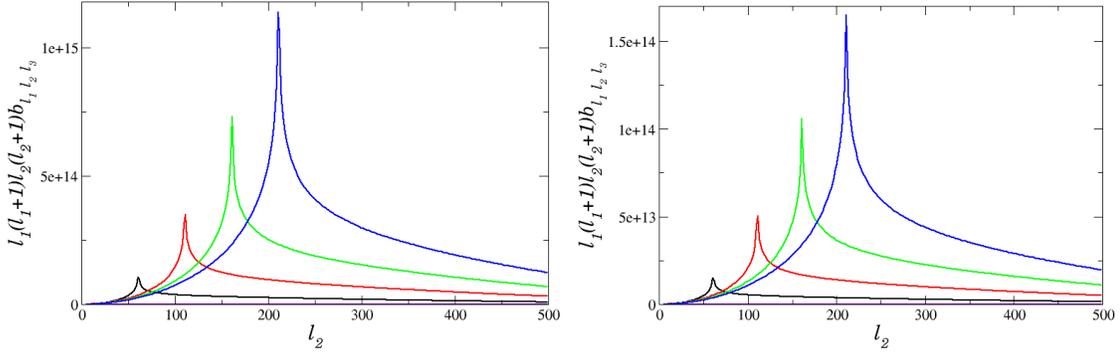

(a) Reduced bispectrum of compensated PMFs for $n = -3/2$.

(b) Reduced bispectrum of compensated PMFs for $n = -1{,}9$.

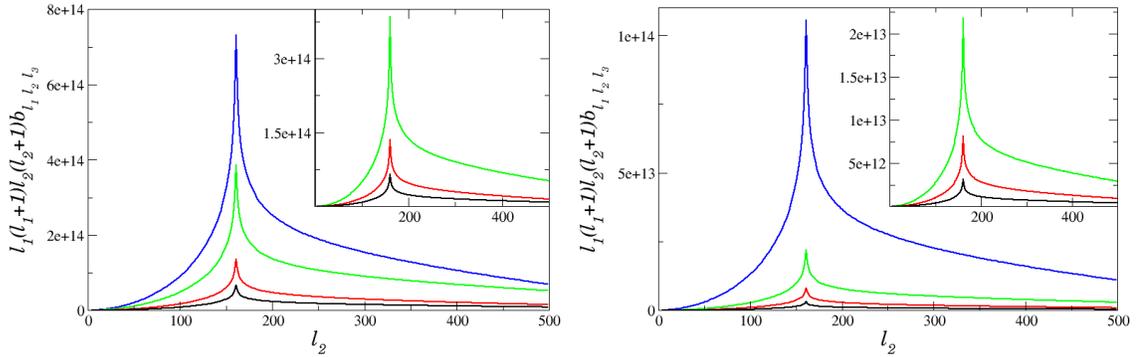

(c) Effects of infrared cut-off on the reduced bispectrum with $l_1 = 161$ for $n = -3/2$.

(d) Effects of infrared cut-off on the reduced bispectrum with $l_1 = 161$ for $n = -1{,}9$.

**Figure 7-12**.: Absolute values of reduced bispectrum seeded by compensated PMFs using the squeezed collinear configuration. Figures (a), (b) display the total contribution of the compensated mode of the magnetic field for $n = -3/2$ and $n = -1{,}9$ respectively; here the lines refer to different values of $l_1$, violet($l_1 = 11$), black($l_1 = 61$), red($l_1 = 110$), green($l_1 = 161$) and blue line($l_1 = 210$). The last Figures (c),(d) explain the effects of an infrared cut-off on the reduced bispectrum for difference values of multipolar numbers $l_1$. Blue; green; red; and black lines refer to lower cut-off of $\alpha = 0{,}001$, $\alpha = 0{,}4$, $\alpha = 0{,}6$, $\alpha = 0{,}8$, respectively. The bispectrum is in units of $4\pi 10^{16} A_P/(8\pi^2 \rho_{\gamma,0})^3$.

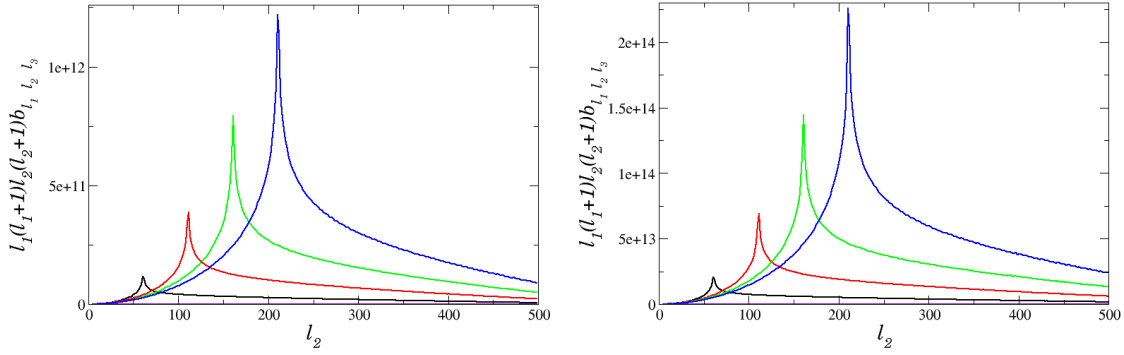

(a) Reduced bispectrum of compensated PMFs for $n = -5/2$.

(b) Reduced bispectrum of compensated PMFs for $n = -3/2$.

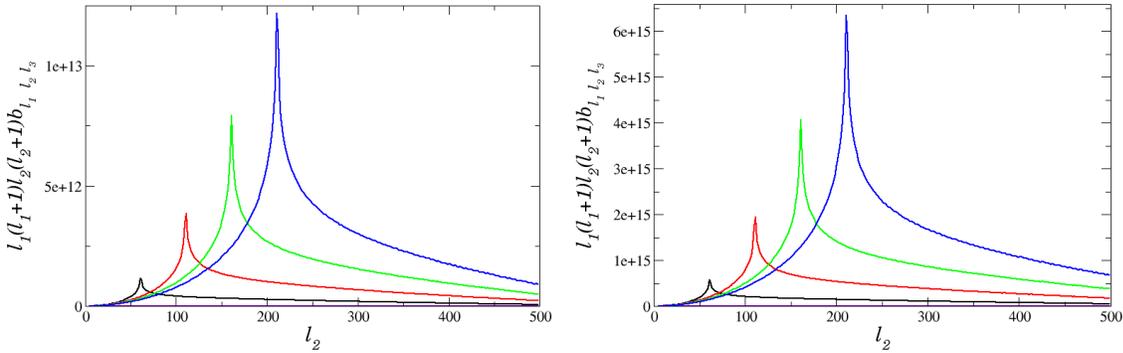

(c) Reduced bispectrum of passive PMFs for $n = -5/2$.

(d) Reduced bispectrum of passive PMFs for $n = -3/2$.

**Figure 7-13**.: Absolute values of reduced bispectrum seeded by passive and compensated PMFs under p-independent approximation. Figures (a), (b) present the total contribution of the compensated mode of the magnetic field for $n = -5/2$ and $n = -3/2$ respectively. Figures (c), (d) explain the total contribution of the passive mode of the magnetic field for $n = -5/2$ and $n = -3/2$ respectively. Here the lines refer to different values of $l_1$, violet($l_1 = 11$), black($l_1 = 61$), red($l_1 = 110$), green($l_1 = 161$) and blue line($l_1 = 210$).

in the study of these non-causal fields and this generates the possibility of determining some important clues in the mechanisms of magnetogenesis.

### 7.5.3. Estimation of the magnetic field amplitude

In fact, it is possible to obtain a rough estimate of $B_\lambda$ using the formula for the primary reduced bispectrum found in [228]

$$b_{lll} \sim l^{-4} \times 2 \times 10^{-17} f_{NL}, \tag{7-50}$$

where the non-Gaussianity (local) is fully specified by a single constant parameter $f_{NL}$. As we mentioned above, the k-dependence on the magnetic bispectrum is similar to the CMB bispectrum arising from the local type NG of curvature perturbations, therefore, by comparing the last equation with (7-49), allow us to express a simple relation between $F_{NL}$ and $B_\lambda$ given by

$$f_{NL} \propto \left(\frac{B_\lambda}{10^{-9}\text{G}}\right)^6. \tag{7-51}$$

In Table **7-2** we present the constant of proportionality of the last expression. Here we use $\frac{\tau_\nu}{\tau_B} = 10^{17}$ which corresponds to the PMF generated at the grand unification energy scale(GUT) scale, $\left(\frac{B_\lambda^2}{8\pi\rho_{\gamma,0}}\right) \sim 10^{-7} \left(\frac{B_\lambda}{10^{-9}\text{G}}\right)^2$, and $R_\gamma \sim 0,6$. In order to constrain the smoothed amplitude of the magnetic field on a scale of 1Mpc ($B_1$), we will use the $f_{NL}$ value reported by Planck Collaboration [12] of $f_{NL} < 5,8$ at $68\%$ CL. The results of $B_1$ are shown in Table **7-3**. Our results for compensated modes lead to upper bounds on the PMF smoothed amplitude which are consistent with the Planck analysis [218], but for passive modes our results are slightly tigher because, these were based on a rough estimation and may involve some uncertainties (except for the causal case where the bound coincides with Planck analysis), however notice that for passive modes the limits are almost 10 times more stringent than the compensated ones as It was reported in [200] for no-helical and scale invariant case, hence CMB-observation are sensitive to the magnetic induced modes. Since our results were obtained under the Sachs-Wolfe approximation, we expected a lower value of $B_1$ for causal fields respect to the non causal fields, and therefore the blue spectra generated by these fields is strongly disfavoured by the CMB bispectrum. On the other hand, we constrain $B_1$ through the helical contribution and we observed an enhance of its amplitude. We see this same effect in the two point correlation as was reported by Planck analysis $B_1 < 5,6$nG at $95\%$ CL [218] for that contribution. In the tables also show the bounds when a high value of the IR cut-off ($\alpha \sim 0,8$) is used. Since the cut-off reduces the amplitude magnetic bispectrum signal, the upper limit of $B_\lambda$ becomes somewhat relaxed and consequently, we are able to illustrate the impact $k_m$ could have on constraints on the PMF amplitude. Although this effect becomes very small for a tiny value of $k_m$, the presence of this scale in the analysis of NG are complementary to the ones found by the two point correlation case and will provide new insight into the nature of primordial magnetic fields.

|            | n=$-\frac{5}{2}$ |         | n=$-1,9$    | n=$-\frac{3}{2}$ |         | n= 2          |               |
|------------|------------------|---------|-------------|------------------|---------|---------------|---------------|
|            | Comp.            | Passive | Comp        | Comp.            | Passive | Comp          | Passive       |
| Helical    | 0.86(1.92)       | 0.12    | 0.47(0.80)  | 0.35(0.49)       | 0.008   | 0.021         | 0.0069        |
| N-Helical  | 1.04(2.43)       | 0.13    | 0.38        | 0.33             | 0.067   | 0.018         | 0.0050        |
| Total      | 0.92(1.85)       | 0.14    | 0.40(0.73)  | 0.38(0.44)       | 0.064   | 0.017(0.021)  | 0.0051(0.006) |

**Table 7-2**.: Constant of proportionality of the Eq.(7-51) for different spectral indices and modes(compensated(comp) or passive) of the PMF without considering an IR cut-off. Parentheses are used to represent this value for $\frac{k_m}{k_D} \sim 0,8$.

|            | n=$-\frac{5}{2}$ |         | n=$-1,9$    | n=$-\frac{3}{2}$ |         | n= 2          |                 |
|------------|------------------|---------|-------------|------------------|---------|---------------|-----------------|
|            | Comp.            | Passive | Comp        | Comp.            | Passive | Comp          | Passive         |
| Helical    | 1.15(2.58)       | 0.16    | 0.64(1.07)  | 0.47(0.66)       | 0.098   | 0.029         | 0.0092          |
| N-Helical  | 1.39(3.25)       | 0.17    | 0.52        | 0.44             | 0.089   | 0.024         | 0.0068          |
| Total      | 1.24(2.48)       | 0.19    | 0.54(0.98)  | 0.52(0.59)       | 0.085   | 0.023(0.029)  | 0.0068(0.0083)  |

**Table 7-3**.: Bound on smoothed amplitude of the magnetic field on a scale of 1Mpc ($B_1$ in units of nG) for different spectral indices and modes(compensated(comp) or passive) of the PMF without considering an IR cut-off. Parentheses are used to represent $B_1$ for $\frac{k_m}{k_D} \sim 0,8$. Here we use $f_{NL} < 5,8$ reported by Planck Collaboration, 2016 [12].

In this thesis, we investigate the effects of helical PMFs in the CMB reduced bispectrum. One of the main motivations to introduce the helicity comes from the fact that these fields are good observables to probe parity-violation in the early stages of the Universe. Furthermore, since magnetic fields depend quadratically on the field, it must induce NG signals on CMB anisotropies at lower order instead of the standard inflationary mechanisms where this signal appears only at high orders [213]. We started our work deriving the full even and odd parts of the bispectrum which comes from the helical magnetic fields, thus extending the previous results reported in [202]. We obtained the full expression for the PMF bispectrum but, we did not consider modes that arise from odd intensity-intensity-intensity bispectrum. Although these signals are smaller than the even ones, the evidence of the odd signals would be decisive observable to probe parity-violating processes in the early Universe [227]. We will provide more details of the parity-odd signals in a future paper. Then, through the methodology used in [200], [226], we found that PMFs bispectrum peaks at $k1 \sim k2$ under a squeezed configuration implying that statistical properties of the PMFs are similar to those of the local-type NG of curvature perturbations. By calculating the bispectrum given by PMFs anisotropic stress, we observed that its amplitude is larger than the density one and also has a negative contribution for values less than $k_D$. Through numerical calculations of the intensity-intensity-intensity reduced bispectra of the scalar modes, we also studied the total contributions of the helical PMF bispectrum and the presence of an IR cut-off in the convolution integrals. Here we observe the same behavior seen in the power spectral case and the presence of negative contribution due to helical terms $A_B A_H^2$. Nevertheless, in the computation in the passive modes for causal fields we observed an unusual behavior in the bispectrum. Indeed, in Figure (**7-9**) we have found out that biggest contribution of the bispectrum comes from an IR cut-off near to $\alpha \sim 0{,}5$ instead of $\alpha = 0$ . Since $k_m$ is dependent on PMF generation model, this behavior might set strong limits on PMF amplitude. Finally, we investigated the effects of $k_m$ on the reduced bispectrum for $n < 0$. Due to the fact that the magnetic field intensity can be enhanced when we use passive modes, it is expected that those modes determine a very strong constraints on the amplitude of the magnetic field on a given characteristic scale $\lambda$. We verify this statement by using the primary reduced bispectrum found in [228] and calculating $B_\lambda$. Our results showed in Tables **7-2**, **7-3** reflect the fact that the corresponding bound on the mean amplitude of the field is dependent on strong values of the minimal cut-off and the helical contribution relaxing the constraints of $B_\lambda$. We also found that for passive modes the limits are almost 10 times more stringent than the compensated ones for both helical and hon helical contribution, this result was also reported in [213] for non helical fields. However, we can observe that effect of $k_m$ is more significant in the magnetic bispectrum driven by negative spectral indices. Hence, the presence of $k_m$ plays an important role in the analysis of the signatures that these non-causal fields may leave in cosmological observations.

In conclusion, we have studied the effects of helicity and a minimal cut-off on the constraints of the PMFs amplitude by computing the CMB reduced bispectrum induced on large angular scales by those fields. Even though $k_m$ for causal modes would be important when this scale is larger than the wavenumber of interest, for non-causal modes, It is related to the horizon scale of the beginning of inflation [1,232], and thus, the study of this cut-off on the bispectrum give us information about the PMF generation mechanisms.

# 8. Conclusions

In this thesis, we have presented the effects on the CMB anisotropy due to primordial magnetic fields and analyzed some favorable scenarios of magnetogenesis constrained by those signatures, including limits on the amplitude of the fields from bounds on CMB non-Gaussianity and background models. We started describing the presence of sharp cut-offs at the convolution integrals and showed the integration scheme used for calculating the spectra and bispectra for any stochastic field. The integration domain reported in Sec.4, generalizes the previous results found in literature and thus, they become an original result of this thesis. With this scheme, we calculated the magnetic spectra and bispectra and displayed the effects of these fields on the CMB anisotropies. The integration scheme is quite general and can be applied to numerous convolution integrals.

Additionally, we have investigated the effects of helical PMFs in the CMB reduced bispectrum. One of the main motivations to introduce the helicity comes from the fact that these fields are good observables to probe parity-violation in the early stages of the Universe. Furthermore, since magnetic fields depend quadratically on the field, it must induce NG signals on CMB anisotropies at lower order instead of the standard inflationary mechanisms where this signal appears only at high orders [213]. We derived the full even and odd parts of the bispectrum which comes from the helical magnetic fields, thus extending the previous results reported in [202] and become another original result of this thesis. We obtained the full expression for the PMF bispectrum but, we did not consider modes that arise from odd intensity-intensity-intensity bispectrum. Although these signals are smaller than the even ones, the evidence of the odd signals would be decisive observable to probe parity-violating processes in the early Universe [227]. We will provide more details of the parity-odd signals in a future work.

Then, through the methodology used in [200], [226], we found that PMFs bispectrum peaks at $k1 \sim k2$ under a squeezed configuration implying that statistical properties of the PMFs are similar to those of the local-type NG of curvature perturbations. By calculating the bispectrum given by PMFs anisotropic stress, we observed that its amplitude is larger than the density one and also has a negative contribution for values less than $k_D$. Through numerical calculations of the intensity-intensity-intensity reduced bispectra of the scalar modes, we also studied the total contributions of the helical PMF bispectrum and the presence of an IR cut-off in the convolution integrals. In Fig.**7-9** we have found out that biggest contribution of the bispectrum comes from an IR cut-off near to $\alpha \sim 0,5$ instead of $\alpha = 0$ . Since $k_m$ is dependent on PMF generation model, this behavior might set strong limits on PMF ampli-tude. Finally, we investigated the effects of $k_m$ on the reduced bispectrum for $n < 0$. Due to the fact that the magnetic field intensity can be enhanced when we use passive modes, it is expected that those modes determine a very strong constraints on the amplitude of the magnetic field on a given characteristic scale $\lambda$. We verify this statement by using the primary reduced bispectrum found in [228] and calculating $B_\lambda$. Our results showed in Tables **7-2**, **7-3** reflect the fact that the corresponding bound on the mean amplitude of the field is dependent on strong values of the minimal cut-off and the helical contribution relaxing the constraints of $B_\lambda$. We also found that for passive modes the limits are almost 10 times more stringent than the compensated ones for both helical and hon helical contribution. However, we can observe that effect of $k_m$ is more significant in the magnetic bispectrum driven by negative spectral indices. Hence, the presence of $k_m$ plays an important role in the analysis of the signatures that these non-causal fields may leave in cosmological observations. Even though $k_m$ for causal modes would be important when this scale is larger than the wavenumber of interest, for non-causal modes, It is related to the horizon scale of the beginning of inflation [1, 232], and thus, the study of this cut-off on the bispectrum give us information about the PMF generation mechanisms.

We also studied in detail the cosmological perturbation theory which has become an important tool in theoretical cosmology to link scenarios of the early universe with cosmological data such as CMB-fluctuations. We have analyzed different approaches that have been developed to manage the gauge problem: 1+3 covariant gauge invariant and the gauge invariant approaches. Following some results shown in [129] and [133], we have contrasted these formalisms comparing their gauge invariant variables defined in each case. Using a magnetic scenario, we have shown a strong relation between both formalisms, indeed, we found out that gauge invariant defined by 1+3 covariant approach is related to spatial variations of the magnetic field energy density (variable defined in the invariant gauge formalism) between two closed fundamental observers as it is noticed in Eqs.(3-144), (3-154) and (3-162). Moreover, we have also derived the gauge transformations for electromagnetic potentials, Eqs.(3-113) and (3-114), which are relevant in the study of evolution of primordial magnetic fields in scenarios such as inflation or later phase transitions become an original result of this thesis too. With the description of the electromagnetic potentials, we have expressed the Maxwell's equations in terms of these ones, finding again an important coupling with the gravitational potentials. As a future research, we expect to calculate the vector and tensor contribution modes in the bispectrum and trispectrum and be able to constrain the amplitude of the fields from bounds on CMB non-Gaussianity driven by future CMB experiments. We will also expect to analyze our results including the full transfer functions in order to get more precision in our findings and summit the code which allowed us to calculate the spectra with both IR and UV cutoff's.

# A. General Relativity notation

We consider a 4-dimensional pseudo-Riemannian spacetime and the metric $g$ is given by the ten components of a $4 \times 4$ symmetry tensor

$$ds^2 = g_{\mu\nu} dx^\mu dx^\nu. \tag{A-1}$$

This metric is also used for raising and lowering the indices of any tensor field, e.g.,

$$X^\alpha = g^{\alpha\nu} X_\nu, \quad X_\alpha = g_{\alpha\nu} X^\nu, \quad g_{\alpha\nu} g^{\alpha\mu} = \delta^\mu_\nu. \tag{A-2}$$

The Christoffel symbols are defined as

$$\Gamma^\mu_{\alpha\beta} = \frac{1}{2} g^{\mu\nu} \left( \partial_\alpha g_{\nu\beta} + \partial_\beta g_{\nu\alpha} - \partial_\nu g_{\alpha\beta} \right), \tag{A-3}$$

where $\partial_\alpha$ indicates a partial derivative with respect to the coordinate $x^\alpha$. The covariant derivatives of a tensor field are given by

$$\nabla_\mu X^{\alpha_1 \alpha_2 \cdots \alpha_n}_{\beta_1 \beta_2 \cdots \beta_m} = \partial_\mu X^{\alpha_1 \alpha_2 \cdots \alpha_n}_{\beta_1 \beta_2 \cdots \beta_m} + \Gamma^{\alpha_1}_{\mu\sigma} X^{\sigma \alpha_2 \cdots \alpha_n}_{\beta_1 \beta_2 \cdots \beta_m} + \cdots - \Gamma^\sigma_{\mu\beta_1} X^{\alpha_1 \alpha_2 \cdots \alpha_n}_{\sigma \beta_2 \cdots \beta_m}, \tag{A-4}$$

and any tensor field $X$ is parellel transported along the vector field $\mathcal{X}$ if

$$\mathcal{X}^\mu \nabla_\mu X^{\alpha_1 \alpha_2 \cdots \alpha_n}_{\beta_1 \beta_2 \cdots \beta_m} = 0, \tag{A-5}$$

where $\mathcal{X}$ satisfies the geodesic equation

$$\frac{d^2 \mathcal{X}^\mu}{ds^2} + \Gamma^\mu_{\alpha\beta} \frac{d\mathcal{X}^\alpha}{ds} \frac{d\mathcal{X}^\beta}{ds} = 0. \tag{A-6}$$

The Riemann tensor is defined by

$$R^\alpha_{\beta\mu\nu} = \partial_\mu \Gamma^\alpha_{\nu\beta} - \partial_\nu \Gamma^\alpha_{\mu\beta} + \Gamma^\rho_{\nu\beta} \Gamma^\alpha_{\mu\rho} - \Gamma^\rho_{\mu\beta} \Gamma^\alpha_{\nu\rho}, \tag{A-7}$$

and we can write the Ricci tensor $R_{\mu\nu}$ and Ricci scalar $R$ as

$$R_{\mu\nu} \equiv R^\alpha_{\mu\alpha\nu} = \partial_\sigma \Gamma^\sigma_{\nu\mu} - \partial_\nu \Gamma^\sigma_{\mu\sigma} + \Gamma^\rho_{\rho\beta} \Gamma^\beta_{\mu\nu} - \Gamma^\rho_{\mu\beta} \Gamma^\beta_{\nu\rho}, \quad R \equiv R_{\mu\nu} g^{\mu\nu}. \tag{A-8}$$

With these conventions, we can define the Einstein tensor $G_{\mu\nu}$ as

$$G_{\mu\nu} \equiv R_{\mu\nu} - \frac{1}{2} R g_{\mu\nu}, \tag{A-9}$$

and write the Einstein's equation as we can see in Eq.(2-11). Using the FLRW metric (2-3), the Christoffel symbols are

$$\Gamma^0_{00} = \frac{a'}{a}, \tag{A-10}$$

$$\Gamma^0_{i0} = \Gamma^0_{0i} = 0, \tag{A-11}$$

$$\Gamma^0_{ij} = \frac{a'}{a}\delta_{ij}, \tag{A-12}$$

$$\Gamma^i_{00} = 0, \tag{A-13}$$

$$\Gamma^i_{k0} = \Gamma^i_{0k} = \frac{a'}{a}\delta^i_k, \tag{A-14}$$

$$\Gamma^i_{jk} = 0. \tag{A-15}$$

While non-vanishing components of the Ricci tensor and scalar are respectively

$$R_{00} = -3\left(\frac{a''}{a} - \left(\frac{a'}{a}\right)^2\right), \tag{A-16}$$

$$R_{ij} = \delta_{ij}\left(\frac{a''}{a} + \left(\frac{a'}{a}\right)^2\right), \tag{A-17}$$

$$R = 6a^{-2}\left(\frac{a''}{a}\right). \tag{A-18}$$

# B. Framework for PMFs generation during inflation

In this Appendix we show some details related to generation of helical PMFs during inflation.

## B.1. Quantisation of the vector potential

Firstly, we promote the vector potential $A_i$ to a quantum operator and impose the commutation relation 5-11. Now, It is convenient to make a decomposition of $A_i$ in terms of plane waves with comoving wave vector $k^i$ on a linear polarization basis defined as [166]

$$\epsilon_0^\mu = \left(\frac{1}{a}, 0\right), \quad \epsilon_\lambda^\mu = \left(0, \frac{\tilde{\epsilon}_\lambda^i}{a}\right), \quad \epsilon_3^\mu = \left(0, \frac{1}{a}\frac{k^i}{k}\right), \tag{B-1}$$

with $\lambda = 1, 2$ and by definition, $\delta_{ij}\tilde{\epsilon}_\lambda^i \tilde{\epsilon}_\lambda^j \equiv 1$ (no summation on $\lambda$). The third vector $\lambda = 3$ is constructed such that its spatial component points in the direction $\hat{k}$ [172]. So the completeness relation can be written as

$$\sum_{\lambda=1}^{2} \epsilon_\lambda^i(k)\epsilon_{j\lambda}(k) + \delta_{j\ell}\frac{k^i k^\ell}{k^2} = \delta^i{}_j, \tag{B-2}$$

and we can impose the conditions [172]

$$\hat{k}^i \tilde{\epsilon}(k)_{\lambda i} = 0, \quad \tilde{\epsilon}(k)_{\lambda i}\tilde{\epsilon}(k)^i_{\lambda'} = \delta_{\lambda\lambda'}, \quad \epsilon_{ijl}\tilde{\epsilon}(k)^j_1 \tilde{\epsilon}(k)^l_2 = \hat{k}_i, \quad |\tilde{\epsilon}(k)_{\lambda i}|^2 = 1,$$
$$\tilde{\epsilon}(-k)_{1i} = -\tilde{\epsilon}(k)_{1i}, \quad \tilde{\epsilon}(-k)_{2i} = \tilde{\epsilon}(k)_{2i}. \tag{B-3}$$

We can also work in the circular polarization basis (helicity vectors) defined as

$$\tilde{\epsilon}(k)_{\pm, i} \equiv \frac{1}{\sqrt{2}}(\tilde{\epsilon}(k)_{1i} \pm i\tilde{\epsilon}(k)_{2i}), \tag{B-4}$$

with the following properties

$$\tilde{\epsilon}(k)^*_{h,i}\tilde{\epsilon}(k)^i_{h'} = \delta_{hh'}, \quad \tilde{\epsilon}(-k)^*_{\pm,i} = -\tilde{\epsilon}(k)_{\mp,i}, \quad \tilde{\epsilon}(k)^*_{\pm,i} = \tilde{\epsilon}(k)_{\mp,i},$$
$$i\epsilon_{ijl}\hat{k}_j\tilde{\epsilon}(k)_{hl} = h\tilde{\epsilon}(k)_{hi}, \quad \sum_h \tilde{\epsilon}(k)_{hi}\tilde{\epsilon}(k)_{-hj} = \delta_{ij} - \hat{k}_i\hat{k}_j, \tag{B-5}$$

where $h, h' = (+, -)$ and the star denotes complex conjugation. If we write the creation and annihilation operators in the helicity basis

$$b_1 = \frac{1}{\sqrt{2}}(b_+(k) + b_-(k)); \quad b_2 = \frac{i}{\sqrt{2}}(b_+(k) - b_-(k)),$$

we have

$$\begin{aligned} b_1(k)\tilde{\epsilon}(k)^i_1 + b_2(k)\tilde{\epsilon}(k)^i_2 &= \frac{1}{\sqrt{2}}b_+(k)[\tilde{\epsilon}(k)^i_1 + i\tilde{\epsilon}(k)^i_2] + \frac{1}{\sqrt{2}}b_-(k)[\tilde{\epsilon}(k)^i_1 - i\tilde{\epsilon}(k)^i_2] \\ &= b_+(k)\tilde{\epsilon}(k)^i_+ + b_-(k)\tilde{\epsilon}(k)^i_-. \end{aligned} \quad \text{(B-6)}$$

Inserting Eq.(5-13) into conjugate momentum of $A_i$ field (5-10), we can expand $\pi_i$ in terms of plane waves

$$\begin{aligned} \pi_i(\tau, x) &= (aI)^2 \int \frac{d^3k}{(2\pi)^{3/2}} \sum_{h=\pm} \bigg[ \big( \tilde{\epsilon}_{ih}(k) b_h(k) \bar{A}'_h(\tau, k) e^{ik\cdot x} + h.c. \big) \\ &+ \gamma_g a^4 \epsilon_{ilk} g^{ln} g^{ks} \big( ik_n \epsilon_{sh}(k) b_h(k) A_h(\tau, k) e^{ik\cdot x} \big) + h.c. \bigg]. \end{aligned} \quad \text{(B-7)}$$

Putting Eq.(5-13) and Eq.(B-7) into the commutation relation (5-11), we find for the first terms

$$\begin{aligned} \left[ A^i(\tau, x), \pi_j(\tau, y) \right]_1 &\sim e^{ik\cdot x} e^{ik'\cdot y} \underbrace{[b_h(k), b_{h'}(k')]}_{0} + e^{-ik\cdot x} e^{-ik'\cdot y} \underbrace{[b_h^\dagger(k), b_{h'}^\dagger(k')]}_{0} \\ &+ e^{-ik\cdot x} e^{ik'\cdot y} \underbrace{[b_h^\dagger(k), b_{h'}(k')]}_{-\delta(k-k')\delta_{hh'}} + e^{ik\cdot x} e^{-ik'\cdot y} \underbrace{[b_h(k), b_{h'}^\dagger(k')]}_{\delta(k-k')\delta_{hh'}}, \end{aligned} \quad \text{(B-8)}$$

getting the commutation relation for the creation and annihilation operators (5-43). The rest terms corresponding to helical contributions are

$$\begin{aligned} \left[ A^i(\tau, x), \pi_j(\tau, y) \right]_2 &\sim -iA_h(k,\tau) A_h^*(k,\tau) k^l \left[ \epsilon_h^i(k)\epsilon_{-h}^k(k) + \epsilon_{-h}^i(k)\epsilon_h^k(k) \right] \epsilon_{jlk} \\ &\sim -A_h(k,\tau) A_h^*(k,\tau) \left[ \frac{-hk}{\sqrt{2}}\epsilon_h^i(k)\epsilon_{-hj}(k) + \frac{-hk}{\sqrt{2}}\epsilon_{-h}^i(k)\epsilon_{hj}(k) \right] \\ &\sim \frac{hk}{\sqrt{2}}|A_h(k,\tau)|^2 \left[ \epsilon_{-hj}(k) - \epsilon_{-h}^i(k)\epsilon_{hj}(k) \right] = 0, \end{aligned} \quad \text{(B-9)}$$

finding no contribution on such additional terms. The previous results lead to

$$\begin{aligned} \left[ A^i(\tau, x), \pi_j(\tau, y) \right] &= \frac{i}{i} \int \frac{d^3k}{(2\pi)^3} \frac{a^2 I^2}{a^2} [\bar{A}_h \bar{A}_h^{*\prime} - \bar{A}_h^* \bar{A}_h'] \left( \delta_j^i - \delta_{lj}\hat{k}^i\hat{k}^l \right) e^{ik\cdot(x-y)} \\ &= i\delta_{\perp j}^{(3)\,i}(x-y), \quad \text{Iff:} \quad [\bar{A}_h \bar{A}_h^{*\prime} - \bar{A}_h^* \bar{A}_h'] = \frac{i}{I^2}, \end{aligned} \quad \text{(B-10)}$$

obtaining the normalization amplitude similar to that reported in Eq.(5-16). Finally, we present a code made in CADABRA [233] which shows the equation of motion given the action.

Code for finding the equation of motion given an action which depends on $A_\mu$ and $\phi$.

Hector Javier Hortua

GGC-UNAL-2017

```
{\beta,\alpha,\mu,\nu,\rho,\sigma,\kappa,\lambda,\eta,\chi,\sigma,\beta#,\
    alpha#,\mu#,\nu#,\rho#,\lambda#,\eta#,\sigma#}::Indices(fourD, position
    =independent);
{m,n,p,q,r,s,u,v,w,z,m#}::Indices(subspace, position=independent, parent=
    fourD);
#{t,t#}::Indices(subspace2, position=independent, name=oneD, parent=fourD);
{m,n,p,q,r,s,u,v,w,z,m#}::Integer(1..3);
{\beta,\alpha,\mu,\nu,\rho,\sigma,\kappa,\lambda,\eta,\chi,\sigma,\beta#,\
    alpha#,\mu#,\nu#,\rho#,\lambda#,\eta#,\sigma#}::Integer(1..4);
x::Coordinate.
y::Coordinate.
```

Attached property Indices(position=independent) to $[\beta,\ \alpha,\ \mu,\ \nu,\ \rho,\ \sigma,\ \kappa,\ \lambda,\ \eta,\ \chi,\ \sigma,\ \beta\#,\ \alpha\#,\ \mu\#,\ \nu\#,\ \rho\#,\ \lambda\#,\ \eta\#,\ \sigma\#]$.

Attached property Indices(position=independent) to $[m,\ n,\ p,\ q,\ r,\ s,\ u,\ v,\ w,\ z,\ m\#]$.

Attached property Integer to $[m,\ n,\ p,\ q,\ r,\ s,\ u,\ v,\ w,\ z,\ m\#]$.

Attached property Integer to $[\beta,\ \alpha,\ \mu,\ \nu,\ \rho,\ \sigma,\ \kappa,\ \lambda,\ \eta,\ \chi,\ \sigma,\ \beta\#,\ \alpha\#,\ \mu\#,\ \nu\#,\ \rho\#,\ \lambda\#,\ \eta\#,\ \sigma\#]$.

```
\partial{#}::PartialDerivative;
F_{\mu?\nu?}::AntiSymmetric;
F_{m? n?}::AntiSymmetric;
F_{\mu\nu}::Depends(x).
A_{\mu}::Depends(x,\partial{#}).
a::Depends().
ad::Depends().
\phi::Depends(x).
f::Depends().
\delta{#}::KroneckerDelta;
\sqrt{-g}::Depends().
\eth{#}::Accent;
g_{\mu \nu}::Metric(signature=-1);
g^{\mu \nu}::InverseMetric;
g_{\mu? \nu?}::Symmetric;
g^{\mu? \nu?}::Symmetric;
h_{m n}::Metric(signature=1);
```

```
h^{m n}::InverseMetric;
h_{m? n?}::Symmetric;
h^{m? n?}::Symmetric;
\gamma::Depends().
V::Depends().
\epsilon_{\mu? \nu? \alpha? \beta?}::EpsilonTensor(delta=\delta, metric=g_{\
    mu \nu}).
\epsilon^{\mu? \nu? \alpha? \beta?}::EpsilonTensor(delta=\delta, metric=g_{\
    mu \nu}).
\epsilon1_{m? n? p?}::EpsilonTensor(delta=\delta, metric=h_{ m n}).
\epsilon1^{m? n? p?}::EpsilonTensor(delta=\delta, metric=h^{ m n}).
```

Attached property PartialDerivative to $\partial\#$.

Attached property AntiSymmetric to $F_{\mu?\nu?}$.

Attached property AntiSymmetric to $F_{m?n?}$.

Attached property KroneckerDelta to $\delta(\#)$.

Attached property Accent to $\tilde{\partial}\#$.

Attached property Metric to $g_{\mu\nu}$.

Attached property TableauSymmetry to $g^{\mu\nu}$.

Attached property Symmetric to $g_{\mu?\nu?}$.

Attached property Symmetric to $g^{\mu?\nu?}$.

Attached property Metric to $h_{mn}$.

Attached property TableauSymmetry to $h^{mn}$.

Attached property Symmetric to $h_{m?n?}$.

Attached property Symmetric to $h^{m?n?}$.

Let us start with calculating the equation of motion for $(F^2 + F\tilde{F})f(\phi)$ in the Coulomb Gauge

```
S2:= \int{ -\sqrt{-g} f^2\frac{1}{4}F_{\nu\beta}F_{\mu\alpha} g^{\alpha \
    beta} g^{\mu \nu} +\sqrt{-g} f^2 \frac{\gamma}{8}
\epsilon^{\alpha \beta \sigma\eta}F_{\alpha\beta}F_{\sigma\eta}- \sqrt{-g}
    (\frac{1}{2} g^{\mu \nu} \partial_{\mu}{\phi} \partial_{\nu}{\phi}+V)}{
    x};
```

$$\int \left( -\frac{1}{4}\sqrt{-g}f^2 F_{\nu\beta}F_{\mu\alpha}g^{\alpha\beta}g^{\mu\nu} + \frac{1}{8}\sqrt{-g}f^2\gamma\epsilon^{\alpha\beta\sigma\eta}F_{\alpha\beta}F_{\sigma\eta} - \sqrt{-g}\left(\frac{1}{2}g^{\mu\nu}\partial_\mu\phi\partial_\nu\phi + V\right)\right)\mathrm{d}x$$

```
rl3:= F_{\mu?\nu?} = \partial_{\mu?}{A_{\nu?}} - \partial_{\nu?}{A_{\mu?}}
rldelA:= \eth{A_{??}}=1

substitute(S2, rl3)

vary(S2, $A_{\mu?} -> \eth{A_{\mu?}}$)

distribute(S2)
integrate_by_parts(S2, $\eth{A_{\mu?}}$)

substitute(_, $\partial_{\mu?}{A_{\nu?}} -> 1/2 \partial_{\mu?}{A_{\nu?}} +
    1/2 F_{\mu?\nu?} + 1/2 \partial_{\nu?}{A_{\mu?}}$)

distribute(_)
rename_dummies(_)

factor_out(_, $\eth{A_{\beta}}$)

kernel(scalar_backend="mathematica")

rename_dummies(_)
canonicalise(_)
simplify(_)

arg3=(_[0]);
```

$$\eth A_\beta \left( -\frac{1}{2}\partial_\alpha\left(\sqrt{-g}f^2\right)F_{\mu\nu}g^{\beta\mu}g^{\alpha\nu} - \frac{1}{2}\sqrt{-g}f^2\partial_\alpha\left(F_{\mu\nu}g^{\beta\mu}g^{\alpha\nu}\right) + \right.$$

$$\frac{1}{4}\partial_\alpha\left(\sqrt{-g}f^2 F_{\mu\nu}\right)g^{\beta\nu}g^{\alpha\mu} + \frac{1}{4}\sqrt{-g}f^2 F_{\alpha\mu}\partial_\nu\left(g^{\alpha\nu}g^{\beta\mu}\right) -$$

$$\frac{1}{4}\partial_\alpha\left(\sqrt{-g}f^2 F_{\mu\nu}\right)g^{\beta\mu}g^{\alpha\nu} - \frac{1}{4}\sqrt{-g}f^2 F_{\alpha\mu}\partial_\nu\left(g^{\alpha\beta}g^{\mu\nu}\right) -$$

$$\left. \frac{1}{4}\partial_\alpha\left(\sqrt{-g}f^2\gamma\epsilon^{\mu\beta\nu\alpha}\right)F_{\mu\nu} + \frac{1}{4}\partial_\alpha\left(\sqrt{-g}f^2\gamma\epsilon^{\beta\alpha\mu\nu}F_{\mu\nu}\right)\right)$$

```
substitute(arg3, rldelA)
canonicalise(_)
simplify(_);
```

This leads to:

$$-\frac{1}{2}\partial_\alpha\left(\sqrt{-g}f^2\right)F_{\mu\nu}g^{\beta\mu}g^{\alpha\nu} + \frac{1}{2}\sqrt{-g}f^2\partial_\alpha\left(F_{\mu\nu}g^{\alpha\mu}g^{\beta\nu}\right) +$$

$$\frac{1}{4}\partial_\alpha\left(\sqrt{-g}f^2 F_{\mu\nu}\right)g^{\beta\nu}g^{\alpha\mu} + \frac{1}{4}\sqrt{-g}f^2 F_{\alpha\mu}\partial_\nu\left(g^{\alpha\nu}g^{\beta\mu}\right) -$$

$$\frac{1}{4}\partial_\alpha\left(\sqrt{-g}f^2 F_{\mu\nu}\right)g^{\beta\mu}g^{\alpha\nu} - \frac{1}{4}\sqrt{-g}f^2 F_{\alpha\mu}\partial_\nu\left(g^{\alpha\beta}g^{\mu\nu}\right) -$$

$$\frac{1}{4}\partial_\alpha\left(\sqrt{-g}f^2\gamma\epsilon^{\alpha\beta\mu\nu}\right)F_{\mu\nu} - \frac{1}{4}\partial_\alpha\left(\sqrt{-g}f^2\gamma\epsilon^{\alpha\beta\mu\nu}F_{\mu\nu}\right)$$

Let's calculate $\beta = 0$:

```
substitute(_, $\beta -> 4$ );
```

$$-\frac{1}{2}\partial_\alpha\left(\sqrt{-g}f^2\right)F_{\mu\nu}g^{4\mu}g^{\alpha\nu} + \frac{1}{2}\sqrt{-g}f^2\partial_\alpha\left(F_{\mu\nu}g^{\alpha\mu}g^{4\nu}\right) +$$

$$\frac{1}{4}\partial_\alpha\left(\sqrt{-g}f^2 F_{\mu\nu}\right)g^{4\nu}g^{\alpha\mu} + \frac{1}{4}\sqrt{-g}f^2 F_{\alpha\mu}\partial_\nu\left(g^{\alpha\nu}g^{4\mu}\right) -$$

$$\frac{1}{4}\partial_\alpha\left(\sqrt{-g}f^2 F_{\mu\nu}\right)g^{4\mu}g^{\alpha\nu} - \frac{1}{4}\sqrt{-g}f^2 F_{\alpha\mu}\partial_\nu\left(g^{\alpha 4}g^{\mu\nu}\right) -$$

$$\frac{1}{4}\partial_\alpha\left(\sqrt{-g}f^2\gamma\epsilon^{\alpha 4\mu\nu}\right)F_{\mu\nu} - \frac{1}{4}\partial_\alpha\left(\sqrt{-g}f^2\gamma\epsilon^{\alpha 4\mu\nu}F_{\mu\nu}\right)$$

```
split_index(_, $\mu, n, 4$, repeat=True)
canonicalise(_)

substitute(_, $\sqrt{-g}\epsilon^{4 n? m? p?} -> \epsilon1^{m? n? p?}$,
    repeat=True)
substitute(_, $\epsilon^{s? n? m? p?} -> 0$, repeat=True)
substitute(_, $g^{4 4} -> -a^{-2}$ )
substitute(_, $g^{4 m?} -> 0$, repeat=True )
substitute(_, $g^{m? 4} -> 0$, repeat=True )
substitute(_, $g^{m? n?} -> h^{m? n?}*a^{-2}$, repeat=True)
substitute(_, $\sqrt{-g} -> a^{4}$)

distribute(_)
product_rule(_)
distribute(_)
product_rule(_)
canonicalise(_)

substitute(_, $\partial_{?}{\gamma} ->0$, repeat=True)
substitute(_, $\partial_{?}{\epsilon1^{s? m? q?}} ->0$, repeat=True)
substitute(_, $\partial_{m}{f^2}->0$, repeat=True)
```

```
substitute(_, $\partial_{m}{\sqrt{-g}} ->0$, repeat=True)
substitute(_, $\partial_{m}{a^{#}} ->0$, repeat=True)
substitute(_, $\partial_{?}{h^{s? m?}} ->0$, repeat=True)
substitute(_, $a^{??}a^{???} ->a^{???+??}$, repeat=True)
collect_factors(_)
substitute(_, $\sqrt{-g} -> a^{4}$)
canonicalise(_);
```

$$a^0 f^2 \partial_m F_{4n} h^{mn} - \frac{1}{4}\epsilon 1^{mnp} f^2 \gamma \partial_m F_{np}$$

```
substitute(_, rl3)
distribute(_)
product_rule(_)
rename_dummies(_)
canonicalise(_);
```

$$a^0 f^2 \partial_{4m} A_n h^{mn} - a^0 f^2 \partial_{mn} A_4 h^{mn}$$

By applying the Coulomb gauge we get:

```
substitute(_, $A_{4}->0$)
substitute(_, $\partial^{m?}{A_{m?}}->0$)
substitute(_, $h^{m? n?}\partial_{?}{\partial_{m?}{A_{n?}}}->0$)
substitute(_, $h^{m? n?}\partial_{?}{\partial_{n?}{A_{m?}}}->0$);
```

$0$

0 For $\beta = m1$ we have:

```
substitute(_, $\beta -> m1$ )

split_index(_, $\mu, n, 4$, repeat=True)
canonicalise(_)

substitute(_, $\sqrt{-g}\epsilon^{4 n? m? p?} -> \epsilon1^{m? n? p?}$,
    repeat=True)
substitute(_, $\epsilon^{s? n? m? p?} -> 0$, repeat=True)
substitute(_, $g^{4 4} -> -a^{-2}$, repeat=True )
substitute(_, $g^{4 m?} -> 0$, repeat=True )
substitute(_, $g^{m? 4} -> 0$, repeat=True )
substitute(_, $g^{m? n?} -> h^{m? n?}*a^{-2}$, repeat=True)
substitute(_, $\sqrt{-g} -> a^{4}$)

distribute(_)
product_rule(_)
distribute(_)
```

```
product_rule(_)
canonicalise(_)

substitute(_, $\partial_{?}{\gamma} ->0$, repeat=True)
substitute(_, $\partial_{?}{\epsilon1^{s? m? q?}} ->0$, repeat=True)
substitute(_, $\partial_{m}{f^2}->0$, repeat=True)
substitute(_, $\partial_{m}{\sqrt{-g}} ->0$, repeat=True)
substitute(_, $\partial_{m}{a^{#}} ->0$, repeat=True)
substitute(_, $\partial_{?}{h^{s? m?}} ->0$, repeat=True)
substitute(_, $a^{??}a^{???} ->a^{???+??}$, repeat=True)
collect_factors(_)
substitute(_, $\sqrt{-g} -> a^{4}$)
canonicalise(_);
```

$$-\partial_4 a^4 f^2 F_{4m} h^{m1m} a^{-4} - a^0 \partial_4 f^2 F_{4m} h^{m1m} - a^0 f^2 \partial_m F_{np} h^{m1n} h^{mp} -$$

$$a^0 f^2 \partial_4 F_{4m} h^{m1m} - \frac{3}{4} a^2 f^2 F_{4m} \partial_4 a^{-2} h^{m1m} - \frac{5}{4} a^2 f^2 F_{4m} h^{m1m} \partial_4 a^{-2} +$$

$$\frac{1}{2}\epsilon 1^{m1mn} \partial_4 f^2 \gamma F_{mn} - \frac{1}{2}\epsilon 1^{m1mn} f^2 \gamma \partial_m F_{4n} + \frac{1}{4}\epsilon 1^{m1mn} f^2 \gamma \partial_4 F_{mn}$$

```
substitute(_, rl3)
distribute(_)
product_rule(_)
rename_dummies(_)
canonicalise(_);
```

$$-\partial_4 a^4 f^2 \partial_4 A_m h^{m1m} a^{-4} + \partial_4 a^4 f^2 \partial_m A_4 h^{m1m} a^{-4} - a^0 \partial_4 f^2 \partial_4 A_m h^{m1m} +$$
$$a^0 \partial_4 f^2 \partial_m A_4 h^{m1m} - a^0 f^2 \partial_{mn} A_p h^{m1m} h^{np} + a^0 f^2 \partial_{mn} A_p h^{m1p} h^{mn} -$$

$$a^0 f^2 \partial_{44} A_m h^{m1m} + a^0 f^2 \partial_{4m} A_4 h^{m1m} - \frac{3}{4} a^2 f^2 \partial_4 A_m \partial_4 a^{-2} h^{m1m} +$$

$$\frac{3}{4} a^2 f^2 \partial_m A_4 \partial_4 a^{-2} h^{m1m} - \frac{5}{4} a^2 f^2 \partial_4 A_m h^{m1m} \partial_4 a^{-2} + \frac{5}{4} a^2 f^2 \partial_m A_4 h^{m1m} \partial_4 a^{-2} +$$

$$\epsilon 1^{m1mn} \partial_4 f^2 \gamma \partial_m A_n$$

This is the equation of motion written in the Coulomb gauge:

```
substitute(_, $A_{4}->0$)
substitute(_, $\partial^{m?}{A_{m?}}->0$)
substitute(_, $h^{m? n?}\partial_{?}{\partial_{m?}{A_{n?}}}->0$)
substitute(_, $h^{m? n?}\partial_{?}{\partial_{n?}{A_{m?}}}->0$)

substitute(_, $\partial_{m}{f^{#}} ->0$);
```

$$-\partial_4 a^4 f^2 \partial_4 A_m h^{m1m} a^{-4} - a^0 \partial_4 f^2 \partial_4 A_m h^{m1m} + a^0 f^2 \partial_{mn} A_p h^{m1p} h^{mn} -$$

$$a^0 f^2 \partial_{44} A_m h^{m1m} - \frac{3}{4} a^2 f^2 \partial_4 A_m \partial_4 a^{-2} h^{m1m} - \frac{5}{4} a^2 f^2 \partial_4 A_m h^{m1m} \partial_4 a^{-2} +$$
$$\epsilon 1^{m1mn} \partial_4 f^2 \gamma \partial_m A_n$$

Getting the following expression:

```
eliminate_metric(_)
distribute(_)
product_rule(_)
rename_dummies(_)
substitute(_, $a^{??}a^{???} ->a^{???+??}$, repeat=True)
canonicalise(_)
substitute(_, $\partial_{4}{a^{?}}->?*a^{?-1} adot$, repeat=True)
sort_product(_)
collect_factors(_)
simplify(_);
```

$$- \partial_4 A^{m1} \partial_4 f^2 a^0 + \partial^m{}_m A^{m1} a^0 f^2 - \partial_{44} A^{m1} a^0 f^2 + \epsilon 1^{m1mn} \gamma \partial_m A_n \partial_4 f^2$$

Let us now calculate the variation respect to $\phi$

```
rl2:= f^2 = f2'\phi
rl1:= V = V'\phi
rldelph:= \delta{\phi}=1

substitute(S2, rl1)
substitute(S2, rl2);
```

$$\int \left( -\frac{1}{4}\sqrt{-g} f2'\phi F_{\nu\beta} F_{\mu\alpha} g^{\alpha\beta} g^{\mu\nu} + \frac{1}{8}\sqrt{-g} f2'\phi\gamma \epsilon^{\alpha\beta\sigma\eta} F_{\alpha\beta} F_{\sigma\eta} - \right.$$
$$\left. \sqrt{-g}\left(\frac{1}{2} g^{\mu\nu} \partial_\mu\phi \partial_\nu\phi + V'\phi\right) \right) \mathrm{d}x$$

```
vary(S2, $\phi -> \delta{\phi}$)

distribute(S2)
integrate_by_parts(S2, $\delta{\phi}$)

sort_product(_)
factor_out(_, $\delta{\phi}$);
```

$$\int \delta(\phi) \left( -\frac{1}{4} F_{\mu\alpha} F_{\nu\beta} \sqrt{-g} f2' g^{\alpha\beta} g^{\mu\nu} + \frac{1}{8} F_{\alpha\beta} F_{\sigma\eta} \epsilon^{\alpha\beta\sigma\eta} \gamma \sqrt{-g} f2' + \right.$$
$$\left. \frac{1}{2} \partial_\nu\phi \partial_\mu\left(\sqrt{-g} g^{\mu\nu}\right) + \frac{1}{2} \partial_{\mu\nu}\phi \sqrt{-g} g^{\mu\nu} + \frac{1}{2} \partial_\nu\left(\partial_\mu\phi\sqrt{-g} g^{\mu\nu}\right) - V'\sqrt{-g} \right) \mathrm{d}x$$

```
rename_dummies(_)
canonicalise(_)
simplify(_)

arg4=(_[0]);
```

$$\delta\left(\phi\right)\left(-\frac{1}{4}F_{\alpha\beta}F_{\mu\nu}\sqrt{-g}f2'g^{\alpha\mu}g^{\beta\nu}+\frac{1}{8}F_{\alpha\beta}F_{\mu\nu}\epsilon^{\alpha\beta\mu\nu}\gamma\sqrt{-g}f2'+\right.$$
$$\left.\frac{1}{2}\partial_{\alpha}\phi\partial_{\beta}\left(\sqrt{-g}g^{\beta\alpha}\right)+\frac{1}{2}\partial_{\alpha\beta}\phi\sqrt{-g}g^{\alpha\beta}+\frac{1}{2}\partial_{\alpha}\left(\partial_{\beta}\phi\sqrt{-g}g^{\alpha\beta}\right)-V'\sqrt{-g}\right)$$

```
substitute(arg4, rldelph)
rename_dummies(_)
canonicalise(_)
simplify(_);
```

$$-\frac{1}{4}F_{\alpha\beta}F_{\mu\nu}\sqrt{-g}f2'g^{\alpha\mu}g^{\beta\nu}+\frac{1}{8}F_{\alpha\beta}F_{\mu\nu}\epsilon^{\alpha\beta\mu\nu}\gamma\sqrt{-g}f2'+\frac{1}{2}\partial_{\alpha}\phi\partial_{\beta}\left(\sqrt{-g}g^{\beta\alpha}\right)+$$
$$\frac{1}{2}\partial_{\alpha\beta}\phi\sqrt{-g}g^{\alpha\beta}+\frac{1}{2}\partial_{\alpha}\left(\partial_{\beta}\phi\sqrt{-g}g^{\alpha\beta}\right)-V'\sqrt{-g}$$

Where we obtain the Klein-Gordon equation sourced by EM terms!!

# C. Scalar bispectra of the magnetic field

Here we present the results for the even and odd contributions of the magnetic field bispectrum. The expressions for the magnetic bispectrum reported in this thesis were performed with the help of the tensor computer algebra xAct package in Mathematica [204].

**Scalar correlation ($<\rho\rho\rho>$)**

For the energy density of PMF bispectrum ($<\rho\rho\rho>$) we have

$$F^1_{\rho\rho\rho} = \beta^2 + \gamma^2 + \mu^2 - \beta\gamma\mu, \tag{C-1}$$

$$F^2_{\rho\rho\rho} = \beta\gamma + \mu, \tag{C-2}$$

$$F^3_{\rho\rho\rho} = \beta\mu + \gamma, \tag{C-3}$$

$$F^4_{\rho\rho\rho} = \beta + \gamma\mu, \tag{C-4}$$

for the even part, and using definition of eq.(7-16) we have for the odd part

$$\mathcal{B}^{(A)\,ljk}_{\rho_B\rho_B\rho_B} = B^{(A)}_{\rho_B\rho_B\rho_B} \widehat{\mathbf{k1}-\mathbf{p}}^l \widehat{\mathbf{k2}+\mathbf{p}}^j \hat{\mathbf{p}}^k, \tag{C-5}$$

with

$$\begin{aligned} B^{(A)}_{\rho_B\rho_B\rho_B} &= \frac{8}{(2\pi)^3(4\pi)^3}\int d^3p\,(P_H(|\mathbf{p}+\mathbf{k2}|)(P_H(p)P_H(|\mathbf{k1}-\mathbf{p}|) - P_B(p)P_B(|\mathbf{k1}-\mathbf{p}|)\beta) \\ &+ P_B(|\mathbf{p}+\mathbf{k2}|)(P_B(p)P_H(|\mathbf{k1}-\mathbf{p}|)\gamma - P_H(p)P_B(|\mathbf{k1}-\mathbf{p}|)\mu)). \end{aligned} \tag{C-6}$$

**Scalar cross-correlation($<\rho\rho\Pi>$)**

For the three-point cross-correlation ($<\rho\rho\Pi>$) we obtain

$$\begin{aligned} F^1_{\rho\rho\Pi^{(S)}} &= 3(-1 + \alpha_q^2 + \beta_q^2 + \gamma_q^2) \\ &+ \gamma^2 + \beta^2 + \mu^2 - \beta\gamma\mu + 3\beta\beta_q\gamma\gamma_q - 3\alpha_q(\beta\beta_q + \gamma\gamma_q) - 3\mu\beta_q\gamma_q, \end{aligned} \tag{C-7}$$

$$F^2_{\rho\rho\Pi^{(S)}} = 2\beta\gamma - 3\alpha_q\beta_q\gamma - 3\alpha_q\beta\gamma_q - \mu + 3\alpha_q^2\mu, \tag{C-8}$$

$$F^3_{\rho\rho\Pi^{(S)}} = \gamma(-2 + 3\beta_q^2) + 3\alpha_q\gamma_q + \beta\mu - 3\alpha_q\beta_q\mu, \tag{C-9}$$

$$F^4_{\rho\rho\Pi^{(S)}} = \beta(-2 + 3\gamma_q^2) + 3\alpha_q\beta_q + \gamma\mu - 3\alpha_q\gamma_q\mu, \tag{C-10}$$

for the even contribution, and with the definition of the three point correlation eq.(7-16), the odd part is given by

$$\begin{aligned}
\mathcal{B}^{(A)\,ljk}_{\rho_B\rho_B\Pi^{(S)}_B} &= B^{(A1)}_{\rho_B\rho_B\Pi^{(S)}_B}\widehat{\mathbf{k2}+\mathbf{p}}^l\widehat{\mathbf{k3}}^j\hat{\mathbf{p}}^k + B^{(A2)}_{\rho_B\rho_B\Pi^{(S)}_B}\widehat{\mathbf{k1}-\mathbf{p}}^l\widehat{\mathbf{k3}}^j\hat{\mathbf{p}}^k \\
&+ B^{(A3)}_{\rho_B\rho_B\Pi^{(S)}_B}\widehat{\mathbf{k1}-\mathbf{p}}^l\widehat{\mathbf{k2}+\mathbf{p}}^j\hat{\mathbf{k3}}^k + B^{(A4)}_{\rho_B\rho_B\Pi^{(S)}_B}\widehat{\mathbf{k1}-\mathbf{p}}^l\widehat{\mathbf{k2}+\mathbf{p}}^j\hat{\mathbf{p}}^k,
\end{aligned} \tag{C-11}$$

where

$$\begin{aligned}
B^{(A1)}_{\rho_B\rho_B\Pi^{(S)}_B} &= \frac{-12}{(2\pi)^3(4\pi)^3}\int d^3p P_B(|\mathbf{k1}-\mathbf{p}|)\left(P_H(p)P_B(|\mathbf{p}+\mathbf{k2}|)\gamma_q\right. \\
&+ \left. P_B(p)P_H(|\mathbf{p}+\mathbf{k2}|)(\alpha_q - \beta\beta_q)\right),
\end{aligned} \tag{C-12}$$

$$\begin{aligned}
B^{(A2)}_{\rho_B\rho_B\Pi^{(S)}_B} &= \frac{12}{(2\pi)^3(4\pi)^3}\int d^3p P_B(|\mathbf{p}+\mathbf{k2}|)\left(P_H(p)P_B(|\mathbf{k1}-\mathbf{p}|)\beta_q\right. \\
&+ \left. P_B(p)P_H(|\mathbf{k1}-\mathbf{p}|)(\gamma\gamma_q - \alpha_q)\right),
\end{aligned} \tag{C-13}$$

$$\begin{aligned}
B^{(A3)}_{\rho_B\rho_B\Pi^{(S)}_B} &= \frac{12}{(2\pi)^3(4\pi)^3}\int d^3p \left(P_H(|\mathbf{p}+\mathbf{k2}|)P_H(p)P_H(|\mathbf{k1}-\mathbf{p}|)\alpha_q\right. \\
&+ \left. P_B(p)(P_H(|\mathbf{k1}-\mathbf{p}|)P_B(|\mathbf{p}+\mathbf{k2}|)\gamma_q - P_B(|\mathbf{k1}-\mathbf{p}|)P_H(|\mathbf{p}+\mathbf{k2}|)\beta_q)\right),
\end{aligned} \tag{C-14}$$

$$\begin{aligned}
B^{(A4)}_{\rho_B\rho_B\Pi^{(S)}_B} &= \frac{-4}{(2\pi)^3(4\pi)^3}\int d^3p \left(P_H(|\mathbf{p}+\mathbf{k2}|)P_H(p)P_H(|\mathbf{k1}-\mathbf{p}|)\right. \\
&- P_B(p)(P_B(|\mathbf{k1}-\mathbf{p}|)P_H(|\mathbf{p}+\mathbf{k2}|)\beta - P_H(|\mathbf{k1}-\mathbf{p}|)P_B(|\mathbf{p}+\mathbf{k2}|)\gamma \\
&+ \left. P_B(|\mathbf{k1}-\mathbf{p}|)P_H(p)P_B(|\mathbf{p}+\mathbf{k2}|)(3\gamma_q\beta_q - \mu)\right).
\end{aligned} \tag{C-15}$$

### Scalar cross-correlation ($<\rho\Pi\Pi>$)

The result for the even contribution of the three-point cross-correlation ($<\rho\Pi\Pi>$) is given by

$$\begin{aligned}
F^1_{\rho\Pi^{(S)}\Pi^{(S)}} &= -6 + 3(\alpha_p^2 + \beta_p^2 + \gamma_p^2) + 3(\alpha_q^2 + \beta_q^2 + \gamma_q^2) + \beta^2 + \gamma^2 + \mu^2 + 3\beta_q(\beta\gamma\gamma_q + 3\beta_p\gamma_p\gamma_q) \\
&- 3\beta_p\gamma_p\mu - \beta\gamma\mu - 3\alpha_q(\beta\beta_q + \gamma\gamma_q) - 9\theta_{pq}(\beta_q\beta_p + \gamma_p\gamma_q - \theta_{pq}) - 3\beta_q\gamma_q\mu \\
&- 3\alpha_p\beta(\beta_p + 3\beta_q\gamma_p\gamma_q - 3\beta_q\theta_{pq} - \gamma_p\mu) - 3\alpha_p(\gamma\gamma_p - 3\alpha_q\gamma_p\gamma_q + 3\alpha_q\theta_{pq}),
\end{aligned} \tag{C-16}$$

$$\begin{aligned}F^2_{\rho\Pi^{(S)}\Pi^{(S)}} &= -3\alpha_q\beta_q\gamma + 9\gamma_q\beta_q + \beta(2\gamma - 3\alpha_q\gamma_q) + \mu(-7 + 3\alpha_q^2 + 9\theta_{pq}^2) + \alpha_p^2(6\mu - 9\beta_q\gamma_q) \\ &\quad - 9\theta_{pq}(\beta_q\gamma_p + \beta_p\gamma_q) + \alpha_p\beta_p(-6\gamma + 9\alpha_q\gamma_q) + 9\alpha_p\theta_{pq}(\beta_q\gamma - \alpha_q\mu) + 6\beta_p\gamma_p,\end{aligned} \quad (\text{C-17})$$

$$\begin{aligned}F^3_{\rho\Pi^{(S)}\Pi^{(S)}} &= 9\alpha_q\gamma_p(\beta_p\beta_q - \theta_{pq}) - 3\alpha_p(3\gamma_p(\beta_q^2 - 1) + 3\gamma_q\theta_{pq} + \beta_p\mu - 3\beta_q\theta_{pq}\mu) + 6\alpha_q\gamma_q \\ &\quad - 3\beta\beta_p\gamma_p + \gamma(-7 + 3\beta_p^2 + 6\beta_q^2 - 9\beta_p\beta_q\theta_{pq} + 9\theta_{pq}^2) + 2\beta\mu - 6\alpha_q\beta_q\mu,\end{aligned} \quad (\text{C-18})$$

$$\begin{aligned}F^4_{\rho\Pi^{(S)}\Pi^{(S)}} &= 3\alpha_p\beta_p + 3\alpha_q\beta_q - 3\beta_p\gamma\gamma_p + 9\alpha_q\beta_p(\gamma_p\gamma_q - \theta_{pq}) \\ &\quad + \beta(-5 + 3\gamma_p^2 + 3\gamma_q^2 - 9\gamma_p\gamma_q\theta_{pq} + 9\theta_{pq}^2) + \gamma\mu - 3\alpha_q\gamma_q\mu.\end{aligned} \quad (\text{C-19})$$

Using eq.(7-16), the odd contribution can be written as

$$\begin{aligned}\mathcal{B}^{(A)\,ljk}_{\rho_B\Pi^{(S)}_B\Pi^{(S)}_B} &= B^{(A1)}_{\rho_B\Pi^{(S)}_B\Pi^{(S)}_B} \widehat{\mathbf{k2+p}}^l \widehat{\mathbf{k3}}^j \hat{\mathbf{p}}^k \\ &\quad + B^{(A2)}_{\rho_B\Pi^{(S)}_B\Pi^{(S)}_B} \widehat{\mathbf{k2}}^l \widehat{\mathbf{k3}}^j \hat{\mathbf{p}}^k + B^{(A3)}_{\rho_B\Pi^{(S)}_B\Pi^{(S)}_B} \widehat{\mathbf{k2}}^l \widehat{\mathbf{k2+p}}^j \hat{\mathbf{k3}}^k \\ &\quad + B^{(A4)}_{\rho_B\Pi^{(S)}_B\Pi^{(S)}_B} \widehat{\mathbf{k1-p}}^l \widehat{\mathbf{k2+p}}^j \hat{\mathbf{k3}}^k + B^{(A5)}_{\rho_B\Pi^{(S)}_B\Pi^{(S)}_B} \widehat{\mathbf{k1-p}}^l \widehat{\mathbf{k2+p}}^j \hat{\mathbf{p}}^k \\ &\quad + B^{(A6)}_{\rho_B\Pi^{(S)}_B\Pi^{(S)}_B} \widehat{\mathbf{k1-p}}^l \widehat{\mathbf{k3}}^j \hat{\mathbf{p}}^k + B^{(A7)}_{\rho_B\Pi^{(S)}_B\Pi^{(S)}_B} \widehat{\mathbf{k1-p}}^l \widehat{\mathbf{k2}}^j \widehat{\mathbf{k2+p}}^k \\ &\quad + B^{(A8)}_{\rho_B\Pi^{(S)}_B\Pi^{(S)}_B} \widehat{\mathbf{k1-p}}^l \widehat{\mathbf{k2}}^j \hat{\mathbf{k3}}^k + B^{(A9)}_{\rho_B\Pi^{(S)}_B\Pi^{(S)}_B} \widehat{\mathbf{k1-p}}^l \widehat{\mathbf{k2}}^j \hat{\mathbf{p}}^k \\ &\quad + B^{(A10)}_{\rho_B\Pi^{(S)}_B\Pi^{(S)}_B} \widehat{\mathbf{k2}}^l \widehat{\mathbf{k2+p}}^j \hat{\mathbf{p}}^k,\end{aligned} \quad (\text{C-20})$$

with

$$\begin{aligned}B^{(A1)}_{\rho_B\Pi^{(S)}_B\Pi^{(S)}_B} &= \frac{6}{(2\pi)^3(4\pi)^3}\int d^3p P_B(|\mathbf{k1}-\mathbf{p}|)\left(P_B(p)P_H(|\mathbf{p}+\mathbf{k2}|)(\alpha_q - \beta\beta_q)\right. \\ &\quad \left. + P_H(p)P_B(|\mathbf{p}+\mathbf{k2}|)\gamma_q\right),\end{aligned} \quad (\text{C-21})$$

$$B^{(A2)}_{\rho_B\Pi^{(S)}_B\Pi^{(S)}_B} = \frac{18}{(2\pi)^3(4\pi)^3}\int d^3p P_B(|\mathbf{k1}-\mathbf{p}|)P_H(p)P_B(|\mathbf{p}+\mathbf{k2}|)(-\gamma_p\gamma_q + \theta_{pq}), \quad (\text{C-22})$$

$$\begin{aligned}B^{(A3)}_{\rho_B\Pi^{(S)}_B\Pi^{(S)}_B} &= \frac{18}{(2\pi)^3(4\pi)^3}\int d^3p P_H(|\mathbf{p}+\mathbf{k2}|)\left(P_H(p)P_H(|\mathbf{k1}-\mathbf{p}|)(\beta_p\alpha_q - \beta\theta_{pq})\right. \\ &\quad \left. + P_B(|\mathbf{k1}-\mathbf{p}|)P_B(p)(-\alpha_p\alpha_q + \alpha_p\beta\beta_q - \beta_p\beta_q + \theta_{pq})\right),\end{aligned} \quad (\text{C-23})$$

$$B^{(A\,4)}_{\rho_B \Pi_B^{(S)} \Pi_B^{(S)}} = \frac{-6}{(2\pi)^3 (4\pi)^3} \int d^3p \, (P_H(|\mathbf{p}+\mathbf{k2}|) P_H(p) P_H(|\mathbf{k1}-\mathbf{p}|) \alpha_q$$
$$+ \; P_B(p)(P_H(|\mathbf{k1}-\mathbf{p}|) P_B(|\mathbf{p}+\mathbf{k2}|) \gamma_q - P_B(|\mathbf{k1}-\mathbf{p}|) P_H(|\mathbf{p}+\mathbf{k2}|) \beta_q)) \, , \quad \text{(C-24)}$$

$$B^{(A\,5)}_{\rho_B \Pi_B^{(S)} \Pi_B^{(S)}} = \frac{2}{(2\pi)^3 (4\pi)^3} \int d^3p \, (P_B(p) \, (P_B(|\mathbf{p}+\mathbf{k2}|) P_H(|\mathbf{k1}-\mathbf{p}|)(\gamma - 3\alpha_p \gamma_p)$$
$$- \; P_B(|\mathbf{k1}-\mathbf{p}|) P_H(|\mathbf{p}+\mathbf{k2}|) \beta) + P_H(p) \, (P_H(|\mathbf{k1}-\mathbf{p}|) P_H(|\mathbf{p}+\mathbf{k2}|)$$
$$+ \; P_B(|\mathbf{k1}-\mathbf{p}|) P_B(|\mathbf{p}+\mathbf{k2}|)(3\beta_q \gamma_q - \mu))) \, , \quad \text{(C-25)}$$

$$B^{(A\,6)}_{\rho_B \Pi_B^{(S)} \Pi_B^{(S)}} = \frac{6}{(2\pi)^3 (4\pi)^3} \int d^3p \, P_B(|\mathbf{p}+\mathbf{k2}|) \, (-P_H(p) P_B(|\mathbf{k1}-\mathbf{p}|) \beta_q$$
$$+ \; P_H(|\mathbf{k1}-\mathbf{p}|) P_B(p)(\alpha_q - \gamma \gamma_q + 3\alpha_p \gamma_p \gamma_q - 3\alpha_p \theta_{pq})) \, , \quad \text{(C-26)}$$

$$B^{(A\,7)}_{\rho_B \Pi_B^{(S)} \Pi_B^{(S)}} = \frac{-6}{(2\pi)^3 (4\pi)^3} \int d^3p \, P_B(p) \, (P_B(|\mathbf{p}+\mathbf{k2}|) P_H(|\mathbf{k1}-\mathbf{p}|) \gamma_p$$
$$+ \; P_B(|\mathbf{k1}-\mathbf{p}|) P_H(|\mathbf{p}+\mathbf{k2}|)(\alpha_p \beta - \beta_p)) \, , \quad \text{(C-27)}$$

$$B^{(A\,8)}_{\rho_B \Pi_B^{(S)} \Pi_B^{(S)}} = \frac{-18}{(2\pi)^3 (4\pi)^3} \int d^3p \, P_B(p) P_B(|\mathbf{p}+\mathbf{k2}|) P_H(|\mathbf{k1}-\mathbf{p}|)(-\gamma_p \gamma_q + \theta_{pq}) \, , \quad \text{(C-28)}$$

$$B^{(A\,9)}_{\rho_B \Pi_B^{(S)} \Pi_B^{(S)}} = \frac{6}{(2\pi)^3 (4\pi)^3} \int d^3p \, P_B(|\mathbf{p}+\mathbf{k2}|) \, (P_B(p) P_H(|\mathbf{k1}-\mathbf{p}|) \alpha_p$$
$$+ \; P_B(|\mathbf{k1}-\mathbf{p}|) P_H(p)(-\beta_p - 3\beta_q \gamma_p \gamma_q + 3\beta_q \theta_{pq} + \gamma_p \mu)) \, . \quad \text{(C-29)}$$

$$B^{(A\,10)}_{\rho_B \Pi_B^{(S)} \Pi_B^{(S)}} = \frac{6}{(2\pi)^3 (4\pi)^3} \int d^3p \, P_H(|\mathbf{p}+\mathbf{k2}|) P_B(p) P_B(|\mathbf{k1}-\mathbf{p}|) \alpha_p$$
$$+ \; P_H(p) \, (P_B(|\mathbf{k1}-\mathbf{p}|) P_B(|\mathbf{k2}+\mathbf{p}|) \beta_p - P_H(|\mathbf{k1}-\mathbf{p}|) P_H(|\mathbf{k2}+\mathbf{p}|) \gamma_p) \, . \quad \text{(C-30)}$$

## Scalar cross-correlation($< \Pi\Pi\Pi >$)

Finally, the even contribution for the three-point cross-correlation of scalar anisotropic stress is written as

$$\begin{aligned}
F^1_{\Pi^{(S)}\Pi^{(S)}\Pi^{(S)}} &= -9 + 3(\alpha_k^2 + \alpha_p^2 + \alpha_q^2) - 9\theta_{pq}(\beta_p\beta_q + \gamma_p\gamma_q - 3\beta_k\beta_q\theta_{kp} + 3\theta_{kp}\theta_{kq}) \\
&+ 3(\beta_q^2 + \beta_p^2 + \beta_k^2) + \beta^2 + \gamma^2 + \mu^2 - \beta\gamma\mu + 3(\gamma_k^2 + \gamma_p^2 + \gamma_q^2) - 3\alpha_q(\beta\beta_q + \gamma\gamma_q) \\
&+ 3\beta_q(\beta\gamma\gamma_q + 3\beta_k\gamma_k\gamma_q + 3\beta_p\gamma_p\gamma_q) - 9\theta_{kp}(\beta_k\beta_p + \gamma_k\gamma_p + 3\beta_q\beta_k\gamma_p\gamma_q - \theta_{kp}) \\
&- 3\alpha_p(\gamma\gamma_p + 3\alpha_q(-\gamma_p\gamma_q + \theta_{pq}) + \beta(\beta_p + 3\beta_q\gamma_p\gamma_q - 3\beta_q\theta_{pq} - \gamma_p\mu)) + 9\theta_{pq}^2 \\
&+ 9\theta_{kq}^2 - 3\alpha_k\left(\beta\beta_k - 3\alpha_q\beta_k\beta_q + \gamma\gamma_k + 3\beta_k\beta_q\gamma\gamma_q + 3\alpha_q\theta_{kq} - 3\gamma\gamma_q\theta_{kq} - \beta_k\gamma\mu \right.\\
&\left. - 3\alpha_p(\gamma_p\gamma_k - \theta_{kp} - 3\gamma_p\gamma_q\theta_{kq} + 3\theta_{kq}\theta_{pq} + \beta_k(\beta_p + 3\beta_q\gamma_p\gamma_q - 3\beta_q\theta_{pq} - \gamma_p\mu)) \right) \\
&- 3\mu(\beta_k\gamma_k + \gamma_p\beta_p + \beta_q\gamma_q - 3\beta_k\gamma_p\theta_{kp}) - 9\theta_{kq}(\beta_k\beta_q + \gamma_k\gamma_q - 3\theta_{kp}\gamma_p\gamma_q), \quad \text{(C-31)}
\end{aligned}$$

$$\begin{aligned}
F^2_{\Pi^{(S)}\Pi^{(S)}\Pi^{(S)}} &= 3\beta_q(3\alpha_k\alpha_q\gamma_k - \alpha_q\gamma + 6\gamma_q - 3\alpha_k^2\gamma_q - 3\gamma_k\theta_{kq}) + \alpha_p^2(-9\beta_q\gamma_q + 6\mu) \\
&+ 6\beta_p(\gamma_p - 3\gamma_k\theta_{kp}) - 9\gamma_q(3\beta_q\theta_{kp}^2 + \beta_k\theta_{kq} - 3\beta_p\theta_{kq}\theta_{kp} + \beta_p\theta_{pq}) \\
&+ \beta(2\gamma - 6\alpha_k\gamma_k - 3\alpha_q\gamma_q + 9\alpha_k\gamma_q\theta_{kq}) + 9\theta_{pq}(3\theta_{kp}\gamma_k\beta_q - \beta_q\gamma_p) + 6\beta_k\gamma_k \\
&+ \mu(-13 + 6\alpha_k^2 + 3\alpha_q^2 + 18\theta_{kp}^2 - 9\alpha_k\alpha_q\theta_{kq} + 9\theta_{kq}^2 - 27\theta_{kp}\theta_{kq}\theta_{pq} + 9\theta_{pq}^2) \\
&+ 3\alpha_p\left(\alpha_k\beta_q\gamma_q\theta_{kp} + \beta_p(-2\gamma + 6\alpha_k\gamma_k + 3\alpha_q\gamma_q - 9\alpha_k\gamma_q\theta_{kq}) + 3\beta_q\gamma\theta_{pq}\right. \\
&\left. - 9\alpha_k\beta_q\gamma_k\theta_{pq} - 6\alpha_k\theta_{kp}\mu - 3\alpha_q\theta_{pq}\mu + 9\alpha_k\theta_{kq}\theta_{pq}\mu \right), \quad \text{(C-32)}
\end{aligned}$$

$$\begin{aligned}
F^3_{\Pi^{(S)}\Pi^{(S)}\Pi^{(S)}} &= 6\beta_p(\gamma\gamma_p - \alpha_p) - 6\alpha_q(\beta_q + 3\beta_p\gamma_p\gamma_q) - 27\alpha_q\beta_k\theta_{kp}\theta_{pq} - 2\mu(\gamma - 3\alpha_q\gamma_q) \\
&+ 3\beta_k(\gamma\gamma_k - 3\alpha_q\gamma_k\gamma_q + 3\alpha_p\theta_{kp} - 3\gamma\gamma_p\theta_{kp} + 9\alpha_q\gamma_p\gamma_q\theta_{kp} + 3\alpha_q\theta_{kq}) + 18\alpha_q\beta_p\theta_{pq} \\
&- \beta\left(-13 + 3(\gamma_k^2 + 2\gamma_p^2 + 2\gamma_q^2) + 9(\theta_{kp}^2 + \theta_{kq}^2 + 2\theta_{pq}^2)\right) - 9\gamma_k(\gamma_p\theta_{kp} + \gamma_q\theta_{kq}) \\
&+ 9\gamma_p\gamma_q(3\theta_{kp}\theta_{kq} - 2\theta_{pq}) - 27\theta_{kp}\theta_{kq}\theta_{pq}) + 3\alpha_k\left(3\beta_q\theta_{kq} + \beta_p(-3\gamma_k\gamma_p + 3\theta_{kp} - 9\theta_{kq}\theta_{pq} \right.\\
&\left. + 9\gamma_p\gamma_q\theta_{kq}) + 3\beta_k(-2 + \gamma_p^2 + \gamma_q^2 - 3\gamma_p\gamma_q\theta_{pq} + 3\theta_{pq}^2) + \mu(\gamma_k - 3\gamma_q\theta_{kq}) \right), \quad \text{(C-33)}
\end{aligned}$$

$$\begin{aligned}
F^4_{\Pi^{(S)}\Pi^{(S)}\Pi^{(S)}} &= -27\alpha_q\beta_k\beta_q\gamma_p\theta_{kp} - 3\beta\beta_p\gamma_p + \mu(2\beta - 3\alpha_p\beta_p - 6\alpha_q\beta_q + 9\alpha_p\beta_q\theta_{pq}) \\
&+ 6\gamma_k(\alpha_k - \beta\beta_k + 3\alpha_q\beta_k\beta_q) + 9\alpha_p\beta_k\beta_p\gamma_k + 9\gamma_p(2\alpha_p - \alpha_p\beta_k^2 + \alpha_q\beta_p\beta_q - \alpha_p\beta_q^2) \\
&+ 6\alpha_q\gamma_q - 9\left(\alpha_k\gamma_p\theta_{kp} - \beta\beta_k\gamma_p\theta_{kp} + 2\alpha_q\gamma_k\theta_{kq} - 3\alpha_q\gamma_p\theta_{kp}\theta_{kq} + \alpha_q\gamma_p\theta_{pq} \right.\\
&\left. + \alpha_p\left(3\gamma_p\theta_{kq}^2 - 3\beta_k\beta_q\gamma_p\theta_{kq} + \gamma_q\theta_{pq} + \gamma_k(\theta_{kp} + 3\beta_k\beta_q\theta_{pq} - 3\theta_{kq}\theta_{pq})\right)\right) \\
&+ \gamma\left(-13 + 3(2\beta_k^2 + \beta_p^2 + 2\beta_q^2) + 9(\theta_{kp}^2 + 2\theta_{kq}^2 + \theta_{pq}^2) - 9\theta_{pq}(\beta_p\beta_q + 3\theta_{kp}\theta_{kq})\right. \\
&\left. - 9\beta_k(\beta_p\theta_{kp} + 2\beta_q\theta_{kq} - 3\beta_q\theta_{kp}\theta_{pq})\right), \quad \text{(C-34)}
\end{aligned}$$

for the odd contribution we found

$$\begin{aligned}
B^{(A1)}_{\Pi_B^{(S)}\Pi_B^{(S)}\Pi_B^{(S)}} &= \frac{3}{(2\pi)^3(4\pi)^3}\int d^3p\,(P_H(p)P_H(|\mathbf{k1}-\mathbf{p}|)P_H(|\mathbf{p}+\mathbf{k2}|)(-2\alpha_q + 3\alpha_k\theta_{kq}) \\
&- P_B(p)(P_B(|\mathbf{k1}-\mathbf{p}|)P_H(|\mathbf{p}+\mathbf{k2}|)\beta_q - P_B(|\mathbf{p}+\mathbf{k2}|)P_H(|\mathbf{k1}-\mathbf{p}|)\gamma_q)) \quad \text{(C-35)}
\end{aligned}$$

$$\begin{aligned}
B^{(A2)}_{\Pi_B^{(S)}\Pi_B^{(S)}\Pi_B^{(S)}} &= \frac{1}{(2\pi)^3(4\pi)^3}\int d^3p\,(P_B(p)(P_B(|\mathbf{k1}-\mathbf{p}|)P_H(|\mathbf{p}+\mathbf{k2}|)(\beta - 3\alpha_k\beta_k) \\
&- P_B(|\mathbf{p}+\mathbf{k2}|)P_H(|\mathbf{k1}-\mathbf{p}|)(\gamma - 3\alpha_p\gamma_p)) + P_H(p)\,(2P_H(|\mathbf{k1}-\mathbf{p}|)P_H(|\mathbf{p}+\mathbf{k2}|) \\
&+ P_B(|\mathbf{k1}-\mathbf{p}|)P_B(|\mathbf{p}+\mathbf{k2}|)(-3\beta_q\gamma_q + \mu))) \quad \text{(C-36)}
\end{aligned}$$

$$\begin{aligned}
B^{(A3)}_{\Pi_B^{(S)}\Pi_B^{(S)}\Pi_B^{(S)}} &= \frac{3}{(2\pi)^3(4\pi)^3}\int d^3p\,P_B(|\mathbf{p}+\mathbf{k2}|)\,(P_H(p)P_B(|\mathbf{k1}-\mathbf{p}|)\beta_q \\
&+ P_H(|\mathbf{k1}-\mathbf{p}|)P_B(p)(-\alpha_q + \gamma\gamma_q - 3\alpha_p\gamma_p\gamma_q + 3\alpha_p\theta_{pq})), \quad \text{(C-37)}
\end{aligned}$$

$$\begin{aligned}
B^{(A4)}_{\Pi_B^{(S)}\Pi_B^{(S)}\Pi_B^{(S)}} &= \frac{3}{(2\pi)^3(4\pi)^3}\int d^3p\,P_B(p)\,(P_B(|\mathbf{p}+\mathbf{k2}|)P_H(|\mathbf{k1}-\mathbf{p}|)\gamma_p \\
&+ P_B(|\mathbf{k1}-\mathbf{p}|)P_H(|\mathbf{p}+\mathbf{k2}|)(\alpha_p\beta - 3\alpha_k\alpha_p\beta_k - \beta_p + 3\beta_k\theta_{kp})), \quad \text{(C-38)}
\end{aligned}$$

$$B^{(A5)}_{\Pi_B^{(S)}\Pi_B^{(S)}\Pi_B^{(S)}} = \frac{9}{(2\pi)^3(4\pi)^3}\int d^3p\,P_B(p)P_B(|\mathbf{p}+\mathbf{k2}|)P_H(|\mathbf{k1}-\mathbf{p}|)(-\gamma_p\gamma_q + \theta_{pq}) \quad \text{(C-39)}$$

$$\begin{aligned}
B^{(A6)}_{\Pi_B^{(S)}\Pi_B^{(S)}\Pi_B^{(S)}} &= \frac{-3}{(2\pi)^3(4\pi)^3}\int d^3p\,P_B(|\mathbf{p}+\mathbf{k2}|)\,(P_B(p)P_H(|\mathbf{k1}-\mathbf{p}|)\alpha_p \\
&+ P_B(|\mathbf{k1}-\mathbf{p}|)P_H(p)(-\beta_p - 3\beta_q\gamma_p\gamma_q + 3\beta_q\theta_{pq} + \gamma_p\mu)), \quad \text{(C-40)}
\end{aligned}$$

$$\begin{aligned}
B^{(A7)}_{\Pi_B^{(S)}\Pi_B^{(S)}\Pi_B^{(S)}} &= \frac{-3}{(2\pi)^3(4\pi)^3}\int d^3p\,P_B(|\mathbf{k1}-\mathbf{p}|)\,(P_B(|\mathbf{p}+\mathbf{k2}|)P_H(p)\gamma_q \\
&+ P_B(p)P_H(|\mathbf{p}+\mathbf{k2}|)(\alpha_q - \beta\beta_q + 3\alpha_k\beta_k\beta_q - 3\alpha_k\theta_{kq})), \quad \text{(C-41)}
\end{aligned}$$

$$B^{(A8)}_{\Pi_B^{(S)}\Pi_B^{(S)}\Pi_B^{(S)}} = \frac{-9}{(2\pi)^3(4\pi)^3}\int d^3p\,P_H(p)P_B(|\mathbf{p}+\mathbf{k2}|)P_B(|\mathbf{k1}-\mathbf{p}|)(-\gamma_p\gamma_q + \theta_{pq}), \quad \text{(C-42)}$$

$$B^{(A\,9)}_{\Pi_B^{(S)}\Pi_B^{(S)}\Pi_B^{(S)}} = \frac{-9}{(2\pi)^3(4\pi)^3}\int d^3p\, P_H(|\mathbf{p}+\mathbf{k2}|)\,(P_B(p)P_B(|\mathbf{k1}-\mathbf{p}|)\,(-\beta_p\beta_q + 3\beta_k\beta_q\theta_{kp} + \theta_{pq}$$
$$-\ 3\theta_{kp}\theta_{kq} + \alpha_p(-\alpha_q + \beta\beta_q - 3\alpha_k\beta_k\beta_q + 3\alpha_k\theta_{kq})) + P_H(|\mathbf{k1}-\mathbf{p}|)P_H(p)\,(-2\alpha_q\beta_p$$
$$+\ 3\alpha_q\beta_k\theta_{kp} + 3\alpha_k\beta_p\theta_{kq} - 3\beta\theta_{kp}\theta_{kq} + 2\beta\theta_{pq} - 3\alpha_k\beta_k\theta_{pq})) \tag{C-43}$$

$$B^{(A\,10)}_{\Pi_B^{(S)}\Pi_B^{(S)}\Pi_B^{(S)}} = \frac{-3}{(2\pi)^3(4\pi)^3}\int d^3p\,(P_H(|\mathbf{p}+\mathbf{k2}|)P_H(p)P_H(|\mathbf{k1}-\mathbf{p}|)\alpha_k$$
$$-\ P_B(p)\,(P_B(|\mathbf{k1}-\mathbf{p}|)P_H(|\mathbf{p}+\mathbf{k2}|)\beta_k + P_H(|\mathbf{k1}-\mathbf{p}|)P_B(|\mathbf{p}+\mathbf{k2}|)(\alpha_k\gamma$$
$$-\ \gamma_k - 3\alpha_k\alpha_p\gamma_p + 3\gamma_p\theta_{kp})))\,, \tag{C-44}$$

$$B^{(A\,11)}_{\Pi_B^{(S)}\Pi_B^{(S)}\Pi_B^{(S)}} = \frac{-9}{(2\pi)^3(4\pi)^3}\int d^3p\, P_B(p)P_B(|\mathbf{p}+\mathbf{k2}|)P_H(|\mathbf{k1}-\mathbf{p}|)(-\alpha_k\alpha_p + \theta_{kp}) \tag{C-45}$$

$$B^{(A\,12)}_{\Pi_B^{(S)}\Pi_B^{(S)}\Pi_B^{(S)}} = \frac{-9}{(2\pi)^3(4\pi)^3}\int d^3p\, P_B(p)P_B(|\mathbf{p}+\mathbf{k2}|)P_H(|\mathbf{k1}-\mathbf{p}|)\,(-\gamma_k\gamma_q + 3\gamma_p\gamma_q\theta_{kp}$$
$$+\ \theta_{kq} - 3\theta_{kp}\theta_{pq} + \alpha_k(-\alpha_q + \gamma\gamma_q - 3\alpha_p\gamma_p\gamma_q + 3\alpha_p\theta_{pq}))\,, \tag{C-46}$$

$$B^{(A\,13)}_{\Pi_B^{(S)}\Pi_B^{(S)}\Pi_B^{(S)}} = \frac{-9}{(2\pi)^3(4\pi)^3}\int d^3p\, P_H(|\mathbf{p}+\mathbf{k2}|)\,(P_B(p)P_B(|\mathbf{k1}-\mathbf{p}|)(-\beta_k\beta_q + \theta_{kq})$$
$$+\ P_H(|\mathbf{k1}-\mathbf{p}|)P_H(p)(-\alpha_q\beta_k + \beta\theta_{kq}))\,, \tag{C-47}$$

$$B^{(A\,14)}_{\Pi_B^{(S)}\Pi_B^{(S)}\Pi_B^{(S)}} = \frac{-3}{(2\pi)^3(4\pi)^3}\int d^3p\,(-P_H(|\mathbf{p}+\mathbf{k2}|)P_B(p)P_B(|\mathbf{k1}-\mathbf{p}|)\alpha_k$$
$$+\ P_H(p)\,(P_H(|\mathbf{k1}-\mathbf{p}|)P_H(|\mathbf{p}+\mathbf{k2}|)\beta_k + P_B(|\mathbf{k1}-\mathbf{p}|)P_B(|\mathbf{p}+\mathbf{k2}|)(-\gamma_k$$
$$-\ 3\beta_k\beta_q\gamma_q + 3\gamma_q\theta_{kq} + \beta_k\mu)))\,, \tag{C-48}$$

$$B^{(A\,15)}_{\Pi_B^{(S)}\Pi_B^{(S)}\Pi_B^{(S)}} = \frac{9}{(2\pi)^3(4\pi)^3}\int d^3p\, P_H(p)P_B(|\mathbf{p}+\mathbf{k2}|)P_B(|\mathbf{k1}-\mathbf{p}|)(-\beta_k\beta_q + \theta_{kq})\,, \tag{C-49}$$

$$B^{(A\,16)}_{\Pi_B^{(S)}\Pi_B^{(S)}\Pi_B^{(S)}} = \frac{9}{(2\pi)^3(4\pi)^3}\int d^3p\, P_H(|\mathbf{p}+\mathbf{k2}|)\,(P_B(p)P_B(|\mathbf{k1}-\mathbf{p}|)(\alpha_k\alpha_p - \theta_{kp})$$
$$+\ P_H(|\mathbf{k1}-\mathbf{p}|)P_H(p)(\alpha_k\beta_p - \beta\theta_{kp}))\,, \tag{C-50}$$

$$B^{(A\,17)}_{\Pi_B^{(S)}\Pi_B^{(S)}\Pi_B^{(S)}} = \frac{3}{(2\pi)^3(4\pi)^3}\int d^3p\, P_H(p)P_B(|\mathbf{p}+\mathbf{k2}|)P_B(|\mathbf{k1}-\mathbf{p}|)\left(-\gamma_k\gamma_p+\theta_{kp}\right.$$
$$\left.+\ 3\gamma_p\gamma_q\theta_{kq}-3\theta_{kq}\theta_{pq}+\beta_k(-\beta_p-3\beta_q\gamma_p\gamma_q+3\beta_q\theta_{pq}+\gamma_p\mu)\right), \qquad \text{(C-51)}$$

$$B^{(A\,18)}_{\Pi_B^{(S)}\Pi_B^{(S)}\Pi_B^{(S)}} = \frac{-3}{(2\pi)^3(4\pi)^3}\int d^3p\, P_B(p)P_H(|\mathbf{p}+\mathbf{k2}|)P_B(|\mathbf{k1}-\mathbf{p}|)\alpha_p$$
$$+\ P_H(p)\left(P_B(|\mathbf{p}+\mathbf{k2}|)P_B(|\mathbf{k1}-\mathbf{p}|)\gamma_p+P_H(|\mathbf{p}+\mathbf{k2}|)P_H(|\mathbf{k1}-\mathbf{p}|)(2\beta_p-3\beta_k\theta_{kp})\right),$$
$$\text{(C-52)}$$

$$B^{(A\,19)}_{\Pi_B^{(S)}\Pi_B^{(S)}\Pi_B^{(S)}} = \frac{-3}{(2\pi)^3(4\pi)^3}\int d^3p\, P_B(|\mathbf{p}+\mathbf{k2}|)\left(P_H(|\mathbf{k1}-\mathbf{p}|)P_B(p)\alpha_k\right.$$
$$-\ P_H(p)P_B(|\mathbf{k1}-\mathbf{p}|)\beta_k\Big), \qquad \text{(C-53)}$$

where

$$\begin{aligned}
\mathcal{B}^{(A)\,ljk}_{\Pi_B^{(S)}\Pi_B^{(S)}\Pi_B^{(S)}} &= B^{(A1)}_{\Pi_B^{(S)}\Pi_B^{(S)}\Pi_B^{(S)}}\widehat{\mathbf{k1}-\mathbf{p}}^l\widehat{\mathbf{k2}+\mathbf{p}}^j\hat{\mathbf{k3}}^k + B^{(A2)}_{\Pi_B^{(S)}\Pi_B^{(S)}\Pi_B^{(S)}}\widehat{\mathbf{k1}-\mathbf{p}}^l\widehat{\mathbf{k2}+\mathbf{p}}^j\hat{\mathbf{p}}^k \\
&+ B^{(A3)}_{\Pi_B^{(S)}\Pi_B^{(S)}\Pi_B^{(S)}}\widehat{\mathbf{k1}-\mathbf{p}}^l\hat{\mathbf{k3}}^j\hat{\mathbf{p}}^k + B^{(A4)}_{\Pi_B^{(S)}\Pi_B^{(S)}\Pi_B^{(S)}}\widehat{\mathbf{k1}-\mathbf{p}}^l\widehat{\mathbf{k2}}^j\widehat{\mathbf{k2}+\mathbf{p}}^k \\
&+ B^{(A5)}_{\Pi_B^{(S)}\Pi_B^{(S)}\Pi_B^{(S)}}\widehat{\mathbf{k1}-\mathbf{p}}^l\widehat{\mathbf{k2}}^j\hat{\mathbf{k3}}^k + B^{(A6)}_{\Pi_B^{(S)}\Pi_B^{(S)}\Pi_B^{(S)}}\widehat{\mathbf{k1}-\mathbf{p}}^l\widehat{\mathbf{k2}}^j\hat{\mathbf{p}}^k \\
&+ B^{(A7)}_{\Pi_B^{(S)}\Pi_B^{(S)}\Pi_B^{(S)}}\widehat{\mathbf{k2}+\mathbf{p}}^l\hat{\mathbf{k3}}^j\hat{\mathbf{p}}^k + B^{(A8)}_{\Pi_B^{(S)}\Pi_B^{(S)}\Pi_B^{(S)}}\widehat{\mathbf{k2}}^l\widehat{\mathbf{k3}}^j\hat{\mathbf{p}}^k \\
&+ B^{(A9)}_{\Pi_B^{(S)}\Pi_B^{(S)}\Pi_B^{(S)}}\widehat{\mathbf{k2}}^l\widehat{\mathbf{k2}+\mathbf{p}}^j\hat{\mathbf{k3}}^k + B^{(A10)}_{\Pi_B^{(S)}\Pi_B^{(S)}\Pi_B^{(S)}}\widehat{\mathbf{k1}}^l\widehat{\mathbf{k1}-\mathbf{p}}^j\widehat{\mathbf{k2}+\mathbf{p}}^k \\
&+ B^{(A11)}_{\Pi_B^{(S)}\Pi_B^{(S)}\Pi_B^{(S)}}\widehat{\mathbf{k1}}^l\widehat{\mathbf{k1}-\mathbf{p}}^j\hat{\mathbf{k2}}^k + B^{(A12)}_{\Pi_B^{(S)}\Pi_B^{(S)}\Pi_B^{(S)}}\widehat{\mathbf{k1}}^l\widehat{\mathbf{k1}-\mathbf{p}}^j\hat{\mathbf{k3}}^k \\
&+ B^{(A13)}_{\Pi_B^{(S)}\Pi_B^{(S)}\Pi_B^{(S)}}\widehat{\mathbf{k1}}^l\widehat{\mathbf{k2}+\mathbf{p}}^j\hat{\mathbf{k3}}^k + B^{(A14)}_{\Pi_B^{(S)}\Pi_B^{(S)}\Pi_B^{(S)}}\widehat{\mathbf{k1}}^l\widehat{\mathbf{k2}+\mathbf{p}}^j\hat{\mathbf{p}}^k \\
&+ B^{(A15)}_{\Pi_B^{(S)}\Pi_B^{(S)}\Pi_B^{(S)}}\widehat{\mathbf{k1}}^l\widehat{\mathbf{k3}}^j\hat{\mathbf{p}}^k + B^{(A16)}_{\Pi_B^{(S)}\Pi_B^{(S)}\Pi_B^{(S)}}\widehat{\mathbf{k1}}^l\widehat{\mathbf{k2}}^j\widehat{\mathbf{k2}+\mathbf{p}}^k \\
&+ B^{(A17)}_{\Pi_B^{(S)}\Pi_B^{(S)}\Pi_B^{(S)}}\widehat{\mathbf{k1}}^l\widehat{\mathbf{k2}}^j\hat{\mathbf{p}}^k + B^{(A18)}_{\Pi_B^{(S)}\Pi_B^{(S)}\Pi_B^{(S)}}\widehat{\mathbf{k2}}^l\widehat{\mathbf{k2}+\mathbf{p}}^j\hat{\mathbf{p}}^k + B^{(A19)}_{\Pi_B^{(S)}\Pi_B^{(S)}\Pi_B^{(S)}}\widehat{\mathbf{k1}}^l\widehat{\mathbf{k1}-\mathbf{p}}^j\hat{\mathbf{p}}^k.
\end{aligned}$$
$$\text{(C-54)}$$

Without helical contributions of the field($A_H = 0$), our results are in agreement with the ones found in [202], however there is an aditional factor of 3 in the eq.(C-18) in three terms.

# D. Integration domain

The angular part of the integrals must be written in spherical coordinates $d^3p = 2\pi p^2 dp d\alpha_k$, where $2\pi$ comes from of the integration of $\theta$. Since we consider an upper cut-off $k_D$ that corresponds to the damping scale at the spectrum, we must introduce the $(k1, k2)$-dependence on the angular integration domain. This implies that we should split the integral domain in different regions such that

$$|\mathbf{k1} - \mathbf{p}| \leq k_D, \quad |\mathbf{k2} + \mathbf{p}| \leq k_D, \tag{D-1}$$

obtaining that region of the wave vectors where $0 < k1, k2 < 2k_D$. Since we expect that most important contribution comes from $\hat{\mathbf{k1}} \to -\hat{\mathbf{k2}}$ and using the above constraints we get the following integration domain in a squeezed configuration

$k_D > k2 > 0$

$$k2 > k1 > 0 \quad \int_0^{k_D-k2} dp \int_{-1}^{1} d\alpha_k + \int_{k_D-k2}^{k_D} dp \int_{\frac{k2^2+p^2-k_D^2}{2k2p}}^{1} d\alpha_k$$

$$k_D > k1 > k2 \quad \int_0^{k_D-k1} dp \int_{-1}^{1} d\alpha_k + \int_{k_D-k1}^{k_D} dp \int_{\frac{k1^2+p^2-k_D^2}{2k1p}}^{1} d\alpha_k$$

$$2k_D > k1 > k_D \quad \int_{k1-k_D}^{k_D} dp \int_{\frac{k1^2+p^2-k_D^2}{2k1p}}^{1} d\alpha_k$$

$2k_D > k2 > k_D$

$$k2 > k1 > 0 \quad \int_{k2-k_D}^{k_D} dp \int_{\frac{k2^2+p^2-k_D^2}{2k2p}}^{1} d\alpha_k$$

$$2k_D > k1 > k2 \quad \int_{k1-k_D}^{k_D} dp \int_{\frac{k1^2+p^2-k_D^2}{2k1p}}^{1} d\alpha_k. \tag{D-2}$$

The above integration domain was used to calculate the bispectrum for causal fields shown in figures (**7-4**) and (**7-5**). However, for the case of non-causal primordial magnetic fields (negative spectral indices) we can approximate the above result by selecting only regions where we can get the biggest contribution to the bispectrum (in fact, in [234] they claimed that the biggest contribution comes from the poles of the integral). Then, we can work with the approximation made in [200, 213, 216] where $k2 < k1 < k_D$ and the angular part is neglected, finding that scheme of integration is reduced to

$k_D > k2 > 0$

$$k_D > k1 > k2 \quad \int_0^{k_D} dp, \tag{D-3}$$

and therefore the bispectrum can be approximated in following way: The wave vector can be expressed in the basis defined in figure **7-2** as follows

$$\begin{aligned} \hat{\mathbf{k1}} &= \hat{\mathbf{e}}_z, \quad \hat{\mathbf{p}} = \sin\theta\cos\phi\hat{\mathbf{e}}_x + \sin\theta\sin\phi\hat{\mathbf{e}}_y + \cos\theta\hat{\mathbf{e}}_z, \\ \hat{\mathbf{k2}} &= -\sin\theta'\hat{\mathbf{e}}_x - \cos\theta'\hat{\mathbf{e}}_z, \quad \hat{\mathbf{k3}} = \sin\theta''\hat{\mathbf{e}}_x + \cos\theta''\hat{\mathbf{e}}_z, \end{aligned} \tag{D-4}$$

being $\theta$, $\theta'$, $\theta''$ the polar angle of **p**, **k2** and **k3** respectively. With these formulas we can find the inner product between different wave vectors

$$\begin{aligned} \hat{\mathbf{p}} \cdot \hat{\mathbf{k2}} &= -\sin\theta\cos\phi\sin\theta' - \cos\theta\cos\theta' \\ \hat{\mathbf{p}} \cdot \hat{\mathbf{k3}} &= \sin\theta''\sin\theta\cos\phi + \cos\theta\cos\theta'' \\ \hat{\mathbf{k2}} \cdot \hat{\mathbf{k3}} &= -\sin\theta''\sin\theta' - \cos\theta'\cos\theta''. \end{aligned} \tag{D-5}$$

Thus, the expression (7-14) can be written as

$$\begin{aligned} \int p^n |\mathbf{k1} - \mathbf{p}|^n |\mathbf{k2} + \mathbf{p}|^n d^3p &\sim 2\pi \int dp\, p^{n+2} \left( |k1 - p|^n (p^2 + k2^2 - 2pk2\cos\theta')^{n/2} \right. \\ &\quad + \left. |k2 - p|^n (p^2 + k1^2 - 2pk1\cos\theta')^{n/2} \right) \\ &\sim \int dp\, p^{n+2} \left( k1^n \left|1 - \frac{p}{k1}\right|^n k2^n (1 + \left(\frac{p}{k2}\right)^2 - 2\frac{p}{k2}\cos\theta')^{n/2} \right. \\ &\quad + \left. k2^n \left|1 - \frac{p}{k2}\right|^n k1^n (1 + \left(\frac{p}{k1}\right)^2 - 2\frac{p}{k1}\cos\theta')^{n/2} \right) \\ &\sim 2\left( \frac{nk1^n k2^{2n+3}}{(n+3)(2n+3)} + \frac{nk1^{3n+3}}{(2n+3)(3n+3)} + \frac{k_D^{3n+3}}{(3n+3)} \right) \end{aligned} \tag{D-6}$$

where in the last equality we have accounted eq.(D-3) and split into sub-ranges: $0 < q < k2$, $k2 < q < k1$ and $k1 < q < k_D$. This result was derived analytically in [199].

# E. Harmonic space and the Wigner-3j symbol

The observable of the CMB is its intensity as a function of frequency and direction on the sky **n**. This latter variable is parametrized by the polar angle $0 \leq \theta \leq \pi$ defined with respect to the zenith (often aligned with z-axis in the Cartesian coordinate system), and the azimuthal angle $0 \leq \phi < 2\pi$ measured from the $x$-axis in the $xy$-plane (see Figure **E-1**a). The components of **n** written in the Cartesian coordinate system are given by

$$n_x = \sin\theta\cos\phi, \quad n_x = \sin\theta\sin\phi, \quad n_z = \cos\theta. \tag{E-1}$$

One generally describes the temperature fluctuation $\Theta(\mathbf{n}) \equiv \frac{\Delta T}{T}$ as a expansion in spherical harmonics

$$\Theta(\mathbf{n}) = \sum_{l=0}^{\infty} \sum_{m=-l}^{l} (-i)^l \sqrt{\frac{4\pi}{2l+1}} \Theta_{lm} Y_{lm}(\mathbf{n}), \tag{E-2}$$

with

$$\Theta_{lm} = i^l \sqrt{\frac{2l+1}{4\pi}} \int_0^\pi \int_0^{2\pi} d\theta d\phi \sin\theta \Theta(\mathbf{n}) Y_{lm}(\mathbf{n}), \tag{E-3}$$

where the coeficient $\Theta_{lm}$ are called the multipoles of $\Theta$ and do not depend on the direction **n**. The spherical harmonics $Y_{lm}$ are defined as [11]

$$Y_{lm}(\theta, \phi) = \sqrt{\frac{(2l+1)(l-m)!}{4\pi(l+m)!}} P_l^m(\cos\theta) \exp^{im\phi}, \tag{E-4}$$

where $P_l^m$ are the associated Legendre polynomials. The $P_l^m$ oscillate in the $\theta$ direction with a wavelength $\lambda \sim 2\pi/l$ as we can see in Figure **E-1**c; thus, the multipoles $\Theta_{lm}$ estimate the autocorrelation of temperature on angular scales $\sim 2\pi/l$, i.e., smaller angular scales correspond to larger multipolar numbers [152]. Now, we will describe some properties of spherical harmonics [11]:

- Orthonormality

$$\int_0^\pi \int_0^{2\pi} d\theta d\phi \sin\theta Y_{l_1 m_1}(\mathbf{n}) Y_{l_2 m_2}^*(\mathbf{n}) = \delta_{l_1 l_2} \delta_{m_1 m_2}. \tag{E-5}$$

- Parity and conjugation relations

$$Y_{lm}(-\mathbf{n}) = (-1)^l Y_{lm}(\mathbf{n}), \quad Y_{l-m}(\mathbf{n}) = (-1)^m Y_{lm}^*(\mathbf{n}), \tag{E-6}$$

using this property we can extract the relation (E-3).

- Addition theorem

$$\sum_{m=-l}^{l} Y_{lm}^*(\mathbf{n}_1) Y_{lm}(\mathbf{n}_2) = \frac{2l+1}{4\pi} P_l(\mathbf{n}_1 \cdot \mathbf{n}_2), \tag{E-7}$$

where $\delta_{lm}$ is the Kronecker delta and $P_l(\mathbf{n})$ is the Legendre polynomial of degree $l$. Furthermore, we can express the product of two spherical harmonics in terms of the Wigner-3j symbols [9]

$$\begin{aligned} Y_{l_1 m_1}(\mathbf{n}) Y_{l_2 m_2}(\mathbf{n}) &= \sum_{m=-l}^{\infty} \sum_{l=0}^{\infty} \sqrt{\frac{(2l_1+1)(2l_2+1)(2l+1)}{4\pi}} \times \\ &\times \begin{pmatrix} l_1 & l_2 & l \\ 0 & 0 & 0 \end{pmatrix} \begin{pmatrix} l_1 & l_2 & l \\ m_1 & m_2 & m \end{pmatrix} Y_{lm}^*(\mathbf{n}), \end{aligned} \tag{E-8}$$

and using the orthonormality of the spherical harmonics (E-5), we can find the expression (7-43).

## E.1. Wigner-3j symbols

In the angular momentum theory, the Clebsch–Gordan coefficients denoted by $\langle l_1 \, m_1 l_2 \, m_2 | l_3 \, m_3 \rangle$ couple two states characterized by $l_1 \, m_1$ and $l_2 \, m_2$ into a new state with quantum numbers $l_3 \, m_3$ [152]. The numbers $l's$ and their projections, or magnetic quantum numbers $m's$ have integer or half-odd integer values. These coeficients are related to the Wigner-3j symbol as [11]

$$\begin{pmatrix} l_1 & l_2 & l_3 \\ m_1 & m_2 & -m_3 \end{pmatrix} = \frac{(-1)^{l_1-l_2+m_3}}{\sqrt{2l_3+1}} \langle l_1 \, m_1 l_2 \, m_2 | l_3 \, m_3 \rangle, \tag{E-9}$$

where the Wigner-3j symbol

$$\begin{pmatrix} l_1 & l_2 & l_3 \\ m_1 & m_2 & -m_3 \end{pmatrix}, \tag{E-10}$$

encodes the geometrical properties of the system of three vectors fulfilling $\mathbf{l_1} + \mathbf{l_2} + \mathbf{l_3} = 0$. This symbol is non-zero only if [9]

$$|l_i - l_j| \leq l_k \leq l_i + l_j, \quad \sum_i^3 m_i = 0, \quad |m_i| \leq l_i, \quad \sum_i^3 l_i \in \mathbb{Z}. \tag{E-11}$$

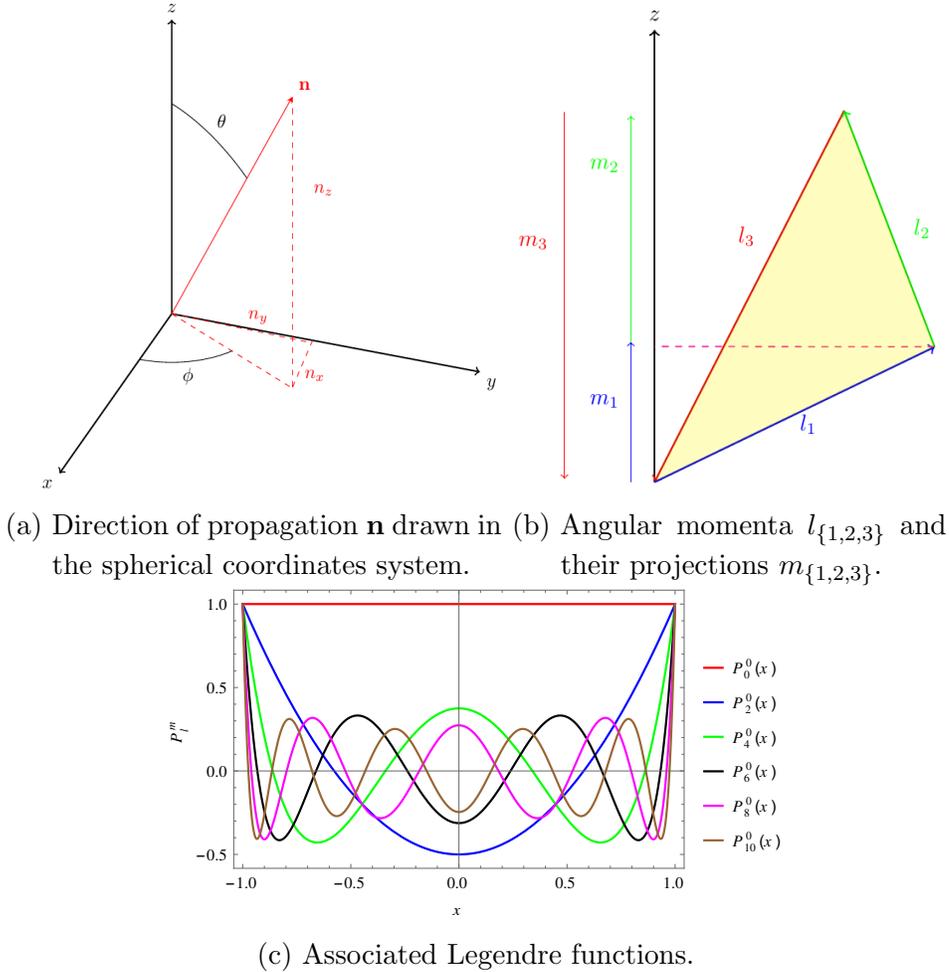

(a) Direction of propagation **n** drawn in the spherical coordinates system.

(b) Angular momenta $l_{\{1,2,3\}}$ and their projections $m_{\{1,2,3\}}$.

(c) Associated Legendre functions.

**Figure E-1**.: In left panel we show the propagation of photons going in a direction **n** drawn in the spherical coordinates system. In middle panel we show a schematic representation of the angular momenta $l_1, l_2, l_3$ in the Wigner-3j symbol and their projective quantum numbers $m_1, m_2, m_3$. Here we observe that $l's$ form the sides of a triangle implying the triangular inequality; this figure was adapted from [11]. Finally, in right panel we plot the associated Legendre functions $P_l^{m=0}$. Notice that peaks of these functions are separated by around $\sim 2\pi/l$.

## E.1.1. Symmetries of the Wigner-3j symbol

An advantage of the Wigner-3j symbol over the coupling coefficients lies in their symmetry properties [11]:

- Even permutation of the columns leaves a 3j-symbol invariant

$$\begin{pmatrix} l_1 & l_2 & l_3 \\ m_1 & m_2 & m_3 \end{pmatrix} = \begin{pmatrix} l_2 & l_3 & l_1 \\ m_2 & m_3 & m_1 \end{pmatrix} = \begin{pmatrix} l_3 & l_1 & l_2 \\ m_3 & m_1 & m_2 \end{pmatrix}. \tag{E-12}$$

- Odd permutation of the columns introduces a phase factor $(-1)^{l_1+l_2+l_3}$

$$\begin{pmatrix} l_1 & l_2 & l_3 \\ m_1 & m_2 & m_3 \end{pmatrix} = (-1)^{l_1+l_2+l_3} \begin{pmatrix} l_2 & l_1 & l_3 \\ m_2 & m_1 & m_3 \end{pmatrix} = \text{etc.} \tag{E-13}$$

- A sign change of $m's$ yields the same phase factor

$$\begin{pmatrix} l_1 & l_2 & l_3 \\ -m_1 & -m_2 & -m_3 \end{pmatrix} = (-1)^{l_1+l_2+l_3} \begin{pmatrix} l_2 & l_1 & l_3 \\ m_2 & m_1 & m_3 \end{pmatrix}, \tag{E-14}$$

which implies that

$$\begin{pmatrix} l_1 & l_2 & l_3 \\ 0 & 0 & 0 \end{pmatrix} = 0, \quad \text{if } l_1 + l_2 + l_3 \text{ is odd.} \tag{E-15}$$

- Orthogonality

$$\sum_{m_1 m_2} (2l_3 + 1) \begin{pmatrix} l_1 & l_2 & l_3 \\ m_1 & m_2 & m_3 \end{pmatrix} \begin{pmatrix} l_1 & l_2 & l'_3 \\ m_1 & m_2 & m'_3 \end{pmatrix} = \delta_{m_3 m'_3} \delta_{l_3 l'_3}, \tag{E-16}$$

$$\sum_{l_3 m_3} (2l_3 + 1) \begin{pmatrix} l_1 & l_2 & l_3 \\ m_1 & m_2 & m_3 \end{pmatrix} \begin{pmatrix} l_1 & l_2 & l_3 \\ m'_1 & m'_2 & m_3 \end{pmatrix} = \delta_{m_1 m'_1} \delta_{m_2 m'_2}, \tag{E-17}$$

$$\sum_{m_1 m_2 m_3} \begin{pmatrix} l_1 & l_2 & l_3 \\ m_1 & m_2 & m_3 \end{pmatrix} \begin{pmatrix} l_1 & l_2 & l_3 \\ m_1 & m_2 & m_3 \end{pmatrix} = 1. \tag{E-18}$$

Therefore, using these symmetries we can obtain some properties of the Gaunt coeficients defined in Eq.(7-43) [152]

- $\mathcal{G}^{l_1 l_2 l_3}_{m_1 m_2 m_3}$ is symmetric with respect to any change of the columns.

- $\mathcal{G}^{l_1 l_2 l_3}_{m_1 m_2 m_3}$ is zero if $l_1 + l_2 + l_3$ is odd.

- $\mathcal{G}^{l_1 l_2 l_3}_{m_1 m_2 m_3} = \mathcal{G}^{l_1 l_2 l_3}_{-m_1 -m_2 -m_3}$.

154                                                                                                  References[41] Bharat Ratra. Cosmological 'seed' magnetic field from inflation. *Astrophys. J.*, 391:L1–L4, 1992.

[42] Francisco D. Mazzitelli and Federico M. Spedalieri. Scalar electrodynamics and primordial magnetic fields. *Phys. Rev. **D***, D52:6694–6699, 1995.

[43] A.-C. Davis and K. Dimopoulos. Primordial magnetic fields in false vacuum inflation. *Phys. Rev. **D***, 55:7398–7414, June 1997.

[44] Antonio L. Maroto. Primordial magnetic fields from metric perturbations. *Phys. Rev. **D***, 64:083006, Sep 2001.

[45] K. Dimopoulos, T. Prokopec, O. Törnkvist, and A. C. Davis. Natural magnetogenesis from inflation. *Phys. Rev. **D***, 65:063505, Feb 2002.

[46] Rajeev Kumar Jain and Martin S. Sloth. Consistency relation for cosmic magnetic fields. *Phys. Rev. **D***, D86:123528, 2012.

[47] M.M. Anber and L. Sorbo. N-flationary magnetic fields. *Journal of Cosmology and Astroparticle Physics*, 10:018, October 2006.

[48] Peng Qian and Zong-Kuan Guo. Model of inflationary magnetogenesis. *Phys. Rev. **D***, 93:043541, Feb 2016.

[49] K. Bamba, C.Q. Geng, and S.H. Ho. Large-scale magnetic fields from inflation due to Chern-Simons-like effective interaction. *Journal of Cosmology and Astroparticle Physics*, 11:013, November 2008.

[50] V. Demozzi, V. Mukhanov, and H. Rubinstein. Magnetic fields from inflation? *Journal of Cosmology and Astroparticle Physics*, 8:025, August 2009.

[51] Ricardo J.Z. Ferreira, Rajeev Kumar Jain, and Martin S. Sloth. Inflationary magnetogenesis without the strong coupling problem. *Journal of Cosmology and Astroparticle Physics*, 2013(10):004, 2013.

[52] C. Caprini and L. Sorbo. Adding helicity to inflationary magnetogenesis. *Journal of Cosmology and Astroparticle Physics*, 10:056, October 2014.

[53] Daniel Green and Takeshi Kobayashi. Constraints on primordial magnetic fields from inflation. *Journal of Cosmology and Astroparticle Physics*, 2016(03):010, 2016.

[54] G. Tasinato. A scenario for inflationary magnetogenesis without strong coupling problem. *Journal of Cosmology and Astroparticle Physics*, 3:040, March 2015.

[55] Massimo Giovannini. Statistical anisotropy from inflationary magnetogenesis. *Phys. Rev. **D***, 93:043543, Feb 2016.